\documentclass[twocolumn]{aastex63}
\usepackage{natbib}

\usepackage{color, colortbl}
\usepackage [english]{babel}
\usepackage [autostyle, english = american]{csquotes}
\MakeOuterQuote{"}

\usepackage{multirow} 

\received{June 29, 2020}
\revised{January 25, 2021}
\accepted{February 8, 2021}

\submitjournal{ApJ}

\shorttitle{Refining the E+A Galaxy}
\shortauthors{Greene et al.}

\graphicspath{{./}{figures/}}

\begin{document}

\title{Refining the E+A Galaxy: A Spatially Resolved Spectrophotometric Sample of Nearby Post-Starburst Systems in SDSS-IV MaNGA (MPL-5)}

\correspondingauthor{Olivia A. Greene}
\email{oliviaallegragreene@gmail.com}

\author{Olivia A. Greene}
\affiliation{Department of Physics \& Astronomy, Vanderbilt University, 
PMB 401807, 2301 Vanderbilt Place, Nashville, TN 37206, USA}

\author{Miguel R. Anderson}
\affiliation{Department of Software Engineering, Bloomberg LP, 731 Lexington Avenue, New York, NY 10022, USA}

\author{Mariarosa Marinelli}
\affiliation{Department of Planetarium Education, Science Museum of Virginia, 2500 West Broad Street, Richmond, VA 23220, USA}

\author{Kelly Holley-Bockelmann}
\affiliation{Department of Physics \& Astronomy, Vanderbilt University, 
PMB 401807, 2301 Vanderbilt Place, Nashville, TN 37206, USA}
\affiliation{Department of Life \& Physical Sciences, Fisk University, 1000 17th Avenue North, Nashville, TN 37208, USA}

\author{Lauren E. P. Campbell}
\affiliation{Department of Physics \& Astronomy, Vanderbilt University, 
PMB 401807, 2301 Vanderbilt Place, Nashville, TN 37206, USA}
\affiliation{Department of Life \& Physical Sciences, Fisk University, 1000 17th Avenue North, Nashville, TN 37208, USA}

\author{Charles T. Liu}
\affiliation{Department of Physics {\&} Astronomy, City University of New York, College of Staten Island, 2800 Victory Blvd, \\ \ Staten Island, NY 10314, USA}
\affiliation{Department of Astrophysics {\&} Hayden Planetarium, American Museum of Natural History, Central Park West at 79th Street,\\ \ New York, NY 10024-5192, USA}
\affiliation{Physics Program, The Graduate Center, CUNY, New York, NY 10016, USA}


\begin{abstract}

Post-starburst galaxies are crucial to disentangling the effect of star formation and quenching on galaxy demographics.  They comprise, however, a heterogeneous population of objects, described in numerous ways.  To obtain a well-defined and uncontaminated sample, we take advantage of spatially-resolved spectroscopy to construct an unambiguous sample of E+A galaxies - post-starburst systems with no observed ongoing star formation. Using data from the Mapping Nearby Galaxies at Apache Point Observatory (MaNGA) Survey, in the fourth generation of the Sloan Digital Sky Survey (SDSS-IV), we have identified 30 E+A galaxies that lie within the green valley of color--stellar mass space. We first identified E+A candidates by their central, single-fiber spectra and (u-r) color from SDSS DR15, and then further required each galaxy to exhibit E+A properties throughout the entirety of the system to 3 effective radii. We describe our selection criteria in detail, note common pitfalls in E+A identification, and introduce the basic characteristics of the sample. We will use this E+A sample, which has been assembled with stringent criteria and thus re-establishes a well-defined sub-population within the broader category of post-starburst galaxies, to study the evolution of galaxies and their stellar populations in the time just after star formation within them is fully quenched.

\end{abstract}



\vspace{0.5cm}
\section{Introduction} \label{sec:intro}
\vspace{0.5cm}

E+A galaxies are systems whose star formation has been completely quenched for approximately $10^9$ years, and whose spectra contain the earmarks of an intermediate age stellar population. As their name implies, the spectra of E+A galaxies resemble those of early-type galaxies, but have a bluer continuum than a typical elliptical, as well as strong stellar hydrogen Balmer absorption lines indicating the substantial presence of A-type stars. 
Notably, $H\alpha$ $\lambda$6563 {\AA} and [O\thinspace II] $\lambda$3727 {\AA} emission lines, the primary indicators at optical wavelengths of ongoing star formation, are absent or weak in E+A galaxies.

Although the term "E+A galaxy" did not appear in the literature for several years afterward \citep{kf89,newberry90,ohh91}, these objects were first recognized as post-starburst systems in distant galaxy clusters in the 1980s \citep{Dressler1,Dressler2,sharples85}. It was subsequently shown \citep{lavery88,cald93,caldwell99} that E+A galaxies appeared in lower-redshift galaxy clusters too; meanwhile, the discovery of an archetypal E+A galaxy in the field at $z =$ 0.088 \citep{ohh91,liu07} was soon followed by systematic studies of low-redshift E+A field galaxies, which showed that a large fraction of them were merging or interacting systems\citep{lk95a,lk95b,zab96}.

Extensive stellar population synthesis work revealed that these uncommon objects, which comprise less than $1\%$ of the galaxy population overall, may represent valuable milestones in the galaxy lifecycle \citep{charlot94,lg96,barger96,liu07,Wild2009,pawlik18,french18}. E+A galaxies have experienced significant evolution in the relatively recent past, and are therefore still harboring important information about how galaxies grow and evolve. Large galaxy surveys (e.g., \citealt{zab96,blake04,goto2007,yang}) have since yielded $\sim 10^3$ E+As at $z \leq 0.2$. In particular, \citet{goto2007}, curated a sample of 564 E+As selected from SDSS DR5, which became the standard catalog for many years.

These rare galaxies have garnered substantial scientific attention as potential sites of AGN feedback and enhanced stellar tidal disruption \citep[e.g.]{guillochon, mockler, baron18,frenchzab}. To this end, multiwavelength work has been done to better understand the global SED and to quantify the interstellar medium, neighbor frequency and AGN fraction \citep[e.g.][]{quillen99,galaz00,yamauchi08,nielsen12,zwaan13,spogs,alatalo16,klitsch17,ardila18,wei18,smercina18,li19}.  Integral field spectroscopy (IFS) has further been applied to study the kinematics of E+A galaxies \citep{goto08,pracy13,pracy14} as well as the spatial distribution of post-starburst substructure \citep{chen}.

Over these several decades of work, the original nomenclature, \textit{E+A galaxy}, has grown to represent a wide range of post-starburst systems. This expansion of meaning is largely understandable, simply because our understanding of galaxy evolution has grown; star formation quenching is thought to be a complex and lengthy process, resulting in a broad set of observed phenomena. In an attempt to organize the taxonomy of E+As, many groups, such as \citet{goto2007}, \citet{Wilkinson}, \citet{Meusinger}, and \citet{chen} to name a few, have identified sub-categories of post-starbursts in an attempt to refine category classifications, carefully comparing the characteristics of each subgroup.  Compounding the challenge of refining such a definition, post-starburst systems in general and E+A galaxies in particular are usually identified only by the equivalent widths of small-aperture spectra; since the quenching of star formation in any given galaxy can exhibit great spatial variation \citep{chen}, mapping the spectral properties over the galaxy is likely to be key to ensure apples-to-apples comparisons of similar galaxy types.

In this paper, we seek to refine and clarify the E+A galaxy class by invoking its original observational intent: a galaxy that, throughout its spatial extent, exhibits a bluer-than-usual, early-type spectrum, has no ongoing star formation, and contains strong hydrogen Balmer absorption. We use the Mapping Nearby Galaxies at Apache Point Observatory (MaNGA) IFS Survey \citep{Law, Law2015, westfall, Belfioredata, Gunn, 2015ApJ...798....7B, 2015AJ....149...77D, 2015AJ....150...19L, 2016AJ....151....8Y,2016AJ....152..197Y, 2016AJ....152...83L,2017AJ....154...86W}, which has observed 10,000 nearby galaxies chosen to ensure a statistically representative galaxy population, as our parent sample. MaNGA provides a detailed view of the content, kinematics, history, and dynamics of galaxies across their entire face, and also generates a sample unbiased by galaxy size, mass, morphology, or environment \citep{Yanoverview, Yan2016, Drory, Smee}.  From the 2,777 galaxies available in the 5th Product Launch (MPL-5) of the MaNGA survey in the fourth generation of the Sloan Digital Sky Survey (SDSS-IV), Data Release 15 (DR15) ~\citep{2017AJ....154...28B, 2018arXiv181202759A}, we identify a sample of objects that retain E+A characteristics throughout the majority of the galaxy.  This carefully curated and confirmed set of E+A galaxies can serve as the basis of a series of investigations of galaxy evolution -- in particular, studies of what may be the last stage of transition between actively star-forming and fully quenched galaxies.



\begin{figure}
	\centering
	{\includegraphics[width=0.47\textwidth]{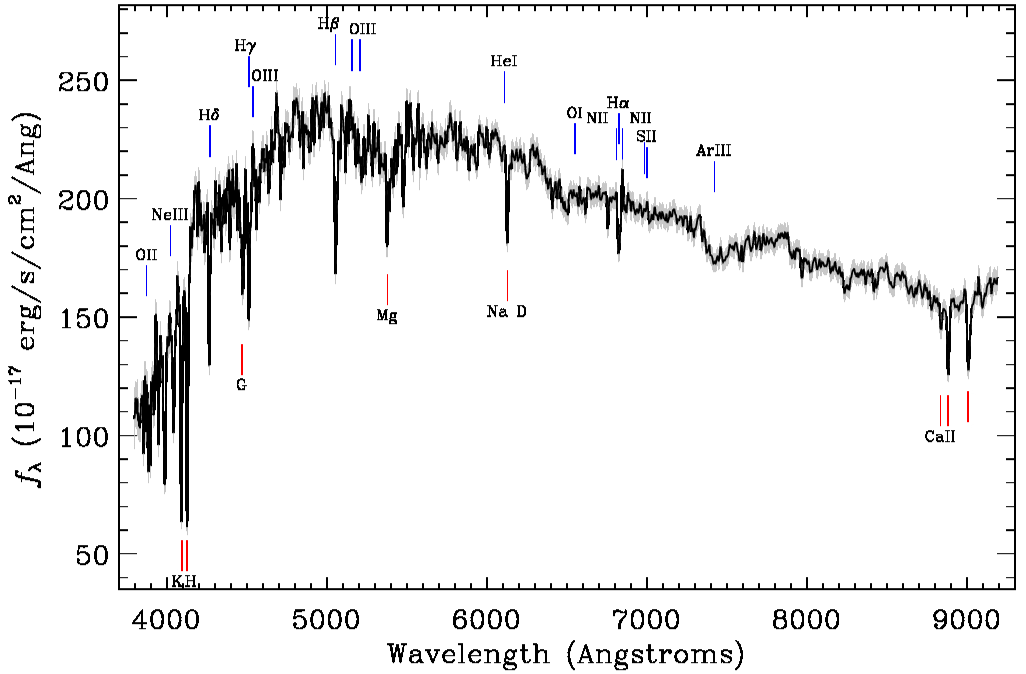}}
    \caption{The single-fiber spectrum for galaxy 8979-1902, obtained from the SDSS SkyServer. This galaxy exhibits a prototypical E+A spectrum. Along with the lack of H${\alpha}$ and [OII] emission, which indicates that there is no ongoing star formation, note the strong Balmer absorption lines from an A-type stellar population and a substantial starburst in the recent past, as well as a $D_n$4000 break {$>$} 1.5, indicating an old stellar population. }
    \label{fig:8979-1902spec}
    \vspace{1cm}
\end{figure}



\begin{table*}
\centering
\tablenum{1}                               
\caption{E+A Galaxy Sample Basic Demographics}
\label{tab:geninfo}
\begingroup
\tablewidth{0pt}
\renewcommand{\arraystretch}{1.3} 
\begin{tabular}{|c|c|c| c c c c c c c c|} 
    
		\hline
		& \textbf{MaNGA ID} & \textbf{PLATE-IFU} & \textbf{RA} & \textbf{DEC} & 	\textbf{z} & \textbf{Mass} & \textbf{g-r} & \textbf{u-r} & \textbf{$R_e$} & \textbf{Morph}\\
		\cline{2-11}
		
		\null & \null & \null & J2000 deg & J2000 deg & \null & \textbf{$log M_*(M_{\odot})$} & mag & mag & arcsec & \null\\
        \cline{1-11}
    \hline
    \hline

\multirow{11}{0.2in}{\rotatebox[origin=c]{90}{Bluer in the center}} & 12-98126 & 7443-12701	& 230.5075	& 43.5323	& 0.021	& 9.96	  & 0.58	& 2.07	& 3.82 & E?\\

 & 1-201180 & 8145-6102	& 116.5535	& 26.9230	& 0.016	    & 9.28	  & 0.61	& 2.28	& 5.20 & E\\

 & 1-235582 & 8326-3704	& 214.8502	& 45.9008	& 0.039	& 10.08  & 0.63	& 2.2	& 4.02 & SO-a\\

 & 1-209078  & 8486-3702	& 235.7749	& 48.3788	& 0.028	& 9.77	  & 0.67	& 2.27	& 1.96 & S?\\

 & 1-92638 & 8548-1901	& 242.2839	& 47.6361	& 0.019	& 9.69	  & 0.64	& 2.26	& 1.33 & N/A\\

 & 1-456505 & 8950-3702	& 193.9990	& 27.9551	& 0.027	    & 9.84	  & 0.62	& 2.14	& 3.73 & E?\\

 & 1-456744 & 8950-3704	& 194.3316	& 27.6139	& 0.026	    & 9.29	  & 0.58	& 2.06	& 3.54 & E?\\

 & 1-456309 & 8950-6101 & 194.7694  & 26.9582   & 0.027     & 10.47   & 0.6     & 1.98  & 5.01 & E?\\

 & 1-210114 & 8979-1902	& 242.5853	& 41.8549	& 0.040	    & 10.84  & 0.63	    & 2.24	& 3.33 & SBbc\\
 
 & 1-135235 & 9029-1901$^{\star}$	& 246.6681	& 42.0268	& 0.031	   & 9.73	  & 0.74	& 2.46	& 2.97 & N/A\\
 
 & 1-178823 & 8623-9102	& 311.7638	& 0.4368	& 0.013	    & 9.77	  & 0.38	& 1.35	& 7.57 & SO-a\\ \cline{1-10}

\hline
\hline

\multirow{4}{0.2in}{\rotatebox[origin=c]{90}{\shortstack{Redder in\\ the center}}} & 1-266298 & 8333-1901	& 215.7042	& 42.3956	& 0.026	    & 9.43	  & 0.72	& 2.45	& 2.13 & N/A\\

\null & 1-489884 & 8338-9102	& 171.2515	& 22.5142	& 0.049	    & 10.41  & 0.69	& 2.33	& 1.69 & E?\\

\null & 1-29809 & 8655-1902$^{\dagger}$	& 358.4688	& -0.0987	& 0.022	    & 9.55	  & 0.71	& 2.37	& 2.71 & E?$^{\diamond}$\\

\null & 1-456380 & 8934-3704	& 194.1990	& 27.4209	& 0.026	    & 9.914	  & 0.69	& 2.27	& 3.59 & SO\\ \cline{1-10}

\hline
\hline

\multirow{15}{0.2in}{\rotatebox[origin=c]{90}{Normal Progression}} & 12-49536 & 7443-1902	& 231.9911	& 42.9712	& 0.076	& 9.69	  & 0.67	& 2.19	& 4.36 & SO-a\\

\null & 1-24124 & 7991-3703	& 258.5303	& 57.4774	& 0.024	& 9.671	  & 0.68	& 2.31	& 3.64 & SBc\\

\null & 1-37034 & 8077-1901	& 41.3789	& 0.9206	& 0.024	    & 9.76	  & 0.75	& 2.42	& 3.06 & E\\

\null & 1-109112 & 8078-1901    & 41.3609	& 0.9101	& 0.025	    & 9.76	  & 0.77	& 2.51	& 2.13 & E\\

\null & 1-38374 & 8082-3704	& 50.8886	& -0.4385	& 0.039	& 9.956	  & 0.65	& 2.22	& 2.71 & SO-a\\

\null & 1-560826 & 8315-3703$^{\dagger}$	& 236.1657	& 38.4254	& 0.027	& 10.63	  & 0.69	& 2.42	& 4.76 & E?\\

\null & 1-90984 & 8553-3701	& 233.4915	& 56.8473	& 0.011	    & 9.18	  & 0.65	& 2.17	& 3.70 & E\\

\null & 1-90176 & 8553-6101	& 233.3523	& 56.6088	& 0.012	    & 9.28	  & 0.69	& 2.33	& 4.57 & Sm\\

\null & 1-95093 & 8588-3704	& 250.9693	& 39.9834	& 0.030	    & 10.54   & 0.7	    & 2.38	& 4.26 & Sa\\

\null & 1-456935 & 8931-12705	& 195.0063	& 27.7312	& 0.027	    & 9.78	  & 0.66	& 2.17	& 6.31 & E?$^{\diamond}$\\

\null & 1-456434 & 8931-3701	& 192.7884	& 27.4068	& 0.023	    & 9.54	  & 0.6	    & 2.12	& 4.75 & E?\\

\null & 1-457004 & 8934-9101	& 196.2637	& 27.5370	& 0.022	    & 9.38	  & 0.57	& 2.00	 & 6.28 & Sc\\

\null & 1-230177 & 8942-6101$^{\star}$	& 124.8979	& 26.3627	& 0.020	    & 9.60	  & 0.71	& 2.3	& 5.11 & E-SO$^{\diamond}$\\

\null & 1-456635 & 8949-3701	& 195.0840	& 27.8435	& 0.026	    & 9.59   & 0.68	& 2.25	& 3.59 & E-SO\\

\null & 1-264510 & 9041-1902$^{\star}$	& 237.0254   & 29.2023	& 0.033	    & 10.21  & 0.70    & 2.38  & 2.54 & N/A\\
		\hline

	\end{tabular}
    \endgroup
    
\bigskip    
\tablecomments{Summary of demographic data for our E+A galaxies sample. Galaxies are identified by MaNGA IDs and PLATE-IFU numbers. The table also includes RA and DEC coordinates, redshift (\textit{z}), total stellar mass from the Mendel Mass Catalog ($log M_*(M_{\odot})$) \citep{mendel}, color, effective radius ($R_e$), and HyperLEDA morphology designation of each galaxy, all (save mass and morphology) obtained from DR15 version of the SDSS-MaNGA pipeline. We group our sample by the radial color gradient: 1) bluer inside $R_e$; 2) redder inside $R_e$; and  3) no color gradient.  The $\star$ superscript denotes galaxies with visible emission lines in [OII], [NII], and [SII]. The $\dagger$ superscript denotes galaxies that have highest flux between 1-2 $R_e$. Finally, the $\diamond$ superscript designates a multiple galaxy system.  Note the galaxies' colors and masses are almost all in the acceptable range for green valley galaxies \citep{sch}.}

\end{table*}


\vspace{0.5cm}
\section{Initial Sample Selection} \label{sec:initialsample}
\vspace{0.5cm}

Our goal in this paper is to take advantage of the extra diagnostic power of an IFS survey to construct a rigorously defined sample of E+A galaxies. To begin our process, we determined our initial candidate pool using single-fiber spectra from SDSS DR15 in order to be able to compare our results with traditional classification methods. Those galaxy spectra were obtained through circular fiber-optic apertures 3 arcseconds in diameter located at the center of each galaxy.

Crucially, our sample selection strategy was very expansive at the outset.  If the original criteria to include an object were too stringent, a large portion of any potential E+A sample could be eliminated from the start.  Additionally, as we wished to examine the post-starburst galaxy population as a whole to refine the definition of E+A galaxies in particular, it was important to include objects with emission lines other than H$\alpha$, in case non-thermal ionization or active galactic nuclei were present even though star formation has been quenched.

With this in mind, our initial criteria for identifying E+A galaxy candidates from the SDSS DR15, single-fiber spectra were as follows:

\begin{enumerate}
\item A blue slope in the continuum flux between 4700 {\AA} and 8000 {\AA}, such that $F(4700{\AA})/F(8000 {\AA}) > 1$.

\item A $D_n$4000 break {$>$} 1.5 to indicate a stellar population not dominated by hot, young stars. 
\item No H$\alpha$ emission line flux (We allow [NII] $\lambda \lambda$6548, 6583 {\AA} emission).
\item Weak to absent [OII] $\lambda$3727 {\AA} emission, typically with less than  $\sim$ 5 {\AA} equivalent width \footnote{ We want to emphasize the \textit{typical} stated in this item, as our galaxies [OII] equivalent width values range from -0.95 to -15.04 {\AA}}.
\item Strong hydrogen Balmer absorption lines (H$\beta$, H$\gamma$, H$\delta$) of at least 2 {\AA} equivalent width - a value corresponding to the "H$\delta$-strong" galaxies of \citet{barger96}.

\end{enumerate}

Initially, the single-fiber spectra for {\it all} 2,777 galaxies in the MPL-5 were individually visually inspected. Objects that showed some initial correspondence with these five criteria were then further measured manually to confirm emission line and absorption line equivalent widths, primarily using the \textit{splot} tool in IRAF \footnote{IRAF is distributed by the NOIRLab (formerly the National Optical Astronomy Observatories) which is operated by the Association of Universities for Research in Astronomy, Inc. (AURA) under cooperative agreement with the National Science Foundation.}. These visual and manual steps turned out to be crucial in our E+A candidate identifications.  As we discuss later, they significantly distinguish our final E+A galaxy sample from other samples selected by more automated methods, as they helped us avoid model-dependent systematic errors in our selection process.

\begin{figure}
	\centering
    \includegraphics[width=0.5\textwidth]{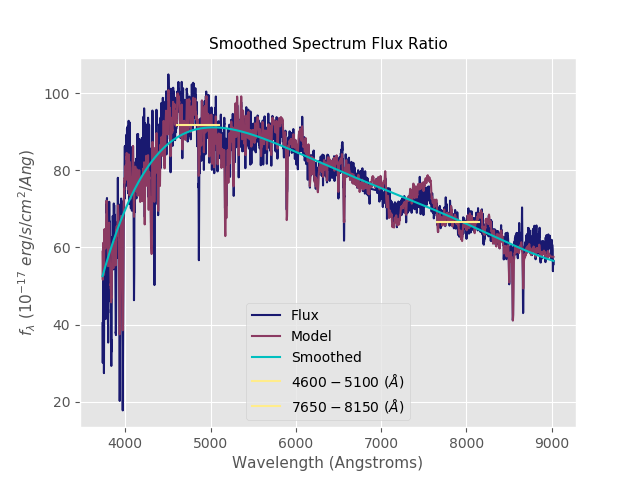}

    \caption{The redshift-corrected single-fiber spectra of 7443-12701. The blue line is the spline-smoothed spectrum. The yellow bars represent the regions used to measure the average flux ratio for our continuum criterion.}
    \label{fig:ratio}
    \vspace{1cm}

\end{figure}

In all, 42 E+A candidates had single-fiber central spectra that satisfied all five criteria. See Figure \ref{fig:8979-1902spec} for a prototypical single-fiber spectrum of these candidates, and Table  \ref{tab:geninfo} for the final sample's galaxy demographics.  Figure \ref{fig:ratio} illustrates the method we used to measure the continuum flux ratio criterion, fitting a spline-smoothed continuum and then comparing the median fluxes in two 500 {\AA} bands at 4700 {\AA} and 8000 {\AA} respectively. The visual inspection of the single-fiber spectra particularly helps us to discern redder galaxies with stronger Balmer lines (sometimes called "k+a" galaxies, whose spectra is that of an older population, dominated by K-giants within a strong A-type stellar population) which can be very difficult to distinguish from E+As using only algorithmic methods.


\begin{table*}
	\centering
	\tablenum{2}
    \caption{WHAN Diagram Spaxel Percentages}
    \vspace{0.3cm}
	\label{tab:WHAN_BPT_info}
    \begingroup
\setlength{\tabcolsep}{10pt} 
\renewcommand{\arraystretch}{1.3} 
	\begin{tabular}{c c| c c>{\bfseries}c}  
	 
	    \hline
    	&& \multicolumn{3}{c}{\textbf{WHAN}}\\
        \hline
		\textbf{MaNGA-ID} & \textbf{PLATE-IFU} & \textbf{SF} & \textbf{AGN} & \textbf{OLD STARS} \\
        \hline
        12-98126    &7443-12701	& 6.62\%	& 20.75\%	&	72.64\%		\\
        12-49536    &7443-1902	& 0.39\%	& 0.00\%	&	99.61\%		\\
        1-24124     &7991-3703	& 1.29\%	& 8.13\%	&	90.57\%		\\
        1-37034     &8077-1901	& 0.00\%	& 0.00\%	&	100\%		\\
        1-109112    &8078-1901	& 0.00\%	& 0.00\%	&	100\%		\\
        1-38374     &8082-3704	& 0.00\%	& 4.21\%	&	95.79\%		\\
        1-201180    &8145-6102	& 1.79\%	& 10.54\%	&	87.67\%		\\
        1-560826    &8315-3703	& 0.14\%	& 0.43\%	&	99.42\%	    \\
        1-235582    &8326-3704	& 0.79\%	& 1.78\%	&	97.44\%		\\
        1-266298    &8333-1901	& 0.00\%	& 0.00\%	&	100\%		\\
        1-489884    &8338-9102	& 2.84\%	& 21.70\%	&	75.46\%		\\
        1-209078    &8486-3702	& 9.43\%	& 22.55\%	&	68.02\%		\\
        1-92638     &8548-1901	& 1.37\%	& 6.85\%	&	91.78\% 	\\
        1-90984     &8553-3701	& 0.00\%	& 0.21\%	&	99.79\% 	\\
        1-90176     &8553-6101	& 0.00\%	& 0.91\%	&	99.09\%		\\
        1-95093     &8588-3704	& 0.28\%	& 0.42\%	&	99.29\%		\\
        1-178823    &8623-9102	& 0.41\%	& 1.35\%	&	98.24\% 	\\
        1-29809     &8655-1902	& 5.45\%	& 0.78\%	&	93.77\%	    \\
        1-456935    &8931-12705	& 2.03\%	& 15.25\%	&	82.72\%		\\
        1-456434    &8931-3701	& 0.00\%	& 1.34\%	&	98.66\%		\\
        1-456380    &8934-3704	& 4.44\%	& 12.13\%	&	83.43\%	    \\
        1-457004    &8934-9101	& 4.44\%	& 6.67\%	&	79.89\%		\\
        1-230177    &8942-6101	& 4.68\%	& 5.83\%	&	89.48\%		\\
        1-456635    &8949-3701	& 3.15\%	& 7.21\%	&	89.64\%		\\
        1-456505    &8950-3702	& 5.23\%	& 1.53\%	&	93.25\%		\\
        1-456744    &8950-3704	& 3.67\%	& 17.85\%	&	78.48\%		\\
        1-456309    &8950-6101  & 2.94\%    & 16.91\%   &   80.15\%      \\
        1-210114    &8979-1902	& 0.00\%	& 0.00\%	&	100\%		\\
        1-135235    &9029-1901	& 0.00\%	& 0.00\%	&	100\%		\\
        1-264510    &9041-1902	& 1.18\%	& 1.57\%	&	97.25\%		\\
       
		\hline
	\end{tabular}
   \endgroup
   
\bigskip

\tablecomments{Percentages of the spaxels found in each region of the WHAN  diagrams (see \ref{subsec:whandiagram}). The WHAN diagram classifies, on average, 92\% of the spaxels as "old stars" (bolded column) as required by our E+A criteria.}

\end{table*}


\vspace{0.5cm}
\section{Final Sample Selection: MaNGA IFS} \label{sec:spec_analysis}
\vspace{0.5cm}

We next used the integral field spectroscopic data from MaNGA to test the identifications of these E+A candidates found using the central single-fiber spectra. The primary software tool we used to examine these galaxies was Marvin {\bf 2.0} \citep{brian_cherinka_2018_1146705}, a software pipeline integrated with the SDSS Skyserver that is designed to search, access, and visualize MaNGA data. 
 
\begin{figure*}
	\centering
	
    \gridline{\fig{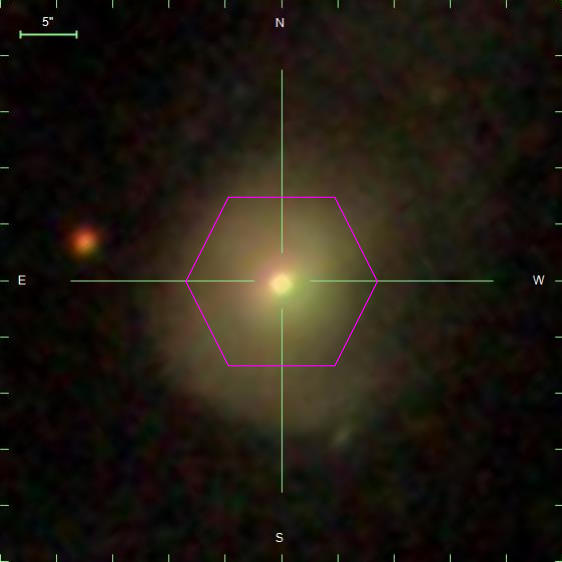}{0.25\textwidth}{8315-3703}
	      \fig{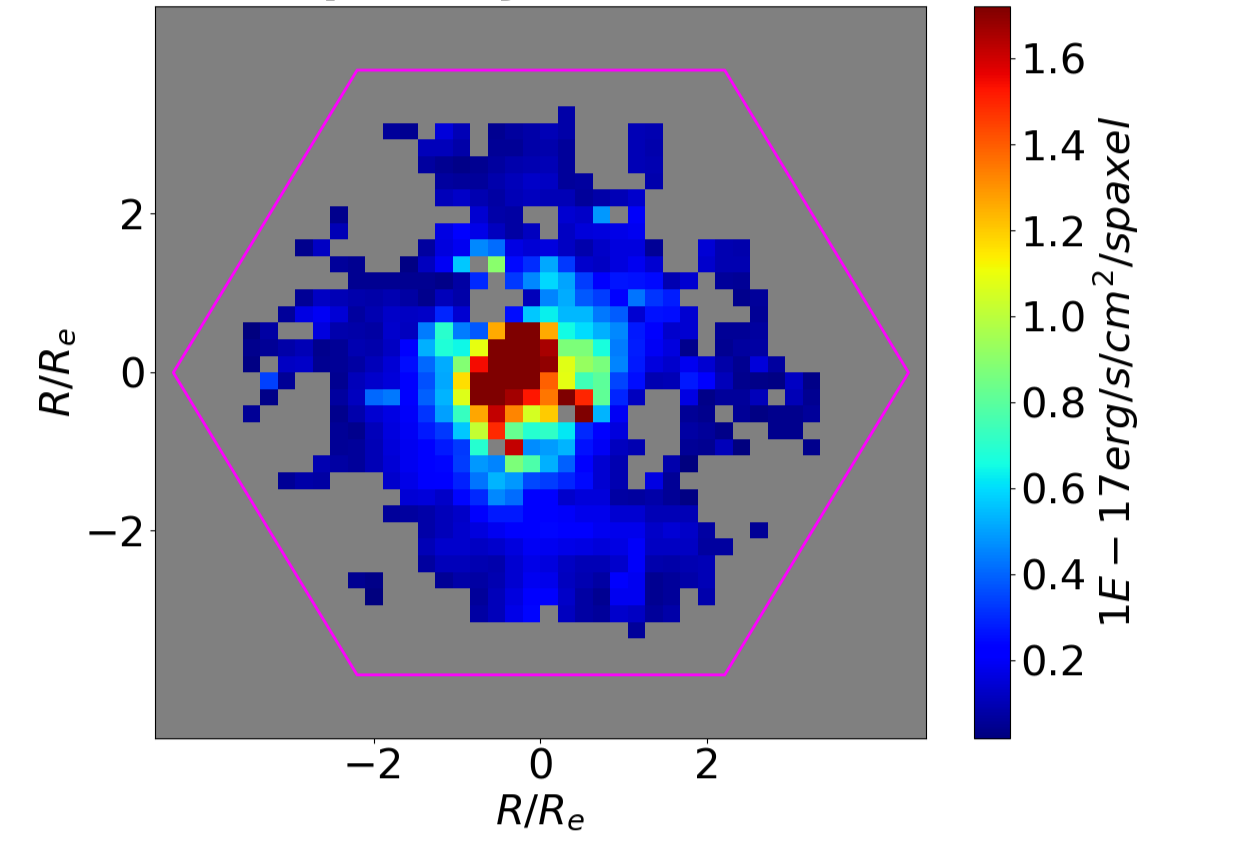}{0.25\textwidth}{[H$\alpha$ $\lambda$6564]}
	      \fig{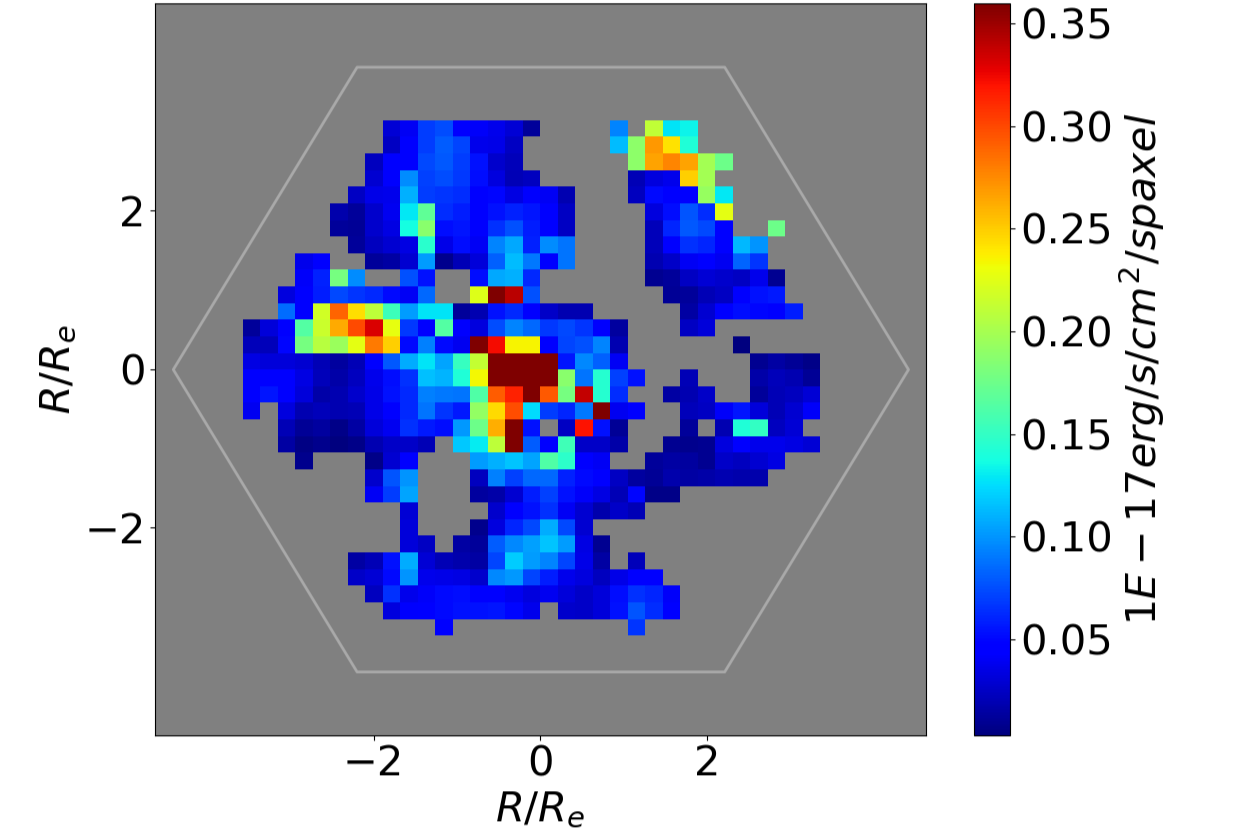}{0.25\textwidth}{[H$\beta$ $\lambda$4826]}
	      \fig{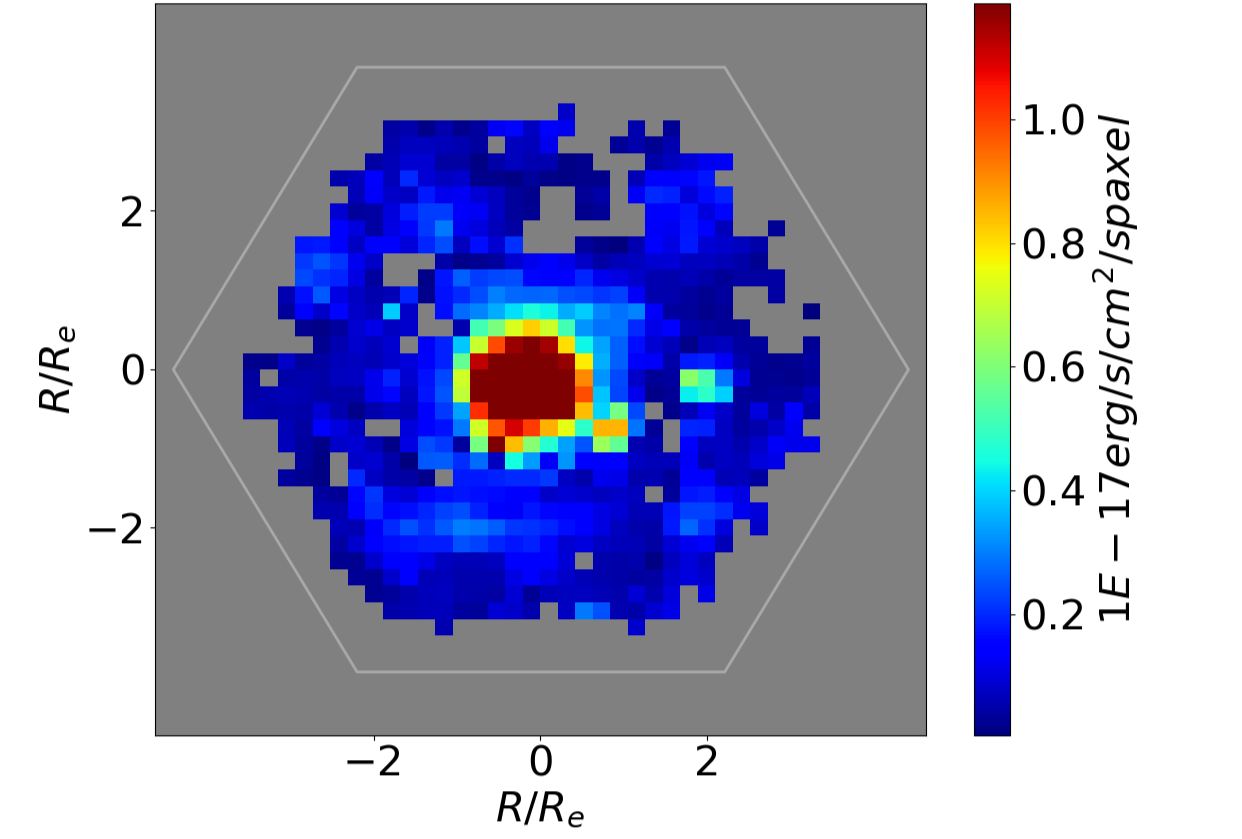}{0.25\textwidth}{[NII $\lambda$6549]}}
	
	\gridline{\fig{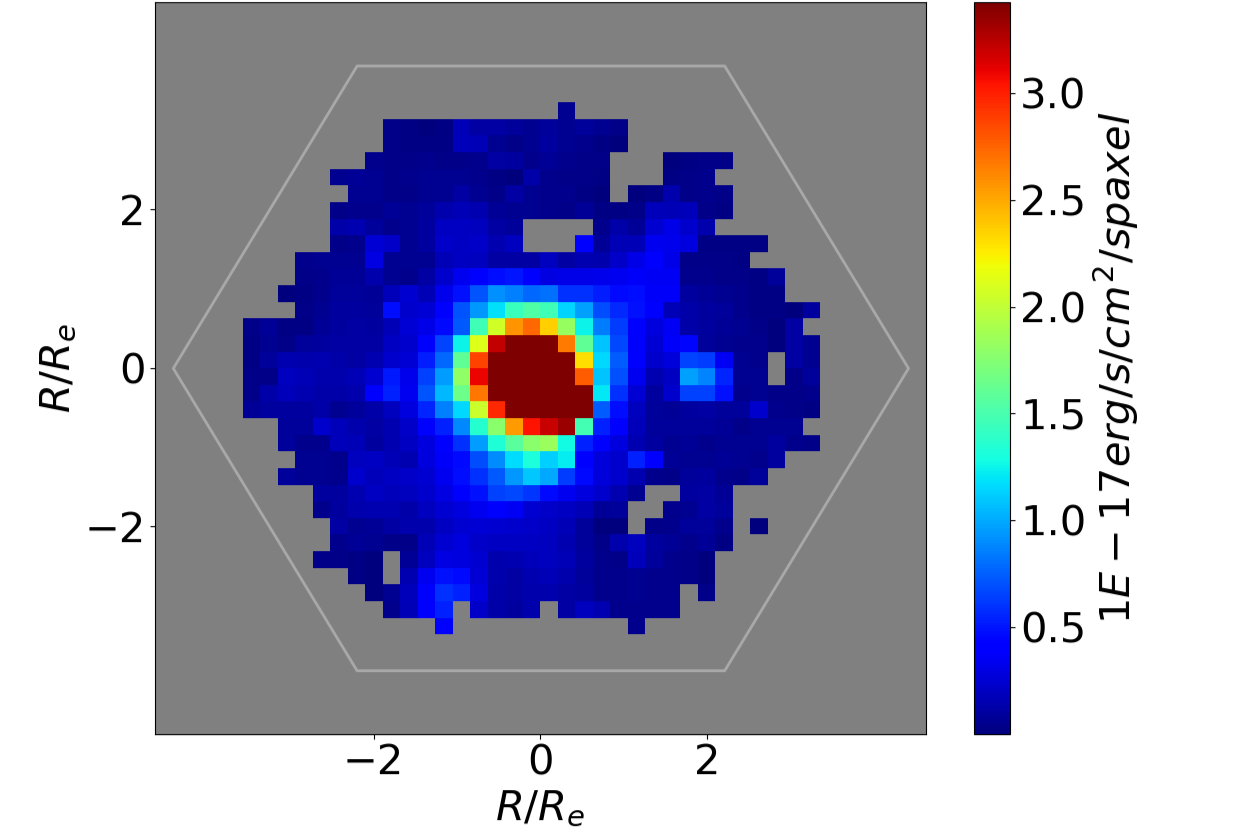}{0.25\textwidth}{[NII $\lambda$6585]}
	      \fig{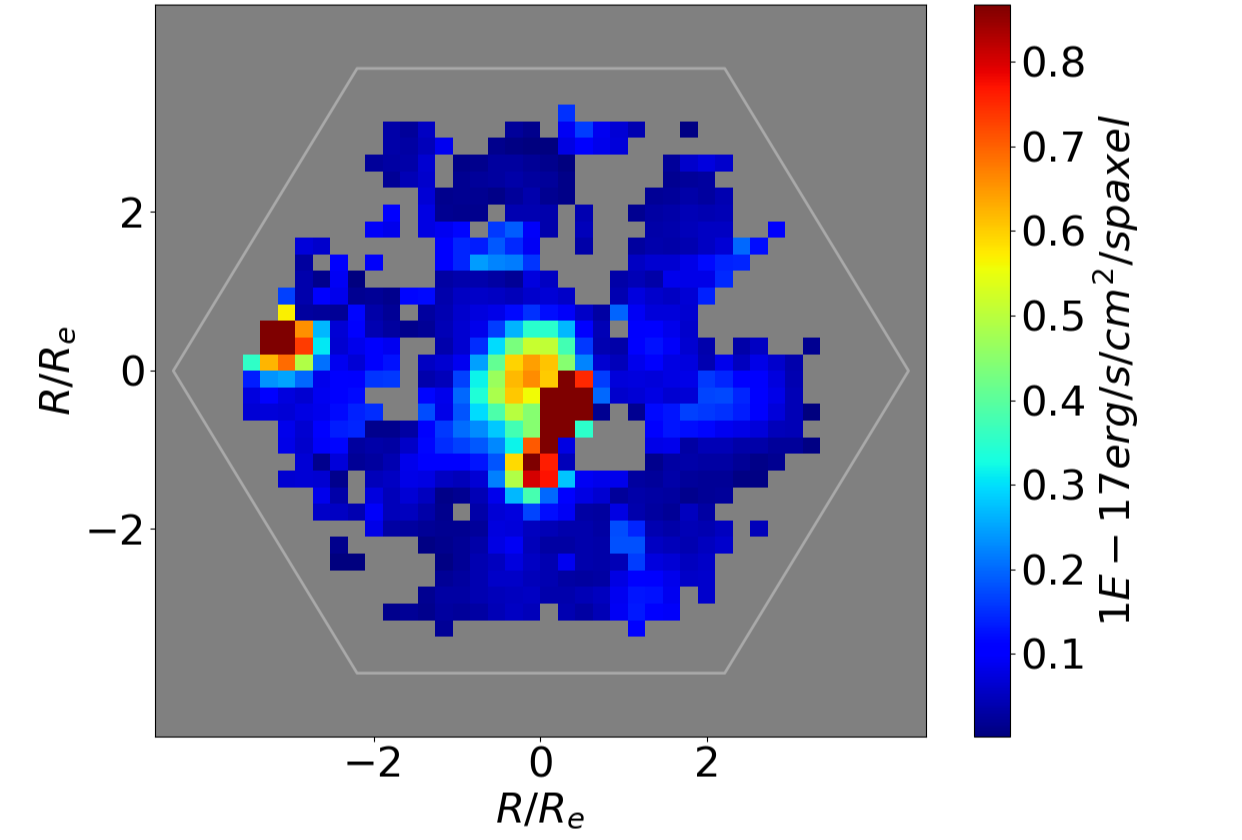}{0.25\textwidth}{[OI $\lambda$6302]}
	      \fig{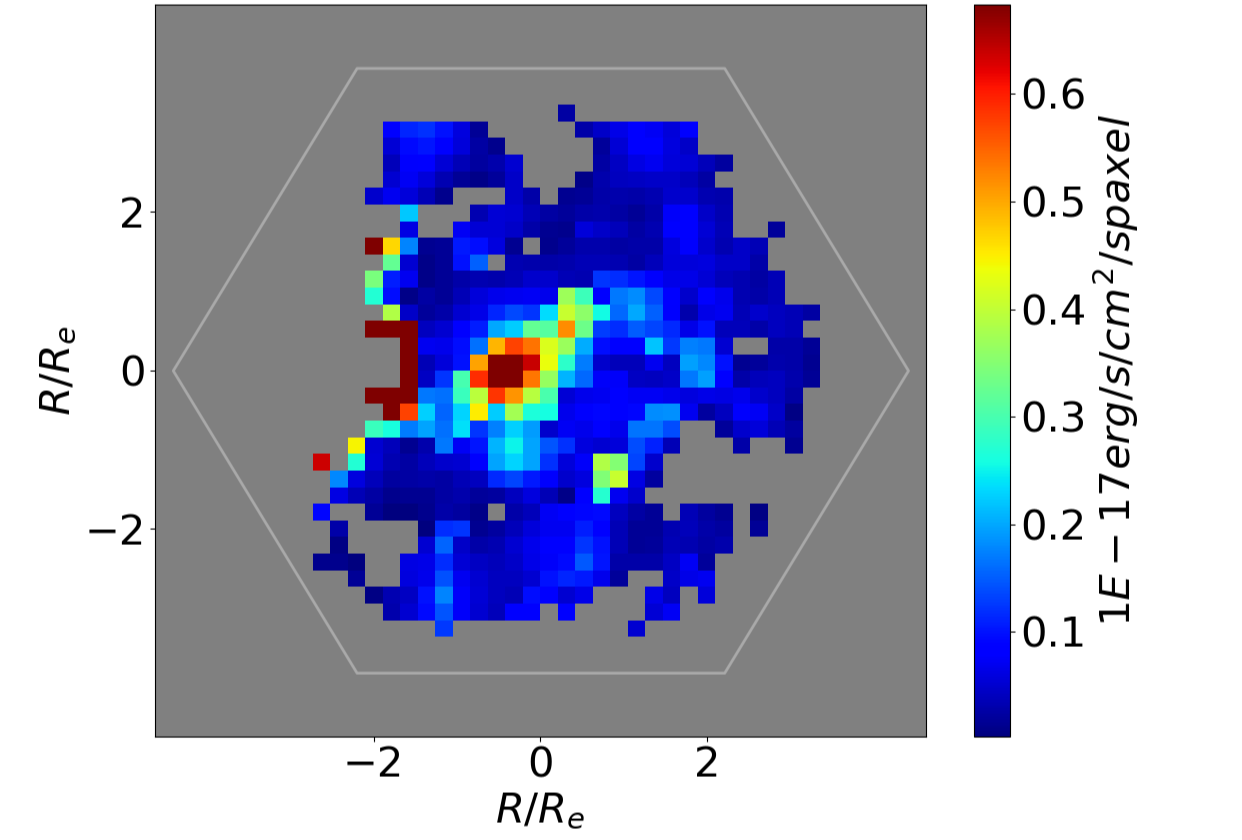}{0.25\textwidth}{[OI $\lambda$6365]}
	      \fig{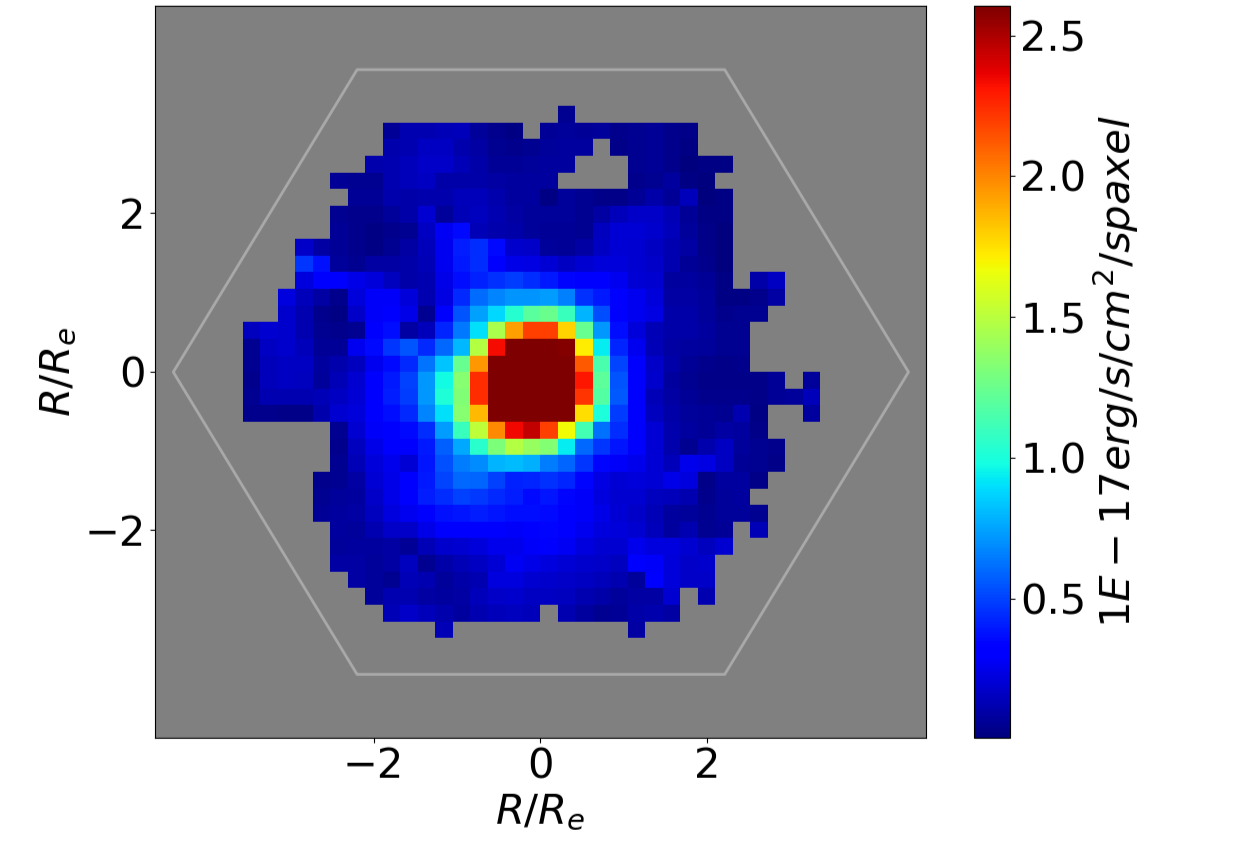}{0.25\textwidth}{[OIId $\lambda$3728]}}
	      
    \gridline{\fig{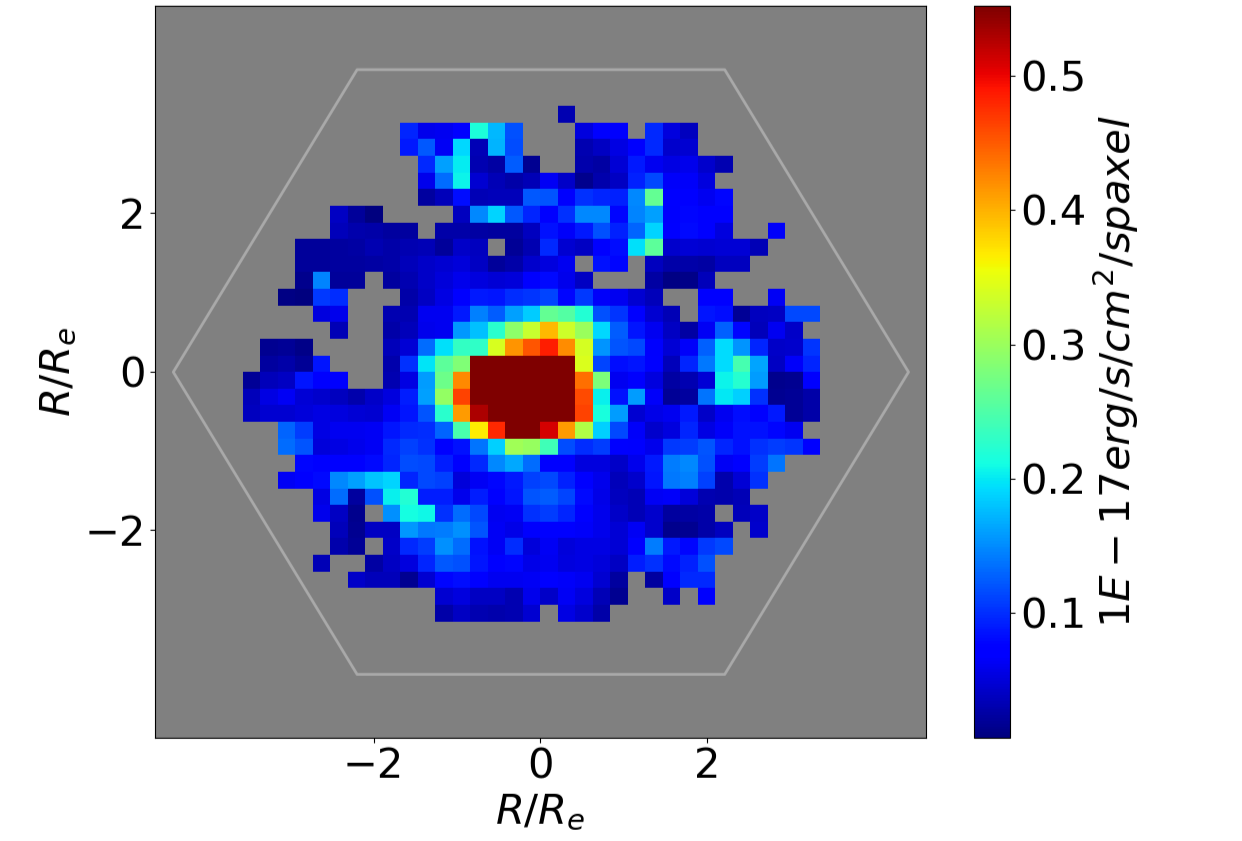}{0.25\textwidth}{[OIII $\lambda$4960]}
	      \fig{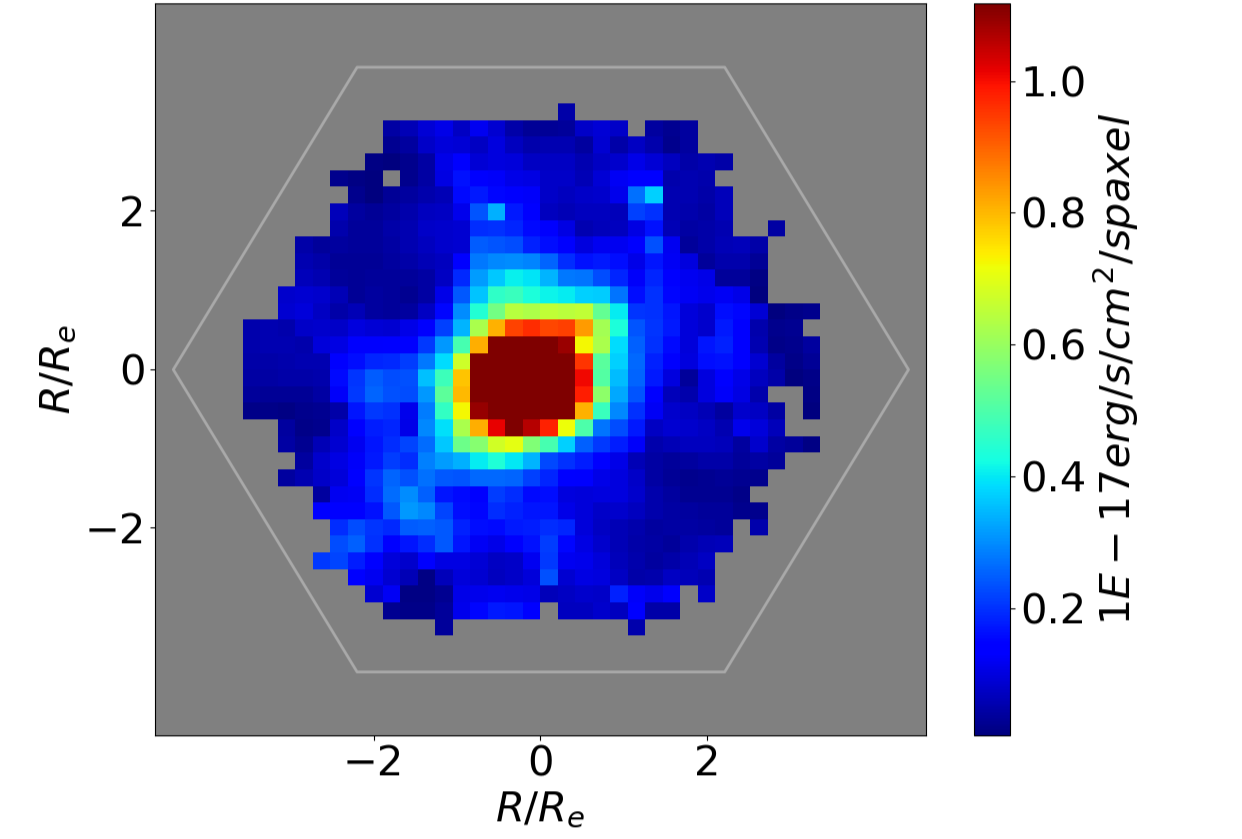}{0.25\textwidth}{[OIII $\lambda$5008]}
	      \fig{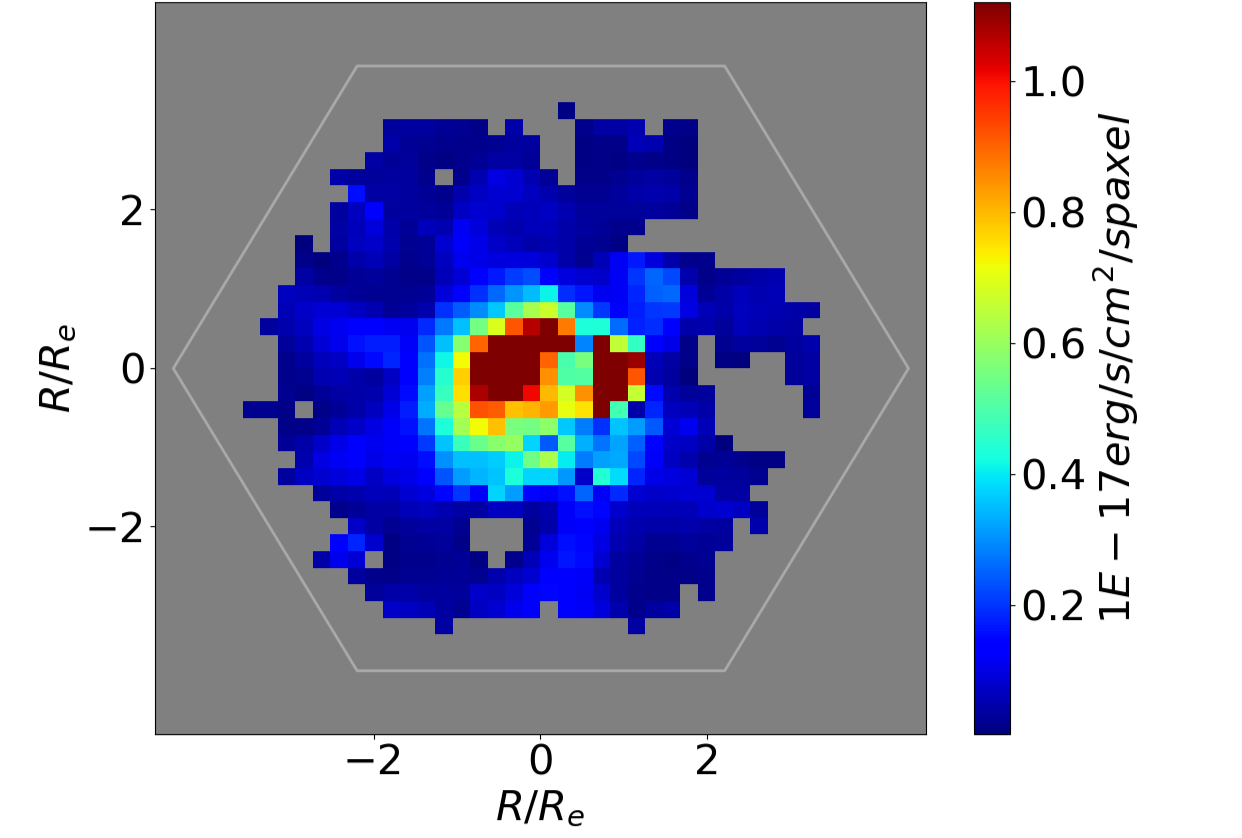}{0.25\textwidth}{[SII $\lambda$6718]}
	      \fig{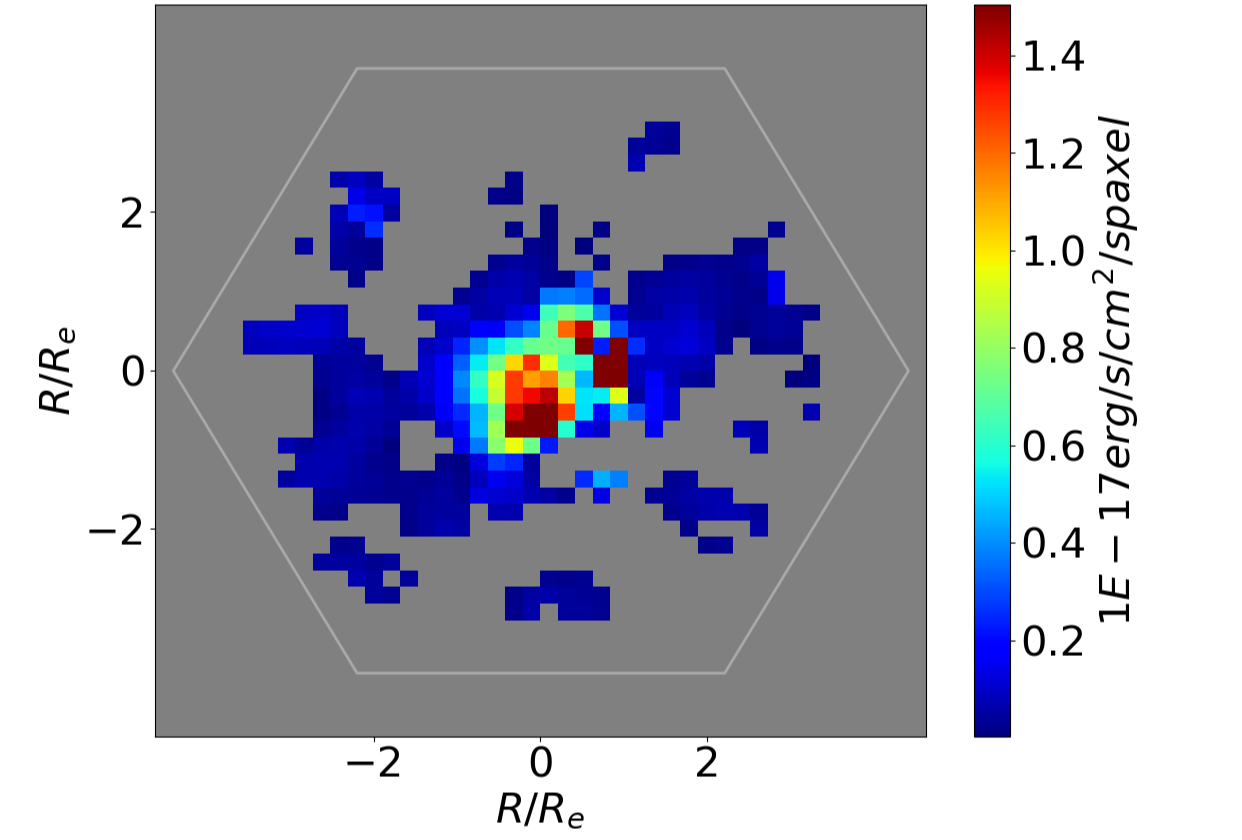}{0.25\textwidth}{[SII $\lambda$6732]}}

    \caption{SDSS image (top left) of galaxy 8315-3703, and the corresponding 11 ionization maps provided from the MaNGA datacube.  These maps are essential in our measurements of stellar population ages and metallicities as a function of distance from the galaxy's center.}
    \label{fig:DAP8315-3703}
    \vspace{1cm}
\end{figure*}

To begin, as a consistency check with our SDSS DR15 single fiber spectra measurements, we used MaNGA data to generate a single composite central spectrum by summing and averaging the spaxels of each galaxy's MaNGA map within one effective radius ({$R_e$}) of the center of each candidate galaxy.  We also collected the non-parametric equivalent widths supplied in 25 {\AA} intervals by the MaNGA Data Analysis Pipeline (DAP), to create, for each galaxy, spatially resolved ionization maps at the wavelengths of 11 well-known optical emission lines (see Figure \ref{fig:DAP8315-3703}).  We compared the MaNGA results to our by-hand EW values measured with \textit{splot}, and confirmed broad consistency between the two sets of values.

We then applied the MaNGA data to complete our E+A galaxy selection process.  This happened in two ways: checking if the spectrophotometry of each galaxy was the same throughout the galaxy, and conducting emission-line diagnostics over the entire galaxy to confirm the lack of star formation. Our multi-parameter selection criteria also allowed us to determine that the nuclear line emission in these E+As is most likely caused by weak AGN rather than centrally concentrated star formation.


\begin{figure}
	\centering
    \includegraphics[width=0.47\textwidth]{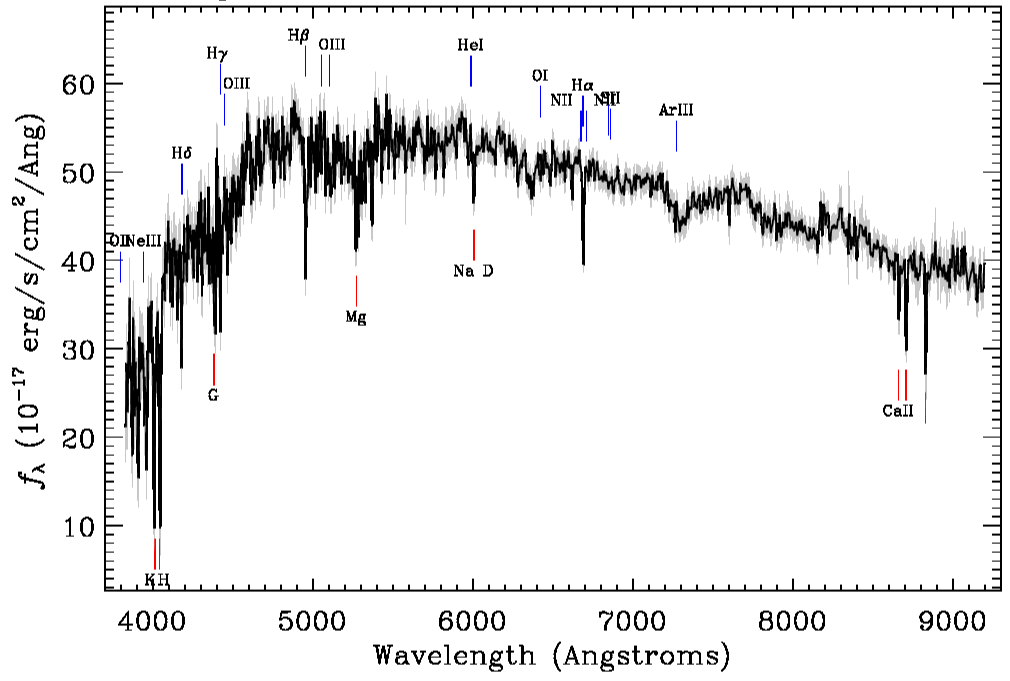}
	\  \includegraphics[width=0.47\textwidth]{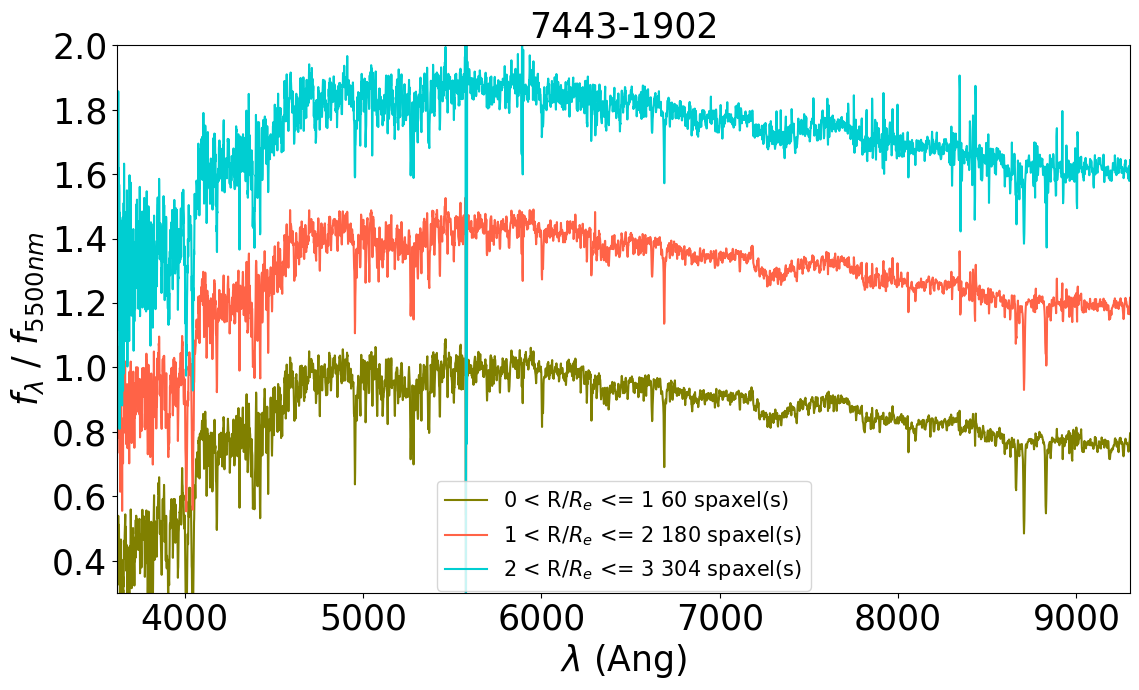}
    \caption{The single-fiber spectra of 7443-1902 (above) and the summed spaxel spectra out to 1, 2 and 3 {$R_e$} (below). The spectra in the bottom plot are normalized to 5500{$\pm$}50 {\AA}, and offset from the 0-1 {$R_e$} aperture by {$R_e$}/2.3. The spaxel count in the legend indicates how many spaxels were summed to produce these spectra. The nearly identical continua in the MaNGA central and annular apertures, compared to the single-fiber spectrum, shows that all three regions could be observed independently and still be categorized as E + A.}
    \label{fig:ReSpec}
\end{figure}


\vspace{0.5cm}
\subsection{Spectrophotometry vs. Radius} \label{subsec:3Re}
\vspace{0.5cm}

To ensure that a candidate galaxy's spectra satisfied our E+A criteria throughout the galaxy, while also increasing the signal-to-noise ratio, we summed the spaxels in three annular regions: from 0-1, 1-2 and 2-3 $R_{e}$ (See Figure \ref{fig:WHANex} for an example). The spaxels in each region were summed to create a single spectrum; then, the three spectra were each normalized to unity at 5500 {\AA} and overlaid to examine the spectrophotometric differences between the three regions.

An example of this process with the summed spectra of one galaxy, 8082-3704, is presented in Figure \ref{fig:ReSpec}.  A comparison to the single-fiber spectrum of the galaxy confirms the same characteristic E + A shape in all three apertures.  By contrast, the bottom panel of Figure \ref{fig:WHANSF} shows another of our E+A candidates, 8077-12704, where the spectra of the three apertures clearly differ.  The former object fulfilled this criterion of being an E+A galaxy by our refined definition, while the latter object did not.


\subsection{WHAN Diagrams} \label{subsec:whandiagram}
\vspace{0.5cm}


\begin{figure}
	\centering
	{\includegraphics[height = 0.21\textwidth]{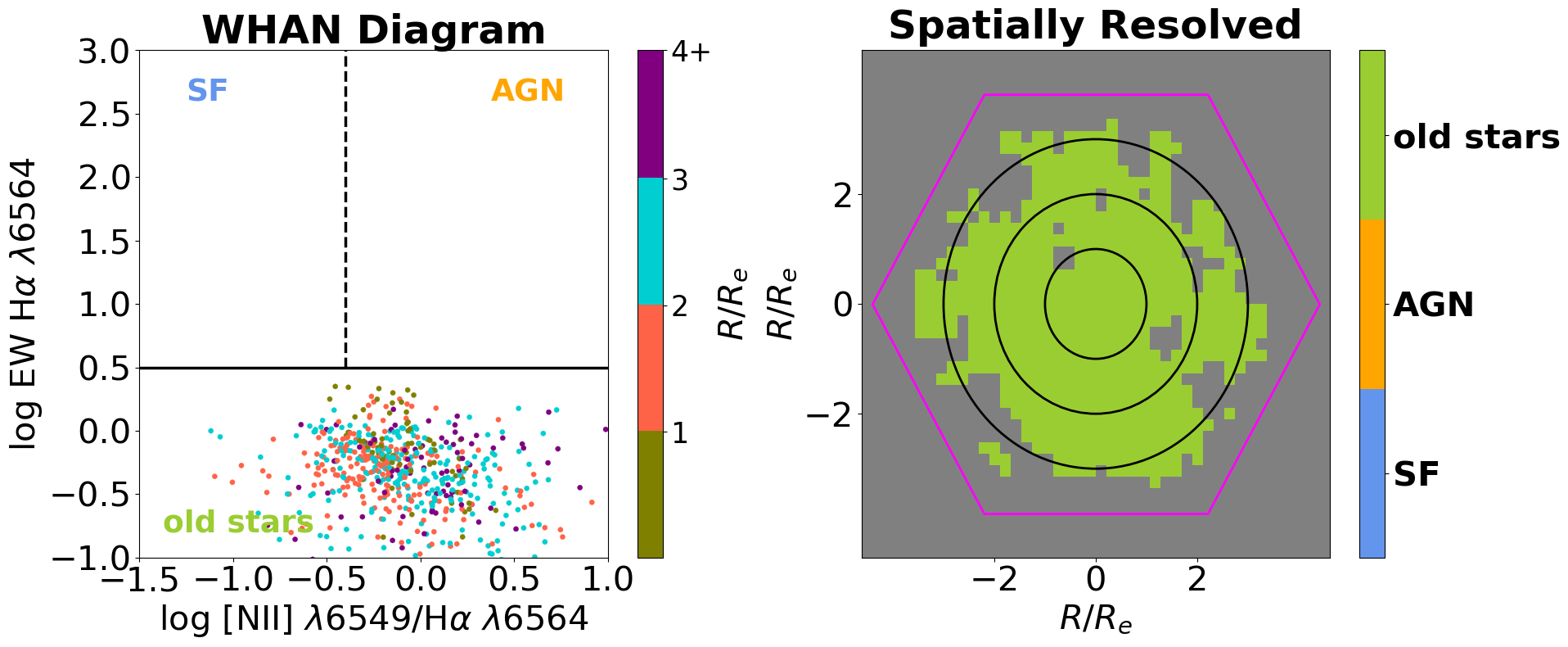}}
    \caption{Line-Ratio and spatially resolved WHAN log [EW H$\alpha$ / ((NII)/H$\alpha$)] diagram for galaxy 8315-3703. Left: the line ratios for each spaxel, color-coded by annuli in units of effective radius. Right: Galaxy image color-coded by sectors in the WHAN diagram: old stars, AGN or star formation (see \ref{subsec:whandiagram}). \citep{Bundy, Belfiore}.}
    \label{fig:WHANex}
    \vspace{1cm}
\end{figure}


Following the approach of \citet{Belfiore}, we plotted spectral line ratios of each spaxel to see how they differ from the single-fiber data, as shown in Figure \ref{fig:WHANex} (left). Specifically, we created BPT (Baldwin-Phillips-Terlevich \citep{bpt} diagrams, which we found to be only somewhat useful as the [OIII], H$\beta$, and [SII] lines in the sample are weak. We also generated WHAN (\textit{log EW H$\alpha$ / (log [NII]/H$\alpha$)}) diagrams in an effort to delineate the regions of the galaxies that would be considered old stars, AGN or star forming. \citet{Belfiore} classifies three regions for the WHAN diagram \footnote {This method is slightly refined from the \citet{fernandes} version that included five categories, two of which would fall into our "old stars" region as Retired Galaxies, and Passive Galaxies.}: 
\\
\\
(1) \textit{Old Stars} if EW(H$\alpha$) {$<$} 3{\AA}. \\
(2) \textit{AGN} if EW(H$\alpha$) {$>$} 3{\AA} and [NII]/H$\alpha$ {$>$} -0.1 \\
(3) \textit{Star Forming} if EW(H$\alpha$) {$>$} 6{\AA} and [NII]/H$\alpha$ {$>$} -0.1 \\
\\

The WHAN diagrams confirm that these galaxies have no current star formation, with, on average, each galaxy having defined more than 90{\%} of the spaxels as an old stellar population (See Figure \ref{fig:WHANex} and Table \ref{tab:WHAN_BPT_info}). This is because H$\alpha$ emission is necessarily low for E + A classification, and is the vertical axis for WHAN diagrams. Because H$\alpha$ emission is possibly the most reliable proxy for ongoing star formation, it gives us confidence that these galaxies are post-starburst. 

\begin{figure*}
   
\gridline{\fig{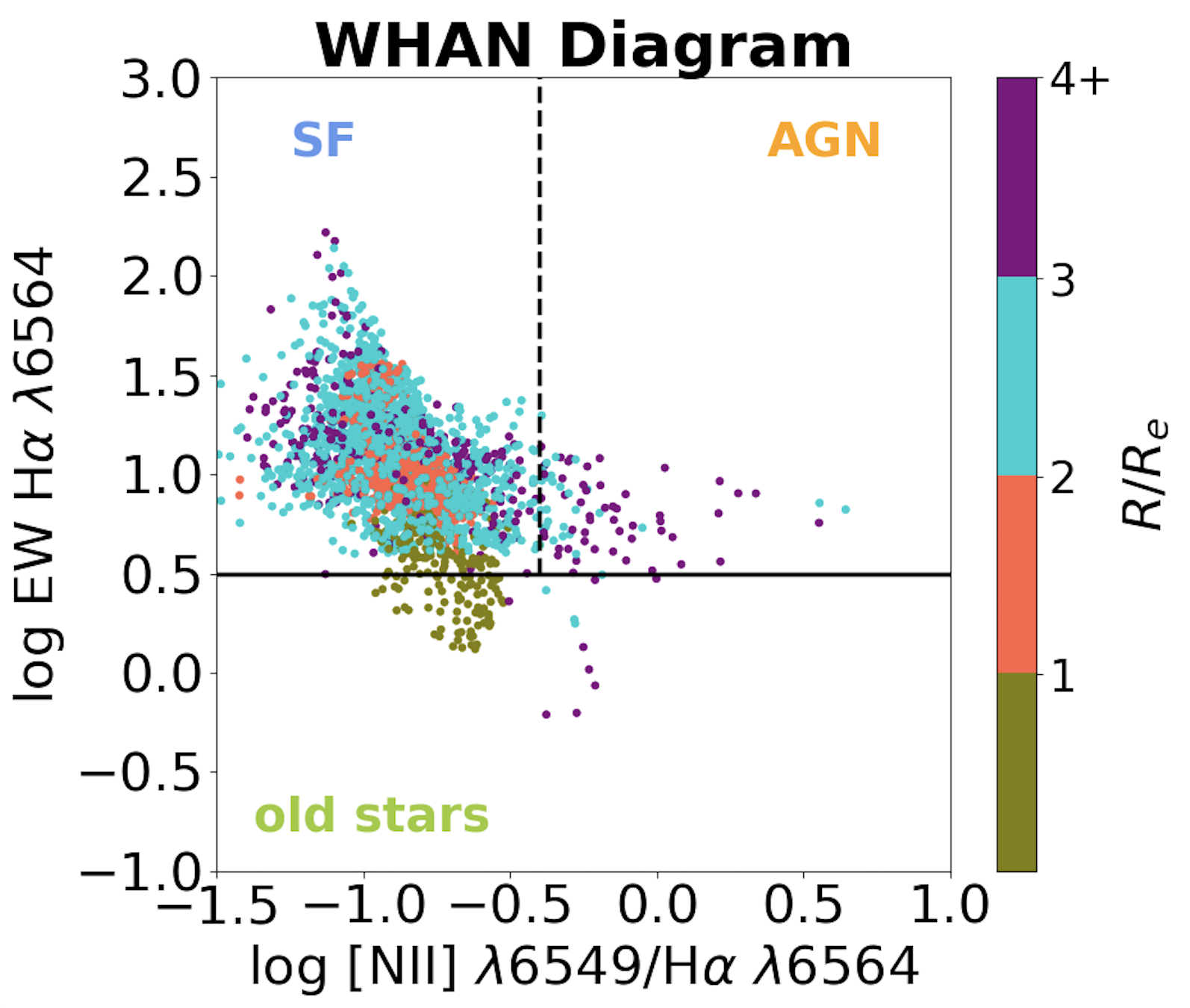}{0.30\textwidth}{Spatially Resolved WHAN Diagram}
	      \fig{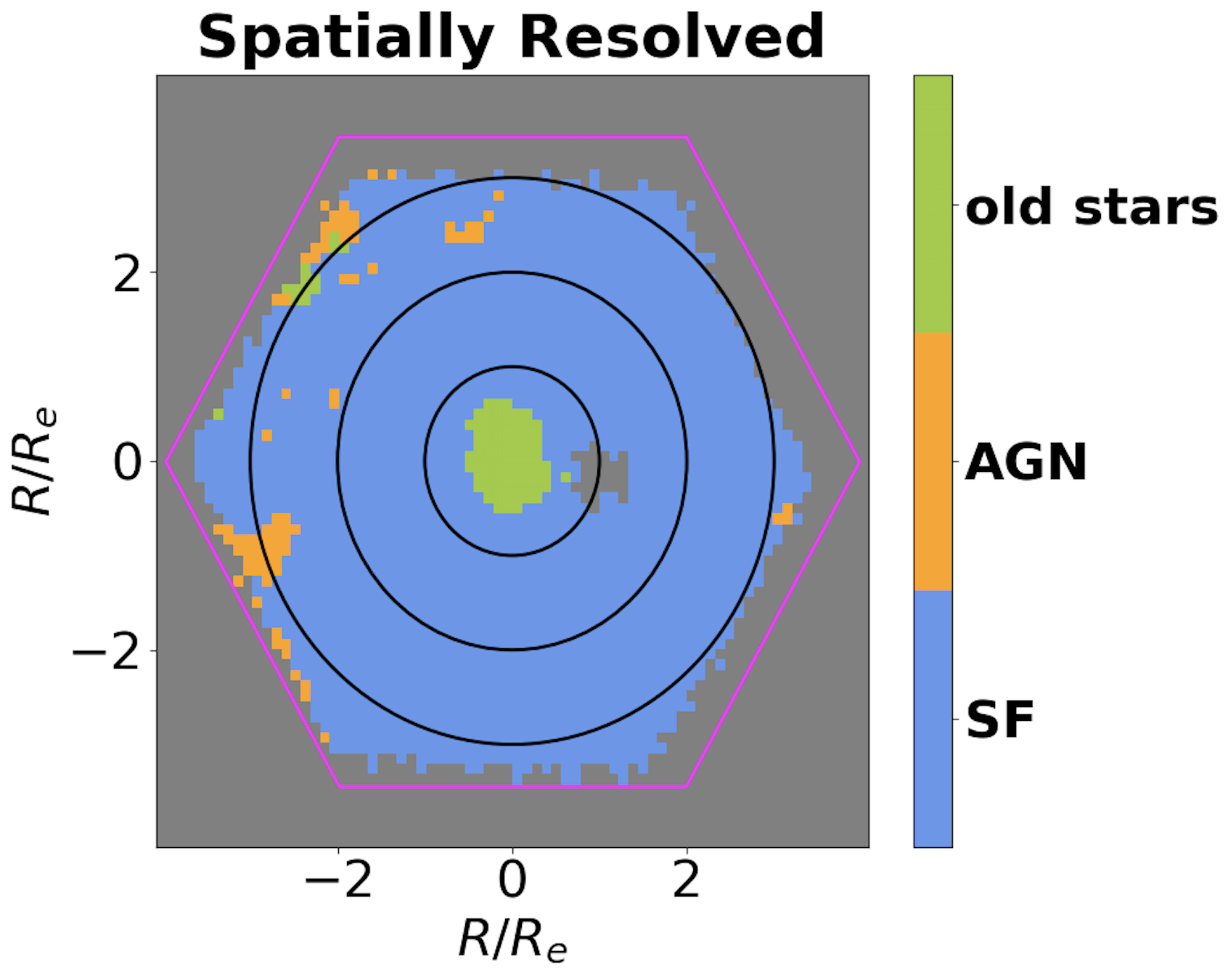}{0.32\textwidth}{}
	      \fig{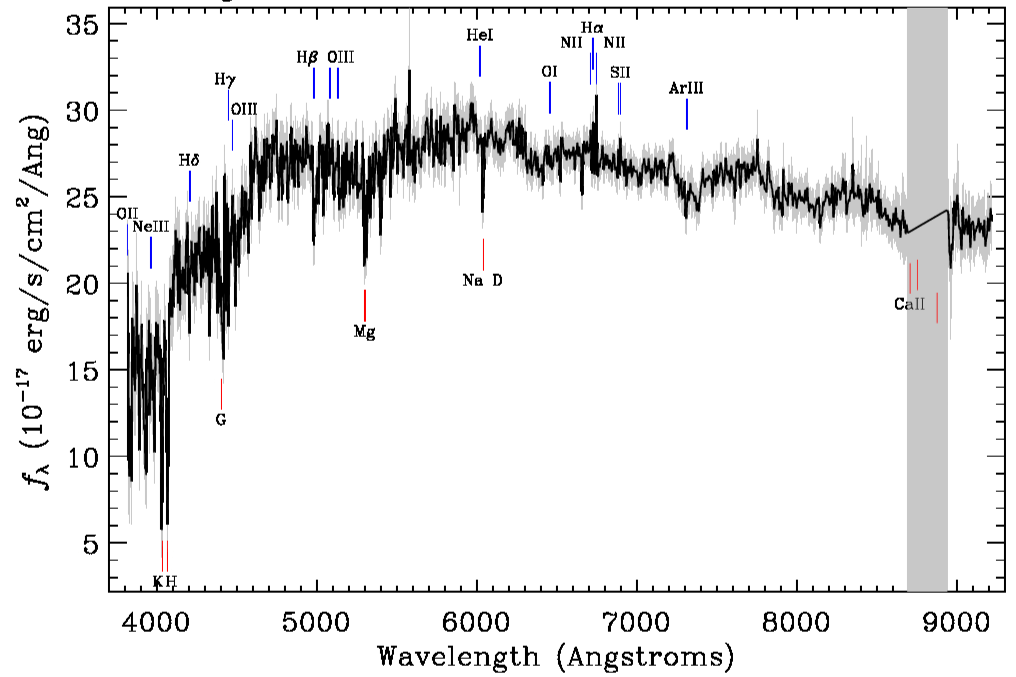}{0.37\textwidth}{Single-Fiber Spectrum obtained from SDSS}}
	      
\gridline{\fig{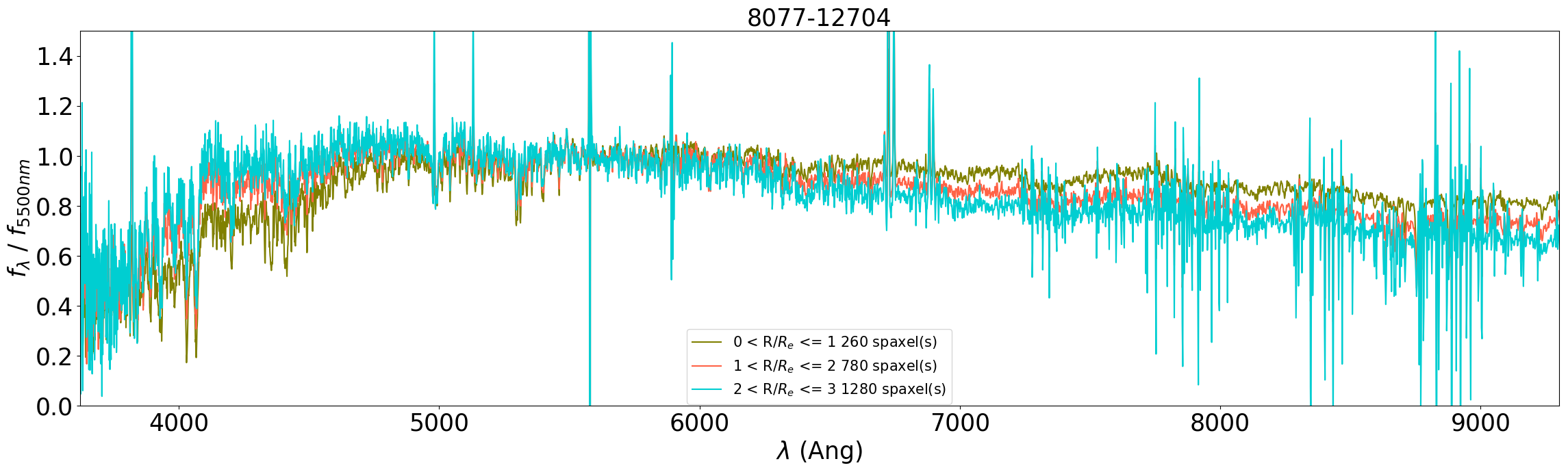}{1\textwidth}{Summed Spaxel Spectrum}}
	       
\caption{The line-ratio \& spatially resolved WHAN diagram (top left), single-fiber spectrum obtained from the SDSS SkyServer (top right), and summed spaxel spectra (bottom) for galaxy 8077-12704. This is a prime example of how the SDSS optical single-fiber spectrum could lead us to believe this galaxy was an E+A; however, the WHAN diagram revealed that the vast majority of the galaxy is actually star-forming, and probably going through an inside-out quenching process. The annular summed spaxel spectra clearly deviates from an E+A galaxy as well, with strong H$\alpha$ evident in the 2nd and 3rd $R_e$ annulii.} 
\label{fig:WHANSF}
\vspace{1cm}
\end{figure*}

The use of these WHAN diagrams confirms the classification of our E+As, across the entirety of the galaxy. Only being able to view a single-fiber region of the galaxy leaves us open for classification mistakes, and outlier data that would muddle the sample.  This is evident in Figure \ref{fig:WHANSF}, which shows the analysis of the galaxy 8077-12704. While its single-fiber spectrum meets the specified criteria we outlined before, this galaxy only retains its E+A properties at the very center, as shown in its WHAN diagram. Only 3.19{\%} of the spaxels in the MaNGA map are considered \textit{"old stars"} whereas 93.9{\%} is highly star forming. Preliminarily, using Marvin's interactive \textit{"Spectrum in Spaxel"} tool, we were able to see that at both two and three $R_e$, the strong Balmer absorption lines disappear, and are replaced with intense H$\alpha$ emission, losing all semblance of E+A classification. The three summed spaxel spectra in the 0-1, 1-2 and 2-3 $R_e$ apertures, as clearly seen in Figure \ref{fig:ReSpec}, have different continuum shapes from one other. Once completed, we were able to note that the spectra did not overlay well; the difference was not as stark as Marvin's interactive spectrum due to our normalization factors, but it was still clearly off by enough that we could confidently say it should not be a part of our sample, and did not meet the specified criteria. This galaxy, had it been included in our sample, would have dramatically skewed results. 

\vspace{0.5cm}
\subsection{Final Sample: 30 Galaxies}\label{comparison}
\vspace{0.5cm}

By applying these two MaNGA-enabled criteria to our 42 candidates, 12 of these candidates were rejected because it was demonstrated either that they were star-forming, or had an inconclusive star-formation history in spaxels beyond their central regions. (As seen below, these twelve rejected candidates provide useful data points in our efforts to understand more clearly the interplay between classification requirements for E+A galaxies.) We were left with a confident final sample of 30 E+A galaxies that fit all seven of our E+A classification criteria -- 5 for one-dimensional spectra, and 2 for the integral field spectra.

\begin{figure}
	\centering
	\includegraphics[height = 0.4\textwidth]{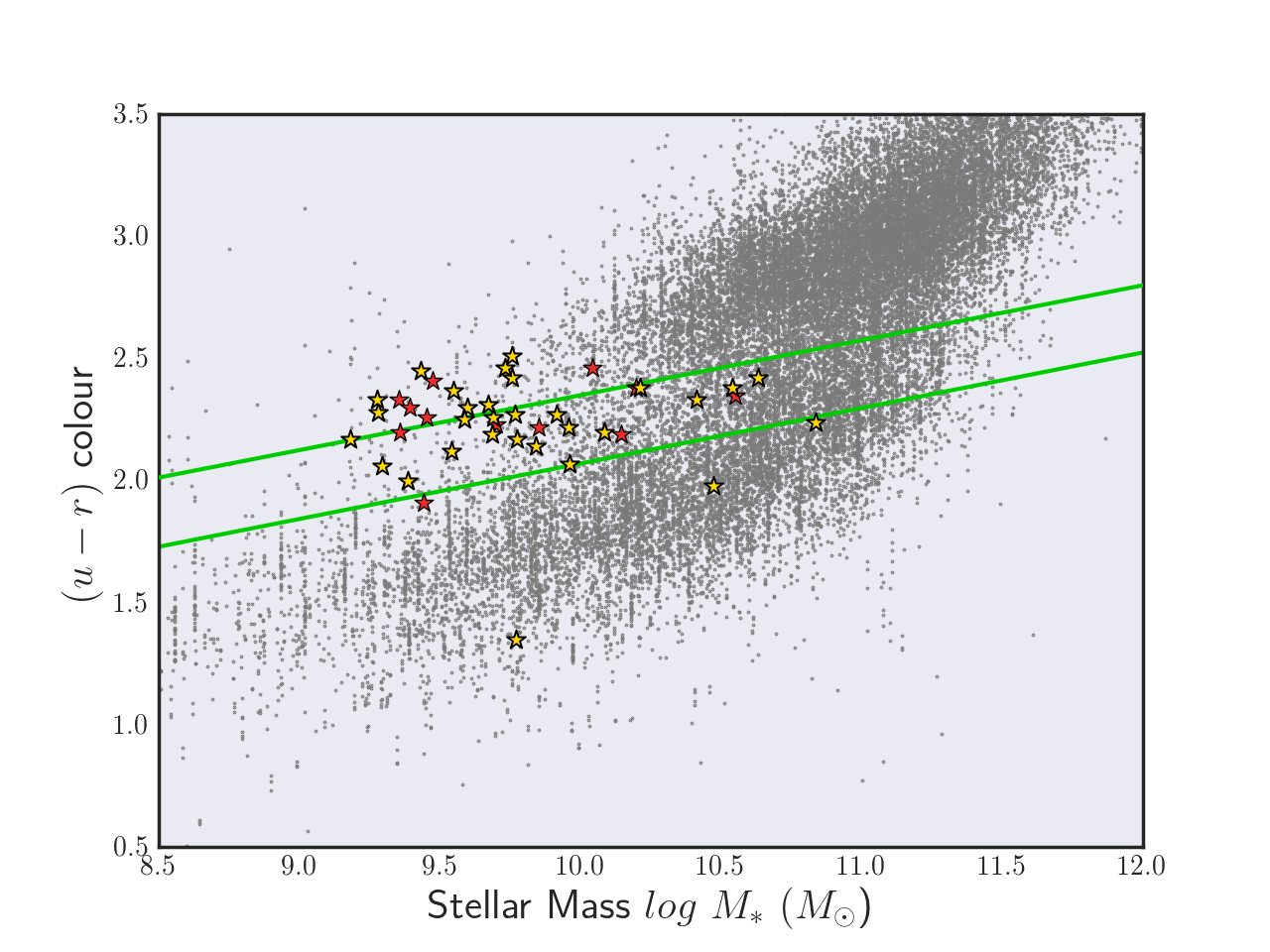}
    \caption{$u-r$ color versus stellar mass. Gold: 30 E + A galaxies from MPL-5. Red: excluded E+A candidates, plotted over a random sample of background SDSS galaxies from the Mendel mass catalog \citep{mendel}. Note that the majority of the E+A galaxies lie squarely in the {\it green valley} \citep{Wild2009}, the region approximately defined by \citealt{sch}, and represented by the two green lines. }
    \label{fig:colormass}
    
\end{figure}



\begin{figure}
	\centering
	{\includegraphics[height = 0.40\textwidth]{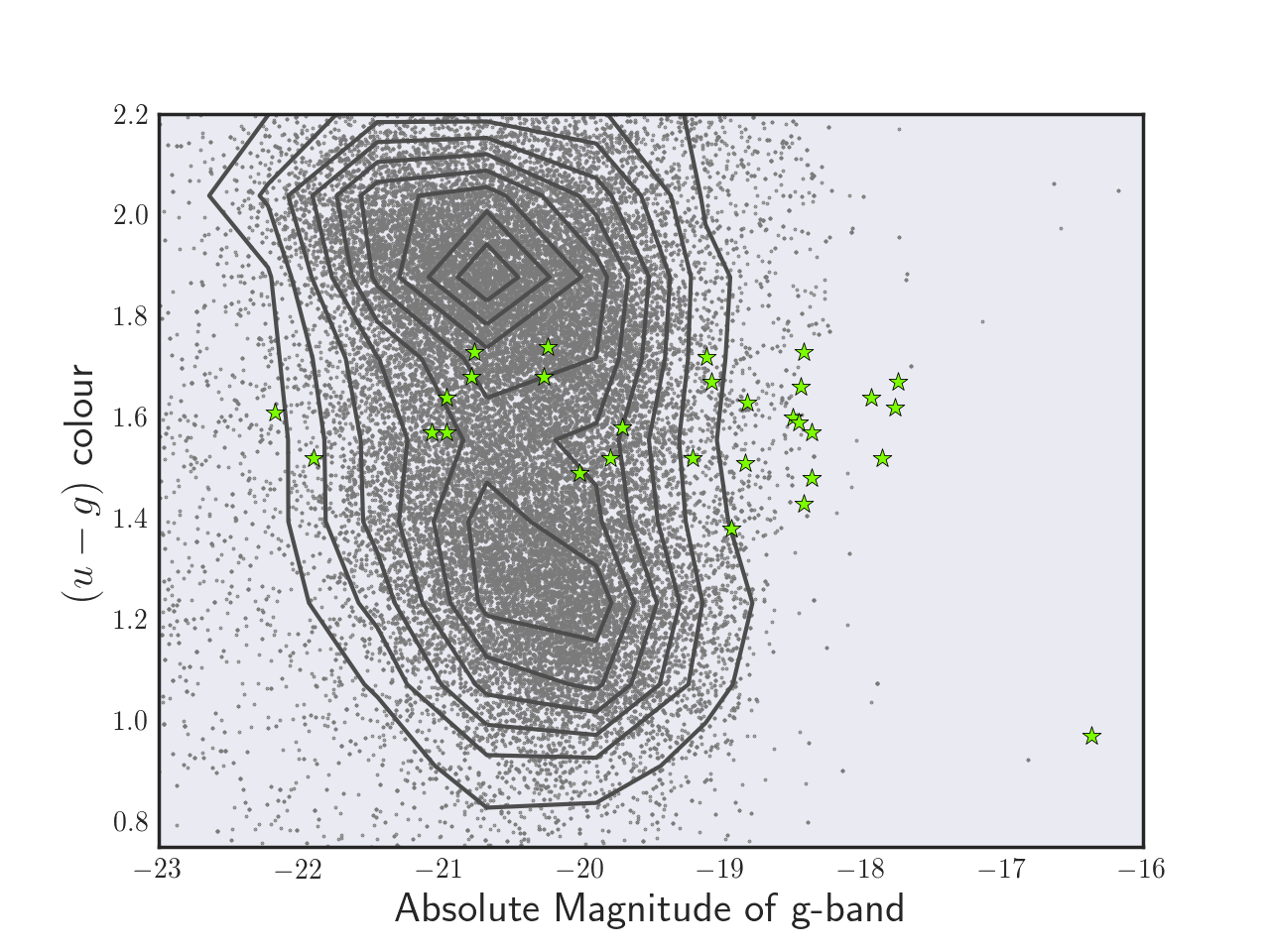}} \
	{\includegraphics[height = 0.40\textwidth]{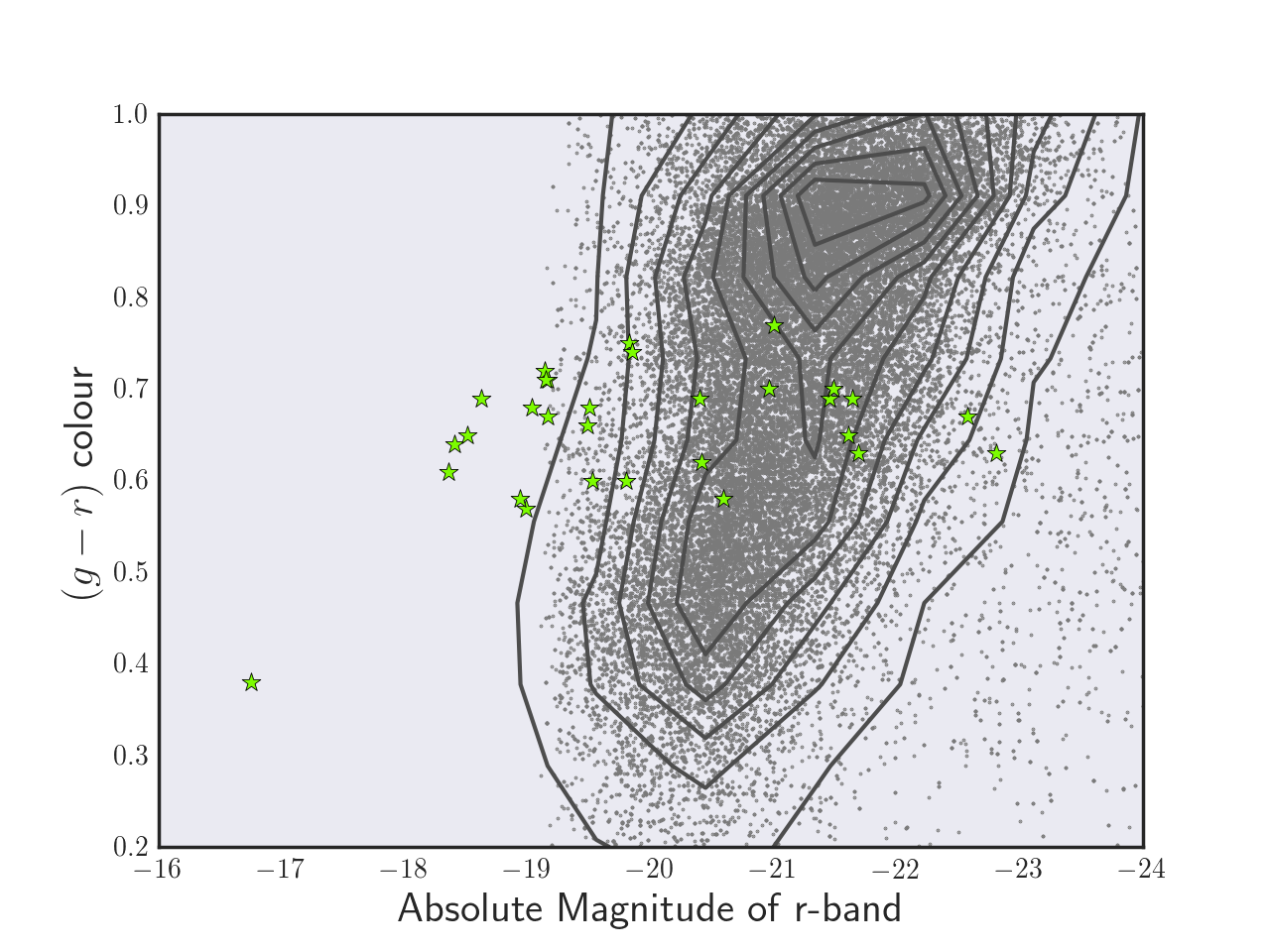}}
    \caption{\textit{(u-g)} (top) \& \textit{(g-r)} (bottom) SDSS color-Magnitude Diagram.}
    \label{fig:colormag}
    \vspace{1cm}
\end{figure} 


\vspace{0.5cm}
\section{Color and Mass} \label{sec:gal-props}
\vspace{0.5cm}

The constraint on the 4700 {\AA} to 8000 {\AA} continuum flux ratios for the E+A galaxies we identified naturally leads to an SDSS $u-r$ color of roughly $< 2.75$. Consequently, our sample is almost entirely comprised of galaxies residing in a region of color-mass and color-magnitude space known as the "green valley" \citep{Wild2009,sch}.  It has been suggested \citep{sch} that this region does not host a uniform population, but rather of two overlapping populations - that of bluish early-type galaxies and of reddish late-type galaxies.  Do E+As, then, represent a distinct galaxy evolutionary stage, or are they just coincidental stragglers in and around the green valley? Put another way, while our criteria provides a sort of red upper bound to E+A galaxies, does it follow that a blue lower bound also exists for E+As?  To test this question, we overlaid our E+A galaxy sample on color-magnitude and color-mass diagrams of galaxies in the local universe to see where they fall in these spaces. 

Our color-magnitude diagrams were created by taking a random sample of 50,000 galaxies from SDSS DR15 to populate the background, all-galaxies space. We limited these background galaxies with a \textit{z} $\leq$ 0.08, a g-band {$<$} 17.77 mag, an absolute g-band magnitude range of  -21 {$<$} $M_g$ {$<$} -16, and an absolute r-band magnitude range of -22 {$<$} $M_g$ {$<$} -14. With such a clean sample of E+A galaxies, we can begin to place this galaxy type within the larger context of galaxy demographics and evolution. 

The result is shown in Figure \ref{fig:colormag}.  Overwhelmingly, our sample, in both the $u-g$ and the $g-r$ diagrams, span the green valley transition zone. While some of the galaxies do sit more in what is considered the upper blue cloud, or lower red sequence (making special note of the one outlying E+A galaxy that is both very bright and very blue in color, and meets all criteria outlined in this paper), it is clear that these E+A galaxies are in some way transitioning from one evolutionary track to the other.

To analyze the color-mass distribution of our E+A galaxy sample, we used the Mendel Mass Catalog \citep{mendel}, which calculates an estimated total mass based on S\'ersic photometry with dust-free models, for the background galaxies in Figure \ref{fig:colormass} \footnote{Although there are masses calculated in SDSS MPL-5, the lower systematic uncertainties in the Mendel catalog were preferred.  The evolutionary trends we find remain intact when we substituted SDSS-MPL-5 masses.}

Figure \ref{fig:colormass} shows the placement of our sample on a color-mass diagram. Again, these E+A galaxies broadly lay in the {\textit green valley}, mirroring our color selection effect. Some important distinctions remain, however; to emphasize this, we overplot the final E+A galaxy sample (green stars) with E+A candidates that were excluded because they failed to meet our criteria throughout the entirety of the galaxy (pink stars). Both the green and pink stars occupy roughly the same color-mass space, but ultimately not all were considered E+A galaxies. 

Figures \ref{fig:colormag} and \ref{fig:colormass} demonstrate that, although our E+A galaxies lie mostly in the green valley range, not all green valley galaxies are E+As.  Had we built our E+A sample only using single-fiber spectra and the colors provided by SDSS, we would have significant contamination from non-E+As. Furthermore, the E+A galaxy that sits very low in the blue cloud, 8623-1902, would definitely have been excluded, despite the evidence that it may be a dwarf quenched by AGN feedback; (see~\citet{penny, marinelli}).  The location of a galaxy in certain regions of color-color space has been used in past studies to narrow the search parameters for E+A galaxies; as we can see, such a color-color cutoff technique does not yield a complete and uncontaminated E+A sample.

\vspace{0.5cm}
\section{Morphology} \label{sec:classification} 
\vspace{0.5cm}

Although the "E" in the name "E+A galaxy" has occasionally been confused to designate an elliptical galaxy instead of a spectral shape, morphology is not a defining feature of an E+A galaxy.  we noted a broad morphological distribution among our sample. Using morphological information listed in HyperLEDA \citep{makarov}, we determine that of the 25 galaxies with morphologies in the database, seven are classified as S0 with possible hints of a bar, six are spirals (mix of both barred and non-barred), four are ellipticals, and eight are indeterminate "E-S0". 

The connections between S0 morphology and star formation quenching have been the subject of considerable study, especially in the context of S0 formation via galaxy-galaxy interactions and mergers (see, e.g., \citep{Eliche-Moral18} and references therein).  Although our results do indicate a slight preponderance of early-type galaxies, the broad overall distribution of morphologies underlines the secondary, less deterministic role that galaxy shape plays in the making of an E+A galaxy.

Depending on whether or not a galaxy is classified as early-type or late-type, \citealt{sch} posit that the different morphological types of galaxies across the color-mass plane are concentrated in their designated sections. In this model, most early-types reside in the red sequence, with about 10\% of the galaxies reaching the blue cloud, representing a rapid transition between stages; late-type galaxies, meanwhile, form a continuous population with no transition zone at all, going directly from the blue cloud to the red sequence without going through a transitional phase. If this were the case, no E+A galaxy could ever be a late-type galaxy. However, \citealt{smethurst} has shown that the green valley is indeed a transitional population between the red sequence and blue cloud, regardless of morphology. Our work supports this picture, as we have at least 4 late-type galaxies, one being a very clear barred spiral, amongst our final sample of 30. Our E+A galaxies are not just blue ellipticals or red spirals; rather, they comprise an evolutionary population in their own right, with their own distinct spectral signature.


\vspace{0.5cm}
\section{Discussion} \label{sec:discussion}

\vspace{0.5cm}
\subsection{E+A Galaxy Sample Comparison} \label{subsec:comparison}
\vspace{0.5cm}

As our primary goal is to refine the definition of an E+A galaxy, and then produce a sample of E+As that fit the refined definition, it is important to compare our results to studies that used other techniques to find E+As. Altogether, we compare to four studies. \citet{goto2007} is used as a standard for identifying E+As, and three post-starburst studies conducted by \citet{cwilkinson}, \citet{Meusinger}, and \citet{chen} lend themselves well to  comparison, as they employ the the SDSS pipeline. Furthermore, \citet{cwilkinson} and \citet{Meusinger} explicitly define an E+A sample, while \citet{chen} also uses MaNGA data.

\begin{table*}
	\centering
	\tablenum{3}
    \caption{Equivalent Width Measurements of Pyraf vs. SDSS}
    \vspace{0.3cm}
	\label{tab:EWpyraf_sdss}
    \begingroup
\setlength{\tabcolsep}{8pt} 
\renewcommand{\arraystretch}{1} 
	\begin{tabular}{c c| >{\bfseries}c c c >{\bfseries}c >{\bfseries}c| >{\bfseries}c c c >{\bfseries}c >{\bfseries}c}  
	   
	    \hline
    	&&& \multicolumn{3}{c}{\textbf{PYRAF}}&&&&{\textbf{SDSS}}\\
        \hline
		\textbf{MaNGA-ID} & \textbf{PLATE-IFU} & \textbf{H$\alpha$} & H$\beta$ & H$\gamma$& \textbf{H$\delta$} & \textbf{[OII]} & \textbf{H$\alpha$} & H$\beta$ & H$\gamma$ &  \textbf{H$\delta$} & \textbf{[OII]}\\
        \hline
       
        1-210114    &8979-1902	&1.95 	&4.79 	&3.41	&4.54   &-2.43   &-0.88   &0.09   &0.28   &0.52  &-0.43\\
        1-109112    &8078-1901	&1.99 	&2.06 	&2.18	&3.27   &-2.22   &-0.07   &0.05   &0.04   &0.27  &0.11\\
        1-38166     &8081-3702	&-1.51 	&4.87 	&2.01	&3.08   &-1.11  &-2.89   &-0.47  &-0.24  &-0.12  &-5.86\\

		\hline
	\end{tabular}
   \endgroup
   
\bigskip

\tablecomments{Measuring EWs using pyraf, three galaxies from our 42 candidates passed all three criteria (bolded values) in \citealt{cwilkinson} and are described in \ref{subsec:comparison}. We compared those three galaxies to the EWs on the SDSS DR16 SkyServer and found significant systematic errors. We also note that galaxy 8081-3702 did not make our final sample because the out annulus of its WHAN diagram did not satisfy our critera (i.e. was too ionized).}

\end{table*}

The following are the E+A definitions from each paper:
\begin{enumerate}

\item \citealt{goto2007} required EW [OII] {$>$} -2.5 {\AA},  EW H{$\alpha$} {$>$} -3 {\AA} and EW H{$\delta$} {$>$} 5 {\AA} in SDSS single-fiber spectra from the DR5 catalog, resulting in a catalog of 564 galaxies.

\item \citealt{cwilkinson} required EW [OII] {$>$} -2.5 {\AA},  EW H{$\alpha$} {$>$} -3 {\AA} and EW H{$\delta$} {$>$} 3 {\AA} in SDSS single-fiber spectra. They curated a total of 25 "pure E+A" galaxies refined from the \citet{goto2007} sample. 

\item \citealt{Meusinger} used similar spectral index limits as \citealt{cwilkinson}, but allowed for a larger EW H{$\alpha$} and EW [OII] range of {$>$} -5 {\AA} and {$>$} -3 {\AA} respectively. They identified 2,665 E+A galaxies under these initial parameters, and 916 galaxies when they lowered their EW parameters to match that of \citet{goto2007}, yielding 0.3\% of their mipmapped sample.

\item \citealt{chen} required an EW H{$\delta$} {$>$} 3 {\AA} as well, but set their EW H{$\alpha$} {$<$} 10 {\AA} from the MaNGA database, and required at least six spaxels. They did not include an [OII] criterion. Their "central post-starburst" sub-category contained 31 objects.  

\end{enumerate}

To better understand the effects of the selection criteria, we subjected our candidate galaxies to the same algorithms each of these studies with the equivalent width measurements for each galaxy listed in the SDSS galSpecLine catalog.  We caution that the emission line catalogs for SDSS contain significant systematic errors and misaligned measurements, especially of weaker spectral features. Indeed, none of our E+A candidates would have satisfied any of the above studies had we used the SDSS EW values.

We found that our E+A sample has no objects in common with the \citet{goto2007} catalog. In particular, although many of our candidates pass the [OII] and H$\alpha$ criteria, none of our sample has H$\delta$ $>$ 5 {\AA}. Only two galaxies in \citet{goto2007} were observed by MaNGA. Of those two, 8083-9101 is a galaxy interaction, and is therefore rejected; furthermore, the minor galaxy in that interaction exhibits a k+a spectrum with strong H{$\alpha$} emission, and would have been rejected as well. The second, 8555-3701, was rejected during our initial search through the MaNGA catalog for having too red a continuum shape that indicates it is more of a transition between E+A and k+a. While we can confidently say these two systems are indeed post-starburst, they do not meet our more stringent criteria for an E+A galaxy.

We found that our E+A sample has no objects in common with the \citet{cwilkinson} catalog. However, it appears that 3 of our E+A galaxies would have appeared in the \citet{cwilkinson} catalog; reanalysis of the EWs revealed that the SDSS DR7 database was in error. These common galaxies are listed in Table \ref{tab:EWpyraf_sdss}, along with their stated and recalculated EW measurements. We believe that although the biggest reason our samples differ is due to our requirement on the continuum shape, we caution that reliance on SDSS spectral values may skew such a sample.

\citealt{Meusinger} used a combination of artificial neural network algorithms, SDSS fits data, and EW measurements to cull their sample from 1 million galaxies. They went further by doing a visual inspection of each returned spectra to eliminate all artifact contamination, leaving only 0.26\% of their original objects. We found that our E+A sample has no objects in common with the \citet{Meusinger} catalog, primarily due to their H$\delta$ {$>$} 3 \AA requirement. However, 7 galaxies in our sample do indeed satisfy these criteria; notably these 7 have erroneous SDSS EWs, and we speculate that this excluded them from their catalog at the onset.

Of these four studies, we had the best chance of overlapping with the \citet{chen} sample because was also selected from MaNGA and used more inclusive criteria for than the other studies. They required at least six spaxels with EW H{$\delta$} {$>$} 3 {\AA} and EW H{$\alpha$} {$<$} 10 {\AA}.  They further subdivided the 74 objects that fit those criteria into different groups; broadly, their group of "central post-starburst" objects were selected in the same way as our sample, as we started with the requirement that the spectrum in each galaxy's central region have E+A characteristics.

We found that of the 31 galaxies in the \citet{chen} central post-starburst subsample, 10 were in our list of 42 candidates. This rate of correspondence is much higher than that of the other three studies, yet still not a majority match to their entire sample.  We think the main discrepancy can be attributed to our requirement that no less than $\sim$ 70\% of the galaxy be classified as post-starburst, which could have been in excess of 900 spaxels per galaxy rather than their six spaxel limit. We note, however, that 12 of our candidate galaxies do satisfy the criteria set forth by \citealt{chen}, and we attribute this very small discrepancy to erroneous EW measurements.

As it turned out, there was little overlap between our final sample and those of the other four studies.  Even after reanalyzing the EWs, at most only 28\%  of our sample of candidate galaxies would have made it into one of these four samples. The greatest overlap was with \citet{chen}. Considering that they also used MaNGA as their parent galaxy sample, that is still somewhat surprisingly low.  As for the \citealt{goto2007}, \citealt{Meusinger} and  \citealt{cwilkinson} samples, our final lists of E+A galaxies did not overlap at all.

A major contributor to this discrepancy is our lower limit of 2 {\AA} EW of Balmer absorption, which is consistent with earlier definitions of E+A galaxies in high-redshift clusters (e.g., \citealt{barger96}) but less stringent than the limit of these four surveys.

The most important difference, however, is that all four of these sample sets based their values of spectral indices on the spectral fits provided in the SDSS database. Our analyses of individual E+A galaxies in our final sample (e.g.,\citealt{marinelli}) showed that those fits frequently failed to measure accurately the equivalent widths of the Balmer absorption lines.  One likely reason for this is that spectral synthesis models do very well with typical galaxy spectra, but are not well-equipped to fit E+A galaxies whose Balmer absorption lines are unusually deep, thus underestimating them significantly.  Like us, \citealt{chen} and \citealt{Meusinger} used visual inspection to confirm their galaxies-- but only on the galaxies returned by their searches, which may have been biased by this systematic effect.

\vspace{0.5cm}
\subsection{E+A Galaxies in IFS} \label{subsec:discussion2}
\vspace{0.5cm}

The detailed, multi-step process that we used to obtain our sample represents a significant refinement of how an E+A galaxy is defined.  By rigorously distinguishing E+A galaxies as a well-defined subset of the heterogeneous population of post-starburst galaxy systems overall, we can use E+As as tracers of galaxy evolution with much greater precision, and investigate the critical process of star formation quenching in its final stages.

With the added diagnostic power of the MaNGA IFS data, we have already noticed some spatially dependent properties of our E+A galaxy sample that bear further investigation.  For example, we noticed several defining features among the galaxies, allowing us to classify them into sub-samples (as shown in Table \ref{tab:geninfo}). 12 objects in the final sample showed almost identical spectra and colors from 0 to 3 {$R_e$}, aligning perfectly with the characteristics we defined in section \ref{sec:initialsample}. 11 of them were significantly bluer in the center than at 2-3 {$R_e$}, while the opposite was true in 4 others that were redder in the center, and bluer at 2-3 {$R_e$}. 3 galaxies exhibited significant forbidden line emission ([OII], [NII], and [SII]) in the central MaNGA aperture that did not appear in their SDSS DR15 single-fiber spectra. Finally, 3 of the galaxies had a significantly higher flux in the annular aperture from 1-2 {$R_e$} than in their central or outer apertures. Further investigation of these traits and what they imply about E+A galaxy evolution will be explained in a forthcoming paper.

Perhaps the most important improvement that MaNGA has allowed in our E+A identifications is that we have been able to confirm that star formation has ceased in our E+As even if some emission lines are present in the spectrum. In particular, whereas automation-dependent searches for E+As almost always exclude galaxies with significant H{$\alpha$} + [NII] emission, we have been able to include E+A galaxies that have [NII] emission lines, but no H{$\alpha$} emission sandwiched between them (Two galaxies in our final sample, in fact, showed very significant line emission).  As a result, we do not exclude the study of weak or nearly-dormant nuclear activity in our E+As.  We are further able to combine that information with the spatially resolved information that integral field spectra provide, such as stellar age and metallicity gradients, gas distribution, and stellar and gas dynamics; in E+A galaxies, that means we can probe not only star formation duty cycles and stellar populations, but also AGN duty cycles, AGN feedback and the starburst-AGN connection in each E+A galaxy's recent history. These topics will be explored in detail in upcoming papers.



\vspace{0.5cm}
\section{Conclusions} \label{sec:conclusion}
\vspace{0.5cm}

We set out to refine a specific set of criteria to be used in the identification of E+A galaxies in the following way:
\begin{enumerate}

\item We mined MaNGA’s MPL-5 sample and selected E+A candidates by their single fiber spectrum housed in the SDSS DR15. We looked for galaxies that presented with strong Balmer absorption lines, with an EW {$>$} 2 {\AA}, No H{$\alpha$} emission, low to no [OII] emission, a S/N $>$ 10, and a significant A-type continuum. We further require that overwhelming majority of spaxels exhibit the same E+A criterion throughout the disk.

\item Using the IFS data from MaNGA, we summed spaxel spectra in annuli of 1, 2, and 3 effective radii to ensure each annular spectrum passed our E+A criteria. This was diagnostic of a dominant E+A spectrum throughout the galaxy.

\item We created spatially resolved WHAN diagrams of each galaxy to confirm an old stellar population. Through this method, we were able to exclude two galaxies that exhibited E+A properties only in the central region, but were highly star forming elsewhere. Overwhelmingly, $> 90\%$ 
of the spaxels in the final E+A sample were post-starburst.

\item Almost all galaxies in our sample have a $u-r$ color between 2 and 2.5, and masses $ < 10^{11}$, which places almost all of our sample in the green valley.

\item Using HyperLEDA for morphology, we find a rich range of morphologies, with over half being S0. 

\item Our final sample of 30 galaxies represents a well-defined population of post-starburst objects, with specific criteria to define them as E+A. The properties of these galaxies in terms of active galactic nuclei, spatial ionization maps, and their diagnostic role in galaxy evolution, will be further explored in future work. 
\end{enumerate}


\vspace{0.5cm}
\section*{Acknowledgements}
\vspace{0.5cm}

This work was supported by the National Science Foundation via Grant No.: 1460939 to the Physical Sciences REU program at the American Museum of Natural History, by the Alfred P. Sloan Foundation via the SDSS-IV Faculty and Student Team (FAST) initiative, ARC Agreement No.SSP483, and by the NSF via Grant AST-1460860, both to the CUNY College of Staten Island, and the CUNY College of Staten Island Academic Affairs Office for the initial work done in the summer of 2016. 
Funding for the Sloan Digital Sky Survey IV has been provided by the Alfred P. Sloan Foundation, the U.S. Department of Energy Office of Science, and the Participating Institutions. SDSS-IV acknowledges
support and resources from the Center for High-Performance Computing at
the University of Utah. The SDSS web site is www.sdss.org.

SDSS-IV is managed by the Astrophysical Research Consortium for the 
Participating Institutions of the SDSS Collaboration including the 
Brazilian Participation Group, the Carnegie Institution for Science, 
Carnegie Mellon University, the Chilean Participation Group, the French Participation Group, Harvard-Smithsonian Center for Astrophysics, 
Instituto de Astrof\'isica de Canarias, The Johns Hopkins University, Kavli Institute for the Physics and Mathematics of the Universe (IPMU) / 
University of Tokyo, the Korean Participation Group, Lawrence Berkeley National Laboratory, 
Leibniz Institut f\"ur Astrophysik Potsdam (AIP),  
Max-Planck-Institut f\"ur Astronomie (MPIA Heidelberg), 
Max-Planck-Institut f\"ur Astrophysik (MPA Garching), 
Max-Planck-Institut f\"ur Extraterrestrische Physik (MPE), 
National Astronomical Observatories of China, New Mexico State University, 
New York University, University of Notre Dame, 
Observat\'ario Nacional / MCTI, The Ohio State University, 
Pennsylvania State University, Shanghai Astronomical Observatory, 
United Kingdom Participation Group,
Universidad Nacional Aut\'onoma de M\'exico, University of Arizona, 
University of Colorado Boulder, University of Oxford, University of Portsmouth, 
University of Utah, University of Virginia, University of Washington, University of Wisconsin, 
Vanderbilt University, and Yale University.

We would also like to acknowledge the undergraduate student collaboration \textit{MOONJAM}: Julia Falcone, Olivia James, Allen Liu, Nicole Wallack, and Muhammad Wally for their dedication in finding a complete sample of E+A galaxies, by eye, in MaNGA. Finally, a special acknowledgement to The Fisk-Vanderbilt Master's to PhD Program, Sam Dunham of Vanderbilt University, Dr. Andrew Johnson of Oxbridge Academy, Eric Vandernoot of Florida Atlantic University, and Shaun Greene for scientific advisement and support throughout this project. 



\bibliography{citations}{}

\begin{thebibliography}{}
\expandafter\ifx\csname natexlab\endcsname\relax\def\natexlab#1{#1}\fi
\providecommand{\url}[1]{\href{#1}{#1}}
\providecommand{\dodoi}[1]{doi:~\href{http://doi.org/#1}{\nolinkurl{#1}}}
\providecommand{\doeprint}[1]{\href{http://ascl.net/#1}{\nolinkurl{http://ascl.net/#1}}}
\providecommand{\doarXiv}[1]{\href{https://arxiv.org/abs/#1}{\nolinkurl{https://arxiv.org/abs/#1}}}

\bibitem[{{Aguado} {et~al.}(2019){Aguado}, {Ahumada}, {Almeida}, {Anderson},
  {Andrews}, {Anguiano}, {Aquino Ort{\'\i}z}, {Arag{\'o}n-Salamanca},
  {Argudo-Fern{\'a}ndez}, {Aubert}, {Avila-Reese}, {Badenes}, {Barboza
  Rembold}, {Barger}, {Barrera-Ballesteros}, {Bates}, {Bautista}, {Beaton},
  {Beers}, {Belfiore}, {Bernardi}, {Bershady}, {Beutler}, {Bird}, {Bizyaev},
  {Blanc}, {Blanton}, {Blomqvist}, {Bolton}, {Boquien}, {Borissova}, {Bovy},
  {Brandt}, {Brinkmann}, {Brownstein}, {Bundy}, {Burgasser}, {Byler}, {Cano
  Diaz}, {Cappellari}, {Carrera}, {Cervantes Sodi}, {Chen}, {Cherinka}, {Choi},
  {Chung}, {Coffey}, {Comerford}, {Comparat}, {Covey}, {da Silva Ilha}, {da
  Costa}, {Dai}, {Damke}, {Darling}, {Davies}, {Dawson}, {de Sainte Agathe},
  {Deconto Machado}, {Del Moro}, {De Lee}, {Diamond-Stanic}, {Dom{\'\i}nguez
  S{\'a}nchez}, {Donor}, {Drory}, {du Mas des Bourboux}, {Duckworth}, {Dwelly},
  {Ebelke}, {Emsellem}, {Escoffier}, {Fern{\'a}ndez-Trincado}, {Feuillet},
  {Fischer}, {Fleming}, {Fraser-McKelvie}, {Freischlad}, {Frinchaboy}, {Fu},
  {Galbany}, {Garcia-Dias}, {Garc{\'\i}a-Hern{\'a}ndez}, {Garma Oehmichen},
  {Geimba Maia}, {Gil-Mar{\'\i}n}, {Grabowski}, {Gu}, {Guo}, {Ha},
  {Harrington}, {Hasselquist}, {Hayes}, {Hearty}, {Hernandez Toledo}, {Hicks},
  {Hogg}, {Holley-Bockelmann}, {Holtzman}, {Hsieh}, {Hunt}, {Hwang},
  {Ibarra-Medel}, {Jimenez Angel}, {Johnson}, {Jones}, {J{\"o}nsson},
  {Kinemuchi}, {Kollmeier}, {Krawczyk}, {Kreckel}, {Kruk}, {Lacerna}, {Lan},
  {Lane}, {Law}, {Lee}, {Li}, {Lian}, {Lin}, {Lin}, {Lintott}, {Long},
  {Longa-Pe{\~n}a}, {Mackereth}, {de la Macorra}, {Majewski}, {Malanushenko},
  {Manchado}, {Maraston}, {Mariappan}, {Marinelli}, {Marques-Chaves},
  {Masseron}, {Masters}, {McDermid}, {Medina Pe{\~n}a}, {Meneses-Goytia},
  {Merloni}, {Merrifield}, {Meszaros}, {Minniti}, {Minsley}, {Muna}, {Myers},
  {Nair}, {Correa do Nascimento}, {Newman}, {Nitschelm}, {Olmstead}, {Oravetz},
  {Oravetz}, {Ortega Minakata}, {Pace}, {Padilla}, {Palicio}, {Pan}, {Pan},
  {Parikh}, {Parker}, {Peirani}, {Penny}, {Percival}, {Perez-Fournon},
  {Peterken}, {Pinsonneault}, {Prakash}, {Raddick}, {Raichoor}, {Riffel},
  {Riffel}, {Rix}, {Robin}, {Roman-Lopes}, {Rose}, {Ross}, {Rossi}, {Rowlands},
  {Rubin}, {S{\'a}nchez}, {S{\'a}nchez-Gallego}, {Sayres}, {Schaefer},
  {Schiavon}, {Schimoia}, {Schlafly}, {Schlegel}, {Schneider}, {Schultheis},
  {Seo}, {Shamsi}, {Shao}, {Shen}, {Shetty}, {Simonian}, {Smethurst}, {Sobeck},
  {Souter}, {Spindler}, {Stark}, {Stassun}, {Steinmetz}, {Storchi-Bergmann},
  {Stringfellow}, {Su{\'a}rez}, {Sun}, {Taghizadeh-Popp}, {Talbot}, {Tayar},
  {Thakar}, {Thomas}, {Tissera}, {Tojeiro}, {Troup}, {Unda-Sanzana},
  {Valenzuela}, {Vargas-Maga{\~n}a}, {V{\'a}zquez-Mata}, {Wake}, {Weaver},
  {Weijmans}, {Westfall}, {Wild}, {Wilson}, {Woods}, {Yan}, {Yang}, {Zamora},
  {Zasowski}, {Zhang}, {Zheng}, {Zheng}, {Zhu}, {Zinn}, \&
  {Zou}}]{2018arXiv181202759A}
{Aguado}, D.~S., {Ahumada}, R., {Almeida}, A., {et~al.} 2019, \apjs, 240, 23,
  \dodoi{10.3847/1538-4365/aaf651}

\bibitem[{{Alatalo} {et~al.}(2016{\natexlab{a}}){Alatalo}, {Cales}, {Rich},
  {Appleton}, {Kewley}, {Lacy}, {Lanz}, {Medling}, \& {Nyland}}]{spogs}
{Alatalo}, K., {Cales}, S.~L., {Rich}, J.~A., {et~al.} 2016{\natexlab{a}},
  \apjs, 224, 38, \dodoi{10.3847/0067-0049/224/2/38}

\bibitem[{{Alatalo} {et~al.}(2016{\natexlab{b}}){Alatalo}, {Lisenfeld}, {Lanz},
  {Appleton}, {Ardila}, {Cales}, {Kewley}, {Lacy}, {Medling}, {Nyland },
  {Rich}, \& {Urry}}]{alatalo16}
{Alatalo}, K., {Lisenfeld}, U., {Lanz}, L., {et~al.} 2016{\natexlab{b}}, \apj,
  827, 106, \dodoi{10.3847/0004-637X/827/2/106}

\bibitem[{{Ardila} {et~al.}(2018){Ardila}, {Alatalo}, {Lanz}, {Appleton},
  {Beaton}, {Bitsakis}, {Cales}, {Falc{\'o}n-Barroso}, {Kewley}, {Medling},
  {Mulchaey}, {Nyland}, {Rich}, \& {Urry}}]{ardila18}
{Ardila}, F., {Alatalo}, K., {Lanz}, L., {et~al.} 2018, \apj, 863, 28,
  \dodoi{10.3847/1538-4357/aad0a3}

\bibitem[{{Baldwin} {et~al.}(1981){Baldwin}, {Phillips}, \& {Terlevich}}]{bpt}
{Baldwin}, J.~A., {Phillips}, M.~M., \& {Terlevich}, R. 1981, \pasp, 93, 5,
  \dodoi{10.1086/130766}

\bibitem[{{Barger} {et~al.}(1996){Barger}, {Aragon-Salamanca}, {Ellis},
  {Couch}, {Smail}, \& {Sharples}}]{barger96}
{Barger}, A.~J., {Aragon-Salamanca}, A., {Ellis}, R.~S., {et~al.} 1996, \mnras,
  279, 1, \dodoi{10.1093/mnras/279.1.1}

\bibitem[{{Baron} {et~al.}(2018){Baron}, {Netzer}, {Prochaska}, {Cai},
  {Cantalupo}, {Martin}, {Matuszewski}, {Moore}, {Morrissey}, \&
  {Neill}}]{baron18}
{Baron}, D., {Netzer}, H., {Prochaska}, J.~X., {et~al.} 2018, \mnras, 480,
  3993, \dodoi{10.1093/mnras/sty2113}

\bibitem[{{Belfiore} {et~al.}(2015){Belfiore}, {Maiolino}, {Bundy}, {Thomas},
  {Maraston}, {Wilkinson}, {S{\'a}nchez}, {Bershady}, {Blanc}, {Bothwell},
  {Cales}, {Coccato}, {Drory}, {Emsellem}, {Fu}, {Gelfand}, {Law}, {Masters},
  {Parejko}, {Tremonti}, {Wake}, {Weijmans}, {Yan}, {Xiao}, {Zhang}, {Zheng},
  {Bizyaev}, {Kinemuchi}, {Oravetz}, \& {Simmons}}]{Belfiore}
{Belfiore}, F., {Maiolino}, R., {Bundy}, K., {et~al.} 2015, \mnras, 449, 867,
  \dodoi{10.1093/mnras/stv296}

\bibitem[{{Belfiore} {et~al.}(2019){Belfiore}, {Westfall}, {Schaefer},
  {Cappellari}, {Ji}, {Bershady}, {Tremonti}, {Law}, {Yan}, {Bundy}, {Shetty},
  {Drory}, {Thomas}, {Emsellem}, \& {S{\'a}nchez}}]{Belfioredata}
{Belfiore}, F., {Westfall}, K.~B., {Schaefer}, A., {et~al.} 2019, \aj, 158,
  160, \dodoi{10.3847/1538-3881/ab3e4e}

\bibitem[{{Blake} {et~al.}(2004){Blake}, {Pracy}, {Couch}, {Bekki}, {Lewis},
  {Glazebrook}, {Baldry}, {Baugh}, {Bland-Hawthorn}, {Bridges}, {Cannon},
  {Cole}, {Colless}, {Collins}, {Dalton}, {De Propris}, {Driver}, {Efstathiou},
  {Ellis}, {Frenk}, {Jackson}, {Lahav}, {Lumsden}, {Maddox}, {Madgwick},
  {Norberg}, {Peacock}, {Peterson}, {Sutherland}, \& {Taylor}}]{blake04}
{Blake}, C., {Pracy}, M.~B., {Couch}, W.~J., {et~al.} 2004, \mnras, 355, 713,
  \dodoi{10.1111/j.1365-2966.2004.08351.x}

\bibitem[{{Blanton} {et~al.}(2017){Blanton}, {Bershady}, {Abolfathi},
  {Albareti}, {Allende Prieto}, {Almeida}, {Alonso-Garc{\'{\i}}a}, {Anders},
  {Anderson}, {Andrews}, \& et~al.}]{2017AJ....154...28B}
{Blanton}, M.~R., {Bershady}, M.~A., {Abolfathi}, B., {et~al.} 2017, \aj, 154,
  28, \dodoi{10.3847/1538-3881/aa7567}

\bibitem[{{Bundy} {et~al.}(2015){Bundy}, {Bershady}, {Law}, {Yan}, {Drory},
  {MacDonald}, {Wake}, {Cherinka}, {S{\'a}nchez-Gallego}, {Weijmans}, {Thomas},
  {Tremonti}, {Masters}, {Coccato}, {Diamond-Stanic}, {Arag{\'o}n-Salamanca},
  {Avila-Reese}, {Badenes}, {Falc{\'o}n-Barroso}, {Belfiore}, {Bizyaev},
  {Blanc}, {Bland-Hawthorn}, {Blanton}, {Brownstein}, {Byler}, {Cappellari},
  {Conroy}, {Dutton}, {Emsellem}, {Etherington}, {Frinchaboy}, {Fu}, {Gunn},
  {Harding}, {Johnston}, {Kauffmann}, {Kinemuchi}, {Klaene}, {Knapen},
  {Leauthaud}, {Li}, {Lin}, {Maiolino}, {Malanushenko}, {Malanushenko}, {Mao},
  {Maraston}, {McDermid}, {Merrifield}, {Nichol}, {Oravetz}, {Pan}, {Parejko},
  {Sanchez}, {Schlegel}, {Simmons}, {Steele}, {Steinmetz}, {Thanjavur},
  {Thompson}, {Tinker}, {van den Bosch}, {Westfall}, {Wilkinson}, {Wright},
  {Xiao}, \& {Zhang}}]{2015ApJ...798....7B}
{Bundy}, K., {Bershady}, M.~A., {Law}, D.~R., {et~al.} 2015, \apj, 798, 7,
  \dodoi{10.1088/0004-637X/798/1/7}

\bibitem[{Bundy {et~al.}(2015)Bundy, Bershady, Law, Yan, Drory, MacDonald,
  Wake, Cherinka, S\'{a}nchez-Gallego, Weijmans, Thomas, Tremonti, Masters,
  Coccato, Diamond-Stanic, Arag\'{o}n-Salamanca, Avila-Reese, Badenes,
  Falc\'{o}n-Barroso, Belfiore, Bizyaev, Blanc, Bland-Hawthorn, Blanton,
  Brownstein, Byler, Cappellari, Conroy, Dutton, Emsellem, Etherington,
  Frinchaboy, Fu, Gunn, Harding, Johnston, Kauffmann, Kinemuchi, Klaene,
  Knapen, Leauthaud, Li, Lin, Maiolino, Malanushenko, Malanushenko, Mao,
  Maraston, McDermid, Merrifield, Nichol, Oravetz, Pan, Parejko, Sanchez,
  Schlegel, Simmons, Steele, Steinmetz, Thanjavur, Thompson, Tinker, van~den
  Bosch, Westfall, Wilkinson, Wright, Xiao, \& Zhang}]{Bundy}
Bundy, K., Bershady, M.~A., Law, D.~R., {et~al.} 2015, The {A}stro{p}hysical
  {J}ournal ({ApJ}), 798, 24

\bibitem[{{Caldwell} {et~al.}(1999){Caldwell}, {Rose}, \& {Dendy}}]{caldwell99}
{Caldwell}, N., {Rose}, J.~A., \& {Dendy}, K. 1999, \aj, 117, 140,
  \dodoi{10.1086/300679}

\bibitem[{{Caldwell} {et~al.}(1993){Caldwell}, {Rose}, {Sharples}, {Ellis}, \&
  {Bower}}]{cald93}
{Caldwell}, N., {Rose}, J.~A., {Sharples}, R.~M., {Ellis}, R.~S., \& {Bower},
  R.~G. 1993, \aj, 106, 473, \dodoi{10.1086/116656}

\bibitem[{{Charlot} \& {Silk}(1994)}]{charlot94}
{Charlot}, S., \& {Silk}, J. 1994, \apj, 432, 453, \dodoi{10.1086/174584}

\bibitem[{{Chen} {et~al.}(2019){Chen}, {Shi}, {Wild}, {Tremonti}, {Rowlands},
  {Bizyaev}, {Yan}, {Lin}, \& {Riffel}}]{chen}
{Chen}, Y.-M., {Shi}, Y., {Wild}, V., {et~al.} 2019, \mnras, 489, 5709,
  \dodoi{10.1093/mnras/stz2494}

\bibitem[{Cherinka {et~al.}(2018)Cherinka, Sánchez-Gallego, Andrews, \&
  Brownstein}]{brian_cherinka_2018_1146705}
Cherinka, B., Sánchez-Gallego, J., Andrews, B., \& Brownstein, J. 2018,
  sdss/marvin: Marvin Beta 2.2.0, \dodoi{10.5281/zenodo.1146705}

\bibitem[{{Cid Fernandes} {et~al.}(2011){Cid Fernandes}, {Stasi{\'n}ska},
  {Mateus}, \& {Vale Asari}}]{fernandes}
{Cid Fernandes}, R., {Stasi{\'n}ska}, G., {Mateus}, A., \& {Vale Asari}, N.
  2011, \mnras, 413, 1687, \dodoi{10.1111/j.1365-2966.2011.18244.x}

\bibitem[{{Dressler} \& {Gunn}(1982)}]{Dressler1}
{Dressler}, A., \& {Gunn}, J.~E. 1982, \apj, 263, 533, \dodoi{10.1086/160524}

\bibitem[{{Dressler} \& {Gunn}(1983)}]{Dressler2}
---. 1983, \apj, 270, 7, \dodoi{10.1086/161093}

\bibitem[{{Drory} {et~al.}(2015{\natexlab{a}}){Drory}, {MacDonald}, {Bershady},
  {Bundy}, {Gunn}, {Law}, {Smith}, {Stoll}, {Tremonti}, {Wake}, {Yan},
  {Weijmans}, {Byler}, {Cherinka}, {Cope}, {Eigenbrot}, {Harding}, {Holder},
  {Huehnerhoff}, {Jaehnig}, {Jansen}, {Klaene}, {Paat}, {Percival}, \&
  {Sayres}}]{2015AJ....149...77D}
{Drory}, N., {MacDonald}, N., {Bershady}, M.~A., {et~al.} 2015{\natexlab{a}},
  \aj, 149, 77, \dodoi{10.1088/0004-6256/149/2/77}

\bibitem[{{Drory} {et~al.}(2015{\natexlab{b}}){Drory}, {MacDonald}, {Bershady},
  {Bundy}, {Gunn}, {Law}, {Smith}, {Stoll}, {Tremonti}, {Wake}, {Yan},
  {Weijmans}, {Byler}, {Cherinka}, {Cope}, {Eigenbrot}, {Harding}, {Holder},
  {Huehnerhoff}, {Jaehnig}, {Jansen}, {Klaene}, {Paat}, {Percival}, \&
  {Sayres}}]{Drory}
---. 2015{\natexlab{b}}, \aj, 149, 77, \dodoi{10.1088/0004-6256/149/2/77}

\bibitem[{{Eliche-Moral} {et~al.}(2018){Eliche-Moral},
  {Rodr{\'\i}guez-P{\'e}rez}, {Borlaff}, {Querejeta}, \&
  {Tapia}}]{Eliche-Moral18}
{Eliche-Moral}, M.~C., {Rodr{\'\i}guez-P{\'e}rez}, C., {Borlaff}, A.,
  {Querejeta}, M., \& {Tapia}, T. 2018, \aap, 617, A113,
  \dodoi{10.1051/0004-6361/201832911}

\bibitem[{{French} {et~al.}(2018{\natexlab{a}}){French}, {Yang}, {Zabludoff},
  \& {Tremonti}}]{french18}
{French}, K.~D., {Yang}, Y., {Zabludoff}, A.~I., \& {Tremonti}, C.~A.
  2018{\natexlab{a}}, \apj, 862, 2, \dodoi{10.3847/1538-4357/aacb2d}

\bibitem[{{French} {et~al.}(2018{\natexlab{b}}){French}, {Zabludoff}, {Yoon},
  {Shirley}, {Yang}, {Smercina}, {Smith}, \& {Narayanan}}]{frenchzab}
{French}, K.~D., {Zabludoff}, A.~I., {Yoon}, I., {et~al.} 2018{\natexlab{b}},
  \apj, 861, 123, \dodoi{10.3847/1538-4357/aac8de}

\bibitem[{{Galaz}(2000)}]{galaz00}
{Galaz}, G. 2000, \aj, 119, 2118, \dodoi{10.1086/301334}

\bibitem[{Goto(2007)}]{goto2007}
Goto, T. 2007, {M}onthly {N}otices of the {R}oyal {A}stronomical {S}ociety
  ({MNRAS}), 381, 187

\bibitem[{{Goto} {et~al.}(2008){Goto}, {Kawai}, {Shimono}, {Sugai}, {Yagi}, \&
  {Hattori}}]{goto08}
{Goto}, T., {Kawai}, A., {Shimono}, A., {et~al.} 2008, \mnras, 386, 1355,
  \dodoi{10.1111/j.1365-2966.2008.12916.x}

\bibitem[{{Guillochon} {et~al.}(2014){Guillochon}, {Manukian}, \&
  {Ramirez-Ruiz}}]{guillochon}
{Guillochon}, J., {Manukian}, H., \& {Ramirez-Ruiz}, E. 2014, \apj, 783, 23,
  \dodoi{10.1088/0004-637X/783/1/23}

\bibitem[{{Gunn} {et~al.}(2006){Gunn}, {Siegmund}, {Mannery}, {Owen}, {Hull},
  {Leger}, {Carey}, {Knapp}, {York}, {Boroski}, {Kent}, {Lupton}, {Rockosi},
  {Evans}, {Waddell}, {Anderson}, {Annis}, {Barentine}, {Bartoszek}, {Bastian},
  {Bracker}, {Brewington}, {Briegel}, {Brinkmann}, {Brown}, {Carr},
  {Czarapata}, {Drennan}, {Dombeck}, {Federwitz}, {Gillespie}, {Gonzales},
  {Hansen}, {Harvanek}, {Hayes}, {Jordan}, {Kinney}, {Klaene}, {Kleinman},
  {Kron}, {Kresinski}, {Lee}, {Limmongkol}, {Lindenmeyer}, {Long}, {Loomis},
  {McGehee}, {Mantsch}, {Neilsen}, {Neswold}, {Newman}, {Nitta}, {Peoples},
  {Pier}, {Prieto}, {Prosapio}, {Rivetta}, {Schneider}, {Snedden}, \&
  {Wang}}]{Gunn}
{Gunn}, J.~E., {Siegmund}, W.~A., {Mannery}, E.~J., {et~al.} 2006, \aj, 131,
  2332, \dodoi{10.1086/500975}

\bibitem[{{Klitsch} {et~al.}(2017){Klitsch}, {Zwaan}, {Kuntschner}, {Couch},
  {Pracy}, \& {Owers}}]{klitsch17}
{Klitsch}, A., {Zwaan}, M.~A., {Kuntschner}, H., {et~al.} 2017, \aap, 600, A80,
  \dodoi{10.1051/0004-6361/201527922}

\bibitem[{{Kollatschny} \& {Fricke}(1989)}]{kf89}
{Kollatschny}, W., \& {Fricke}, K.~J. 1989, in IAU Symposium, Vol. 134, Active
  Galactic Nuclei, ed. D.~E. {Osterbrock} \& J.~S. {Miller}, 449

\bibitem[{{Lavery} \& {Henry}(1988)}]{lavery88}
{Lavery}, R.~J., \& {Henry}, J.~P. 1988, \apj, 330, 596, \dodoi{10.1086/166496}

\bibitem[{{Law} {et~al.}(2015{\natexlab{a}}){Law}, {Yan}, {Bershady}, {Bundy},
  {Cherinka}, {Drory}, {MacDonald}, {S{\'a}nchez-Gallego}, {Wake}, {Weijmans},
  {Blanton}, {Klaene}, {Moran}, {Sanchez}, \& {Zhang}}]{Law2015}
{Law}, D.~R., {Yan}, R., {Bershady}, M.~A., {et~al.} 2015{\natexlab{a}}, \aj,
  150, 19, \dodoi{10.1088/0004-6256/150/1/19}

\bibitem[{{Law} {et~al.}(2015{\natexlab{b}}){Law}, {Yan}, {Bershady}, {Bundy},
  {Cherinka}, {Drory}, {MacDonald}, {S{\'a}nchez-Gallego}, {Wake}, {Weijmans},
  {Blanton}, {Klaene}, {Moran}, {Sanchez}, \& {Zhang}}]{2015AJ....150...19L}
---. 2015{\natexlab{b}}, \aj, 150, 19, \dodoi{10.1088/0004-6256/150/1/19}

\bibitem[{{Law} {et~al.}(2016{\natexlab{a}}){Law}, {Cherinka}, {Yan},
  {Andrews}, {Bershady}, {Bizyaev}, {Blanc}, {Blanton}, {Bolton}, {Brownstein},
  {Bundy}, {Chen}, {Drory}, {D'Souza}, {Fu}, {Jones}, {Kauffmann}, {MacDonald},
  {Masters}, {Newman}, {Parejko}, {S{\'a}nchez-Gallego}, {S{\'a}nchez},
  {Schlegel}, {Thomas}, {Wake}, {Weijmans}, {Westfall}, \& {Zhang}}]{Law}
{Law}, D.~R., {Cherinka}, B., {Yan}, R., {et~al.} 2016{\natexlab{a}}, \aj, 152,
  83, \dodoi{10.3847/0004-6256/152/4/83}

\bibitem[{{Law} {et~al.}(2016{\natexlab{b}}){Law}, {Cherinka}, {Yan},
  {Andrews}, {Bershady}, {Bizyaev}, {Blanc}, {Blanton}, {Bolton}, {Brownstein},
  {Bundy}, {Chen}, {Drory}, {D'Souza}, {Fu}, {Jones}, {Kauffmann}, {MacDonald},
  {Masters}, {Newman}, {Parejko}, {S{\'a}nchez-Gallego}, {S{\'a}nchez},
  {Schlegel}, {Thomas}, {Wake}, {Weijmans}, {Westfall}, \&
  {Zhang}}]{2016AJ....152...83L}
---. 2016{\natexlab{b}}, \aj, 152, 83, \dodoi{10.3847/0004-6256/152/4/83}

\bibitem[{{Li} {et~al.}(2019){Li}, {French}, {Zabludoff}, \& {Ho}}]{li19}
{Li}, Z., {French}, K.~D., {Zabludoff}, A.~I., \& {Ho}, L.~C. 2019, \apj, 879,
  131, \dodoi{10.3847/1538-4357/ab1f68}

\bibitem[{{Liu} \& {Green}(1996)}]{lg96}
{Liu}, C.~T., \& {Green}, R.~F. 1996, \apjl, 458, L63, \dodoi{10.1086/309925}

\bibitem[{{Liu} {et~al.}(2007){Liu}, {Hooper}, {O'Neil}, {Thompson}, {Wolf}, \&
  {Lisker}}]{liu07}
{Liu}, C.~T., {Hooper}, E.~J., {O'Neil}, K., {et~al.} 2007, \apj, 658, 249,
  \dodoi{10.1086/511328}

\bibitem[{{Liu} \& {Kennicutt}(1995{\natexlab{a}})}]{lk95a}
{Liu}, C.~T., \& {Kennicutt}, Robert~C., J. 1995{\natexlab{a}}, \apjs, 100,
  325, \dodoi{10.1086/192222}

\bibitem[{{Liu} \& {Kennicutt}(1995{\natexlab{b}})}]{lk95b}
---. 1995{\natexlab{b}}, \apj, 450, 547, \dodoi{10.1086/176165}

\bibitem[{{Makarov} {et~al.}(2014){Makarov}, {Prugniel}, {Terekhova},
  {Courtois}, \& {Vauglin}}]{makarov}
{Makarov}, D., {Prugniel}, P., {Terekhova}, N., {Courtois}, H., \& {Vauglin},
  I. 2014, \aap, 570, A13, \dodoi{10.1051/0004-6361/201423496}

\bibitem[{{Marinelli} {et~al.}(2020){Marinelli}, {Greene}, {Riffel},
  {Rowlands}, \& {Liu}}]{marinelli}
{Marinelli}, M., {Greene}, O., {Riffel}, R.~A., {Rowlands}, K., \& {Liu}, C.~T.
  2020, Research Notes of the American Astronomical Society, 4, 110,
  \dodoi{10.3847/2515-5172/aba595}

\bibitem[{{Mendel} {et~al.}(2014){Mendel}, {Simard}, {Palmer}, {Ellison}, \&
  {Patton}}]{mendel}
{Mendel}, J.~T., {Simard}, L., {Palmer}, M., {Ellison}, S.~L., \& {Patton},
  D.~R. 2014, \apjs, 210, 3, \dodoi{10.1088/0067-0049/210/1/3}

\bibitem[{{Meusinger} {et~al.}(2016){Meusinger}, {Bruenecke}, {Schalldach}, \&
  {in der}}]{Meusinger}
{Meusinger}, H., {Bruenecke}, J., {Schalldach}, P., \& {in der}, A.~A. 2016,
  VizieR Online Data Catalog, 359

\bibitem[{{Mockler} {et~al.}(2019){Mockler}, {Guillochon}, \&
  {Ramirez-Ruiz}}]{mockler}
{Mockler}, B., {Guillochon}, J., \& {Ramirez-Ruiz}, E. 2019, \apj, 872, 151,
  \dodoi{10.3847/1538-4357/ab010f}

\bibitem[{{Newberry} {et~al.}(1990){Newberry}, {Boroson}, \&
  {Kirshner}}]{newberry90}
{Newberry}, M.~V., {Boroson}, T.~A., \& {Kirshner}, R.~P. 1990, \apj, 350, 585,
  \dodoi{10.1086/168413}

\bibitem[{{Nielsen} {et~al.}(2012){Nielsen}, {Ridgway}, {De Propris}, \&
  {Goto}}]{nielsen12}
{Nielsen}, D.~M., {Ridgway}, S.~E., {De Propris}, R., \& {Goto}, T. 2012,
  \apjl, 761, L16, \dodoi{10.1088/2041-8205/761/2/L16}

\bibitem[{{Oegerle} {et~al.}(1991){Oegerle}, {Hill}, \& {Hoessel}}]{ohh91}
{Oegerle}, W.~R., {Hill}, J.~M., \& {Hoessel}, J.~G. 1991, \apj, 381, L9,
  \dodoi{10.1086/186184}

\bibitem[{{Pawlik} {et~al.}(2018){Pawlik}, {Taj Aldeen}, {Wild},
  {Mendez-Abreu}, {Lah{\'e}n}, {Johansson}, {Jimenez}, {Lucas}, {Zheng},
  {Walcher}, \& {Rowlands}}]{pawlik18}
{Pawlik}, M.~M., {Taj Aldeen}, L., {Wild}, V., {et~al.} 2018, \mnras, 477,
  1708, \dodoi{10.1093/mnras/sty589}

\bibitem[{Penny {et~al.}(2017)Penny, Masters, Smethurst, Nichol, Krawczyk,
  Bizyaev, Greene, Liu, Marinelli, Rembold, Riffel, da~Silva~Ilha, Wylezalek,
  Andrews, Bundy, Drory, Oravetz, \& Pan}]{penny}
Penny, S.~J., Masters, K.~L., Smethurst, R., {et~al.} 2017,
  \dodoi{10.1093/mnras/sty202}

\bibitem[{{Pracy} {et~al.}(2014){Pracy}, {Owers}, {Zwaan}, {Couch},
  {Kuntschner}, {Croom}, \& {Sadler}}]{pracy14}
{Pracy}, M.~B., {Owers}, M.~S., {Zwaan}, M., {et~al.} 2014, \mnras, 443, 388,
  \dodoi{10.1093/mnras/stu1103}

\bibitem[{{Pracy} {et~al.}(2013){Pracy}, {Croom}, {Sadler}, {Couch},
  {Kuntschner}, {Bekki}, {Owers}, {Zwaan}, {Turner}, \& {Bergmann}}]{pracy13}
{Pracy}, M.~B., {Croom}, S., {Sadler}, E., {et~al.} 2013, \mnras, 432, 3131,
  \dodoi{10.1093/mnras/stt666}

\bibitem[{{Quillen} {et~al.}(1999){Quillen}, {Rieke}, {Rieke}, {Caldwell}, \&
  {Engelbracht}}]{quillen99}
{Quillen}, A.~C., {Rieke}, G.~H., {Rieke}, M.~J., {Caldwell}, N., \&
  {Engelbracht}, C.~W. 1999, \apj, 518, 632, \dodoi{10.1086/307307}

\bibitem[{{Schawinski} {et~al.}(2014){Schawinski}, {Urry}, {Simmons},
  {Fortson}, {Kaviraj}, {Keel}, {Lintott}, {Masters}, {Nichol}, {Sarzi},
  {Skibba}, {Treister}, {Willett}, {Wong}, \& {Yi}}]{sch}
{Schawinski}, K., {Urry}, C.~M., {Simmons}, B.~D., {et~al.} 2014, \mnras, 440,
  889, \dodoi{10.1093/mnras/stu327}

\bibitem[{{Sharples} {et~al.}(1985){Sharples}, {Ellis}, {Couch}, \&
  {Gray}}]{sharples85}
{Sharples}, R.~M., {Ellis}, R.~S., {Couch}, W.~J., \& {Gray}, P.~M. 1985,
  \mnras, 212, 687, \dodoi{10.1093/mnras/212.3.687}

\bibitem[{{Smee} {et~al.}(2013){Smee}, {Gunn}, {Uomoto}, {Roe}, {Schlegel},
  {Rockosi}, {Carr}, {Leger}, {Dawson}, {Olmstead}, {Brinkmann}, {Owen},
  {Barkhouser}, {Honscheid}, {Harding}, {Long}, {Lupton}, {Loomis}, {Anderson},
  {Annis}, {Bernardi}, {Bhardwaj}, {Bizyaev}, {Bolton}, {Brewington}, {Briggs},
  {Burles}, {Burns}, {Castander}, {Connolly}, {Davenport}, {Ebelke}, {Epps},
  {Feldman}, {Friedman}, {Frieman}, {Heckman}, {Hull}, {Knapp}, {Lawrence},
  {Loveday}, {Mannery}, {Malanushenko}, {Malanushenko}, {Merrelli}, {Muna},
  {Newman}, {Nichol}, {Oravetz}, {Pan}, {Pope}, {Ricketts}, {Shelden},
  {Sandford}, {Siegmund}, {Simmons}, {Smith}, {Snedden}, {Schneider},
  {SubbaRao}, {Tremonti}, {Waddell}, \& {York}}]{Smee}
{Smee}, S.~A., {Gunn}, J.~E., {Uomoto}, A., {et~al.} 2013, \aj, 146, 32,
  \dodoi{10.1088/0004-6256/146/2/32}

\bibitem[{{Smercina} {et~al.}(2018){Smercina}, {Smith}, {Dale}, {French},
  {Croxall}, {Zhukovska}, {Togi}, {Bell}, {Crocker}, {Draine}, {Jarrett},
  {Tremonti}, {Yang}, \& {Zabludoff}}]{smercina18}
{Smercina}, A., {Smith}, J.~D.~T., {Dale}, D.~A., {et~al.} 2018, \apj, 855, 51,
  \dodoi{10.3847/1538-4357/aaafcd}

\bibitem[{{Smethurst} {et~al.}(2015){Smethurst}, {Lintott}, {Simmons},
  {Schawinski}, {Marshall}, {Bamford}, {Fortson}, {Kaviraj}, {Masters},
  {Melvin}, {Nichol}, {Skibba}, \& {Willett}}]{smethurst}
{Smethurst}, R.~J., {Lintott}, C.~J., {Simmons}, B.~D., {et~al.} 2015, \mnras,
  450, 435, \dodoi{10.1093/mnras/stv161}

\bibitem[{{Wake} {et~al.}(2017){Wake}, {Bundy}, {Diamond-Stanic}, {Yan},
  {Blanton}, {Bershady}, {S{\'a}nchez-Gallego}, {Drory}, {Jones}, {Kauffmann},
  {Law}, {Li}, {MacDonald}, {Masters}, {Thomas}, {Tinker}, {Weijmans}, \&
  {Brownstein}}]{2017AJ....154...86W}
{Wake}, D.~A., {Bundy}, K., {Diamond-Stanic}, A.~M., {et~al.} 2017, \aj, 154,
  86, \dodoi{10.3847/1538-3881/aa7ecc}

\bibitem[{{Wei} {et~al.}(2018){Wei}, {Gu}, {Brotherton}, {Shi}, \&
  {Chen}}]{wei18}
{Wei}, P., {Gu}, Y., {Brotherton}, M.~S., {Shi}, Y., \& {Chen}, Y. 2018, \apj,
  857, 27, \dodoi{10.3847/1538-4357/aab499}

\bibitem[{{Westfall} {et~al.}(2019){Westfall}, {Cappellari}, {Bershady},
  {Bundy}, {Belfiore}, {Ji}, {Law}, {Schaefer}, {Shetty}, {Tremonti}, {Yan},
  {Andrews}, {Brownstein}, {Cherinka}, {Coccato}, {Drory}, {Maraston},
  {Parikh}, {S{\'a}nchez-Gallego}, {Thomas}, {Weijmans}, {Barrera-Ballesteros},
  {Du}, {Goddard}, {Li}, {Masters}, {Ibarra Medel}, {S{\'a}nchez}, {Yang},
  {Zheng}, \& {Zhou}}]{westfall}
{Westfall}, K.~B., {Cappellari}, M., {Bershady}, M.~A., {et~al.} 2019, \aj,
  158, 231, \dodoi{10.3847/1538-3881/ab44a2}

\bibitem[{Wild {et~al.}(2009)Wild, JakobWalcher, Johansson, Tresse, Charlot,
  Pollo, F\`{o}vre, \& de~Ravel}]{Wild2009}
Wild, V., JakobWalcher, C., Johansson, P.~H., {et~al.} 2009, {M}onthly
  {N}otices of the {R}oyal {A}stronomical {S}ociety ({MNRAS}), 395, 144–159

\bibitem[{{Wilkinson} {et~al.}(2017){Wilkinson}, {Pimbblet}, \&
  {Stott}}]{cwilkinson}
{Wilkinson}, C.~L., {Pimbblet}, K.~A., \& {Stott}, J.~P. 2017, \mnras, 472,
  1447, \dodoi{10.1093/mnras/stx2034}

\bibitem[{Wilkinson {et~al.}(2015)Wilkinson, Maraston, Thomas, Coccato,
  Tojeiro, Cappellari, Belfiore, Bershady, Blanton, \& Bundy}]{Wilkinson}
Wilkinson, D.~M., Maraston, C., Thomas, D., {et~al.} 2015, {M}onthly {N}otices
  of the {R}oyal {A}stronomical {S}ociety ({MNRAS}), 449, 328

\bibitem[{{Yamauchi} {et~al.}(2008){Yamauchi}, {Yagi}, \& {Goto}}]{yamauchi08}
{Yamauchi}, C., {Yagi}, M., \& {Goto}, T. 2008, \mnras, 390, 383,
  \dodoi{10.1111/j.1365-2966.2008.13756.x}

\bibitem[{Yan {et~al.}(2015)Yan, Sánchez-Gallego, Zhang, Bundy, Bershady, Law,
  Drory, MacDonald, Wake, Cherinka, Weijmans, Thomas, \&
  Tremonti}]{Yanoverview}
Yan, R., Sánchez-Gallego, J.~R., Zhang, K., {et~al.} 2015, Physics and
  Astronomy Faculty Publications, 332

\bibitem[{{Yan} {et~al.}(2016{\natexlab{a}}){Yan}, {Tremonti}, {Bershady},
  {Law}, {Schlegel}, {Bundy}, {Drory}, {MacDonald}, {Bizyaev}, {Blanc},
  {Blanton}, {Cherinka}, {Eigenbrot}, {Gunn}, {Harding}, {Hogg},
  {S{\'a}nchez-Gallego}, {S{\'a}nchez}, {Wake}, {Weijmans}, {Xiao}, \&
  {Zhang}}]{2016AJ....151....8Y}
{Yan}, R., {Tremonti}, C., {Bershady}, M.~A., {et~al.} 2016{\natexlab{a}}, \aj,
  151, 8, \dodoi{10.3847/0004-6256/151/1/8}

\bibitem[{{Yan} {et~al.}(2016{\natexlab{b}}){Yan}, {Bundy}, {Law}, {Bershady},
  {Andrews}, {Cherinka}, {Diamond-Stanic}, {Drory}, {MacDonald},
  {S{\'a}nchez-Gallego}, {Thomas}, {Wake}, {Weijmans}, {Westfall}, {Zhang},
  {Arag{\'o}n-Salamanca}, {Belfiore}, {Bizyaev}, {Blanc}, {Blanton},
  {Brownstein}, {Cappellari}, {D'Souza}, {Emsellem}, {Fu}, {Gaulme}, {Graham},
  {Goddard}, {Gunn}, {Harding}, {Jones}, {Kinemuchi}, {Li}, {Li}, {Maiolino},
  {Mao}, {Maraston}, {Masters}, {Merrifield}, {Oravetz}, {Pan}, {Parejko},
  {Sanchez}, {Schlegel}, {Simmons}, {Thanjavur}, {Tinker}, {Tremonti}, {van den
  Bosch}, \& {Zheng}}]{2016AJ....152..197Y}
{Yan}, R., {Bundy}, K., {Law}, D.~R., {et~al.} 2016{\natexlab{b}}, \aj, 152,
  197, \dodoi{10.3847/0004-6256/152/6/197}

\bibitem[{{Yan} {et~al.}(2016{\natexlab{c}}){Yan}, {Tremonti}, {Bershady},
  {Law}, {Schlegel}, {Bundy}, {Drory}, {MacDonald}, {Bizyaev}, {Blanc},
  {Blanton}, {Cherinka}, {Eigenbrot}, {Gunn}, {Harding}, {Hogg},
  {S{\'a}nchez-Gallego}, {S{\'a}nchez}, {Wake}, {Weijmans}, {Xiao}, \&
  {Zhang}}]{Yan2016}
{Yan}, R., {Tremonti}, C., {Bershady}, M.~A., {et~al.} 2016{\natexlab{c}}, \aj,
  151, 8, \dodoi{10.3847/0004-6256/151/1/8}

\bibitem[{Yang {et~al.}(2015)Yang, Luo, Chen, Hou, Zhang, Du, Zhang, Cai, Guo,
  Zhang, Zhao, Wu, Wang, Shen, Yang, Zhang, \& Hou}]{yang}
Yang, H.-F., Luo, A.-L., Chen, X.-Y., {et~al.} 2015, {R}esearch in {A}stronomy
  and {A}strophysics ({RAA}), 15, 1414–1423

\bibitem[{{Zabludoff} {et~al.}(1996){Zabludoff}, {Zaritsky}, {Lin}, {Tucker},
  {Hashimoto}, {Shectman}, {Oemler}, \& {Kirshner}}]{zab96}
{Zabludoff}, A.~I., {Zaritsky}, D., {Lin}, H., {et~al.} 1996, \apj, 466, 104,
  \dodoi{10.1086/177495}

\bibitem[{{Zwaan} {et~al.}(2013){Zwaan}, {Kuntschner}, {Pracy}, \&
  {Couch}}]{zwaan13}
{Zwaan}, M.~A., {Kuntschner}, H., {Pracy}, M.~B., \& {Couch}, W.~J. 2013,
  \mnras, 432, 492, \dodoi{10.1093/mnras/stt496}

\end{thebibliography}
\bibliographystyle{aasjournal}
\clearpage


\appendix \label{appendix}
 
\bigskip
\bigskip   

\section{SPECTRAL SYNTHESIS DIAGRAMS}


The following appendix provides, for each galaxy in our sample, the visual image gathered from Marvin, the single-fiber spectrum from SDSS, and the diagnostic and spatially resolved WHAN diagrams and normalized summed spaxel spectrum created by this team-- each identified by the galaxy's MaNGA ID and PLATE-IFU. We would also like to note how that we are, for a future paper,  investigating the likelihood that the ionization ratios present in the WHAN diagrams of the galaxies are produced by shocks or AGN.

\bigskip
\bigskip
\vfill\eject

\begin{center}[\textbf{MaNGA ID: 12-98126 | PLATE-IFU: 7443-12701}]
\end{center}
\includegraphics[height = 0.197\textwidth]{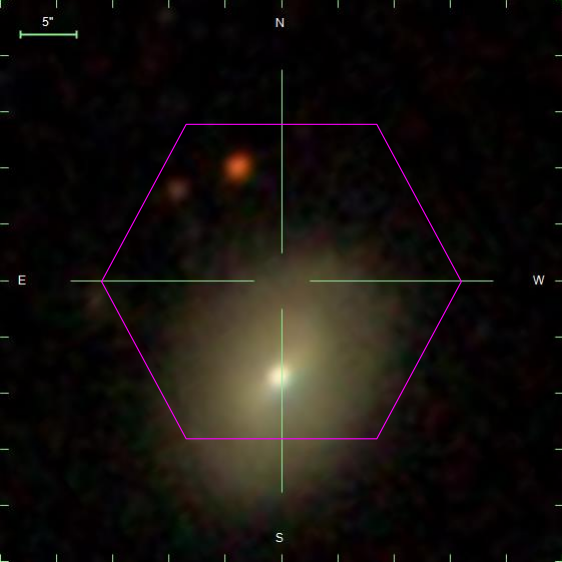}
\includegraphics[height = 0.197\textwidth]{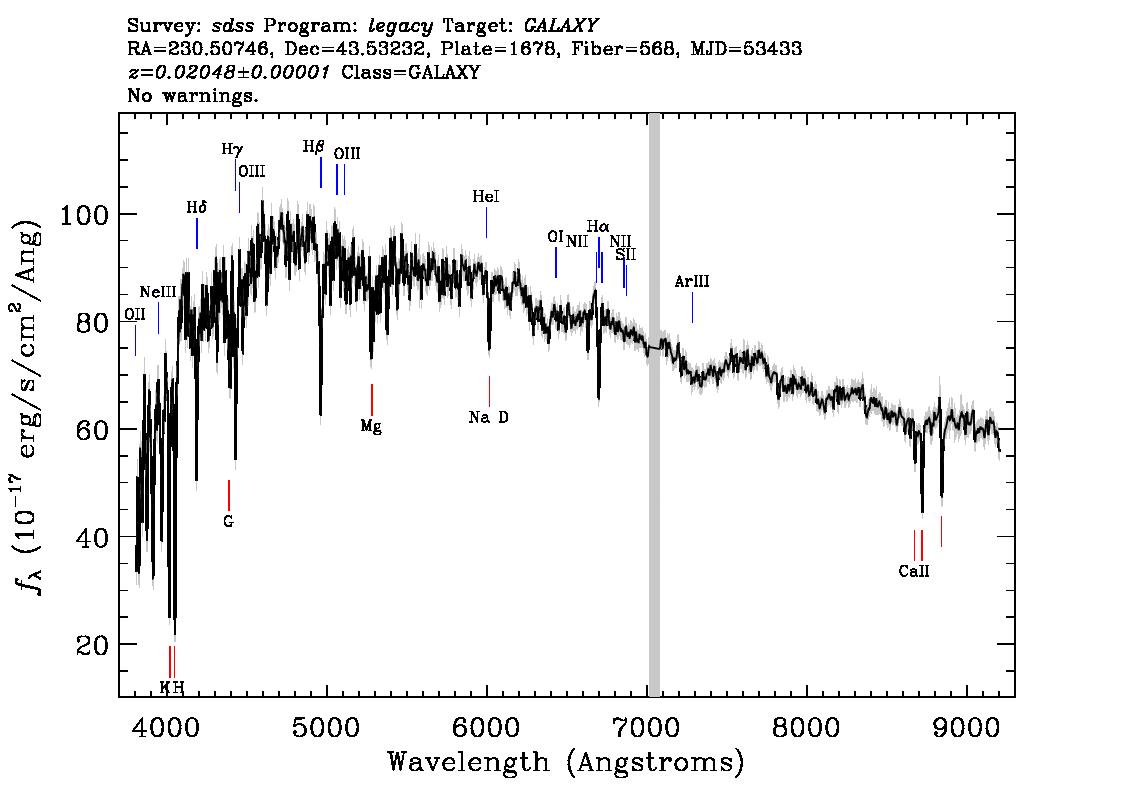}  \includegraphics[height = 0.197\textwidth]{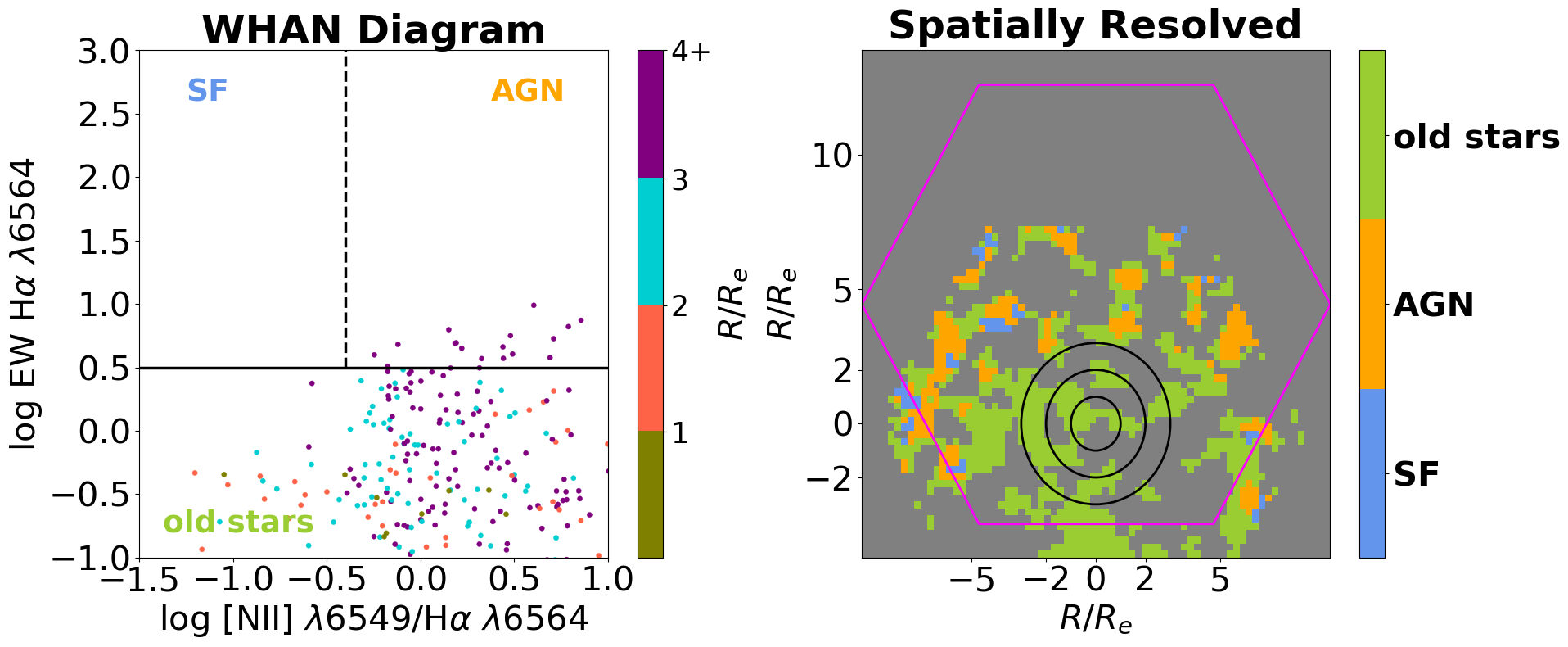} 
\bigskip
\ \includegraphics[height = 0.29\textwidth]{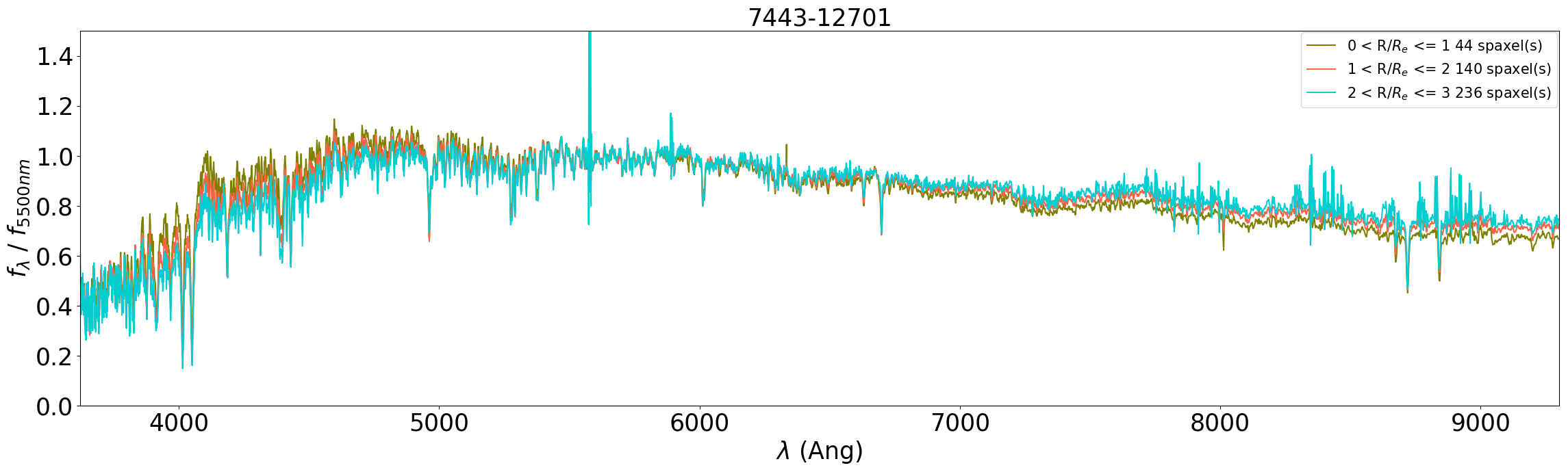}

\bigskip

\begin{center}[\textbf{MaNGA ID: 12-49536 | PLATE-IFU: 7443-1902}]
\end{center}
\includegraphics[height = 0.197\textwidth]{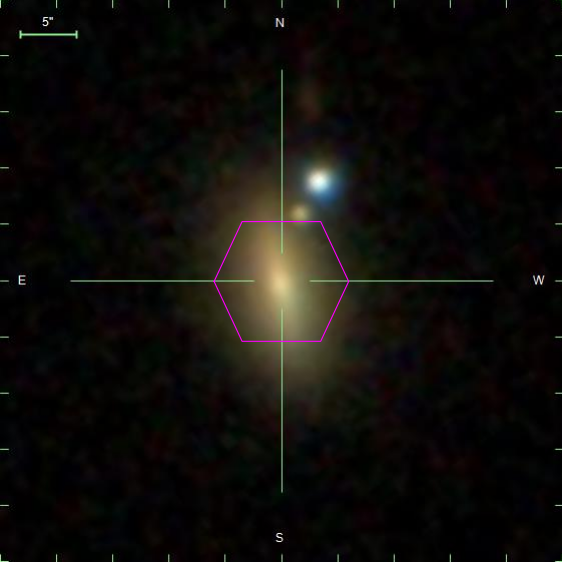}
\includegraphics[height = 0.197\textwidth]{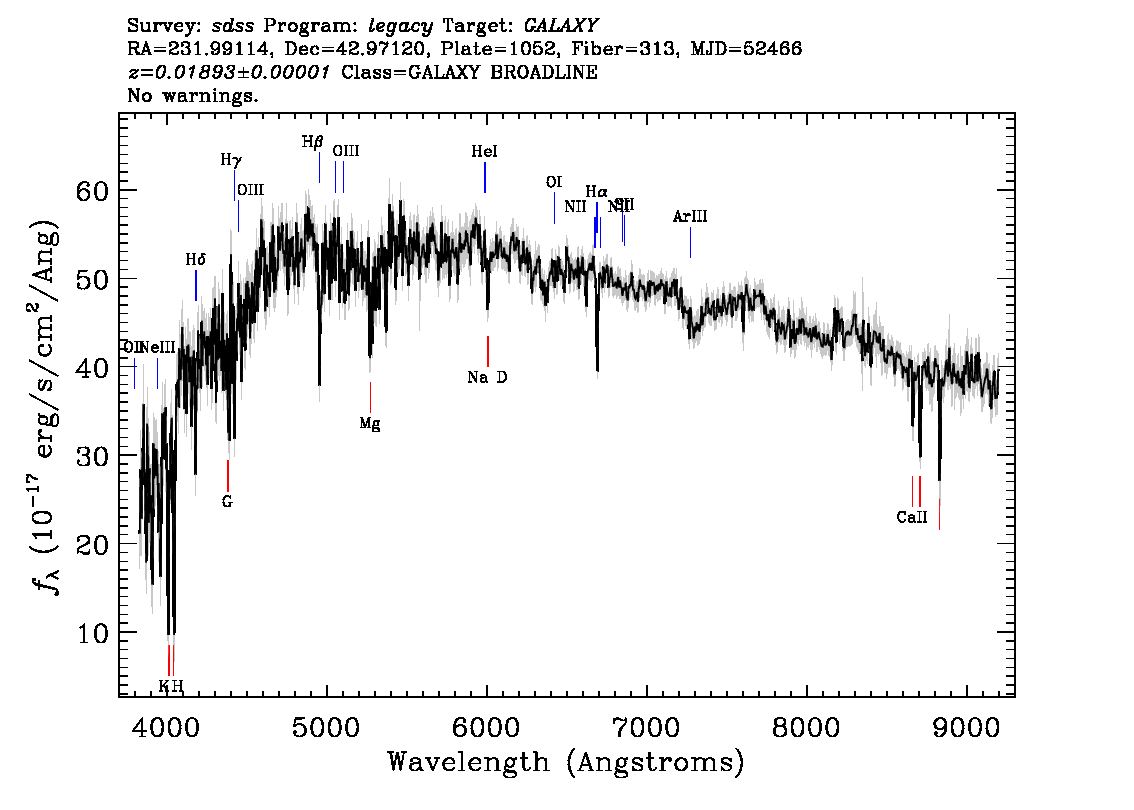}  \includegraphics[height = 0.197\textwidth]{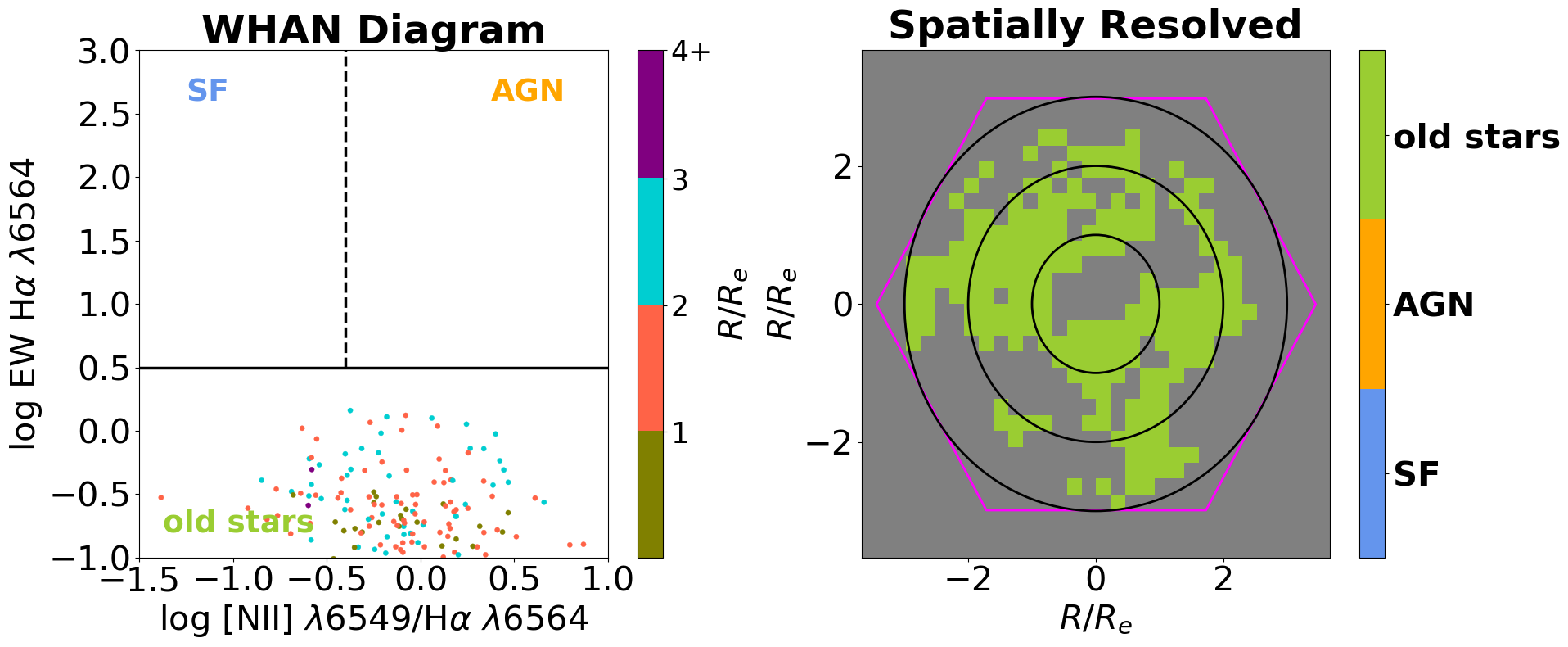} 
\bigskip
\bigskip
\ \includegraphics[height = 0.29\textwidth]{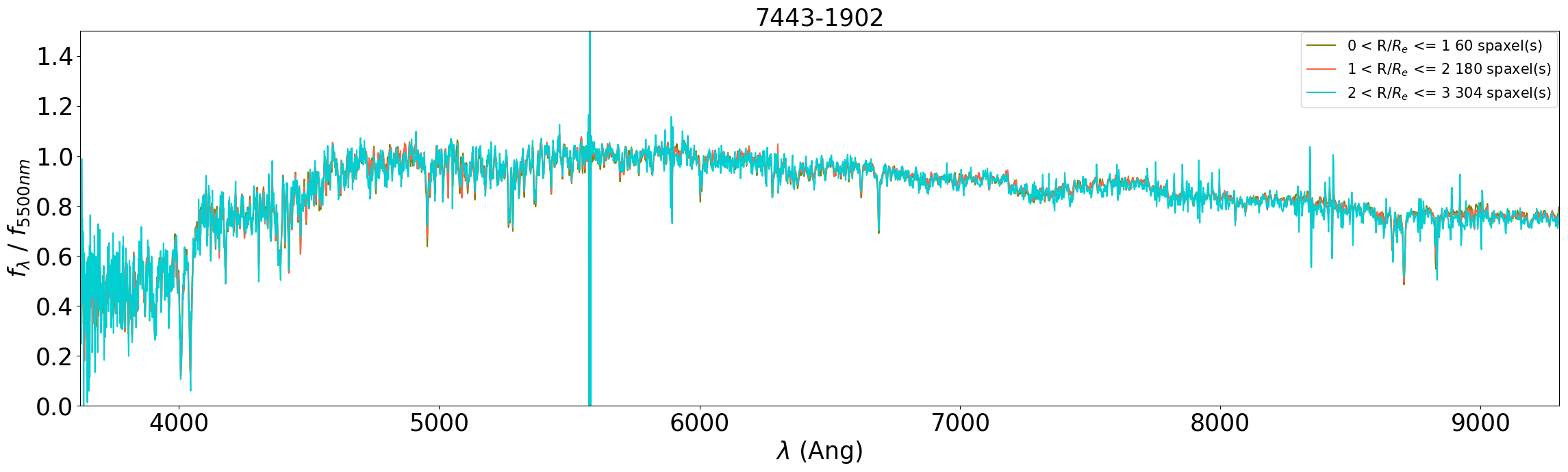}

\bigskip
\vfill\eject

\begin{center}[\textbf{MaNGA ID: 1-24124 | PLATE-IFU: 7991-3703}]
\end{center}
\includegraphics[height = 0.197\textwidth]{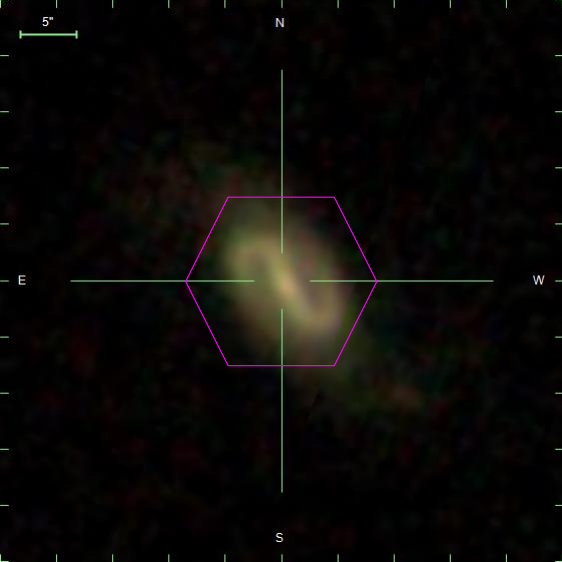}
\includegraphics[height = 0.197\textwidth]{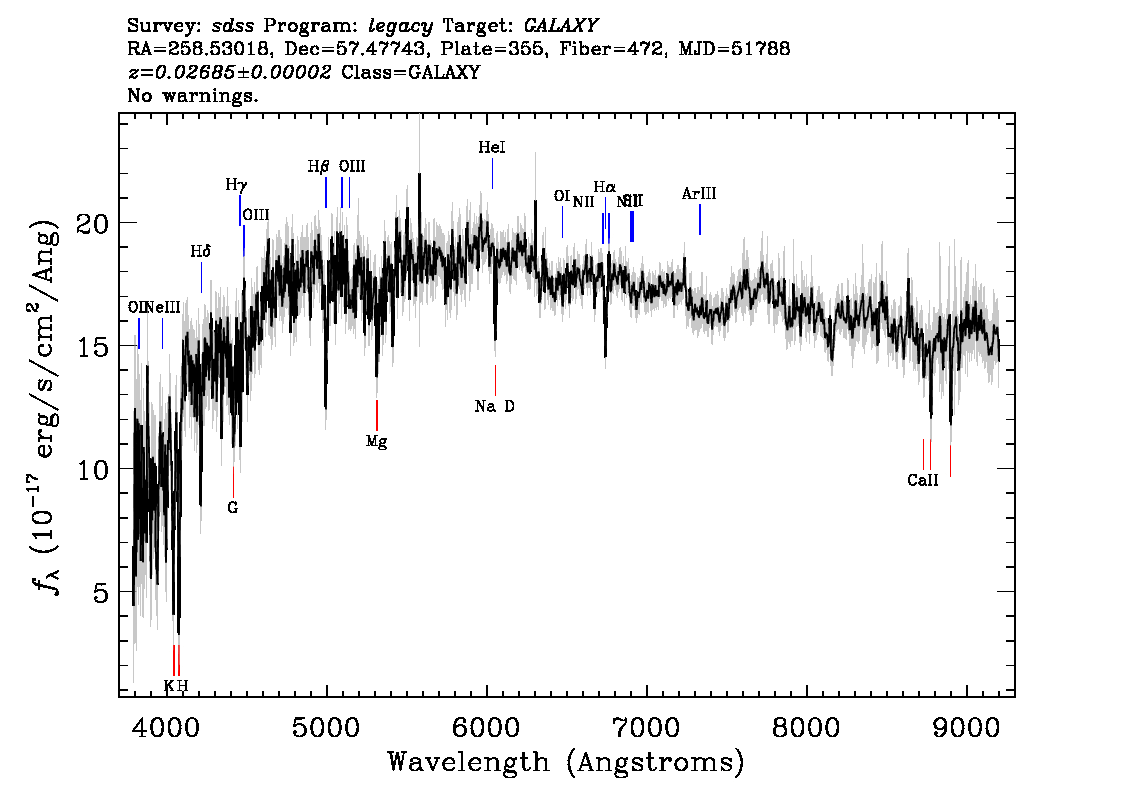}  \includegraphics[height = 0.197\textwidth]{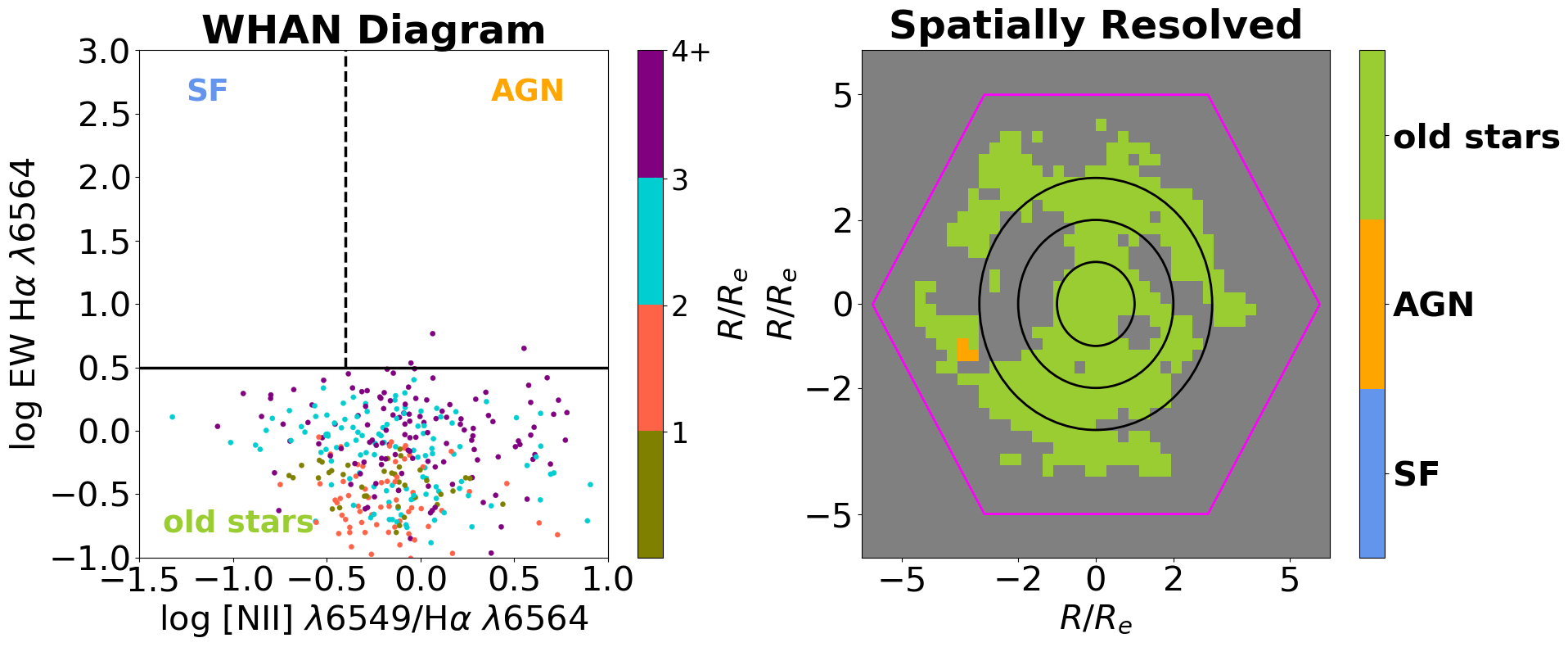} 
\bigskip
\bigskip
\ \includegraphics[height = 0.29\textwidth]{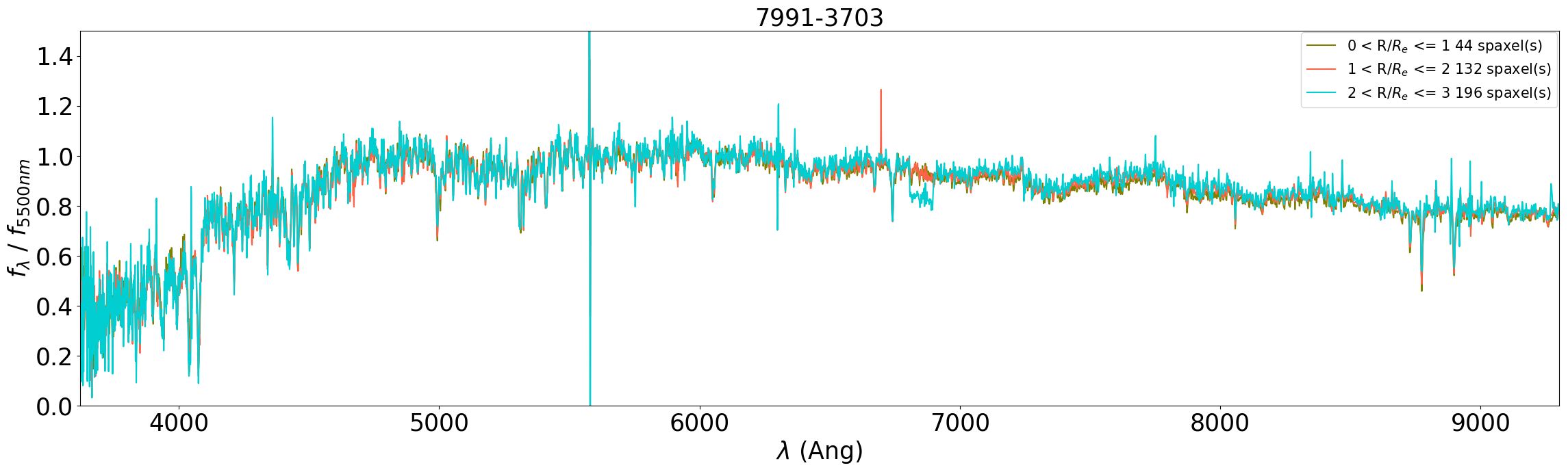}

\bigskip
\bigskip
\bigskip
\bigskip

\begin{center}[\textbf{MaNGA ID: 1-37034 | PLATE-IFU: 8077-1901}]
\end{center}
\includegraphics[height = 0.197\textwidth]{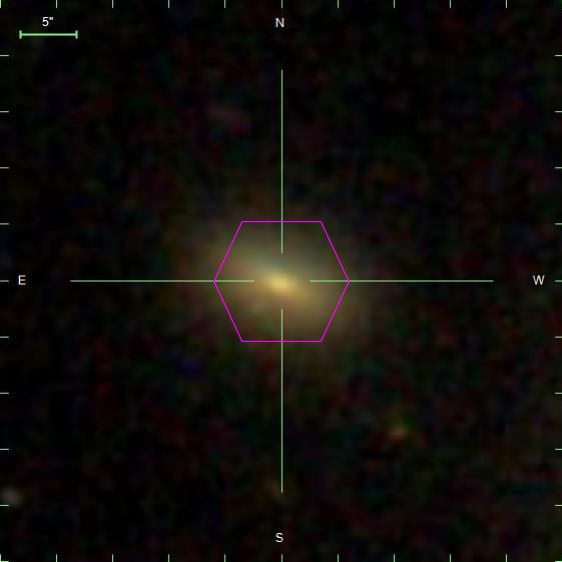}
\includegraphics[height = 0.197\textwidth]{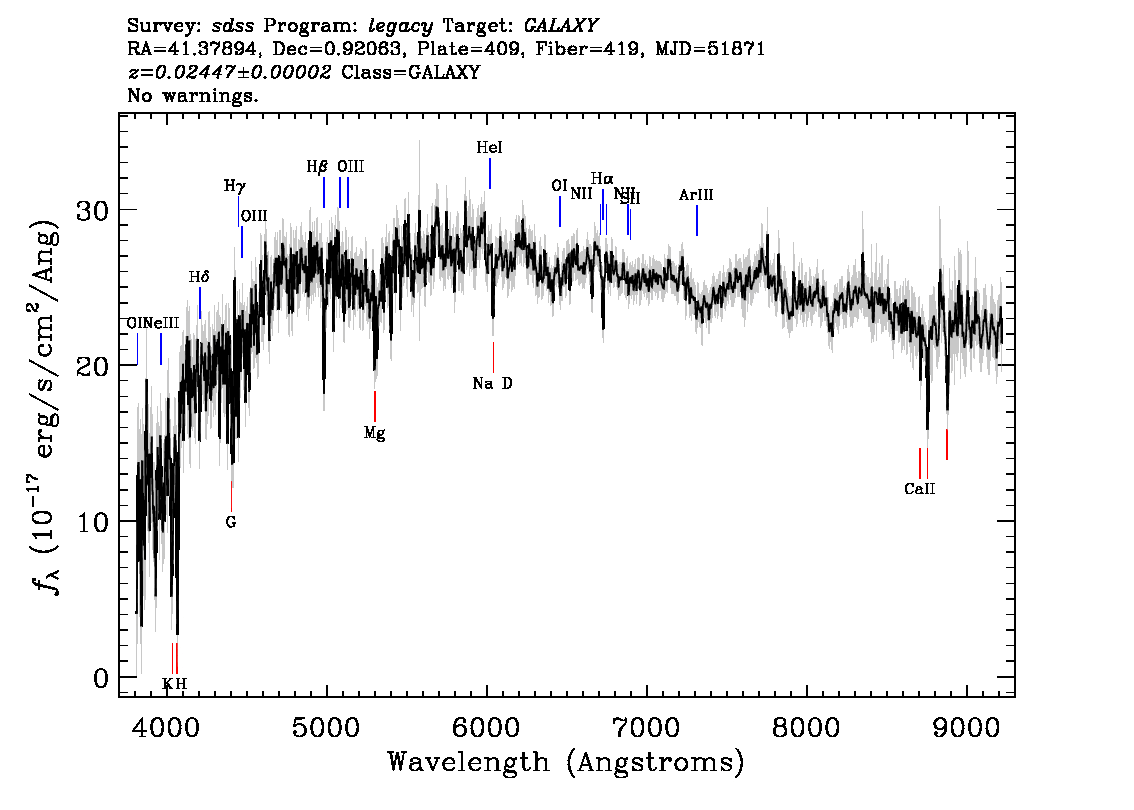}  \includegraphics[height = 0.197\textwidth]{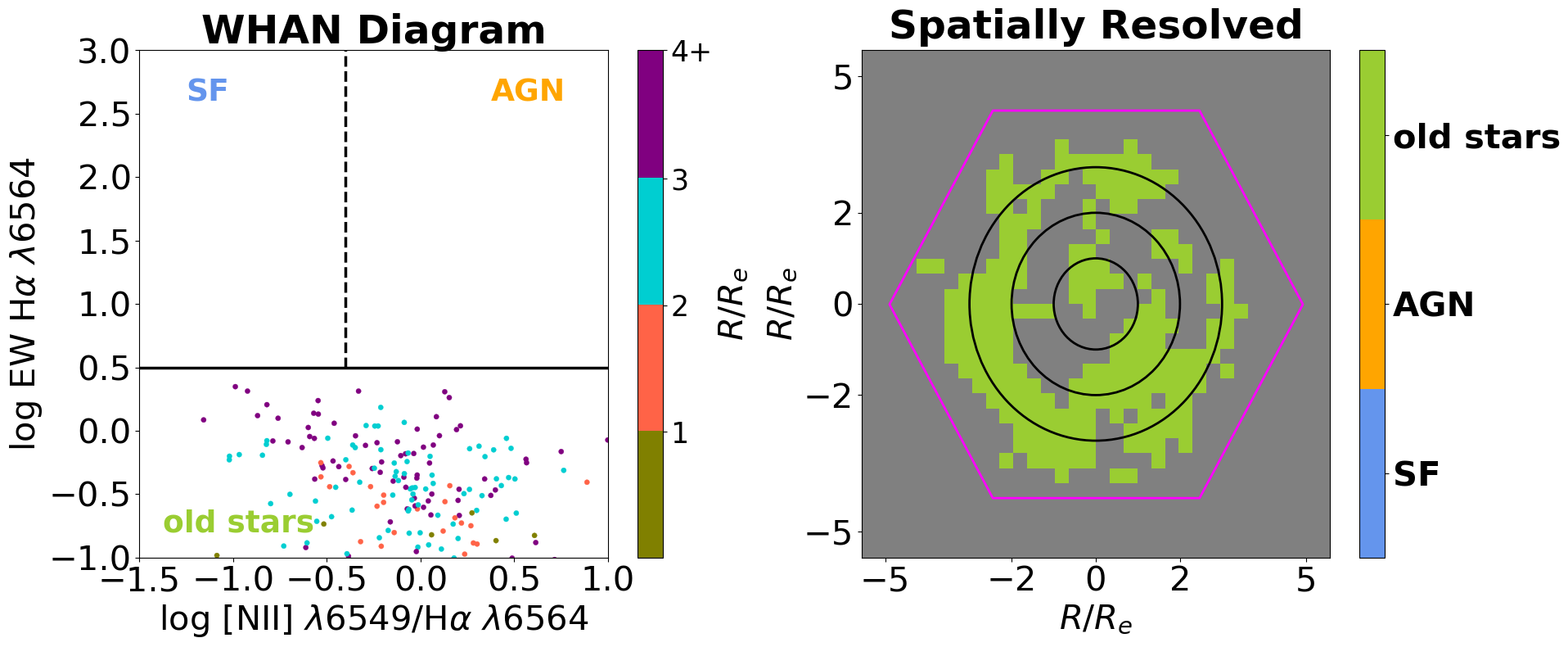} 
\bigskip
\bigskip
\ \includegraphics[height = 0.29\textwidth]{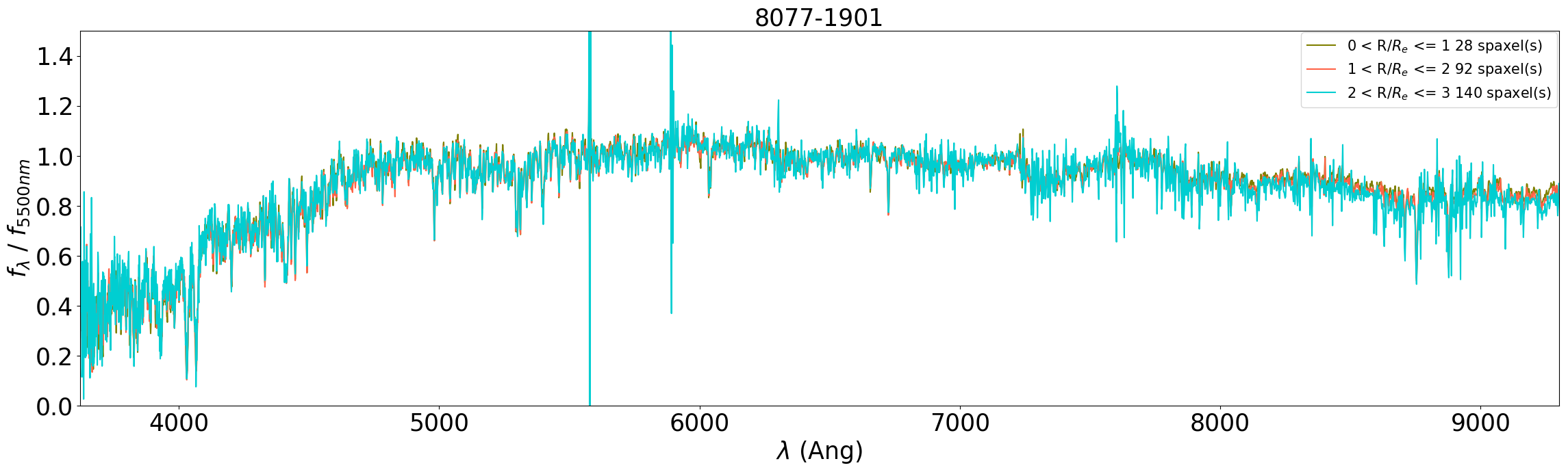}

\bigskip

\vfill\eject
\begin{center}[\textbf{MaNGA ID: 1-109112 | PLATE-IFU: 8078-1901}]
\end{center}
\includegraphics[height = 0.197\textwidth]{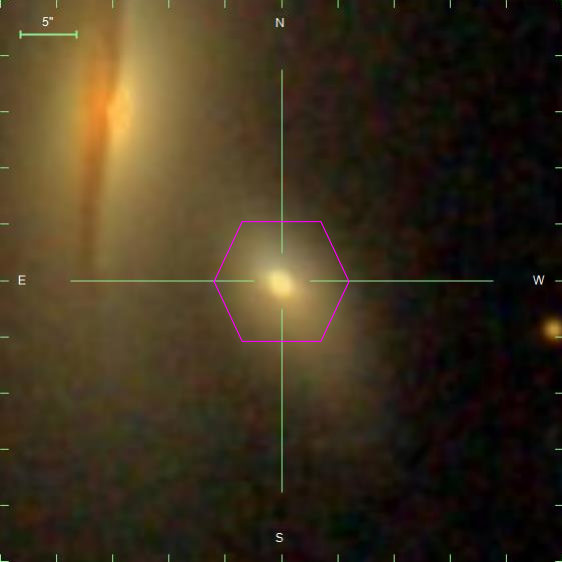}
\includegraphics[height = 0.197\textwidth]{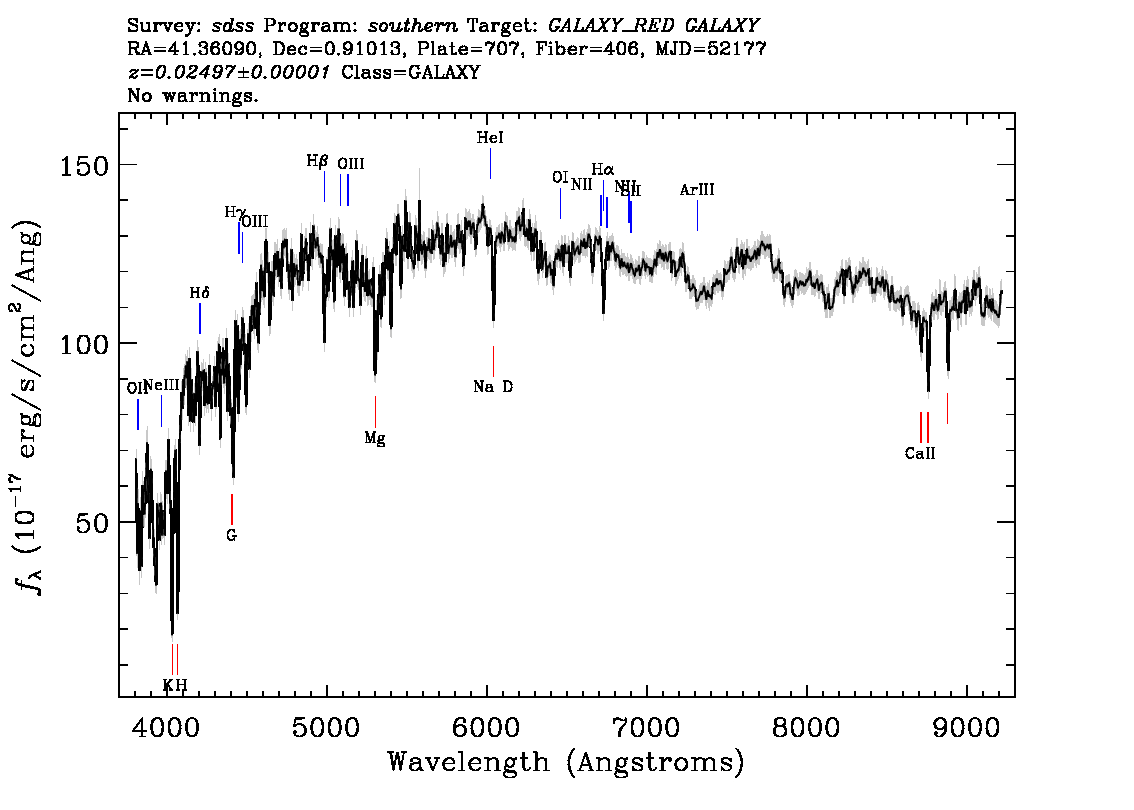}  \includegraphics[height = 0.197\textwidth]{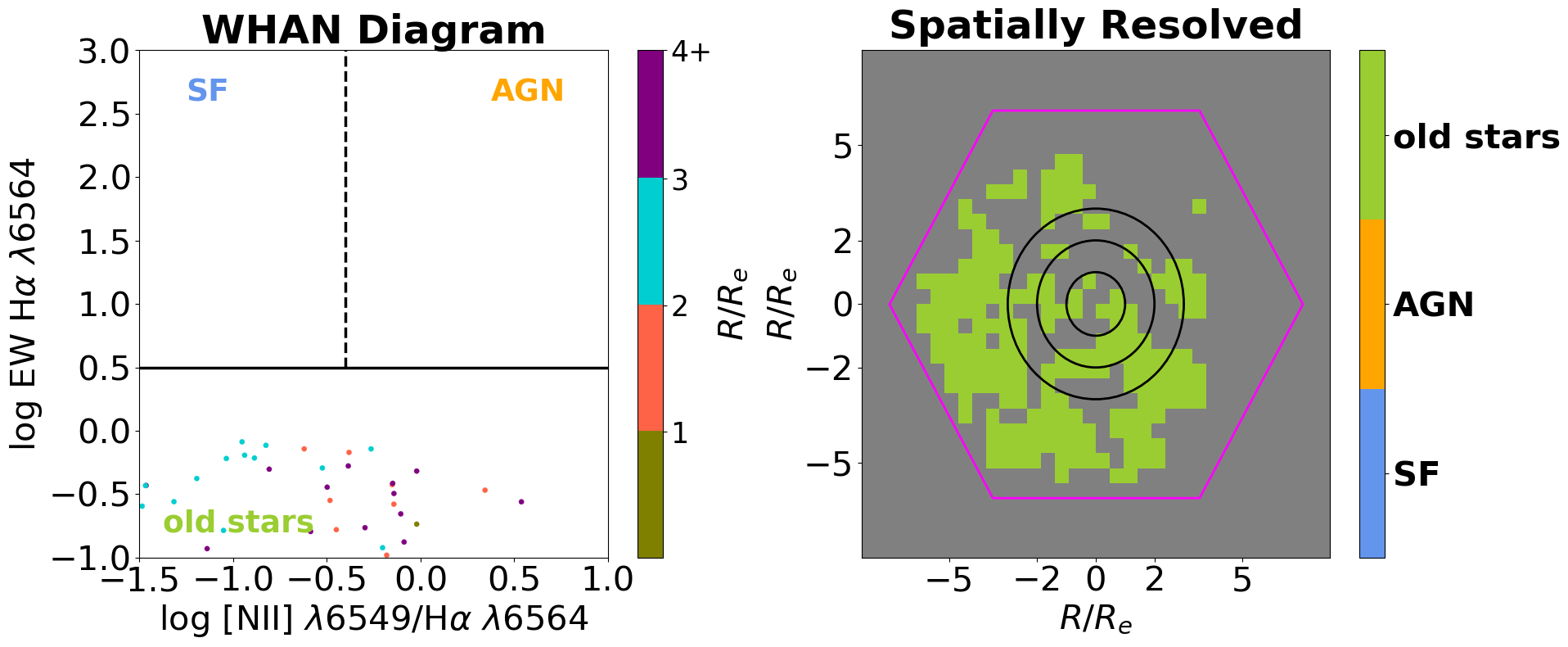} 
\bigskip
\bigskip
\ \includegraphics[height = 0.29\textwidth]{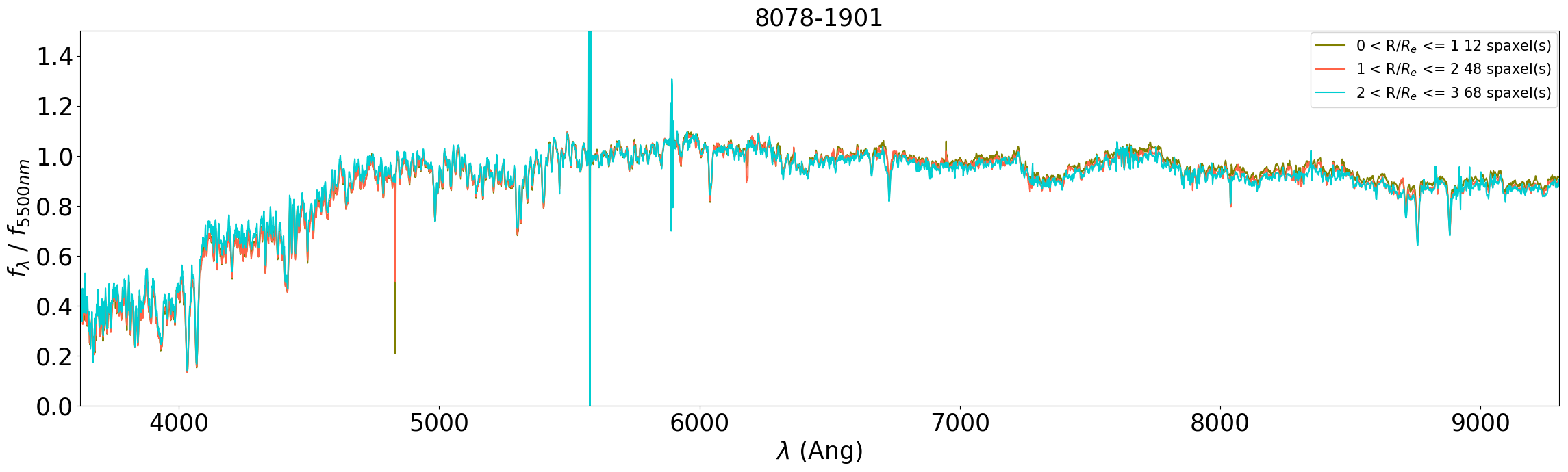}

\bigskip
\bigskip
\bigskip
\bigskip

\begin{center}[\textbf{MaNGA ID: 1-38374 | PLATE-IFU: 8082-3704}]
\end{center}
\includegraphics[height = 0.197\textwidth]{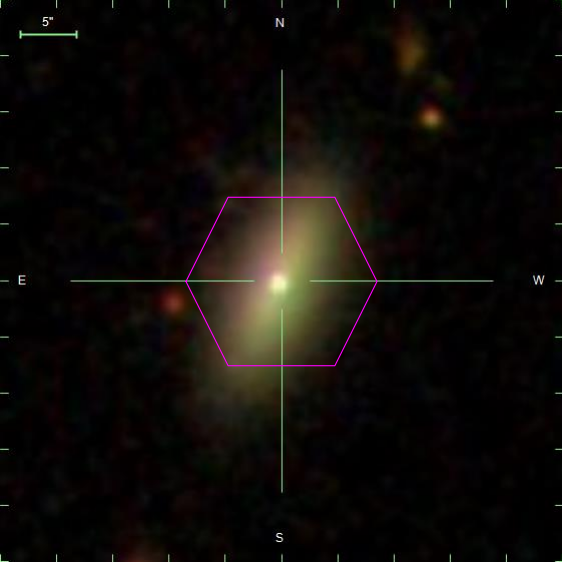}
\includegraphics[height = 0.197\textwidth]{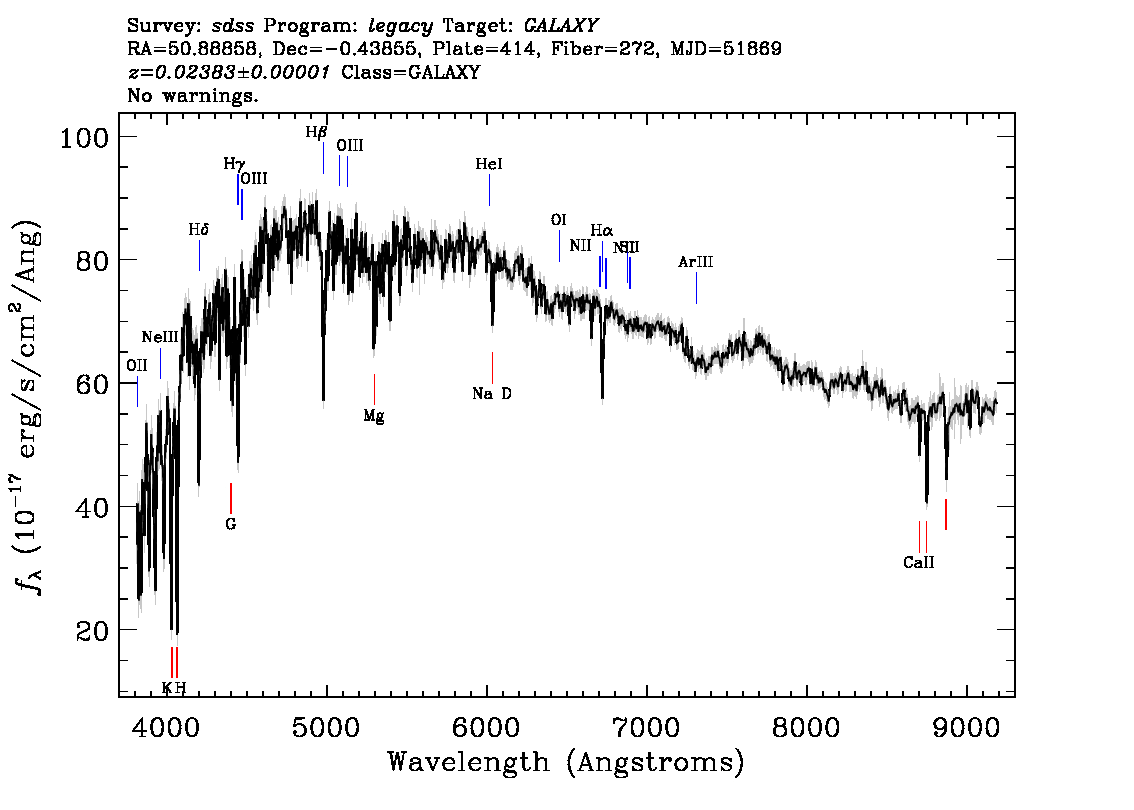}  \includegraphics[height = 0.197\textwidth]{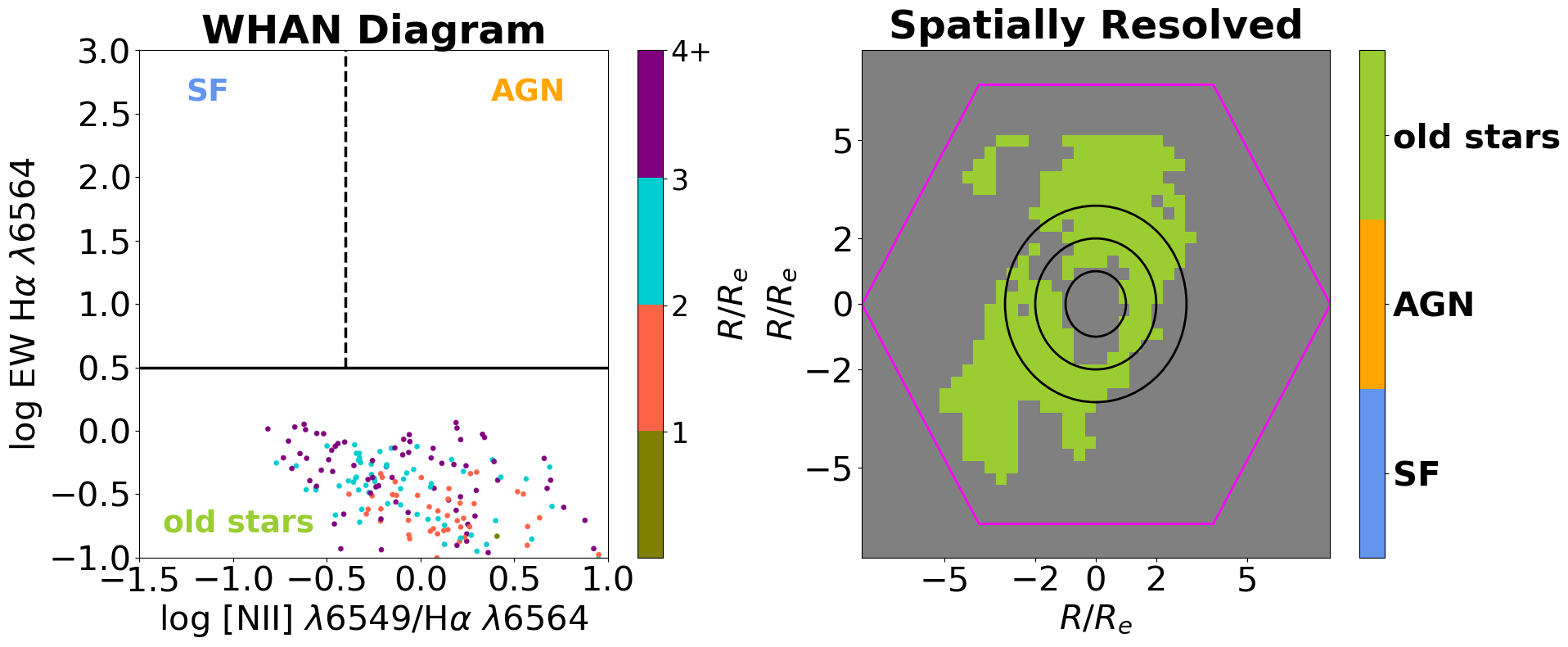}
\bigskip
\bigskip
\ \includegraphics[height = 0.29\textwidth]{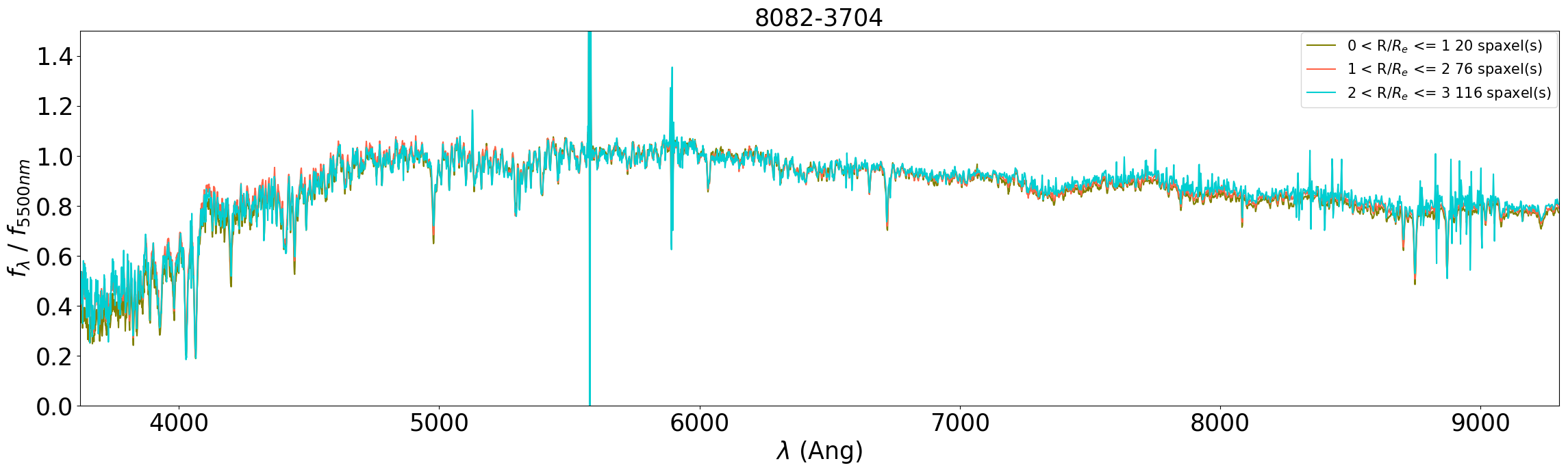}

\bigskip

\vfill\eject
\begin{center}[\textbf{MaNGA ID: 1-201180 | PLATE-IFU: 8145-6102}]
\end{center}
\includegraphics[height = 0.197\textwidth]{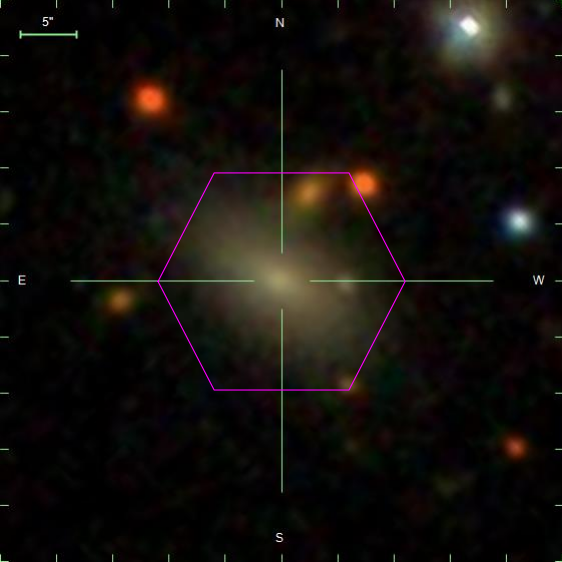}
\includegraphics[height = 0.197\textwidth]{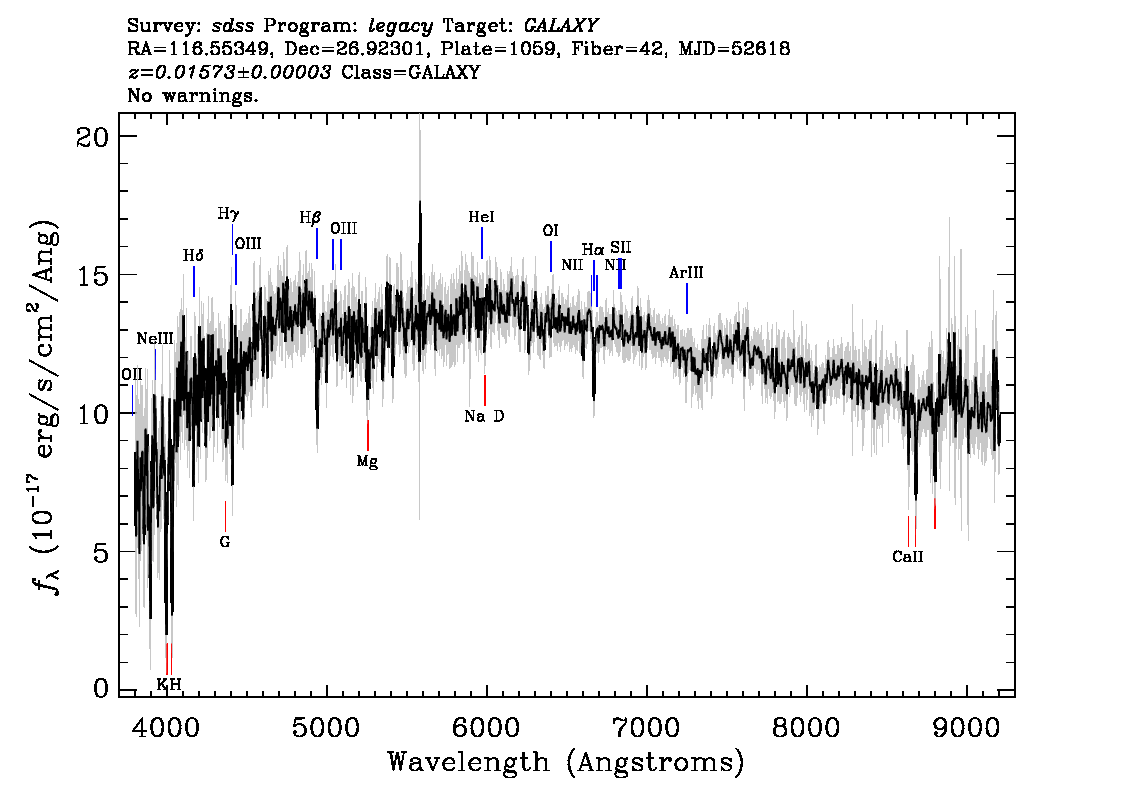}  \includegraphics[height = 0.197\textwidth]{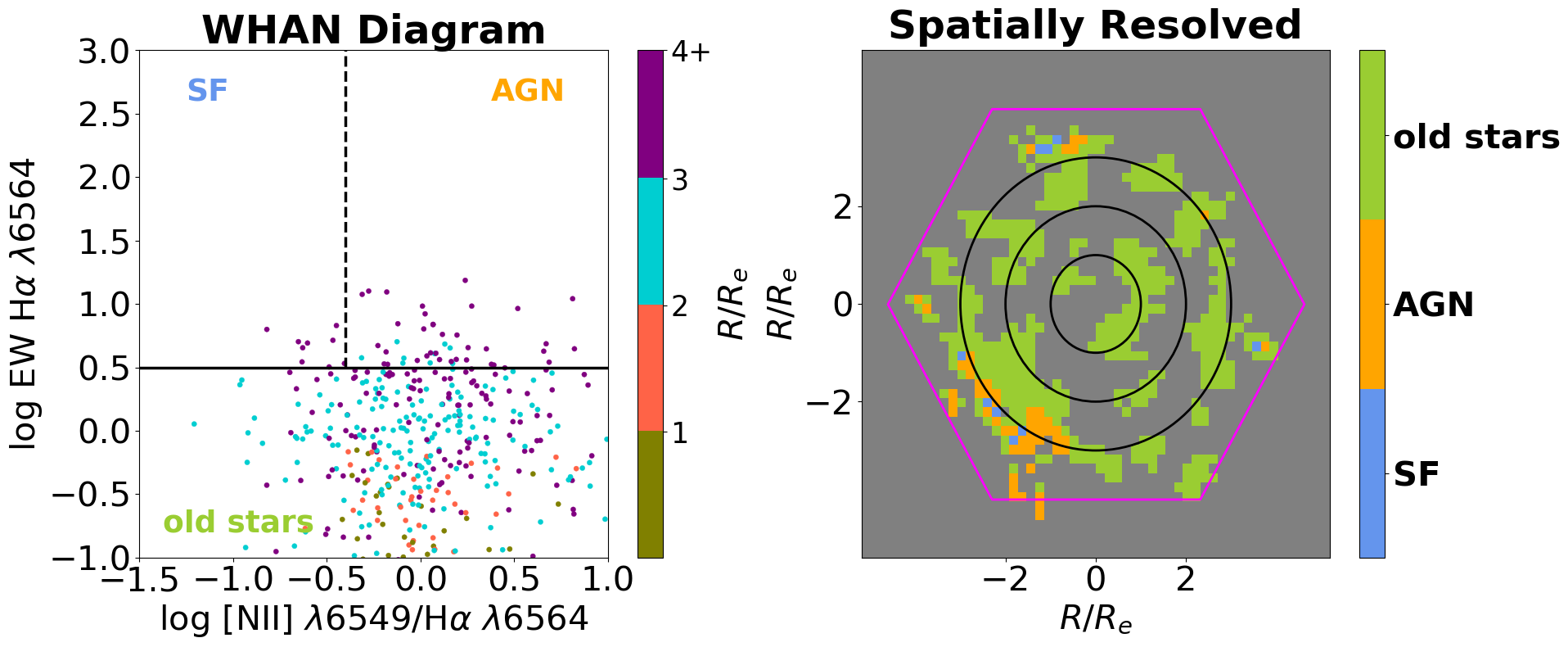} 
\bigskip
\bigskip
\ \includegraphics[height = 0.29\textwidth]{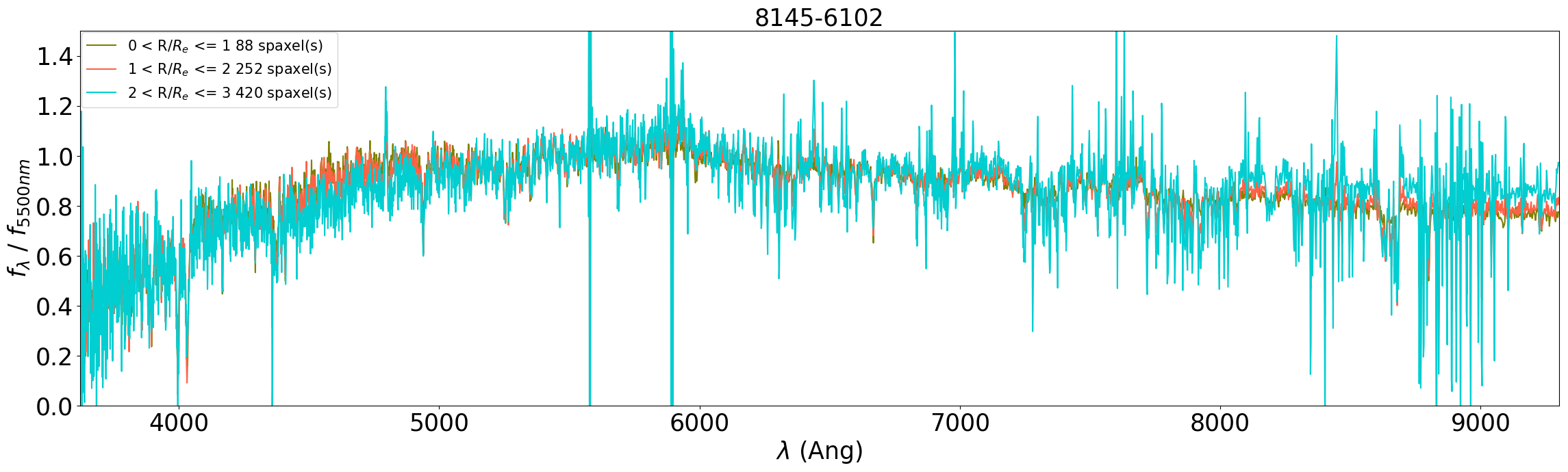}

\bigskip
\bigskip
\bigskip
\bigskip

\begin{center}[\textbf{MaNGA ID: 1-560826 | PLATE-IFU: 8315-3703}]
\end{center}
\includegraphics[height = 0.197\textwidth]{8315-3703vis.png}
\includegraphics[height = 0.197\textwidth]{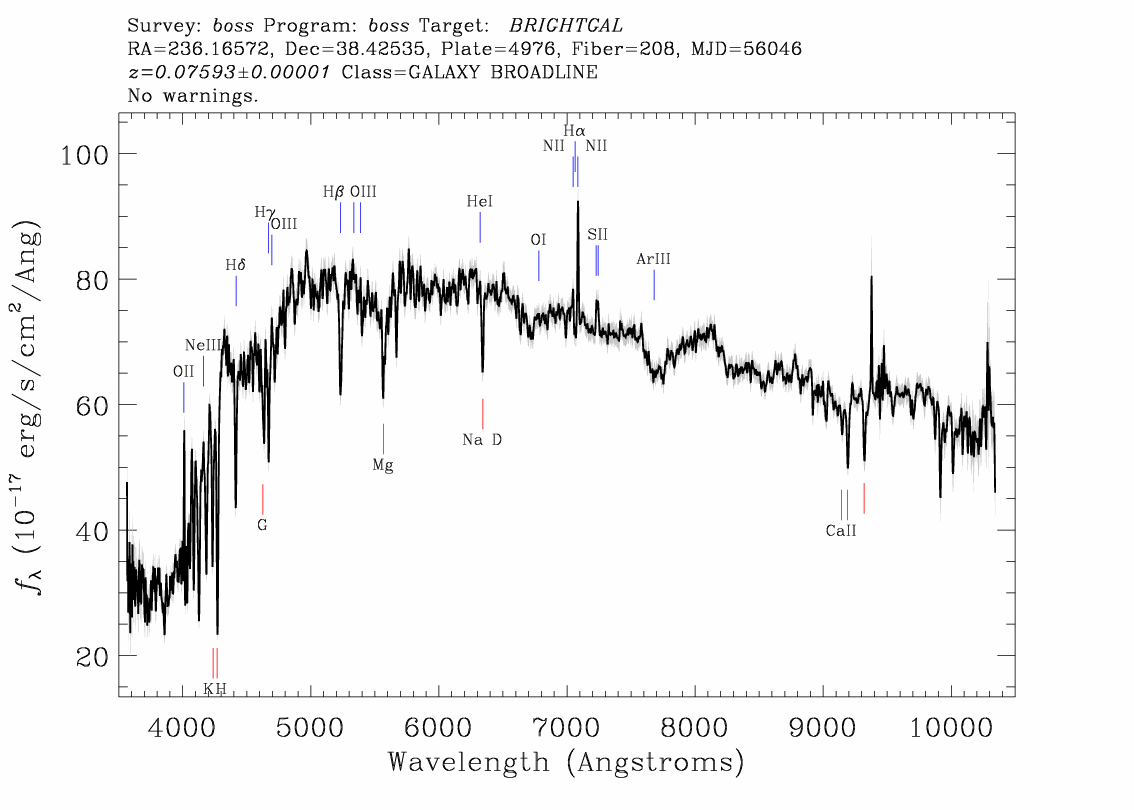}  \includegraphics[height = 0.197\textwidth]{8315-3703_WHAN.png}
\bigskip
\bigskip
\ \includegraphics[height = 0.29\textwidth]{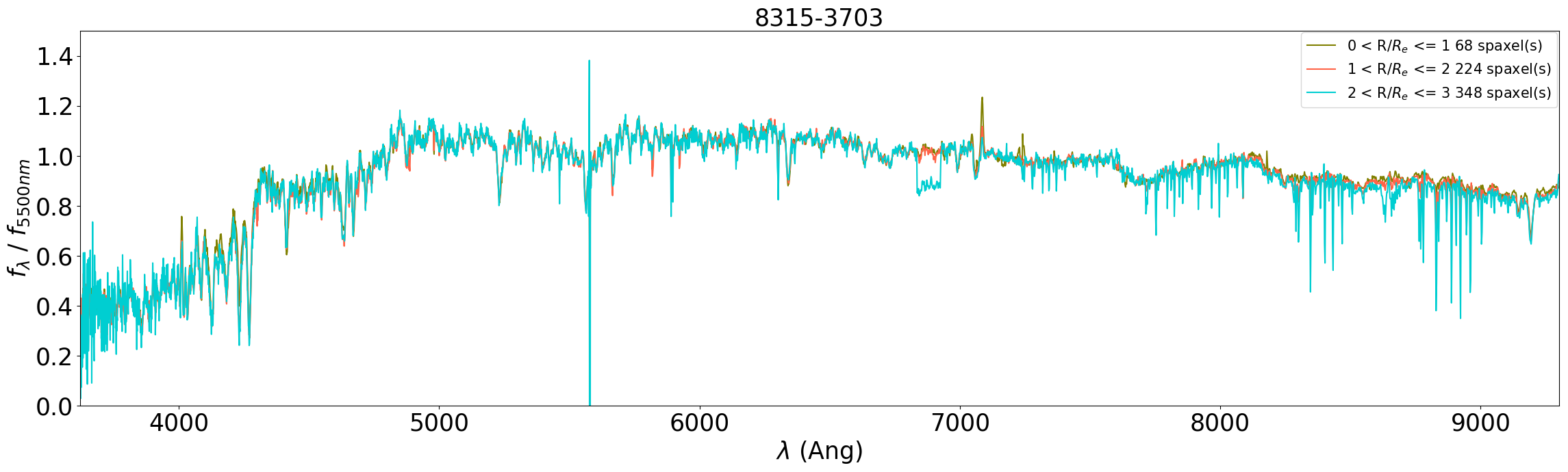}

\bigskip

\vfill\eject
\begin{center}[\textbf{MaNGA ID: 1-235582 | PLATE-IFU: 8326-3704}]
\end{center}
\includegraphics[height = 0.197\textwidth]{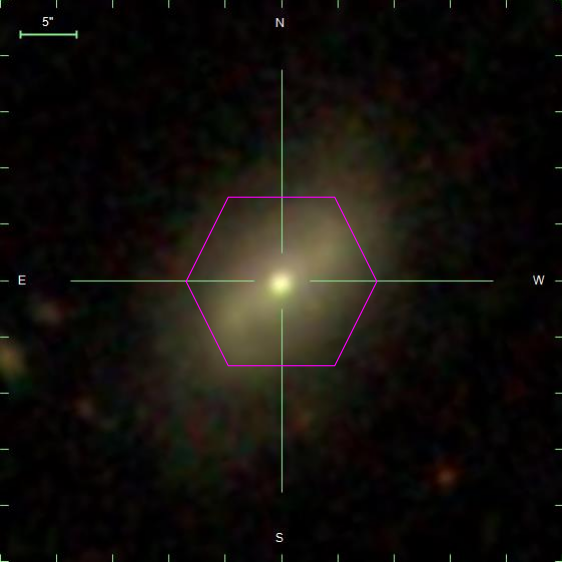}
\includegraphics[height = 0.197\textwidth]{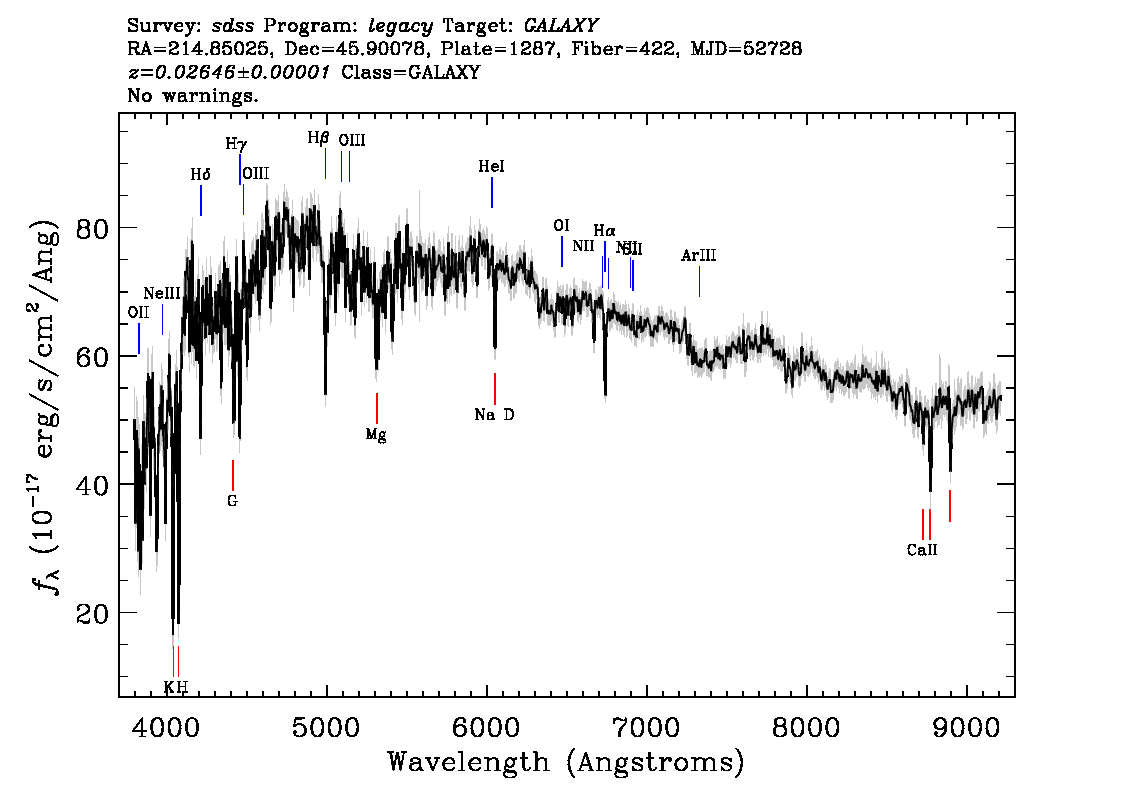}  \includegraphics[height = 0.197\textwidth]{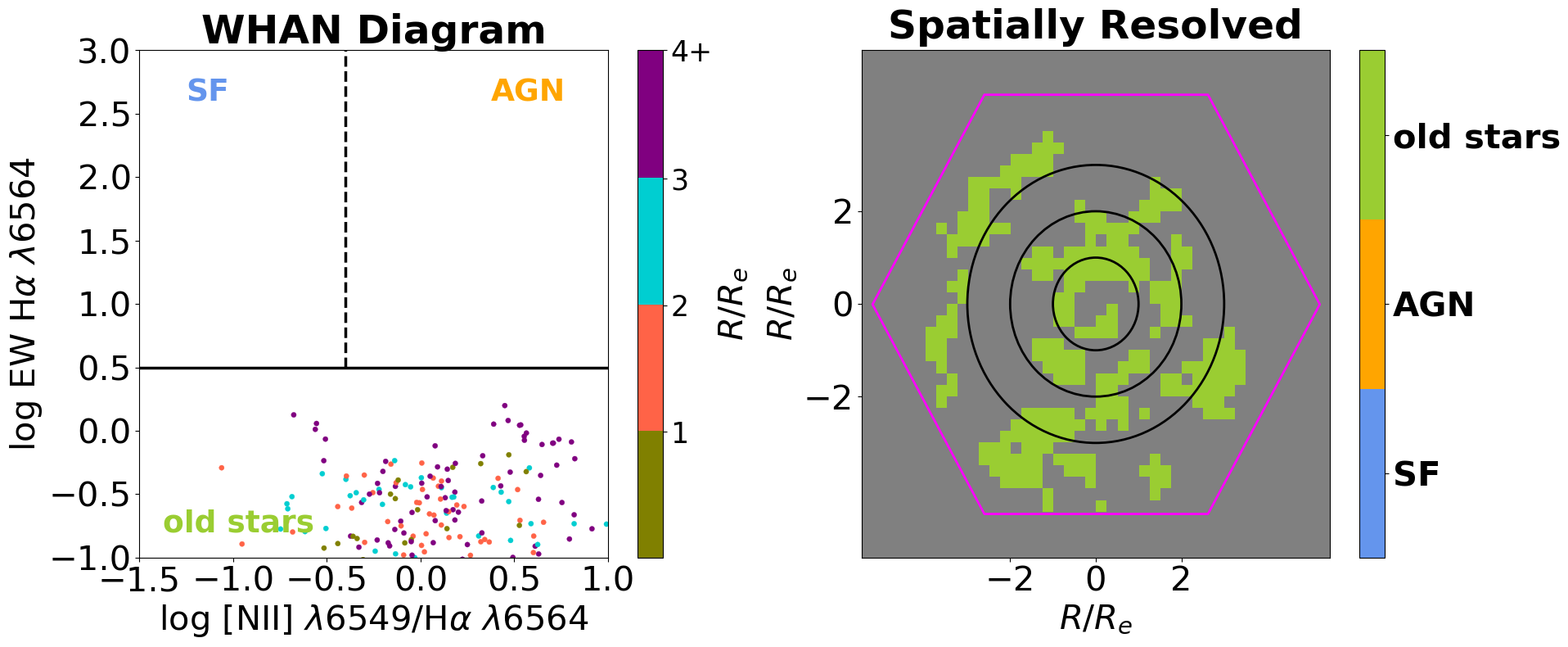} 
\bigskip
\bigskip
\ \includegraphics[height = 0.29\textwidth]{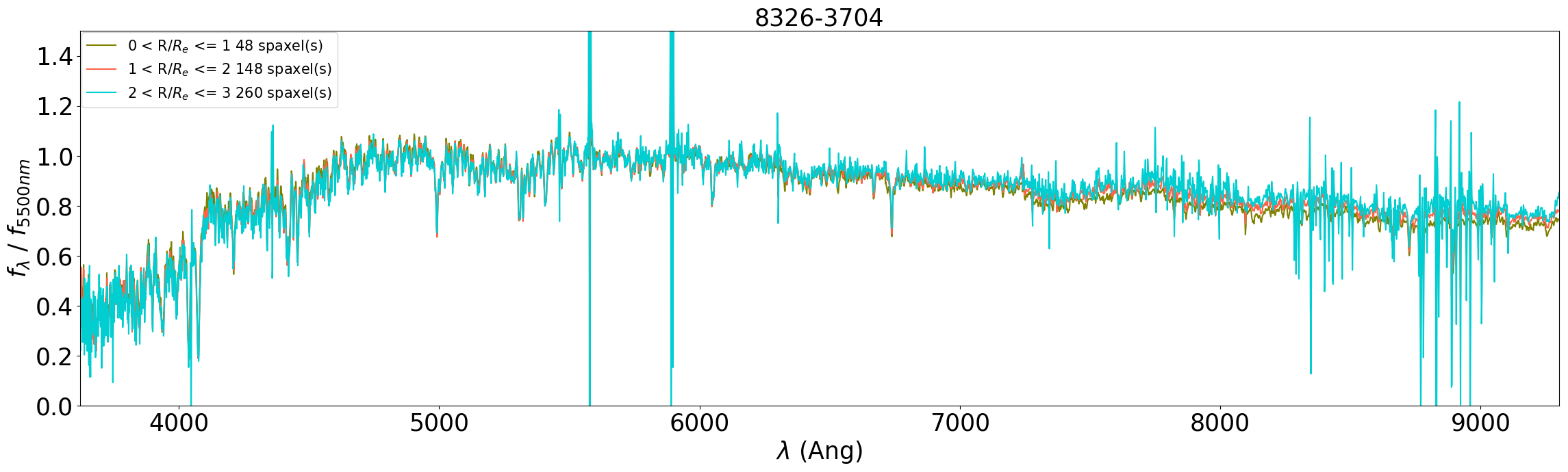}

\bigskip
\bigskip
\bigskip
\bigskip

\begin{center}[\textbf{MaNGA ID: 1-266298 | PLATE-IFU: 8333-1901}]
\end{center}
\includegraphics[height = 0.197\textwidth]{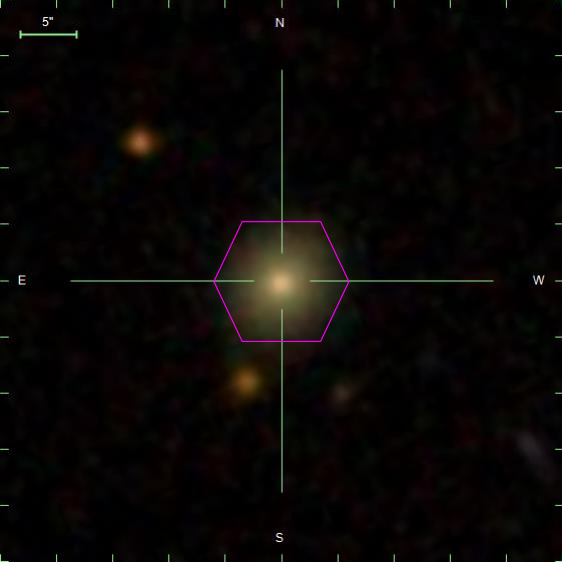}
\includegraphics[height = 0.197\textwidth]{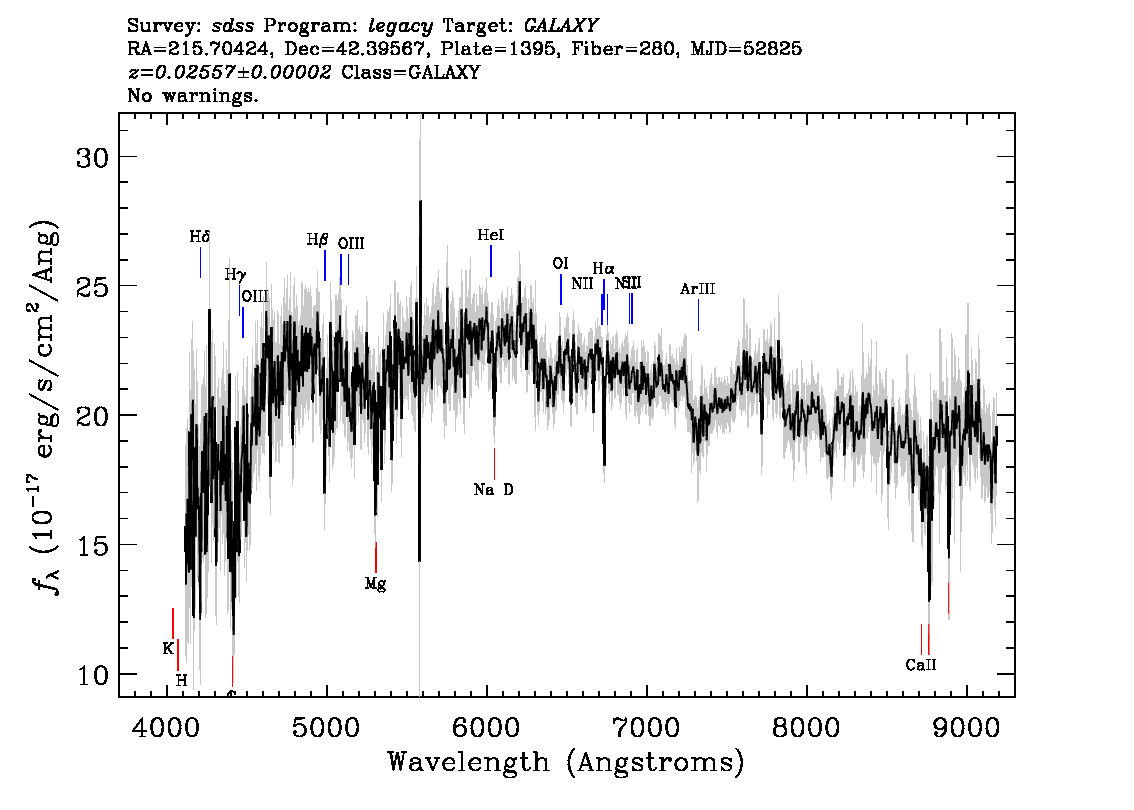}  \includegraphics[height = 0.197\textwidth]{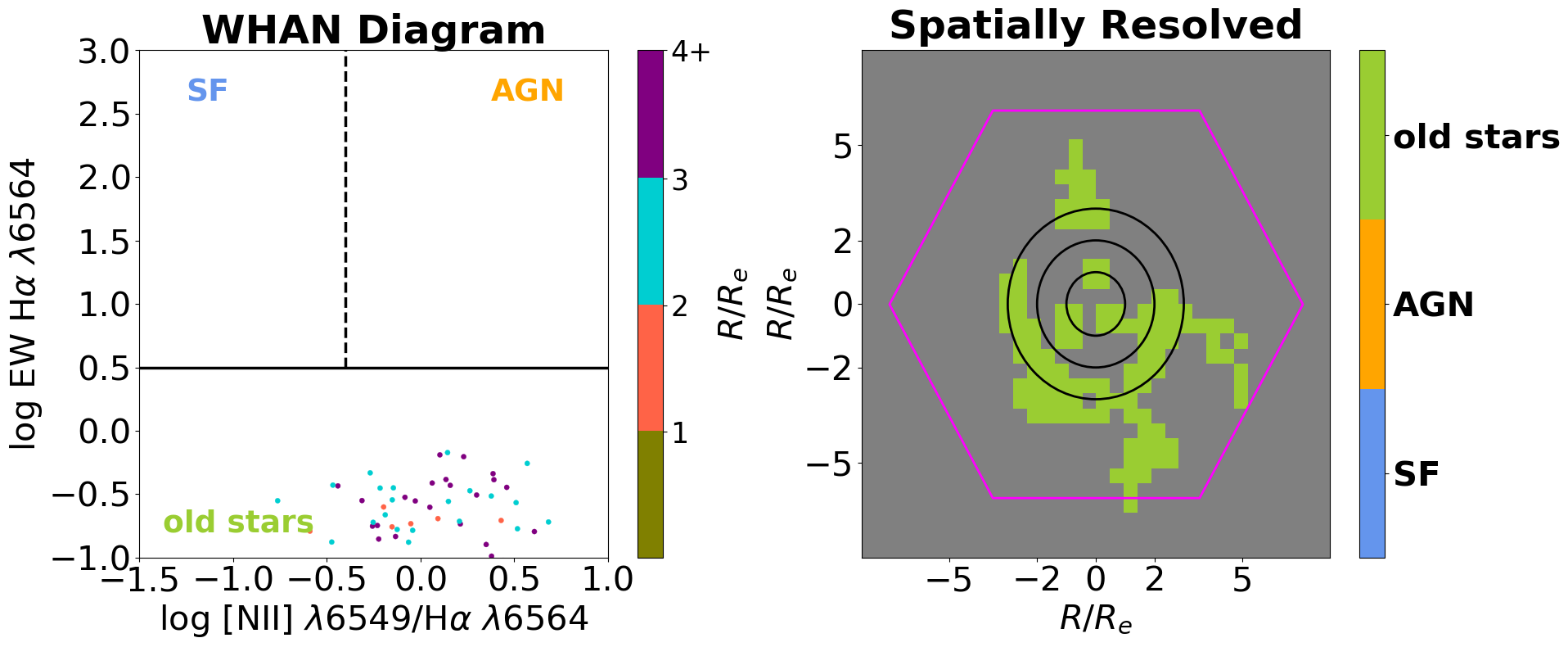} 
\bigskip
\bigskip
\ \includegraphics[height = 0.29\textwidth]{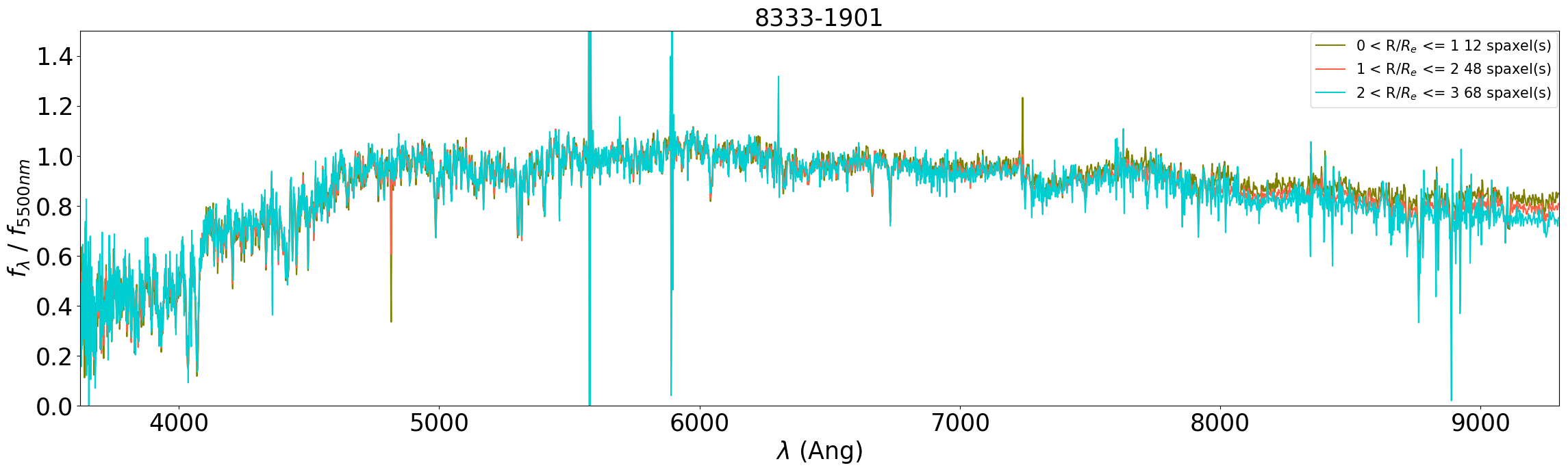}

\bigskip

\vfill\eject
\begin{center}[\textbf{MaNGA ID: 1-489884 | PLATE-IFU: 8338-9102}]
\end{center}
\includegraphics[height = 0.197\textwidth]{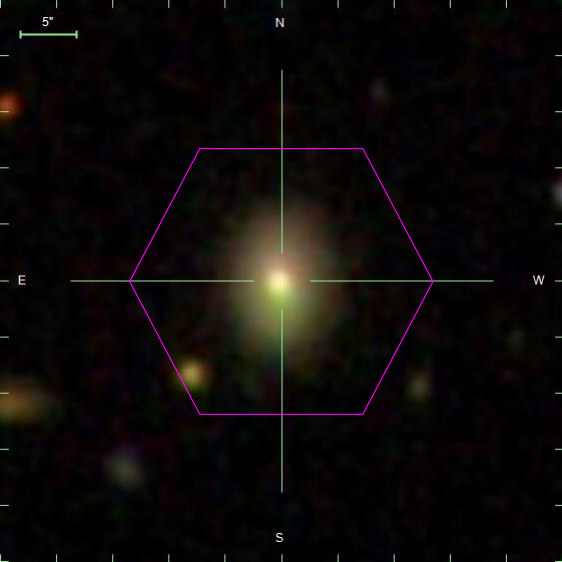}
\includegraphics[height = 0.197\textwidth]{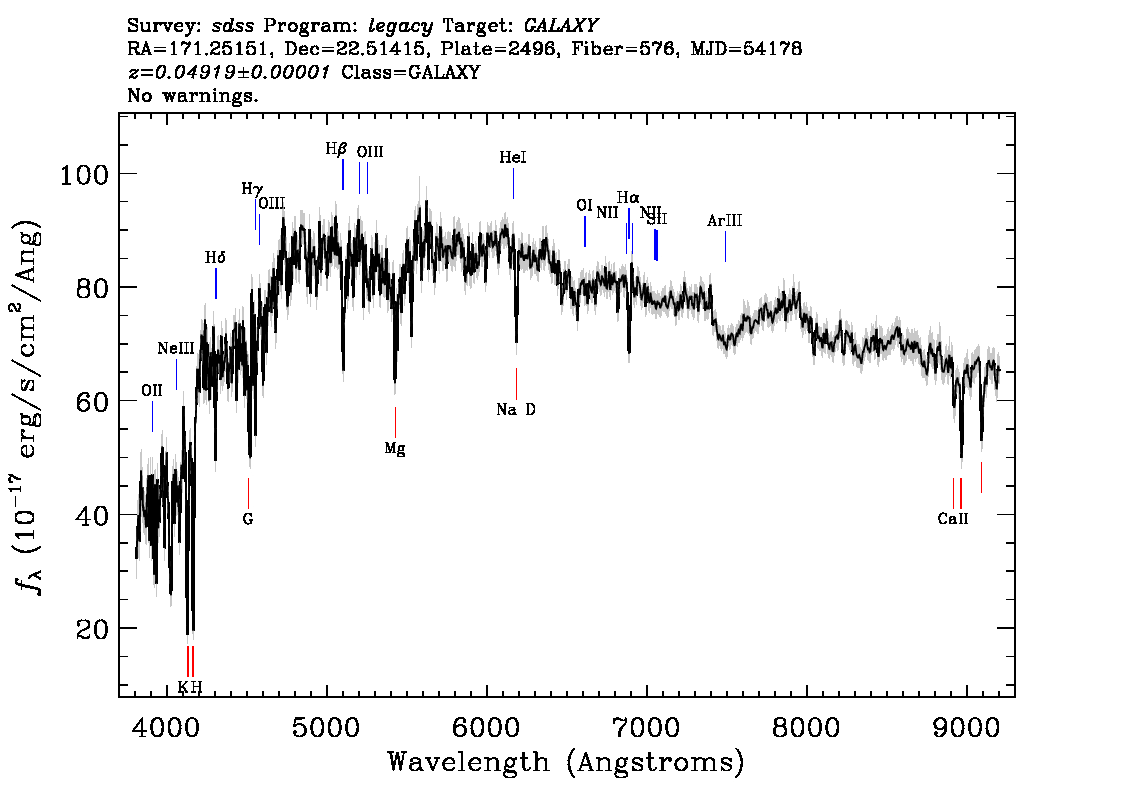}  \includegraphics[height = 0.197\textwidth]{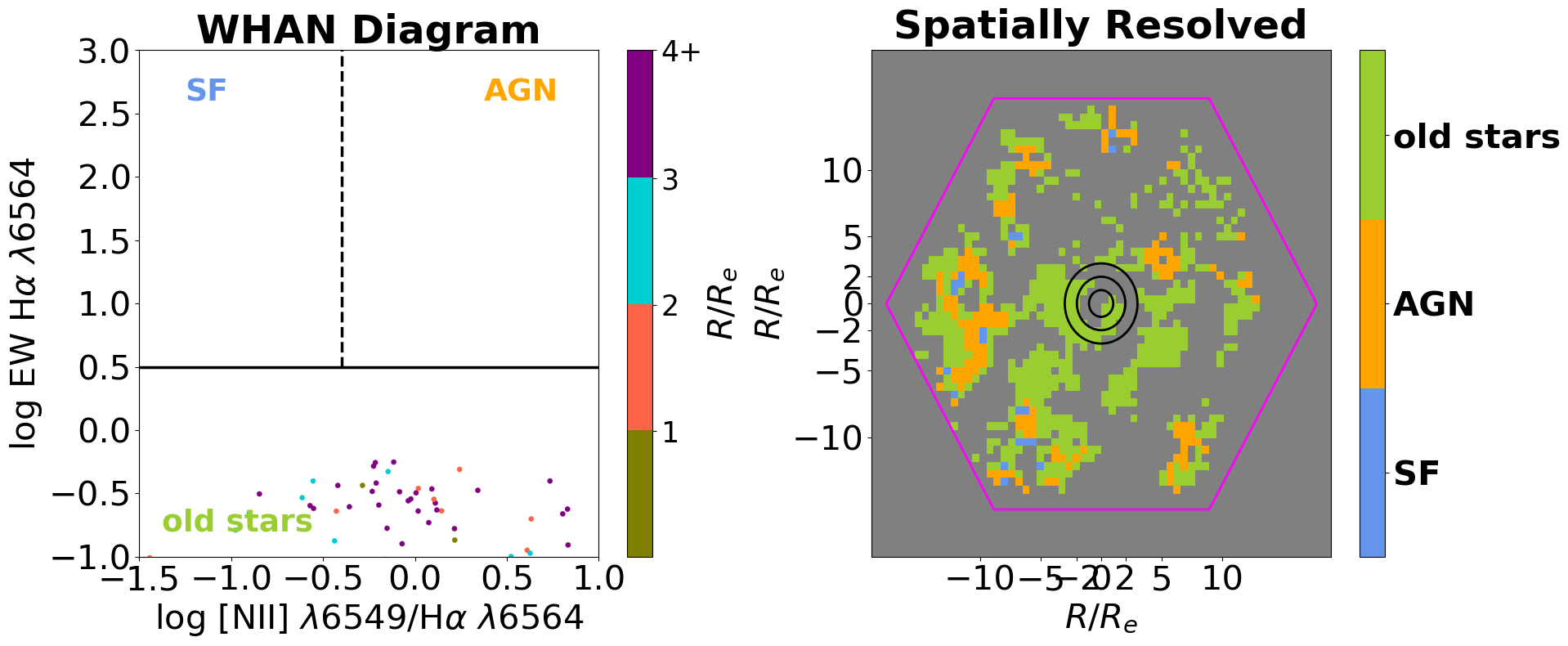} 
\bigskip
\bigskip
\ \includegraphics[height = 0.29\textwidth]{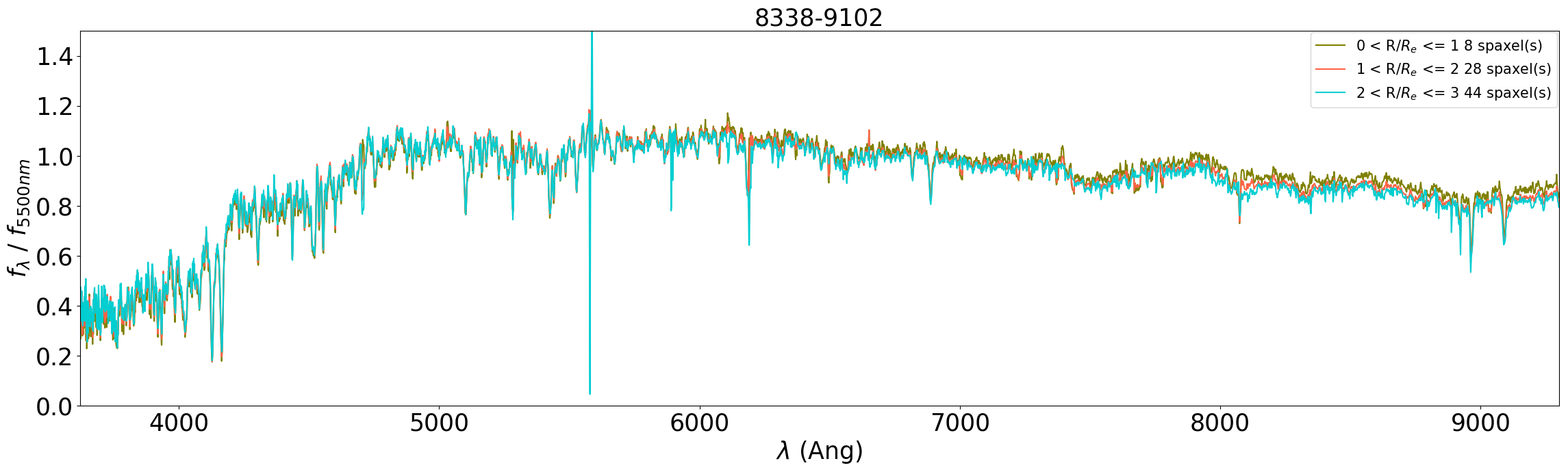}

\bigskip
\bigskip
\bigskip
\bigskip

\begin{center}[\textbf{MaNGA ID: 1-209078 | PLATE-IFU: 8486-3702}]
\end{center}
\includegraphics[height = 0.197\textwidth]{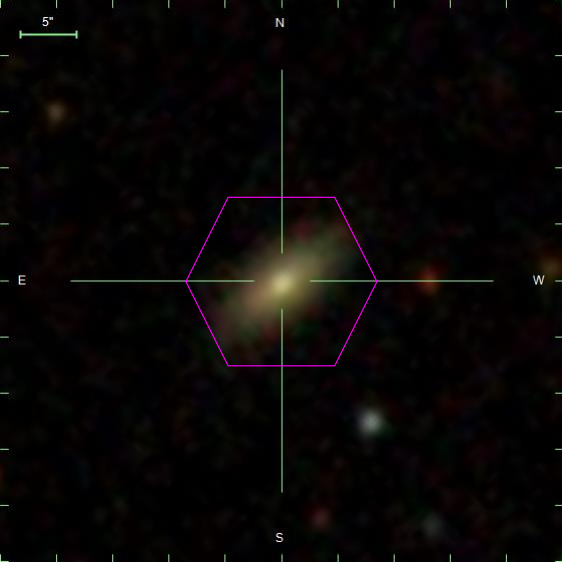}
\includegraphics[height = 0.197\textwidth]{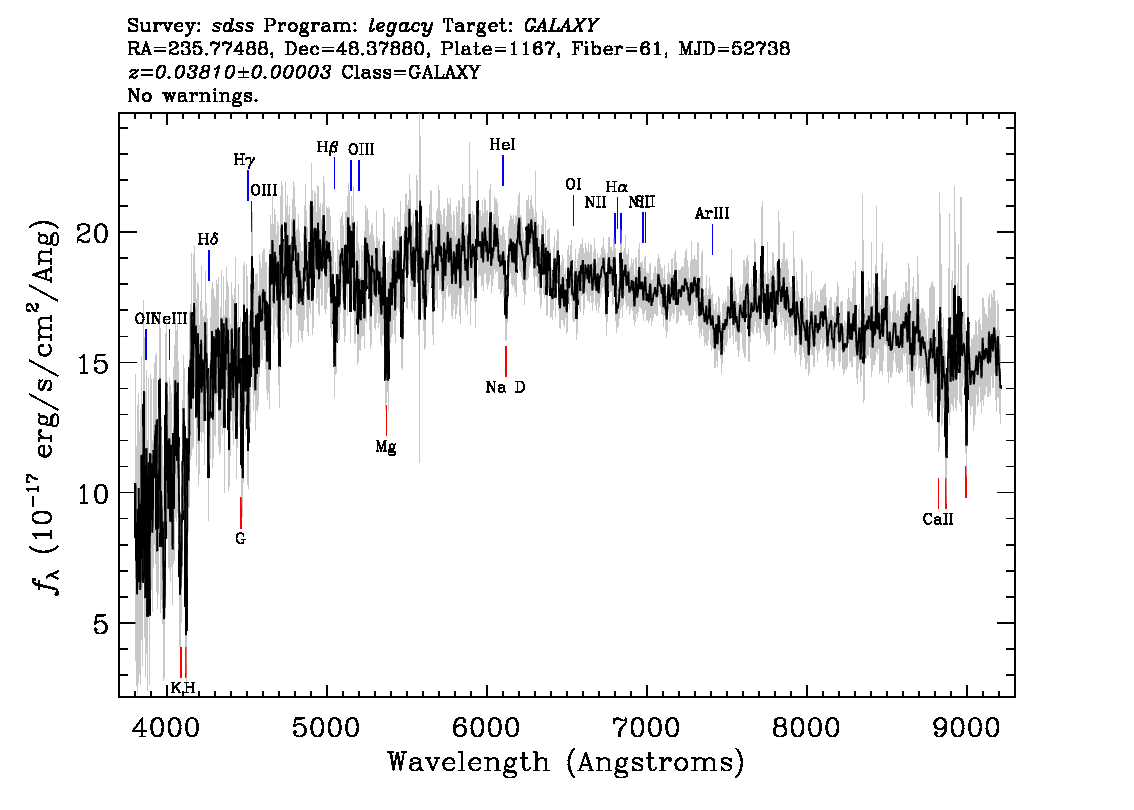}  \includegraphics[height = 0.197\textwidth]{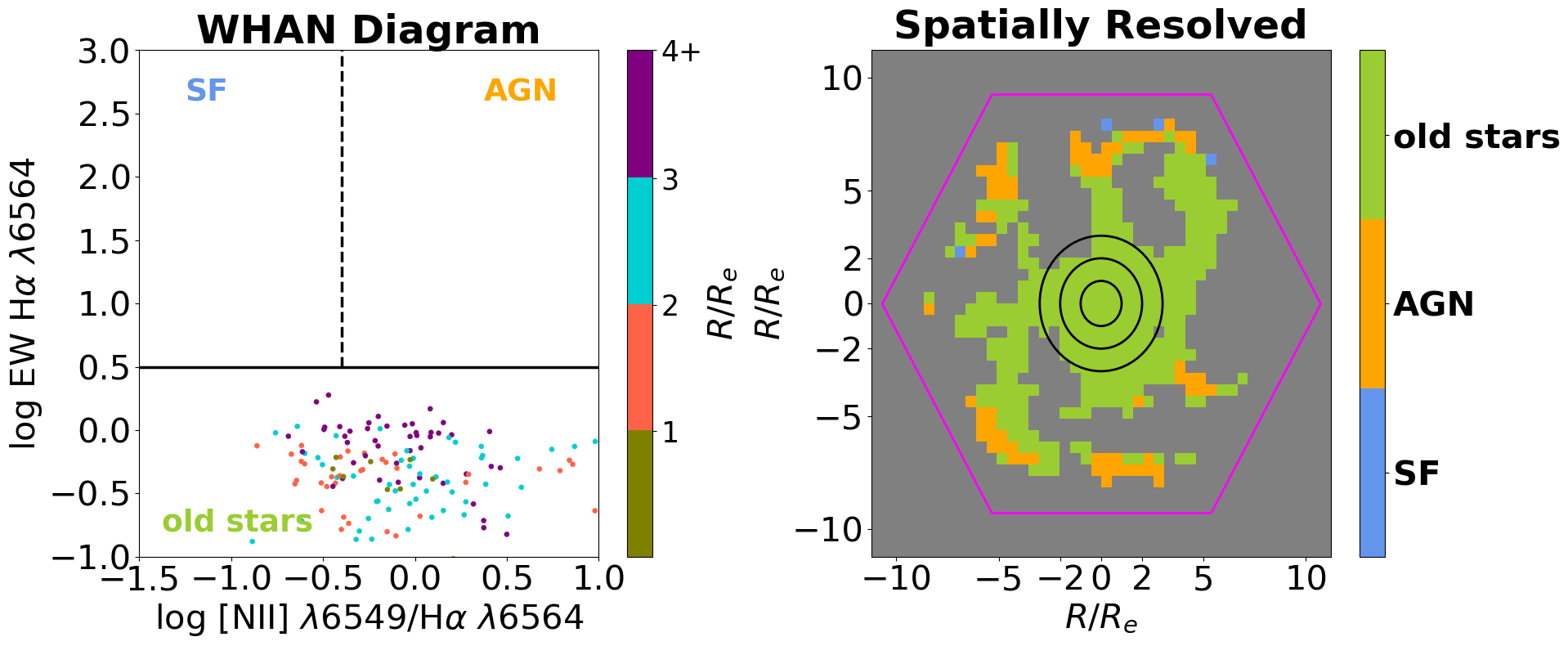} 
\bigskip
\bigskip
\ \includegraphics[height = 0.29\textwidth]{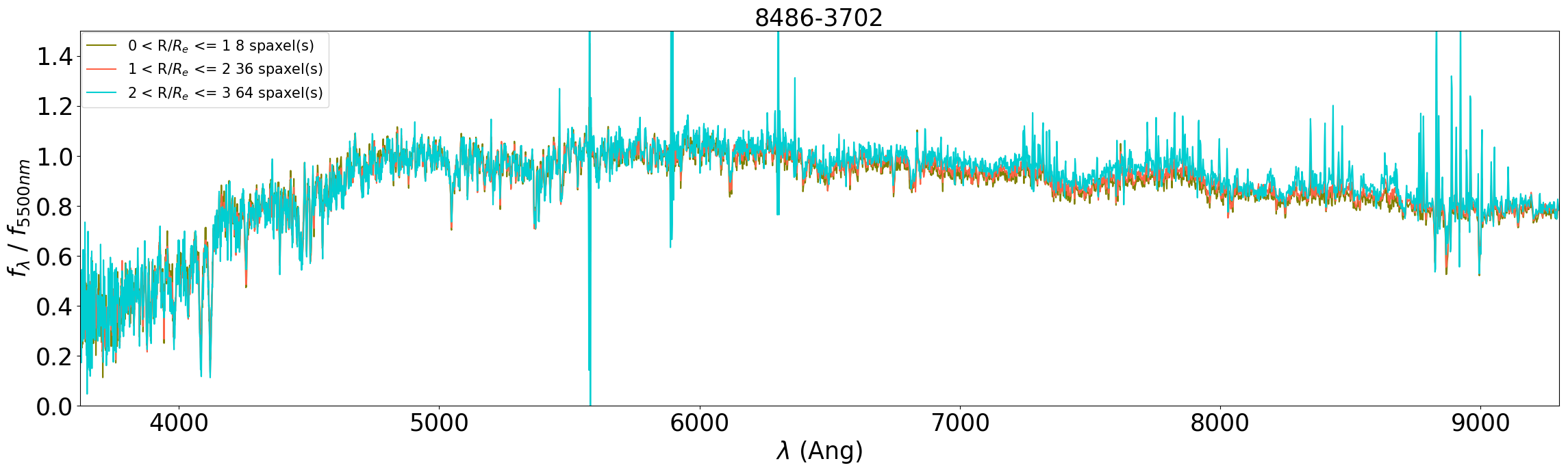}

\bigskip

\vfill\eject
\begin{center}[\textbf{MaNGA ID: 1-92638 | PLATE-IFU: 8548-1901}]
\end{center}
\includegraphics[height = 0.197\textwidth]{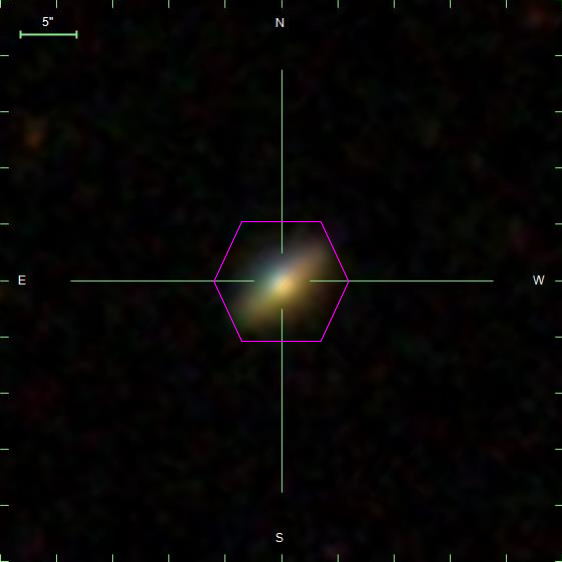}
\includegraphics[height = 0.197\textwidth]{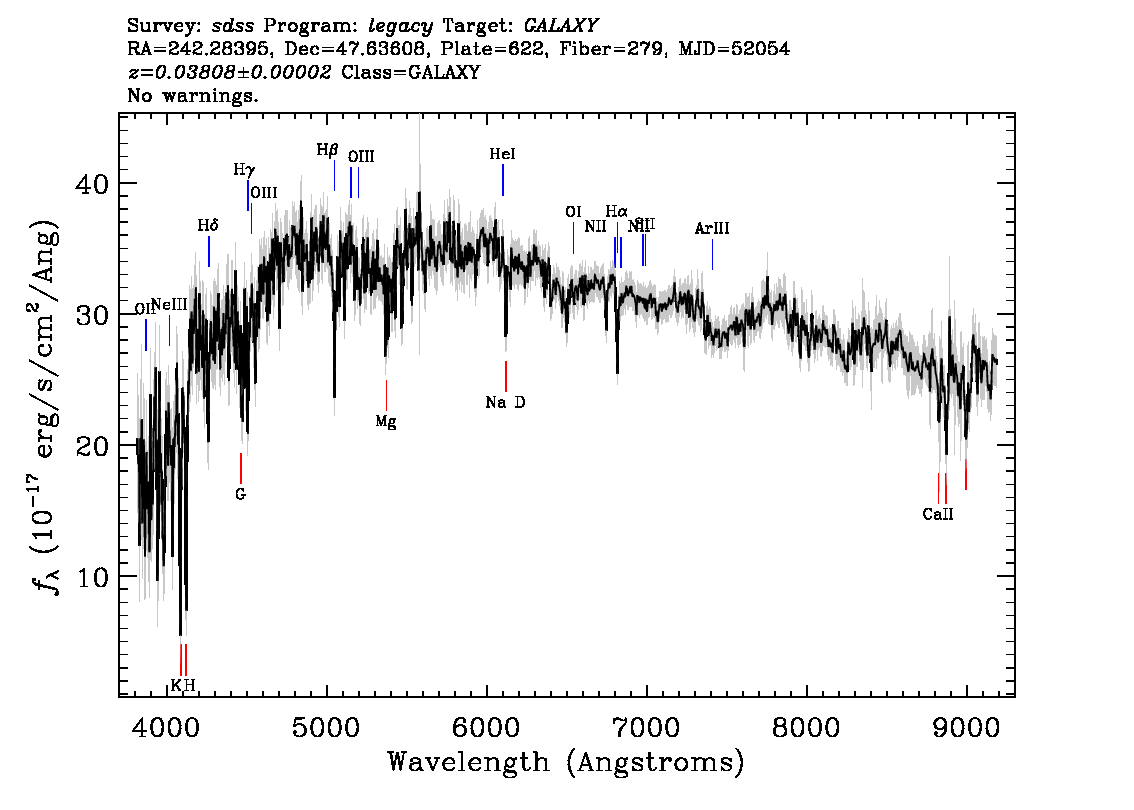}  \includegraphics[height = 0.197\textwidth]{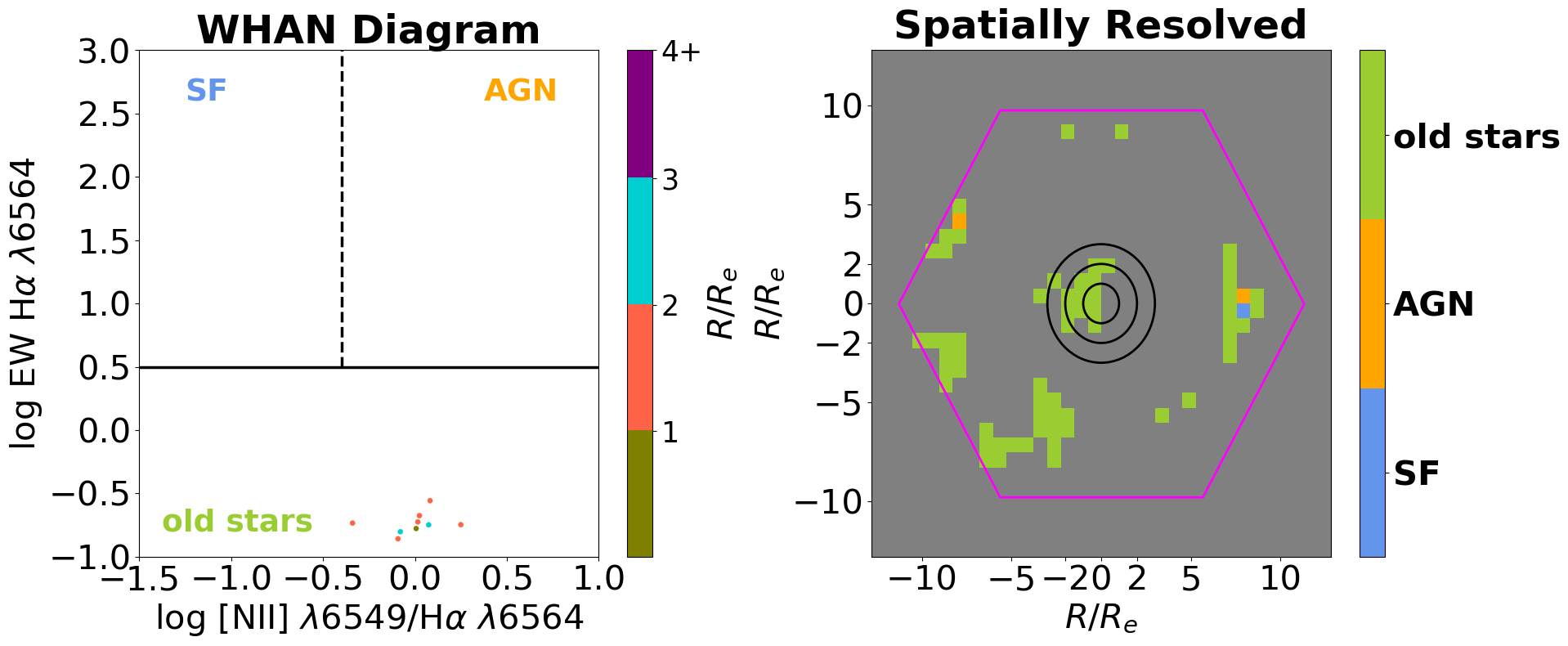}
\bigskip
\bigskip
\ \includegraphics[height = 0.29\textwidth]{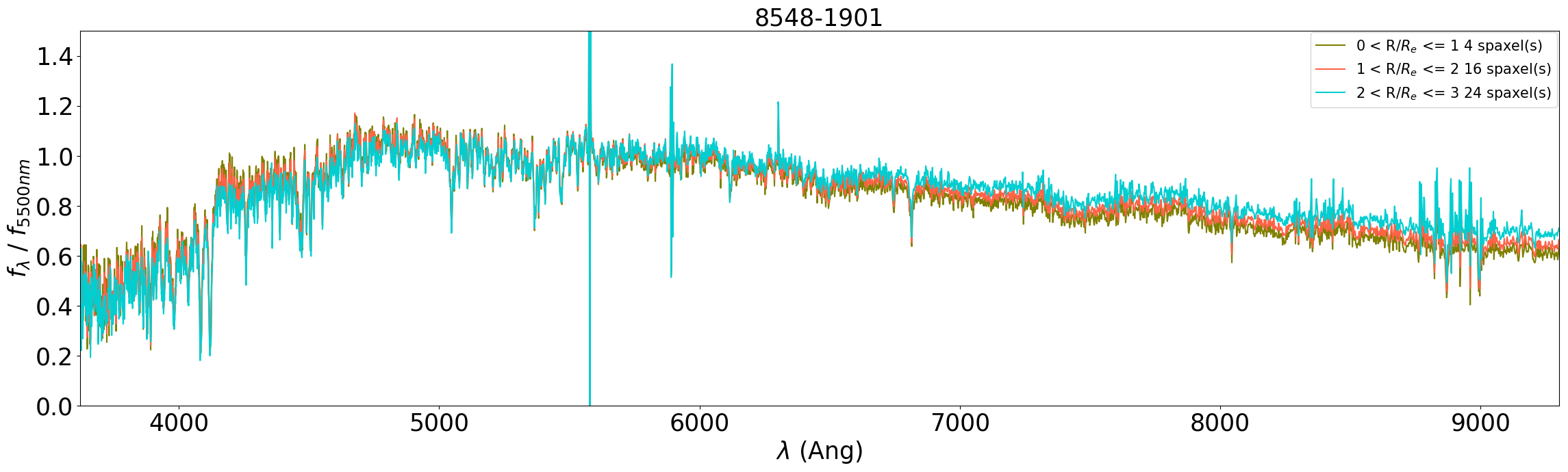}

\bigskip
\bigskip
\bigskip
\bigskip

\begin{center}[\textbf{MaNGA ID: 1-90984 | PLATE-IFU: 8553-3701}]
\end{center}
\includegraphics[height = 0.197\textwidth]{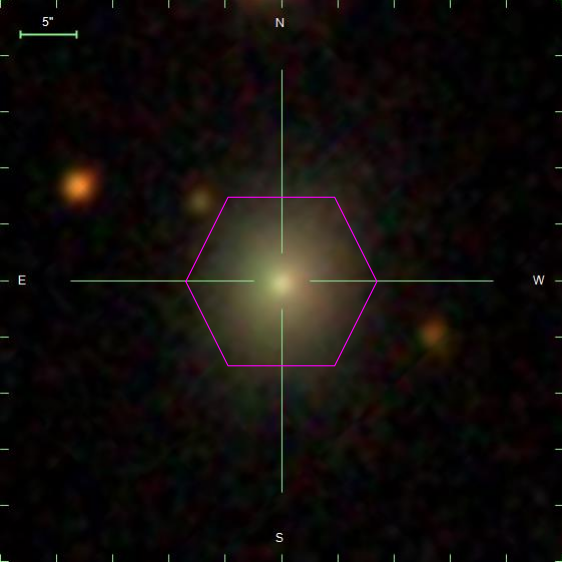}
\includegraphics[height = 0.197\textwidth]{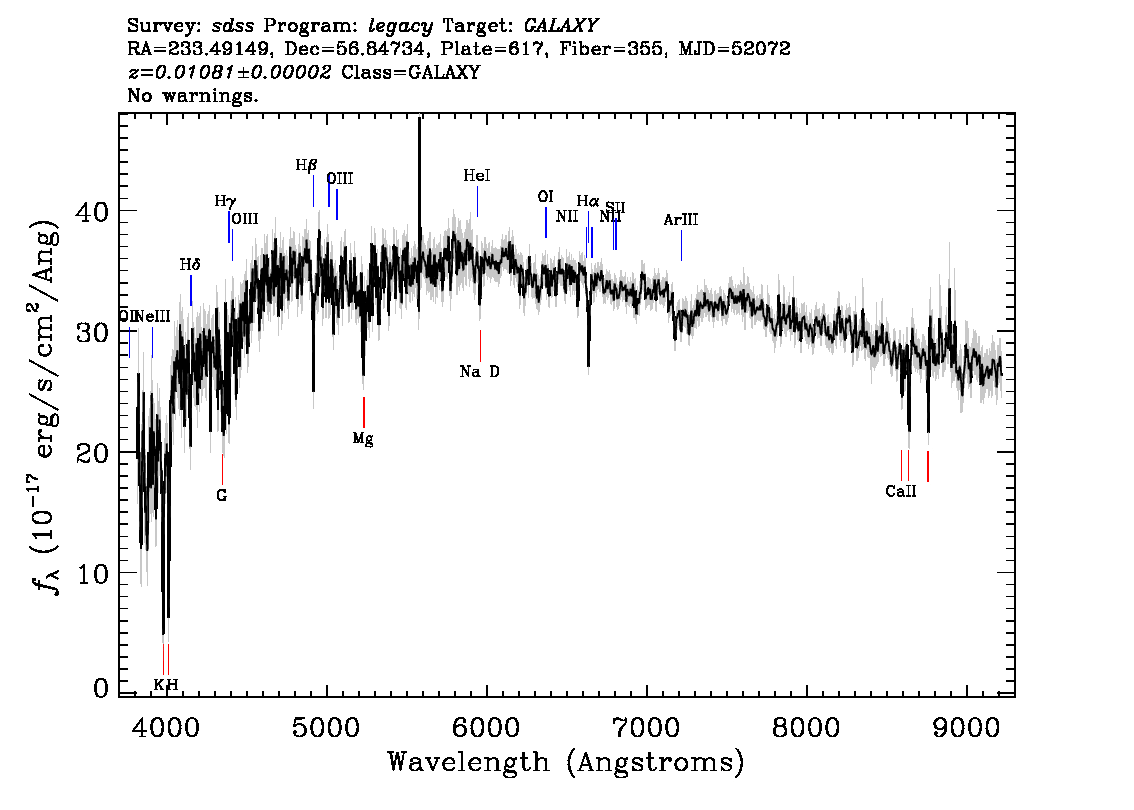}  \includegraphics[height = 0.197\textwidth]{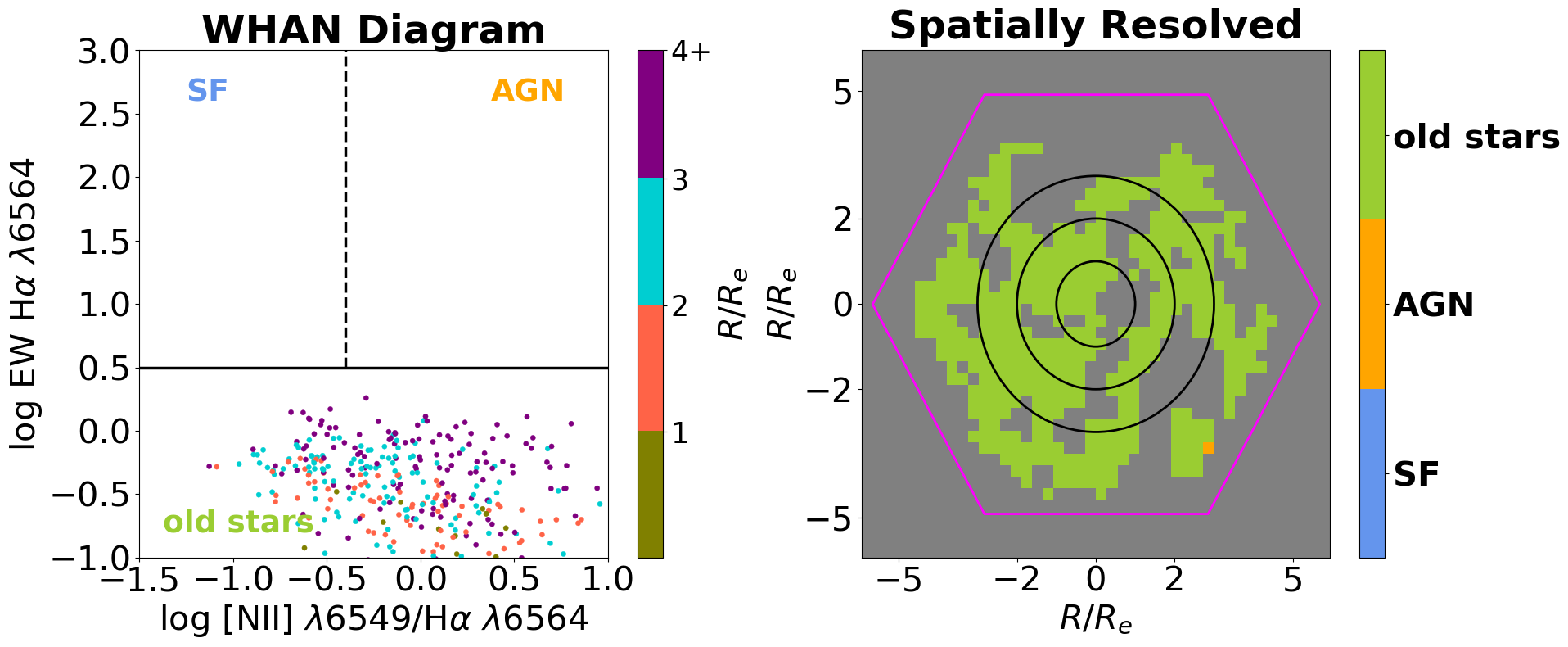} 
\bigskip
\bigskip
\ \includegraphics[height = 0.29\textwidth]{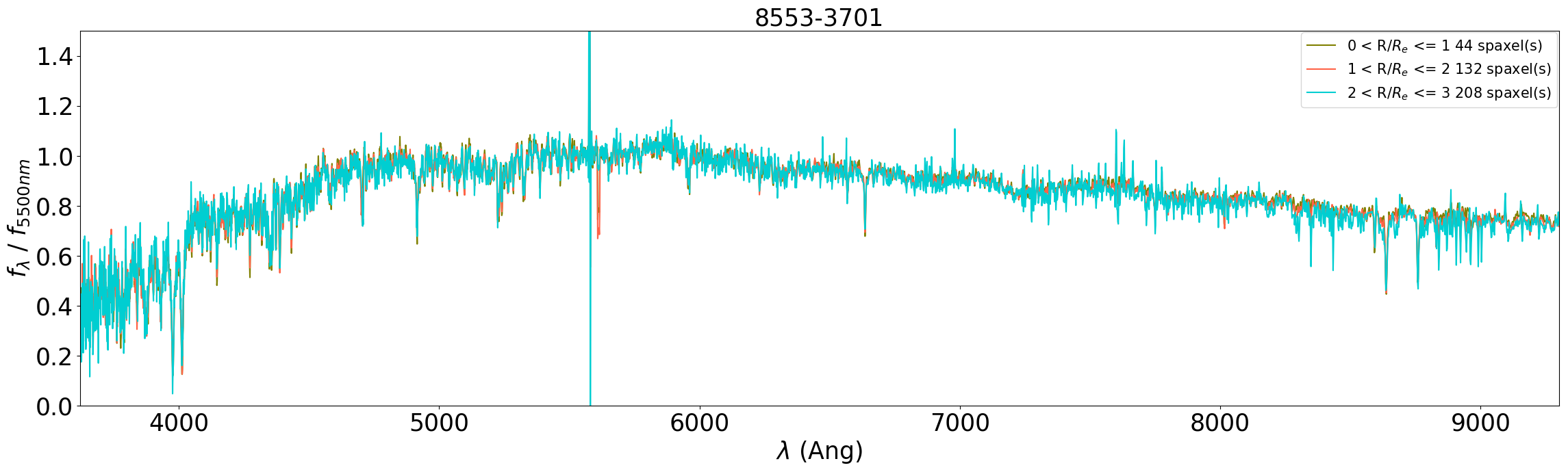}

\bigskip

\vfill\eject
\begin{center}[\textbf{MaNGA ID: 1-90176 | PLATE-IFU: 8553-6101}]
\end{center}
\includegraphics[height = 0.197\textwidth]{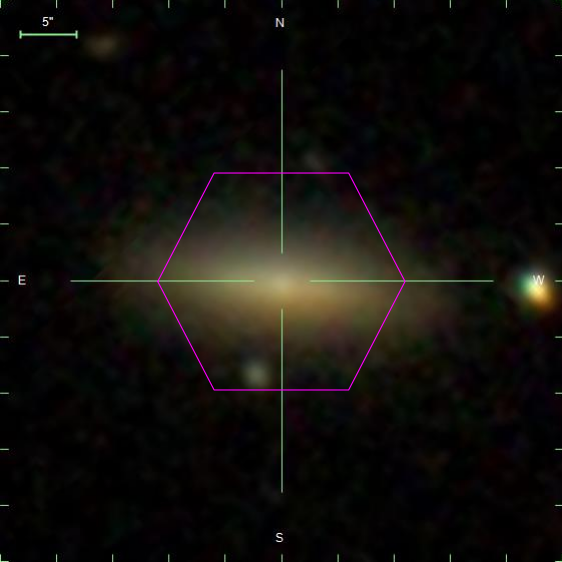}
\includegraphics[height = 0.197\textwidth]{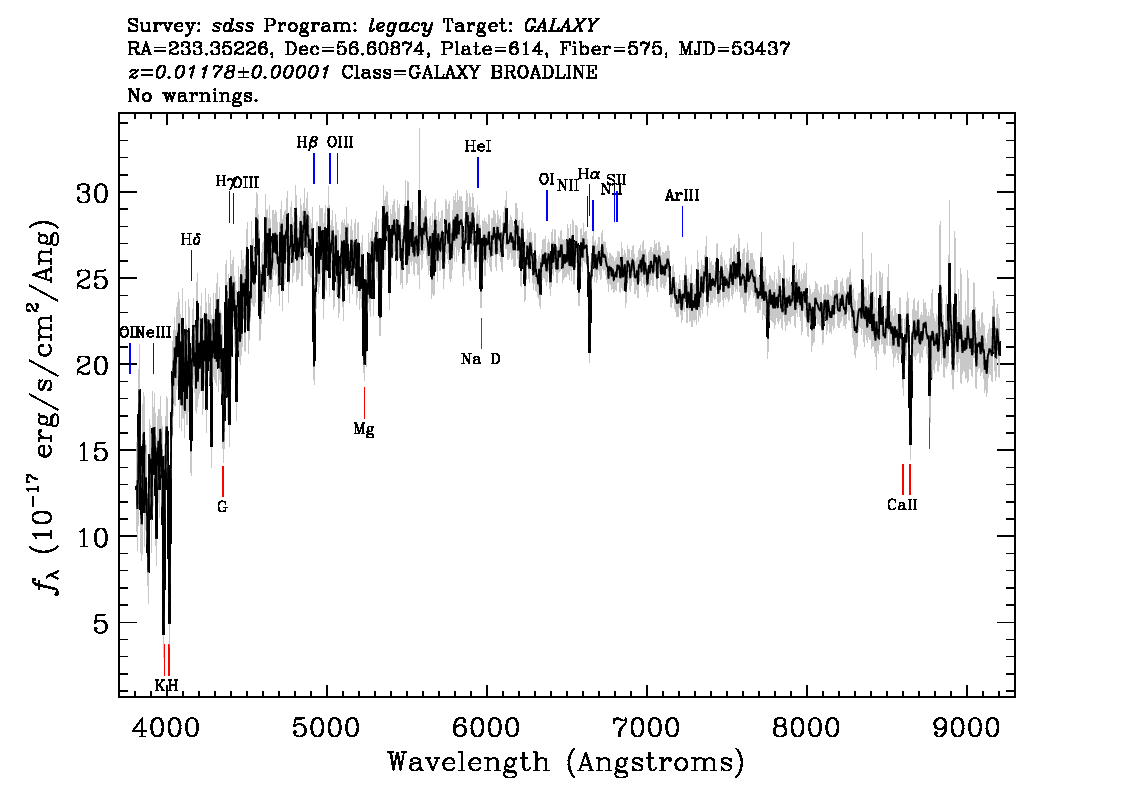}  \includegraphics[height = 0.197\textwidth]{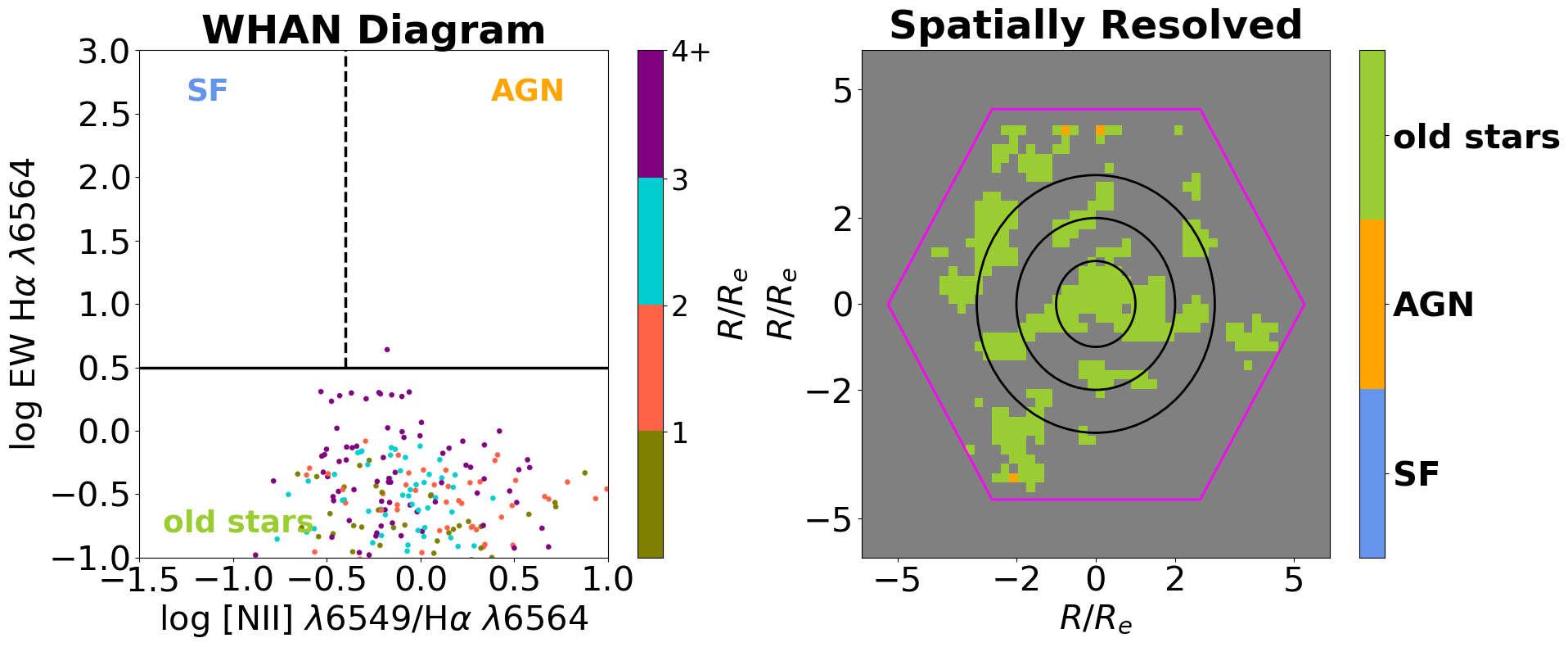} 
\bigskip
\bigskip
\ \includegraphics[height = 0.29\textwidth]{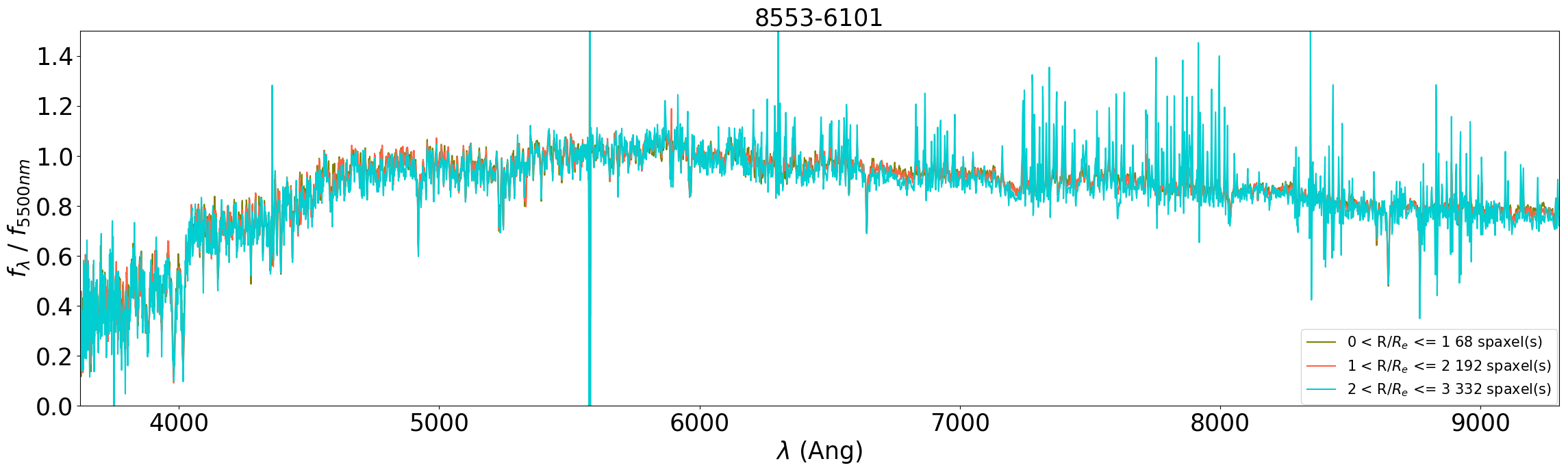}

\bigskip
\bigskip
\bigskip
\bigskip

\begin{center}[\textbf{MaNGA ID: 1-95093 | PLATE-IFU: 8588-3704}]
\end{center}
\includegraphics[height = 0.197\textwidth]{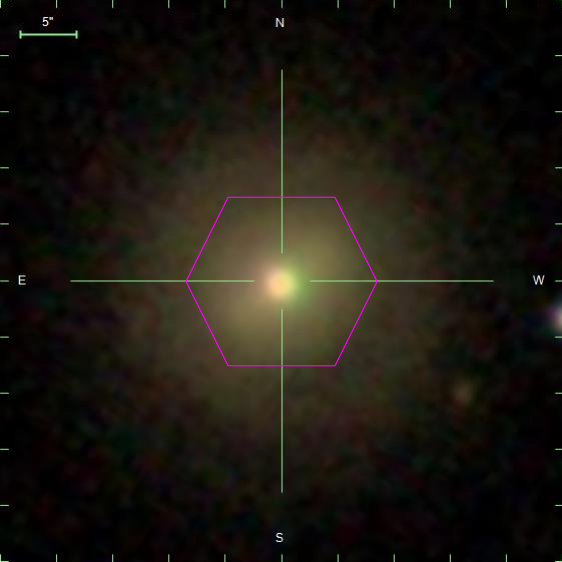}
\includegraphics[height = 0.197\textwidth]{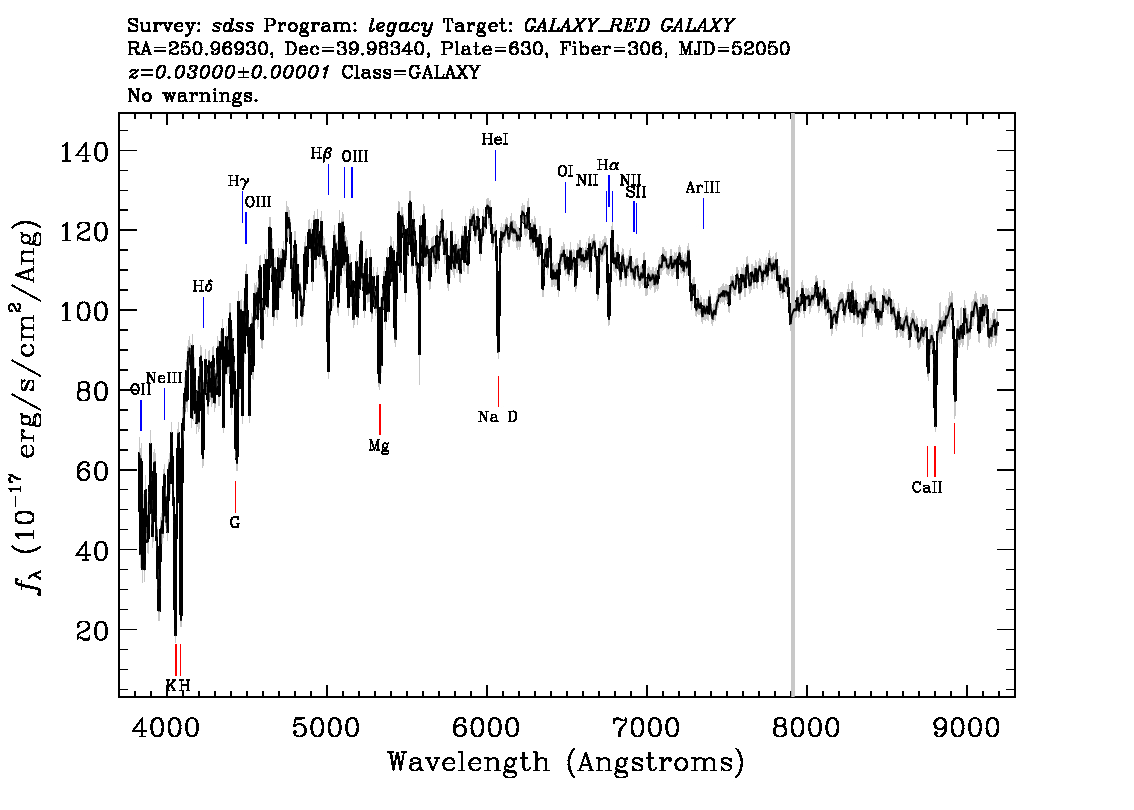}  \includegraphics[height = 0.197\textwidth]{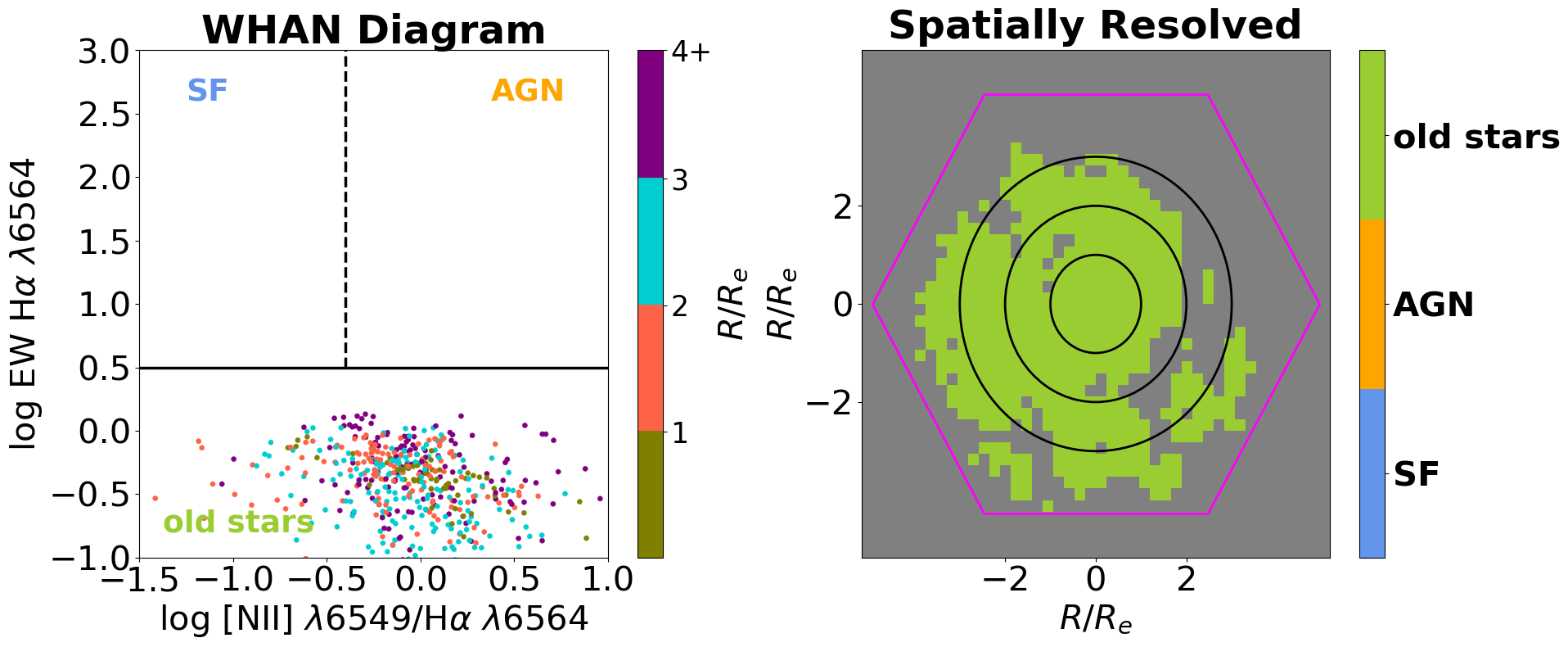} 
\bigskip
\bigskip
\ \includegraphics[height = 0.29\textwidth]{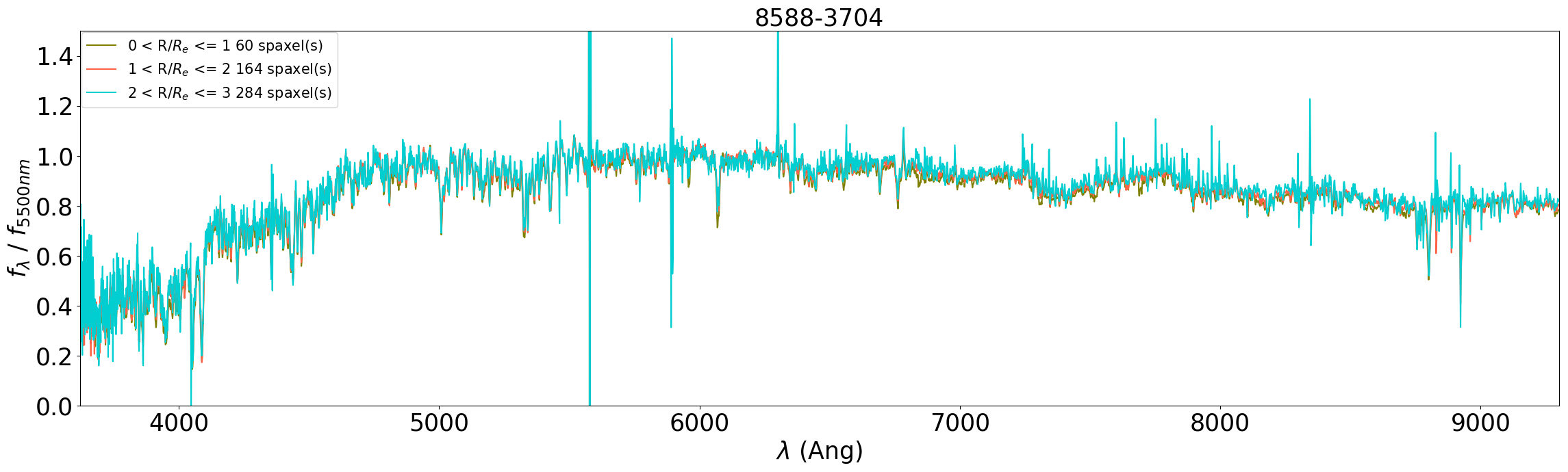}

\bigskip

\vfill\eject
\begin{center}[\textbf{MaNGA ID: 1-178823 | PLATE-IFU: 8623-9102}]
\end{center}
\includegraphics[height = 0.197\textwidth]{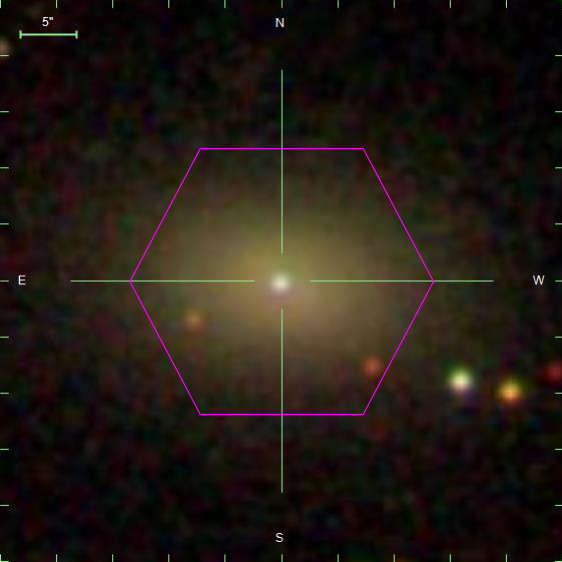}
\includegraphics[height = 0.197\textwidth]{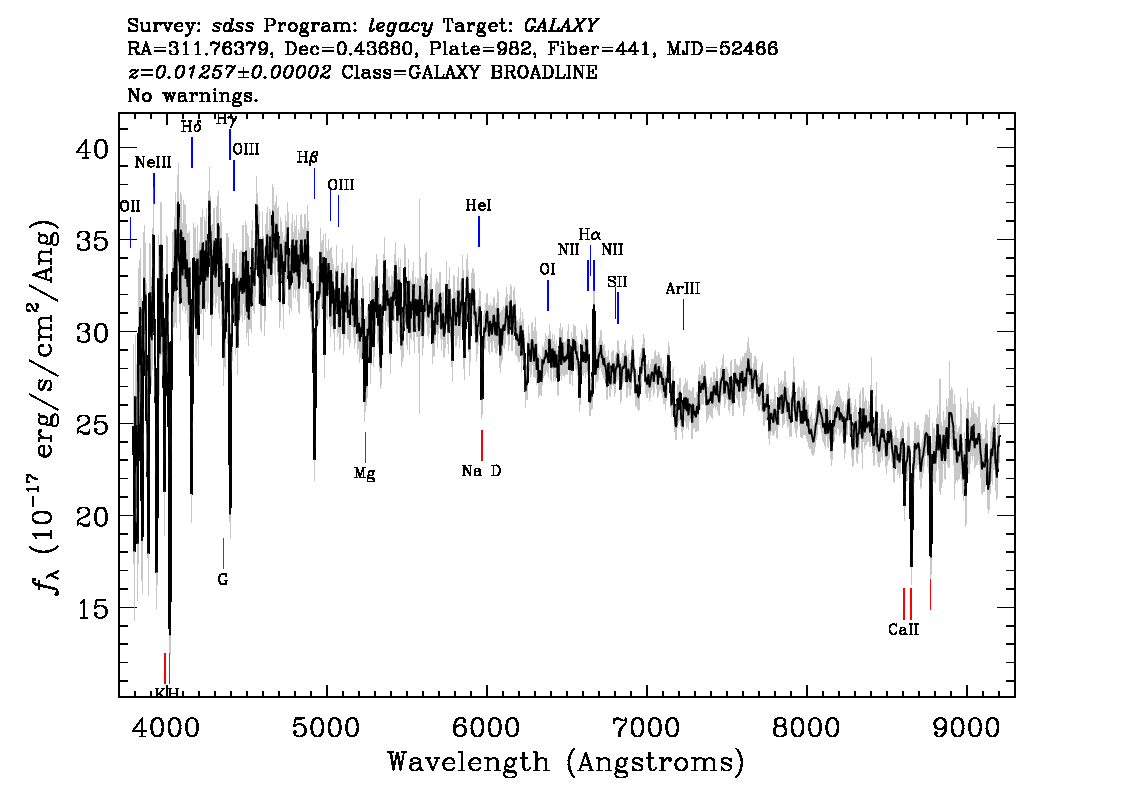}  \includegraphics[height = 0.197\textwidth]{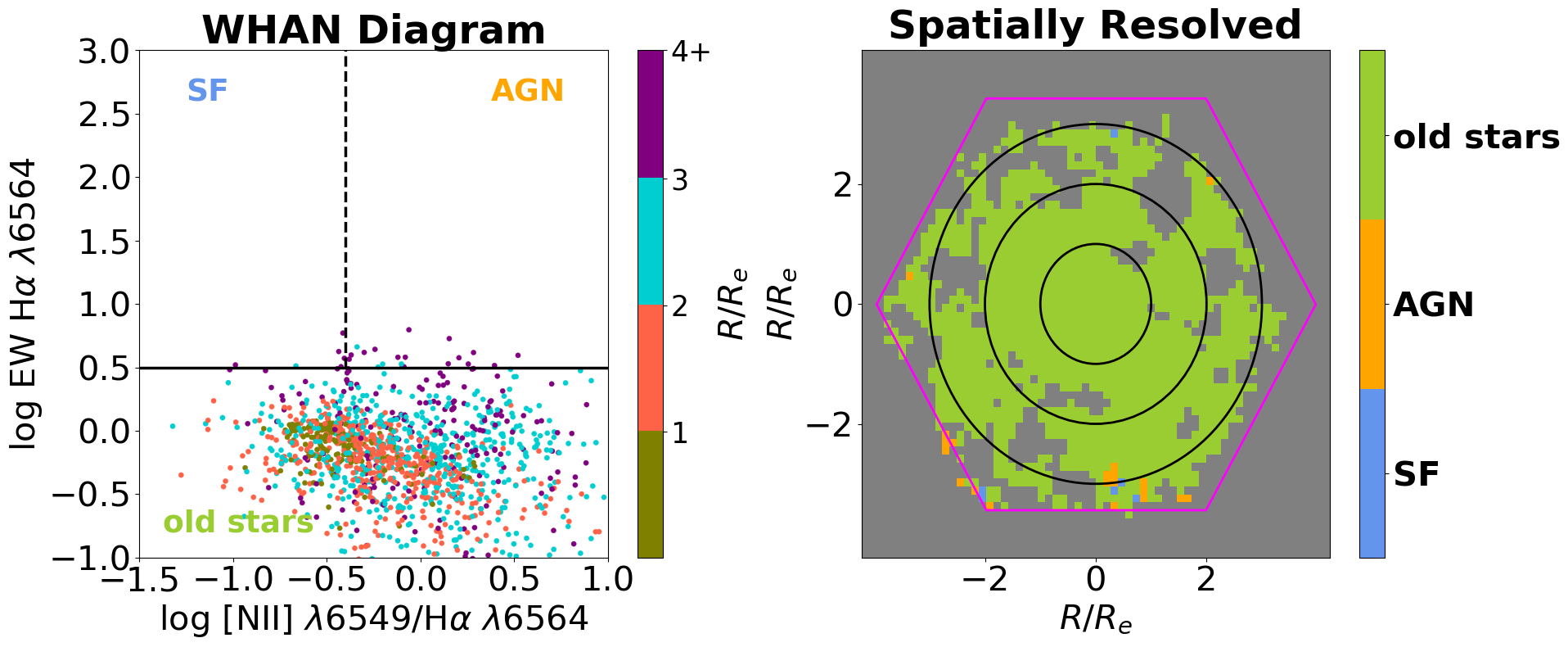} 
\bigskip
\bigskip
\ \includegraphics[height = 0.29\textwidth]{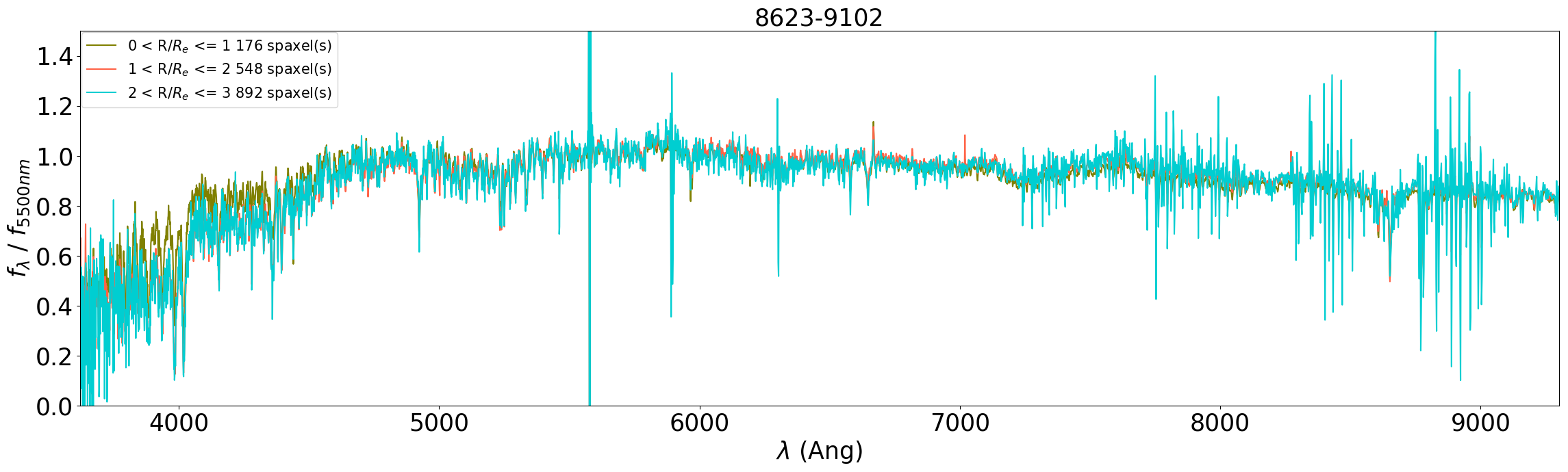}

\bigskip
\bigskip
\bigskip
\bigskip

\begin{center}[\textbf{MaNGA ID: 1-29809 | PLATE-IFU: 8655-1902}]
\end{center}
\includegraphics[height = 0.197\textwidth]{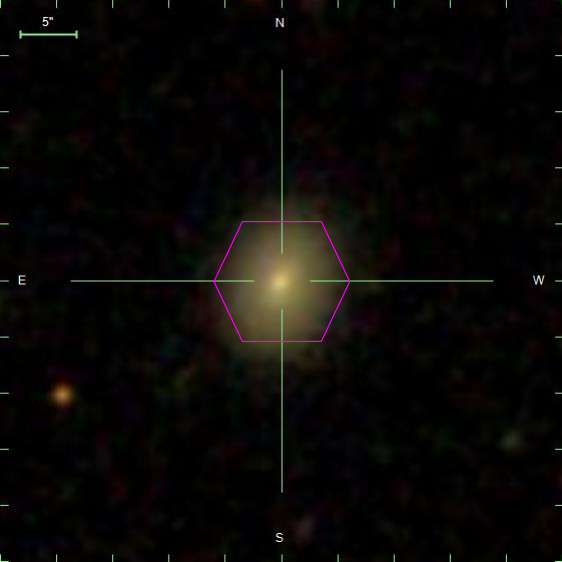}
\includegraphics[height = 0.197\textwidth]{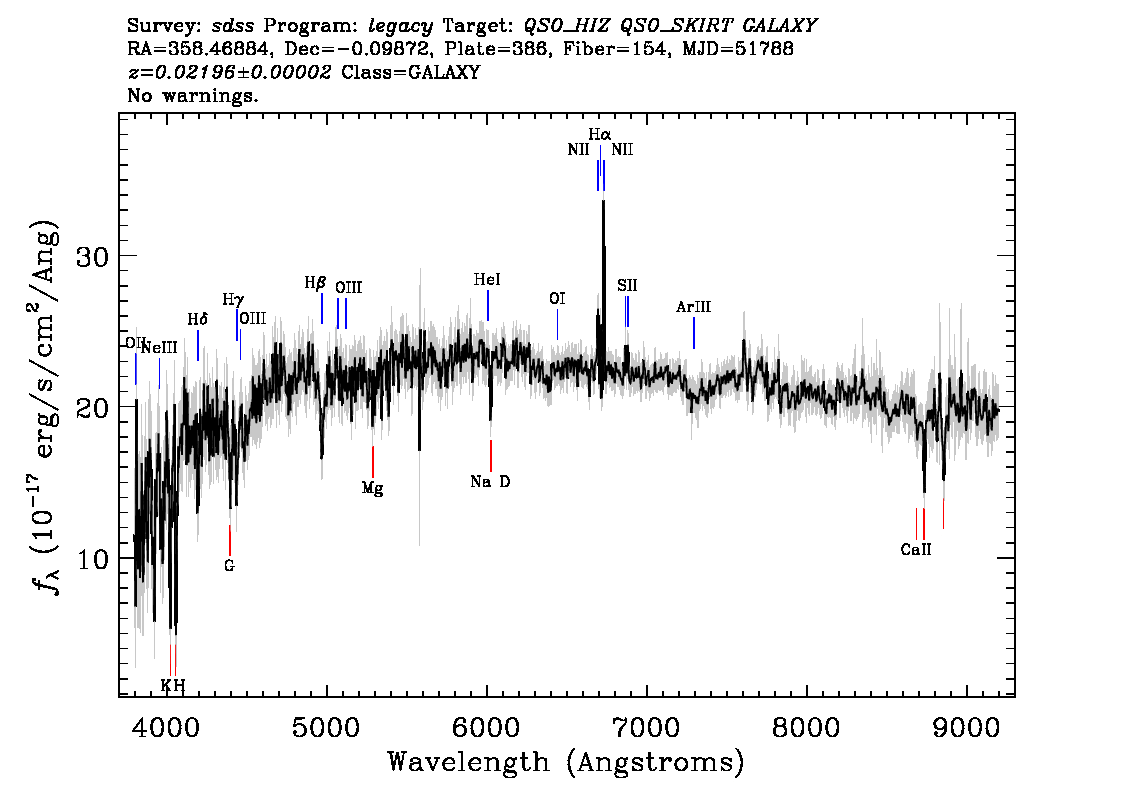}  \includegraphics[height = 0.197\textwidth]{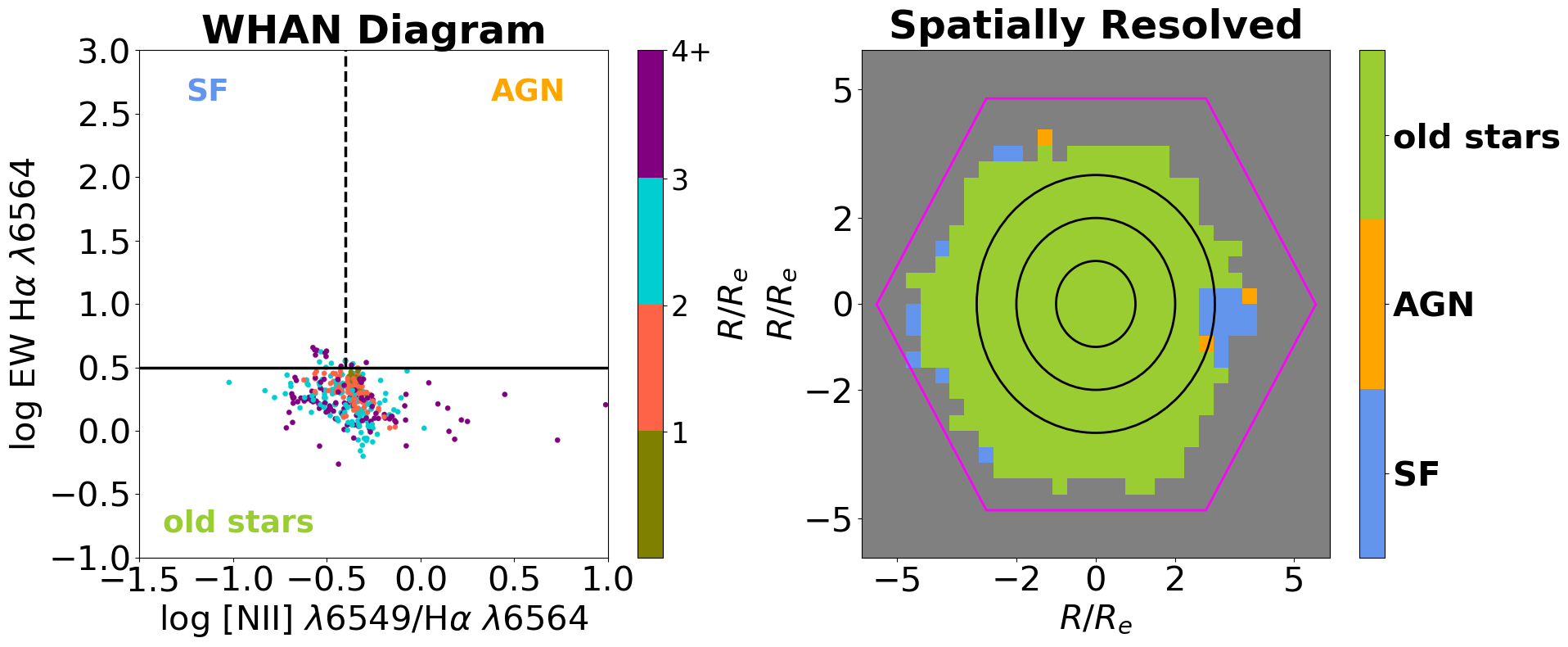} 
\bigskip
\bigskip
\ \includegraphics[height = 0.29\textwidth]{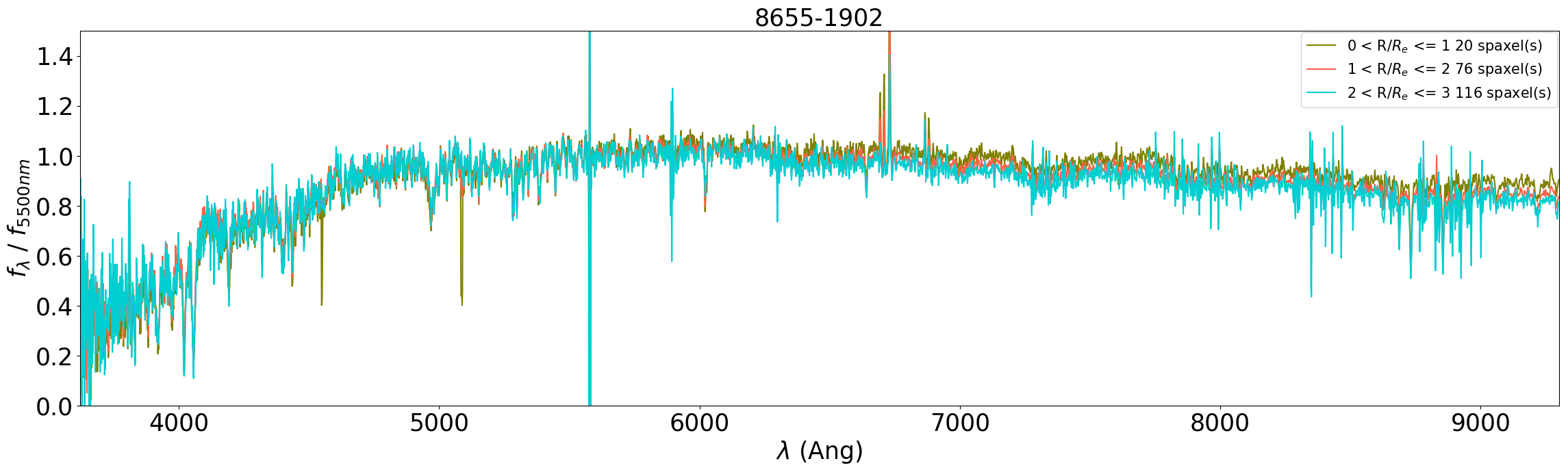}

\bigskip

\vfill\eject
\begin{center}[\textbf{MaNGA ID: 1-456935 | PLATE-IFU: 8931-12705}]
\end{center}
\includegraphics[height = 0.197\textwidth]{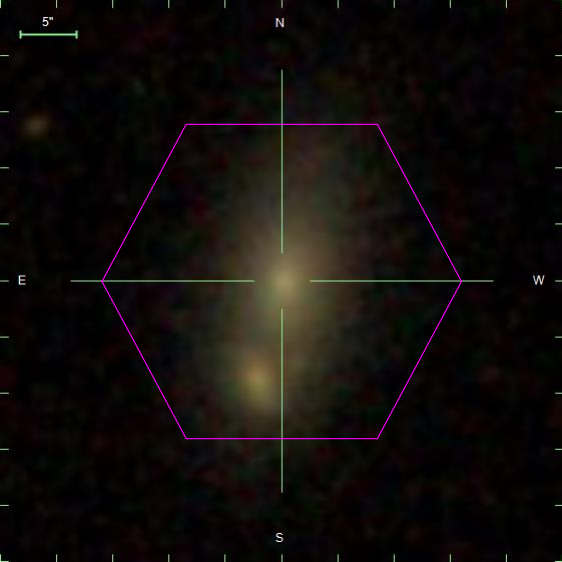}
\includegraphics[height = 0.197\textwidth]{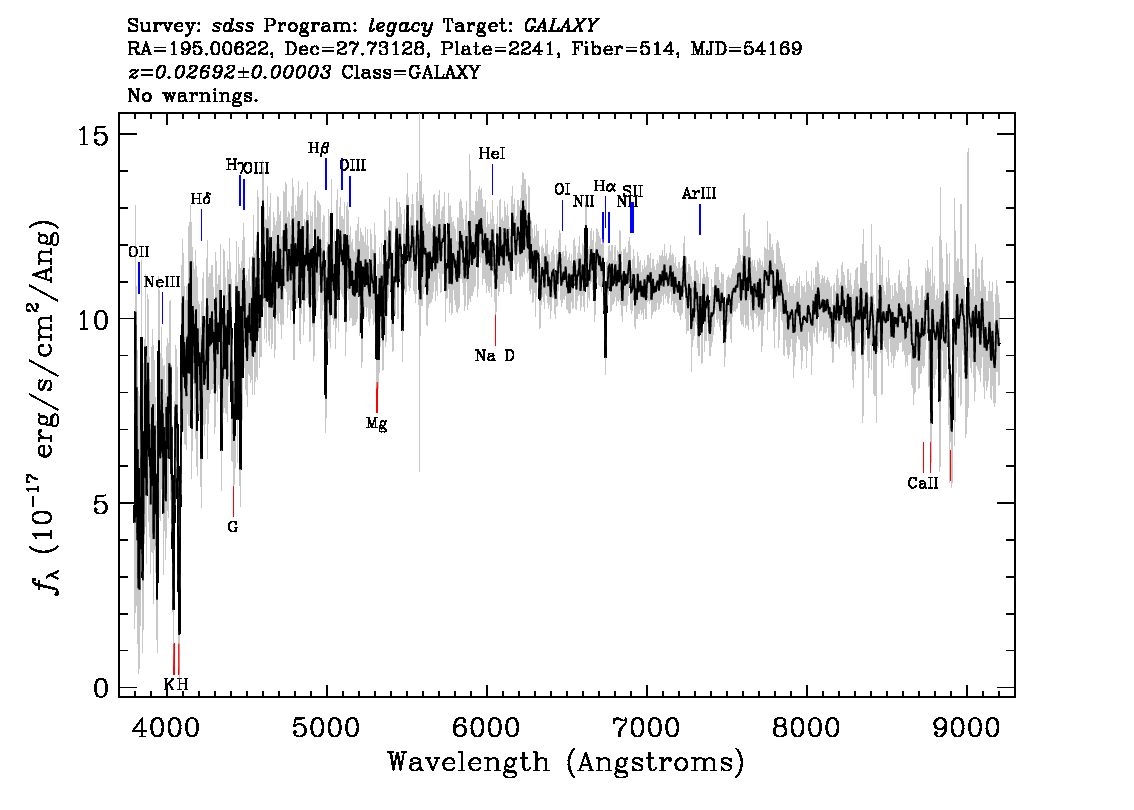}  \includegraphics[height = 0.197\textwidth]{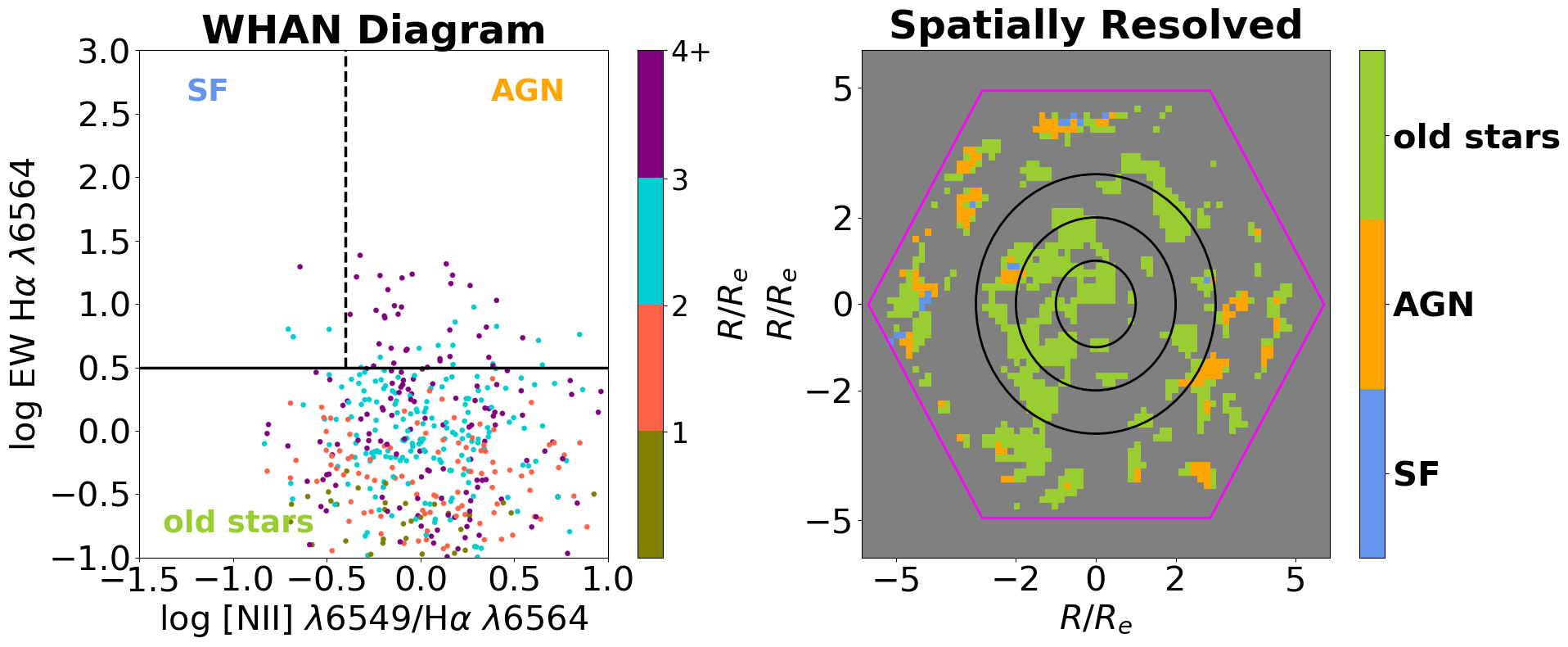} 
\bigskip
\bigskip
\ \includegraphics[height = 0.29\textwidth]{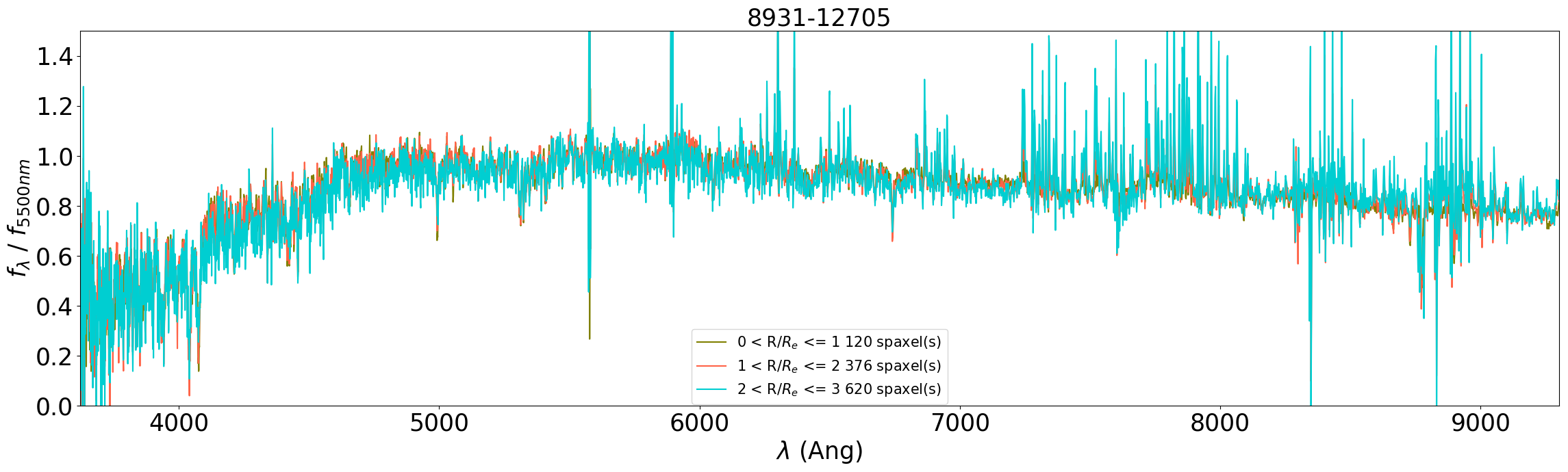}

\bigskip
\bigskip
\bigskip
\bigskip

\begin{center}[\textbf{MaNGA ID: 1-456434 | PLATE-IFU: 8931-3701}]
\end{center}
\includegraphics[height = 0.197\textwidth]{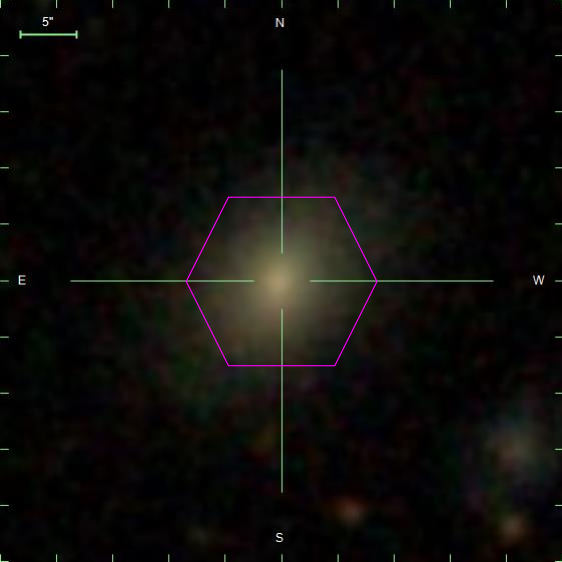}
\includegraphics[height = 0.197\textwidth]{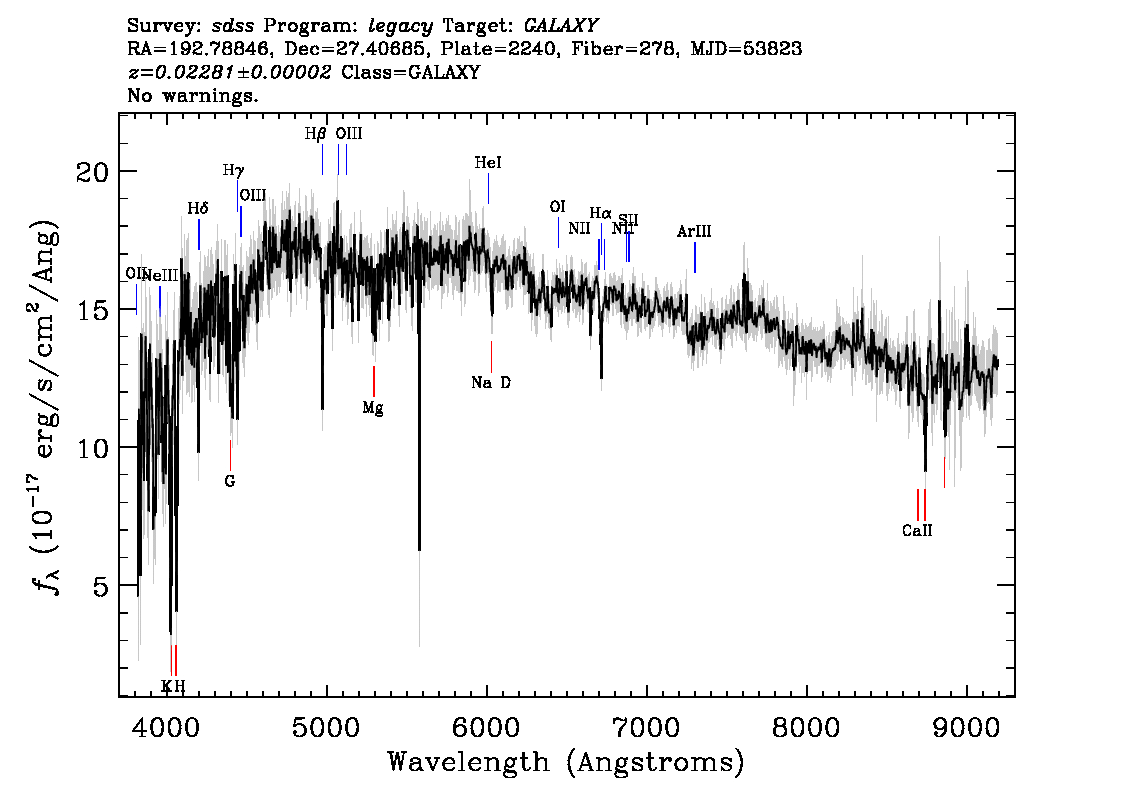}  \includegraphics[height = 0.197\textwidth]{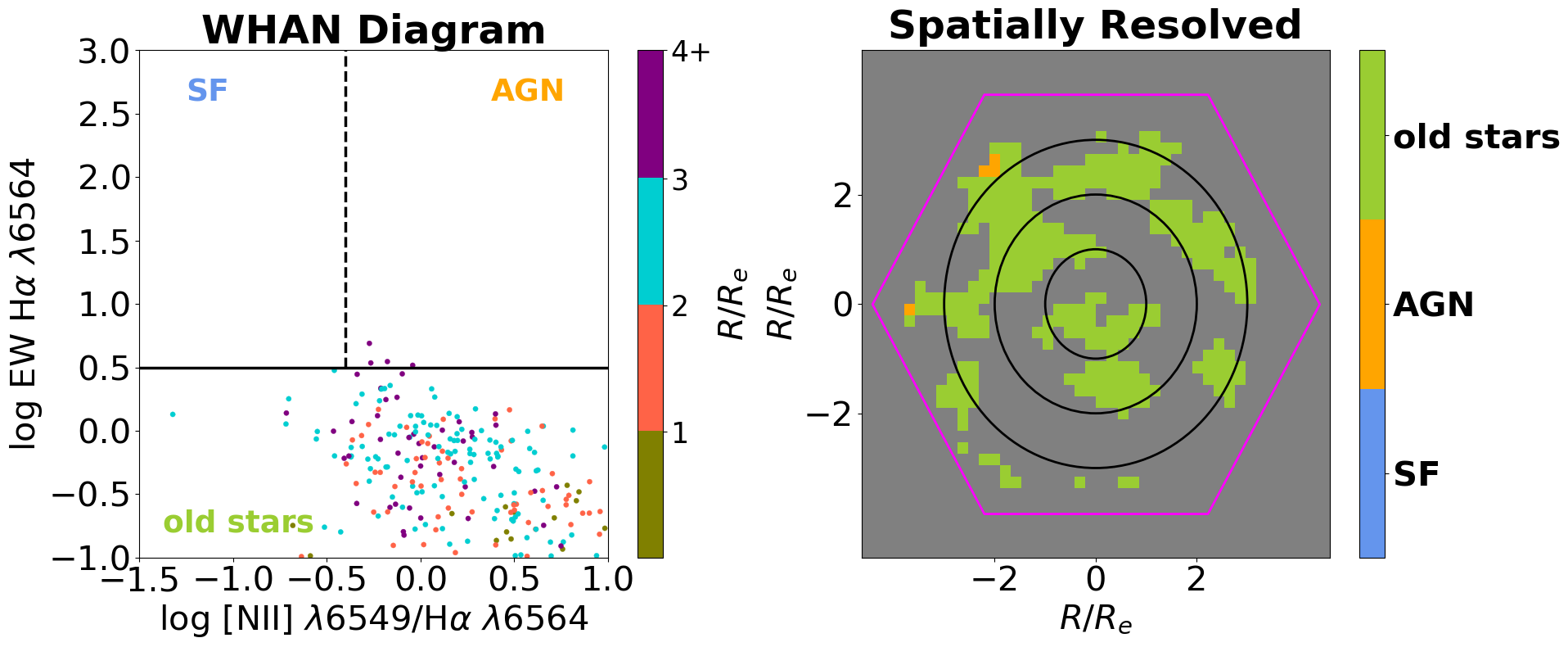} 
\bigskip
\bigskip
\ \includegraphics[height = 0.29\textwidth]{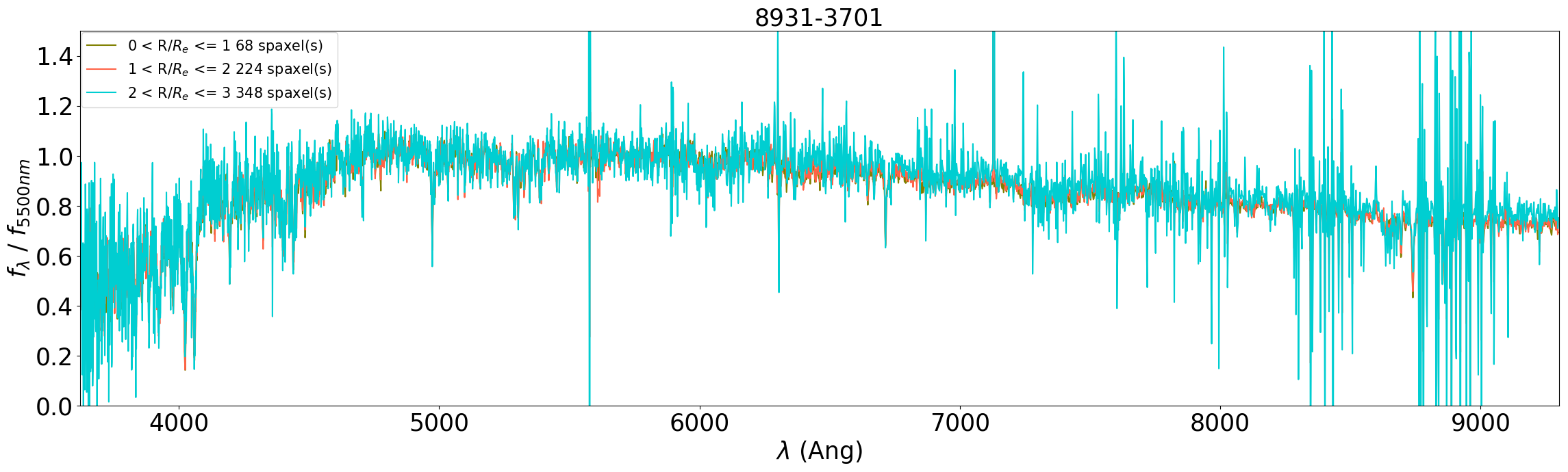}

\bigskip

\vfill\eject
\begin{center}[\textbf{MaNGA ID: 1-456380 | PLATE-IFU: 8934-3704}]
\end{center}
\includegraphics[height = 0.197\textwidth]{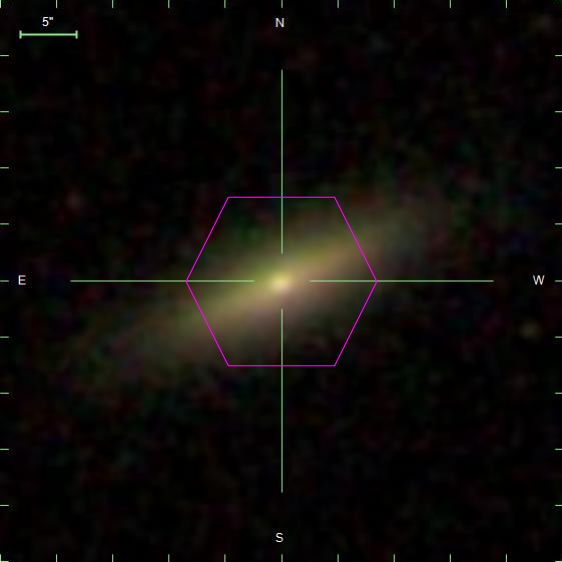}
\includegraphics[height = 0.197\textwidth]{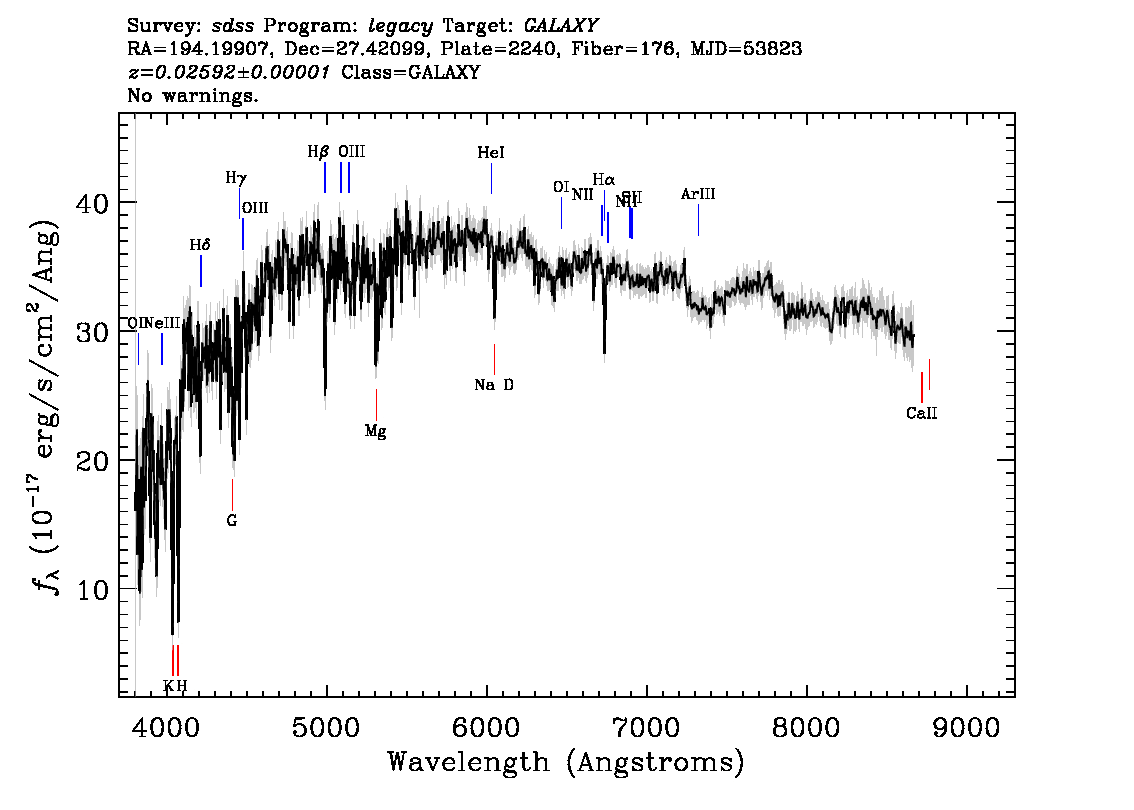}  \includegraphics[height = 0.197\textwidth]{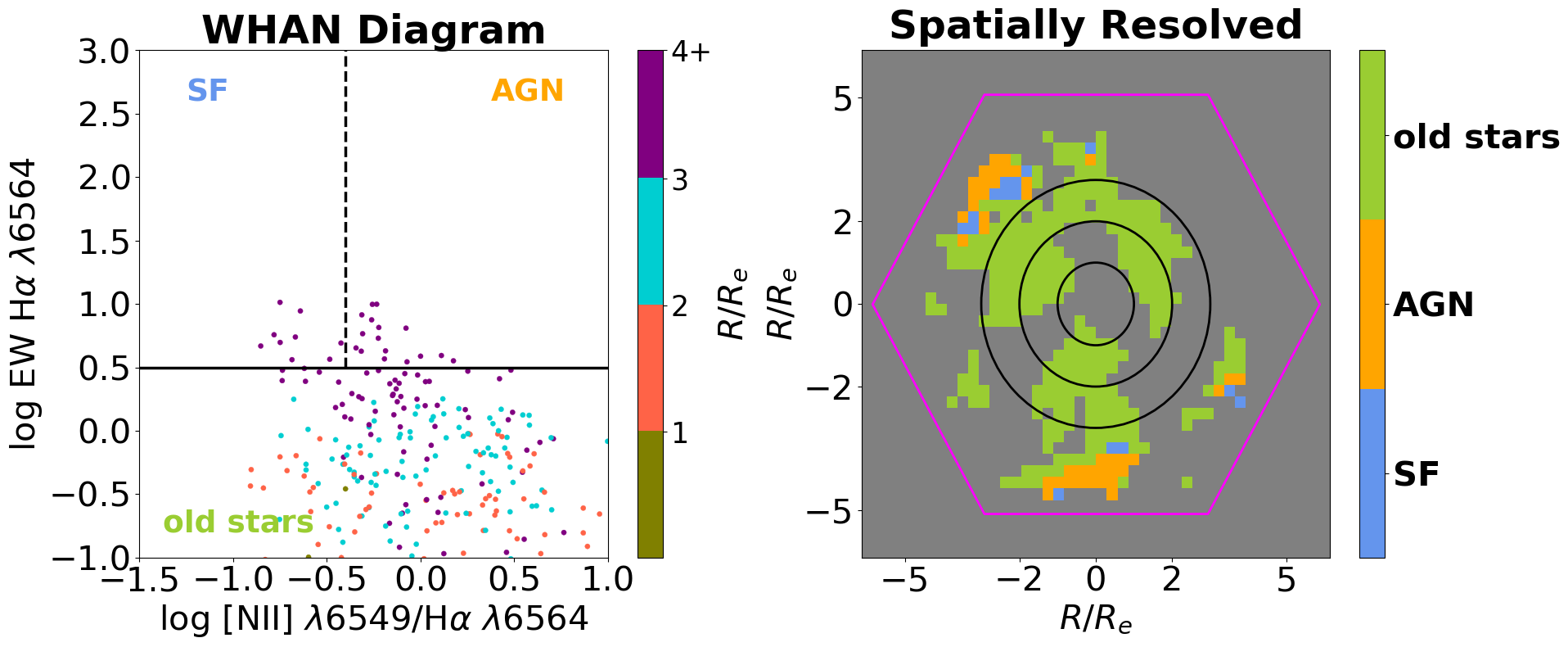} 
\bigskip
\bigskip
\ \includegraphics[height = 0.29\textwidth]{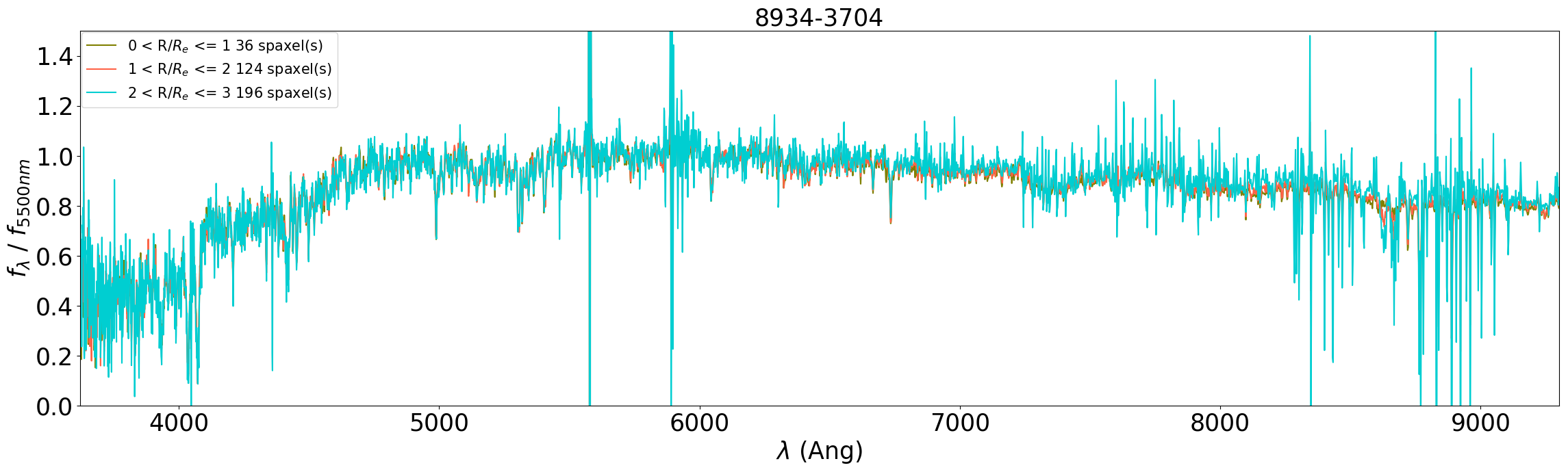}

\bigskip
\bigskip
\bigskip
\bigskip

\begin{center}[\textbf{MaNGA ID: 1-457004 | PLATE-IFU: 8934-9101}]
\end{center}
\includegraphics[height = 0.197\textwidth]{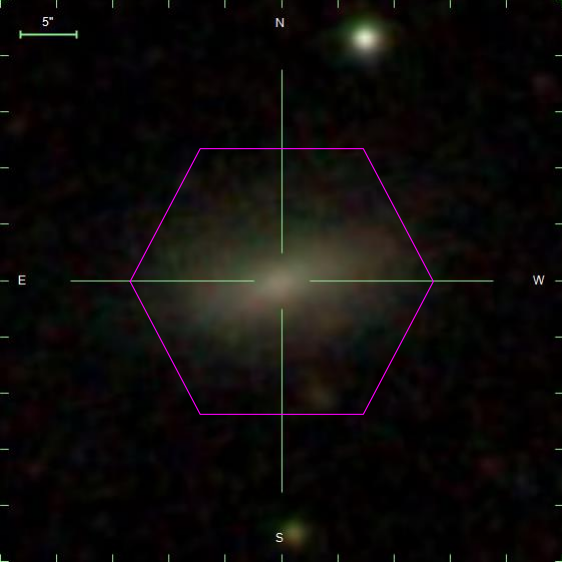}
\includegraphics[height = 0.197\textwidth]{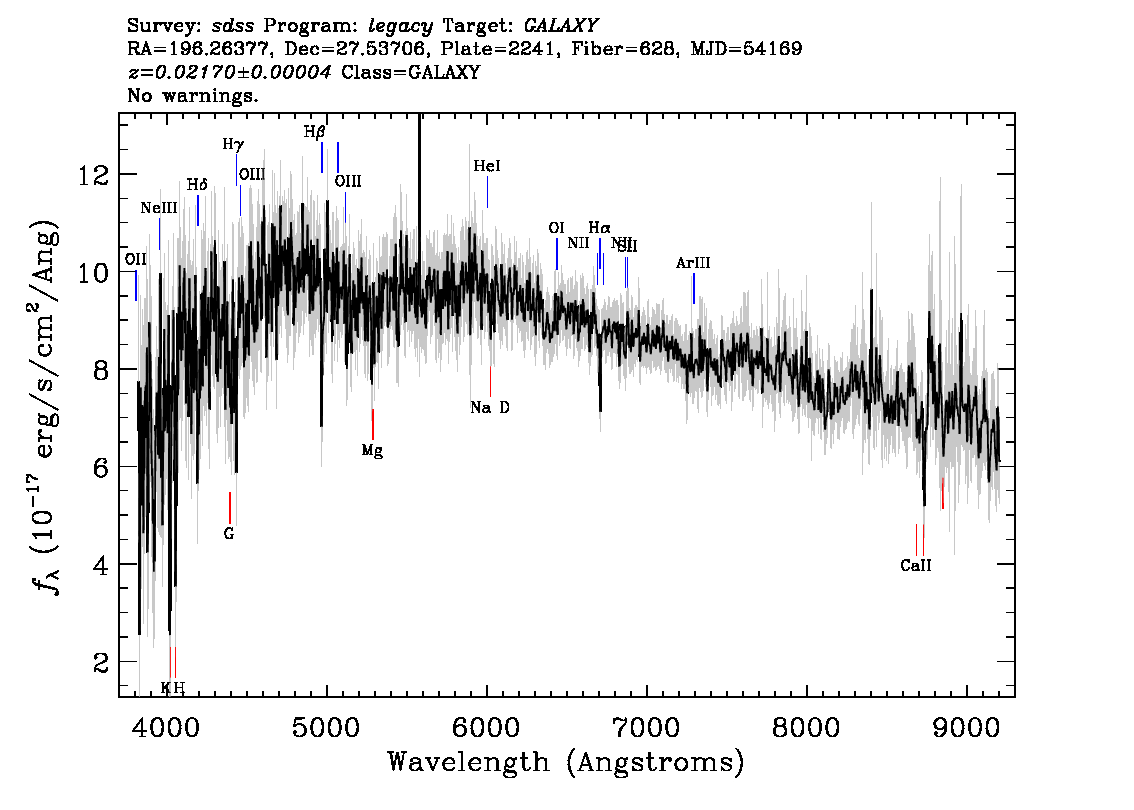}  \includegraphics[height = 0.197\textwidth]{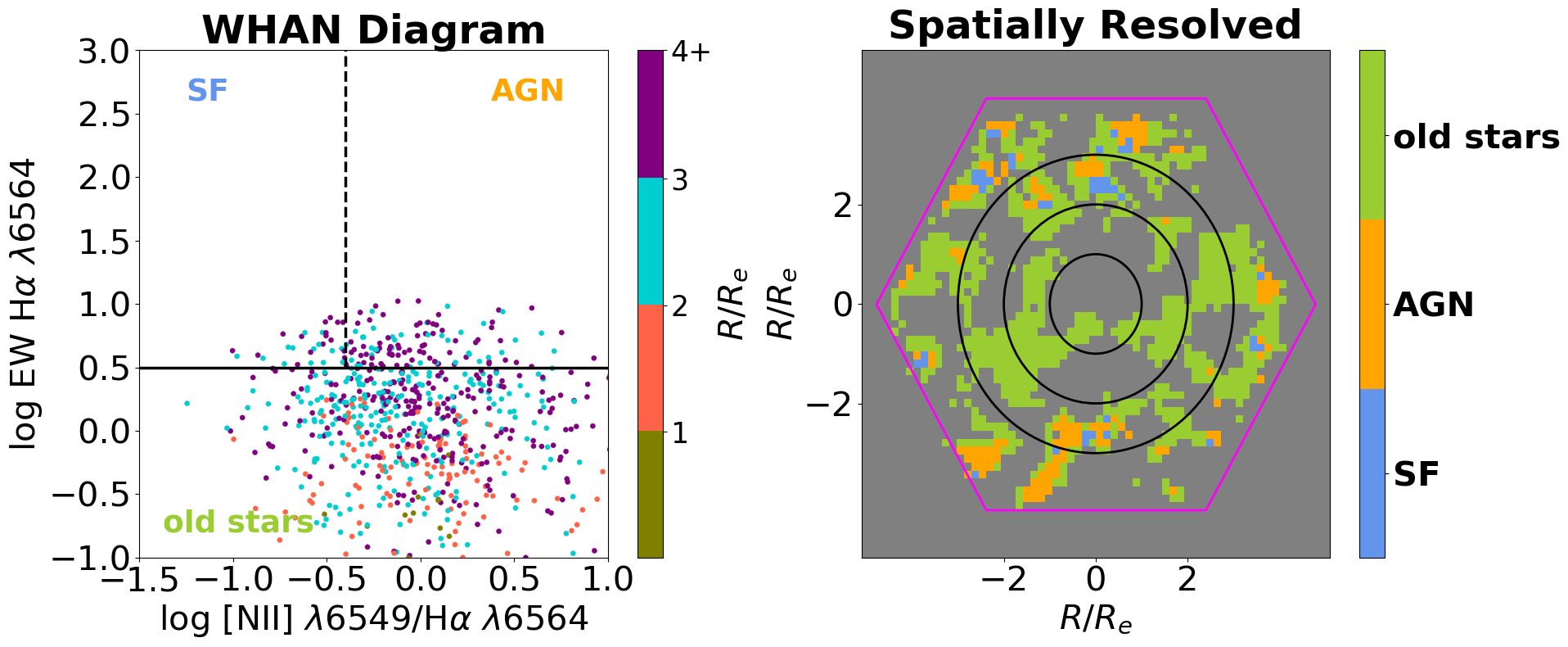} 
\bigskip
\bigskip
\ \includegraphics[height = 0.29\textwidth]{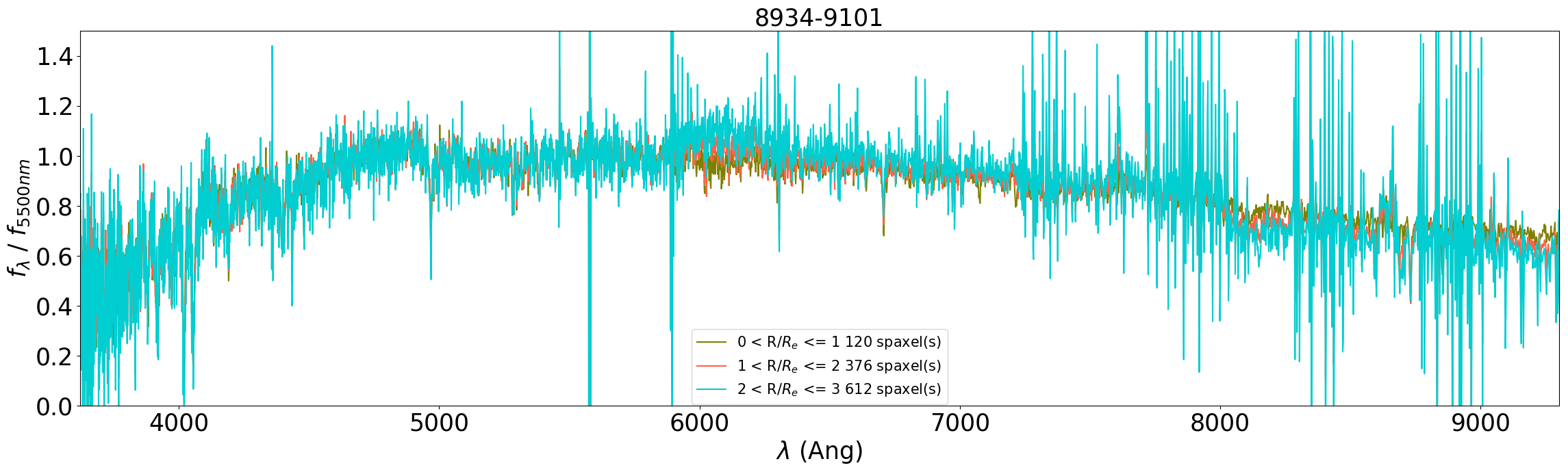}

\bigskip

\vfill\eject
\begin{center}[\textbf{MaNGA ID: 1-230177 | PLATE-IFU: 8942-6101}]
\end{center}
\includegraphics[height = 0.197\textwidth]{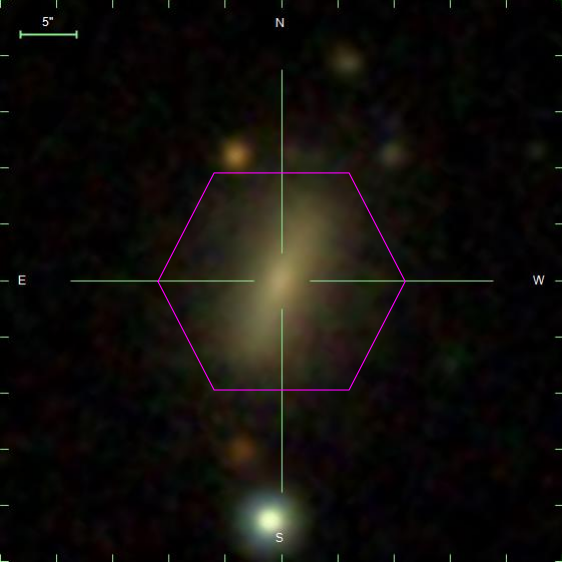}
\includegraphics[height = 0.197\textwidth]{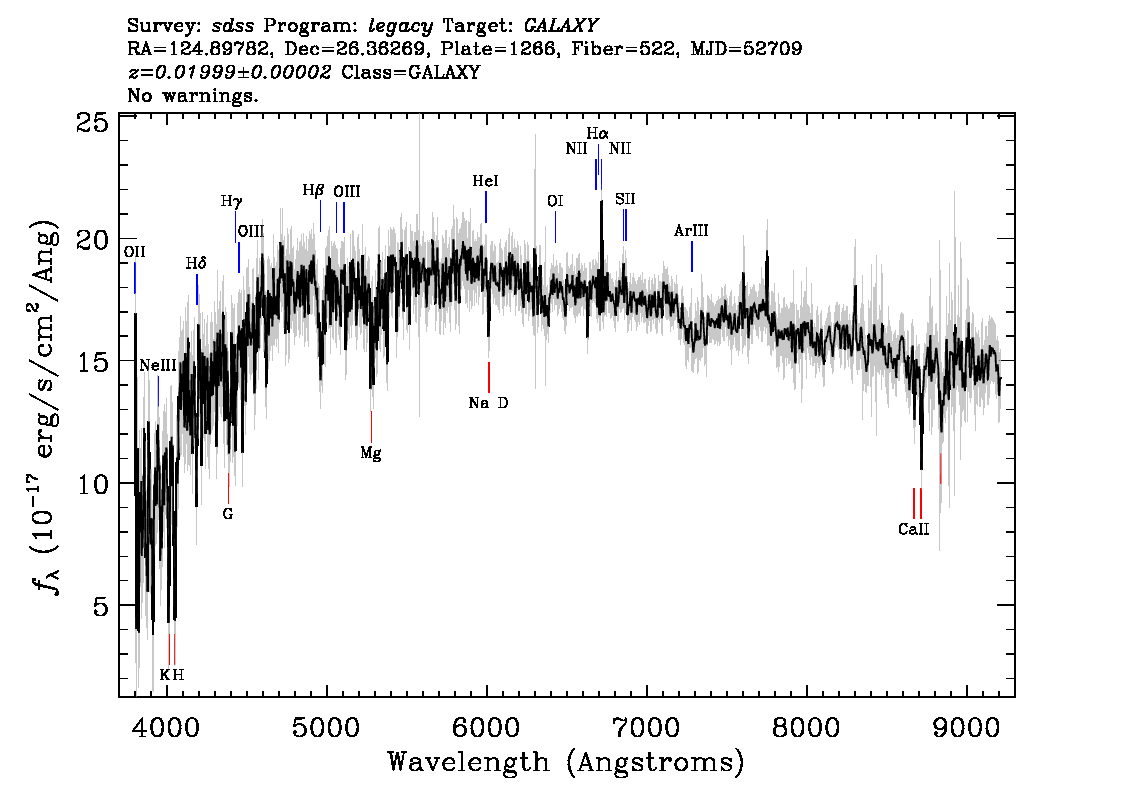}  \includegraphics[height = 0.197\textwidth]{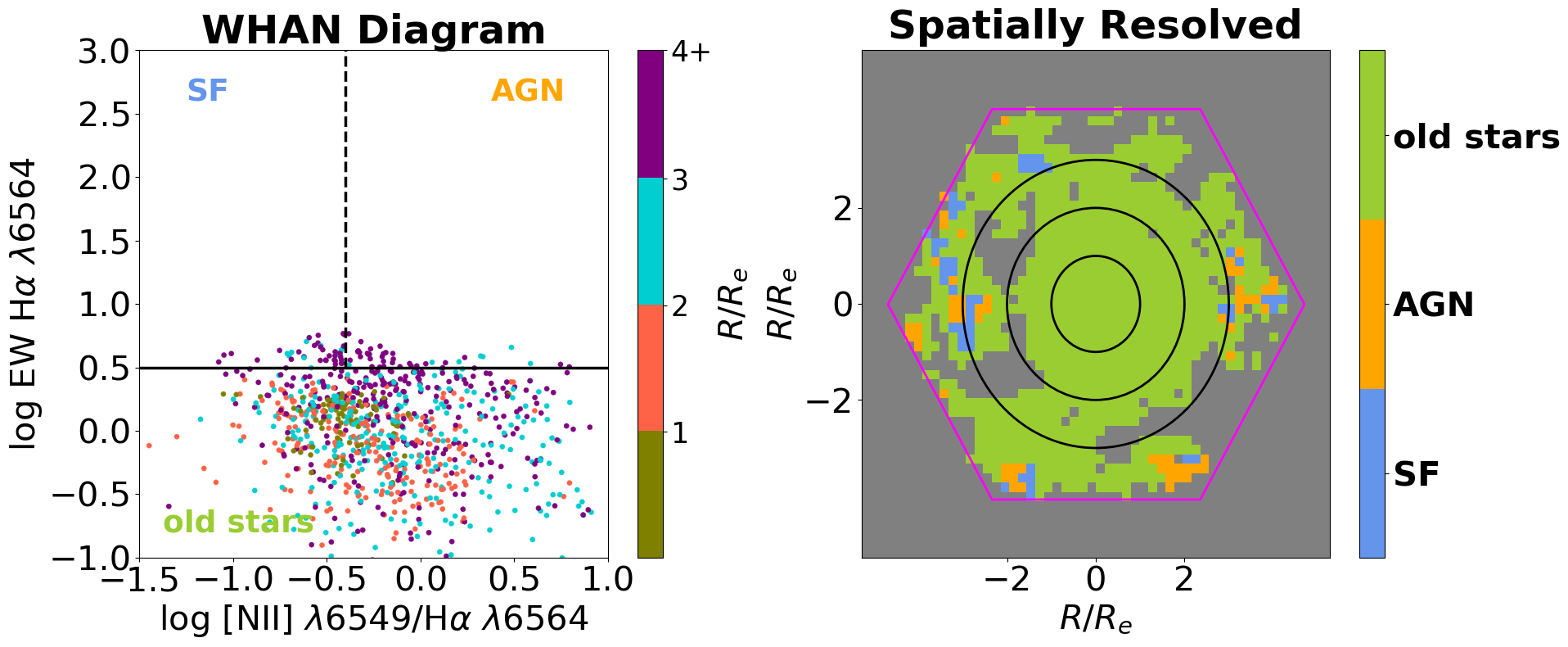} 
\bigskip
\bigskip
\ \includegraphics[height = 0.29\textwidth]{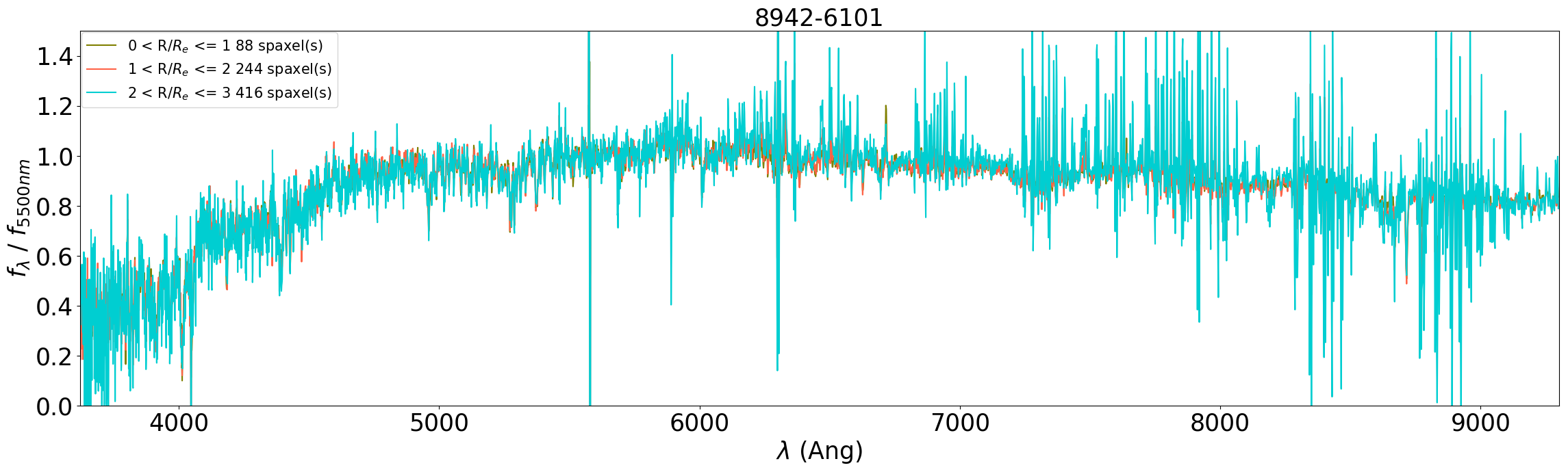}

\bigskip
\bigskip
\bigskip
\bigskip

\begin{center}[\textbf{MaNGA ID: 1-456635 | PLATE-IFU: 8949-3701}]
\end{center}
\includegraphics[height = 0.197\textwidth]{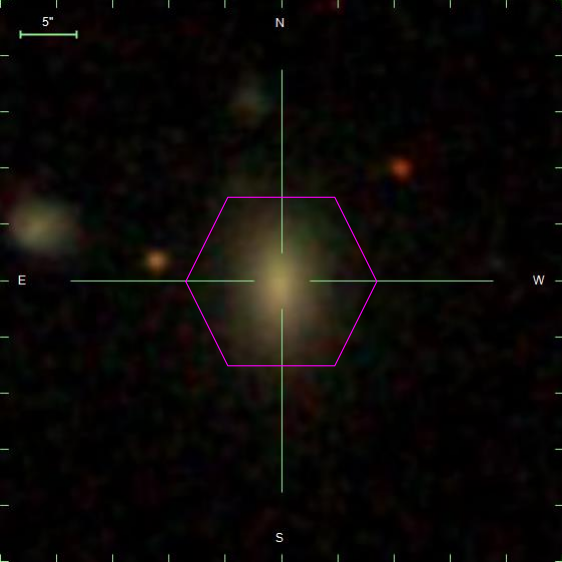}
\includegraphics[height = 0.197\textwidth]{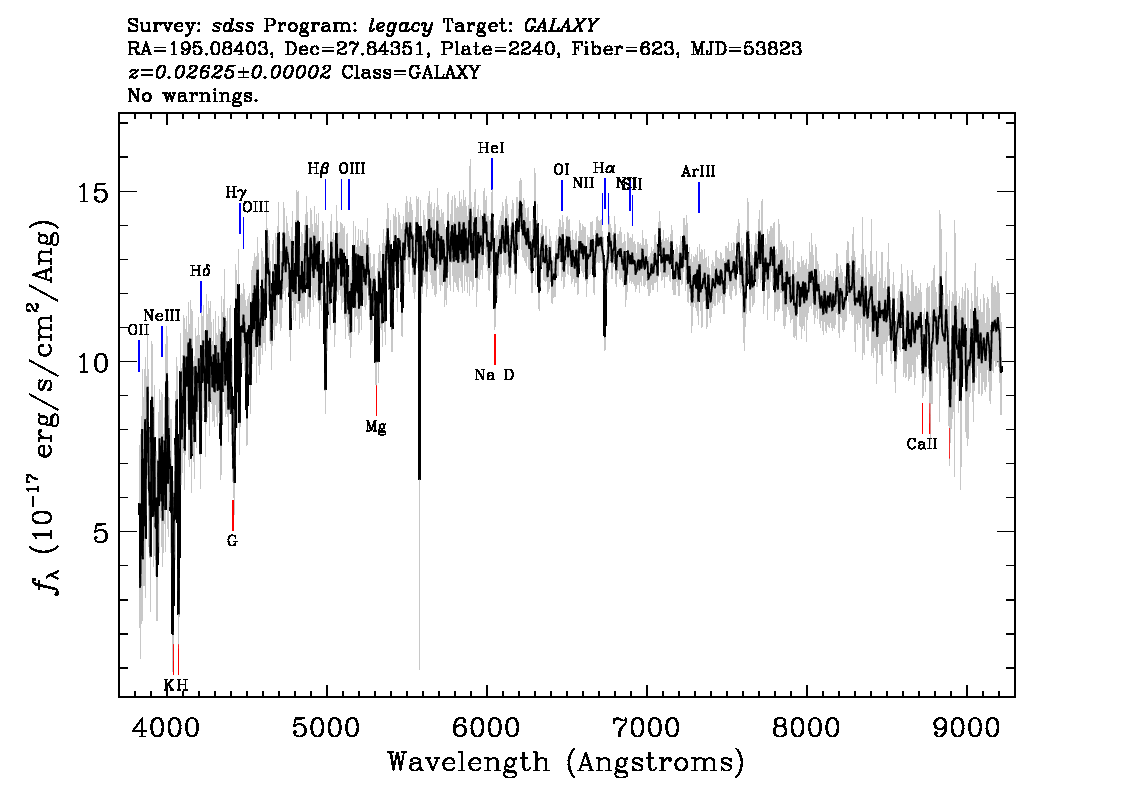}  \includegraphics[height = 0.197\textwidth]{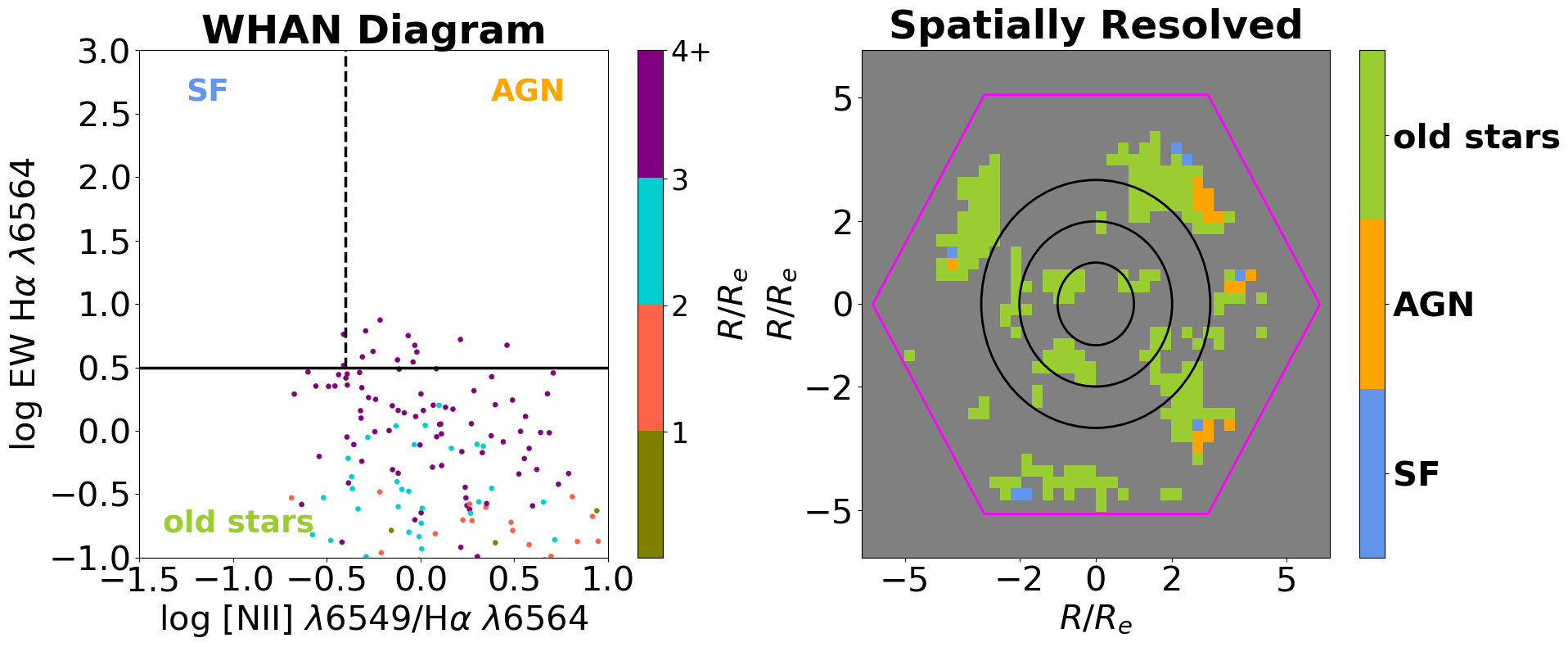} 
\bigskip
\bigskip
\ \includegraphics[height = 0.29\textwidth]{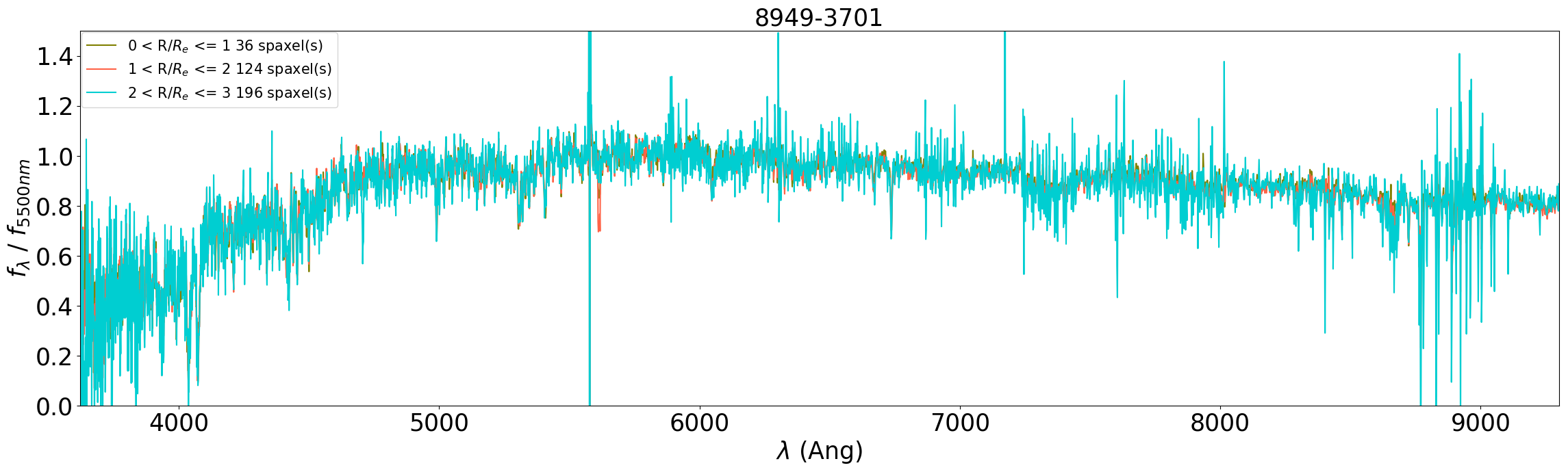}

\bigskip

\vfill\eject
\begin{center}[\textbf{MaNGA ID: 1-456505 | PLATE-IFU: 8950-3702}]
\end{center}
\includegraphics[height = 0.197\textwidth]{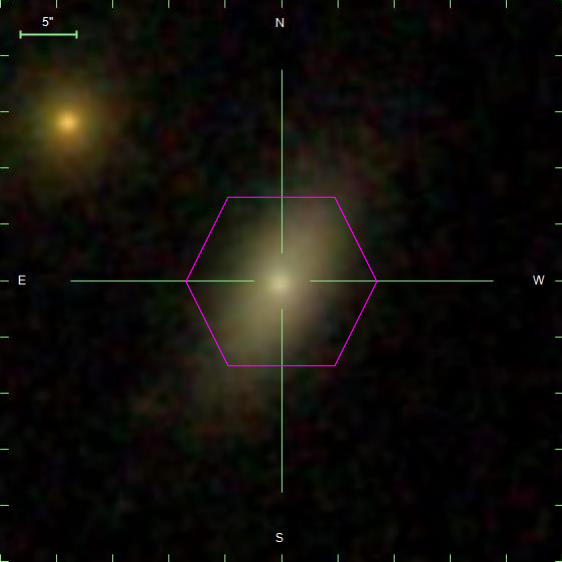}
\includegraphics[height = 0.197\textwidth]{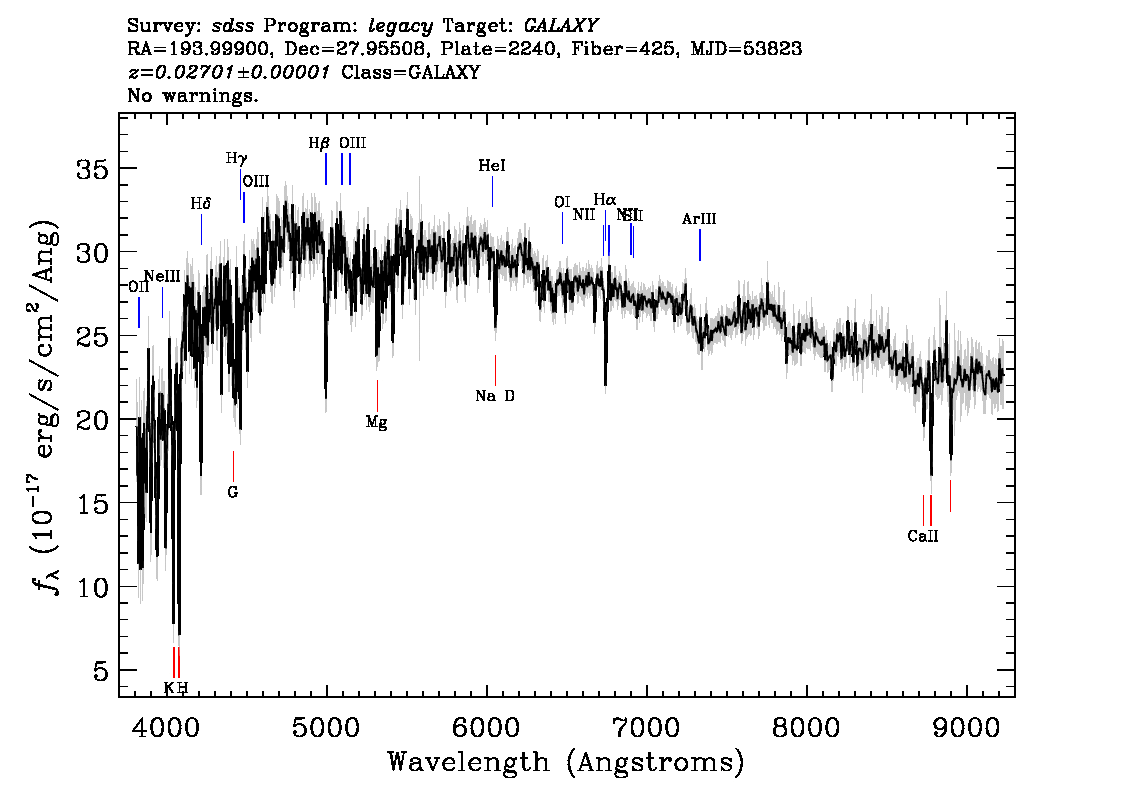}  \includegraphics[height = 0.197\textwidth]{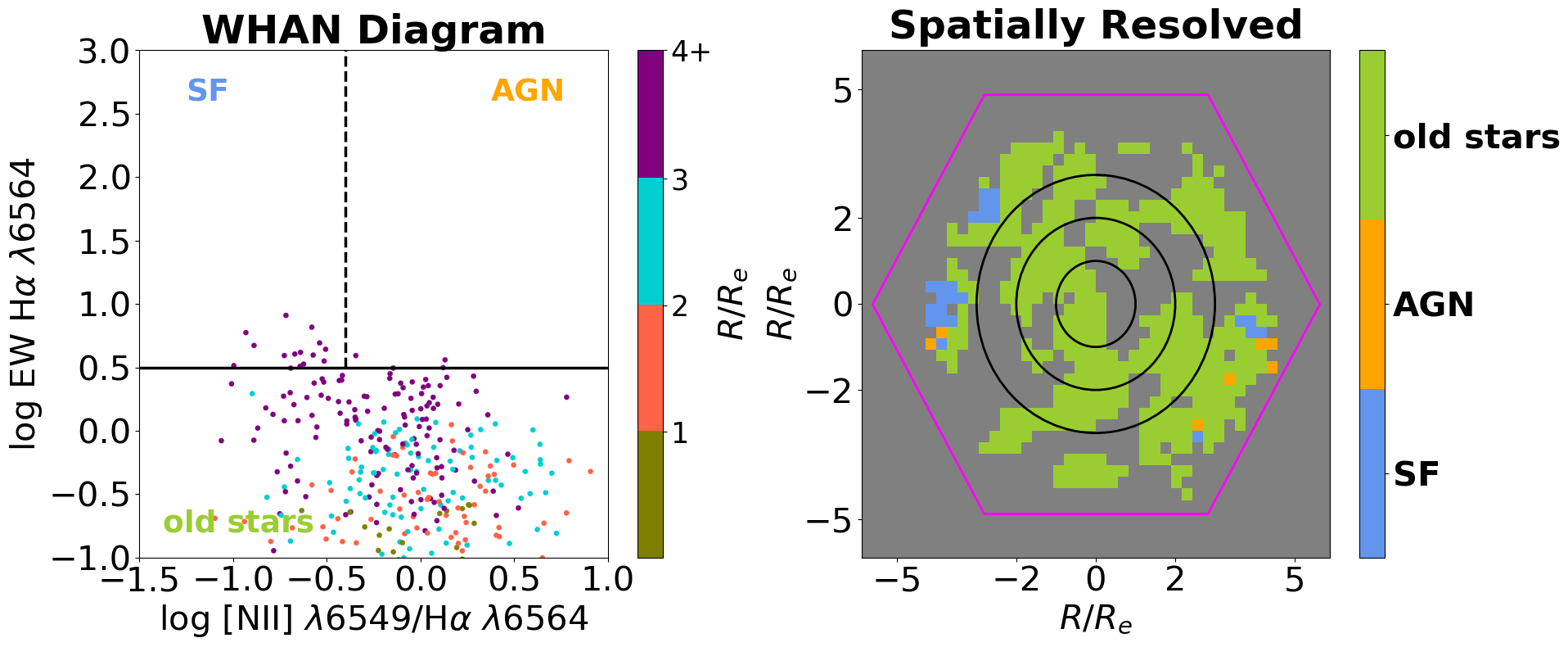} 
\bigskip
\bigskip
\ \includegraphics[height = 0.29\textwidth]{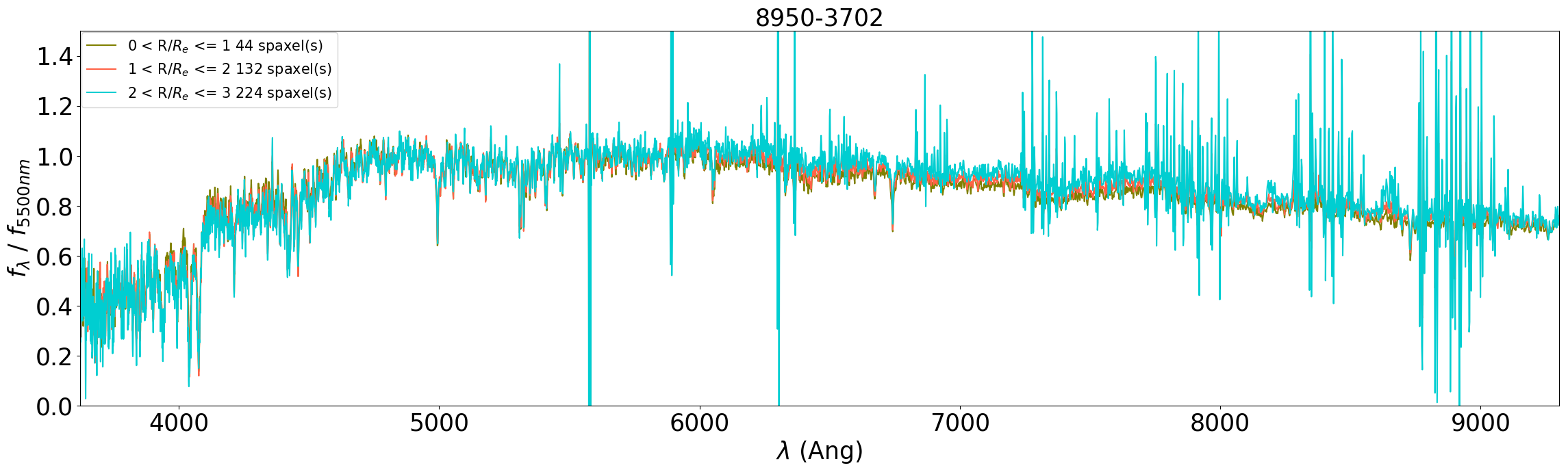}

\bigskip
\bigskip
\bigskip
\bigskip

\begin{center}[\textbf{MaNGA ID: 1-456744 | PLATE-IFU: 8950-3704}]
\end{center}
\includegraphics[height = 0.197\textwidth]{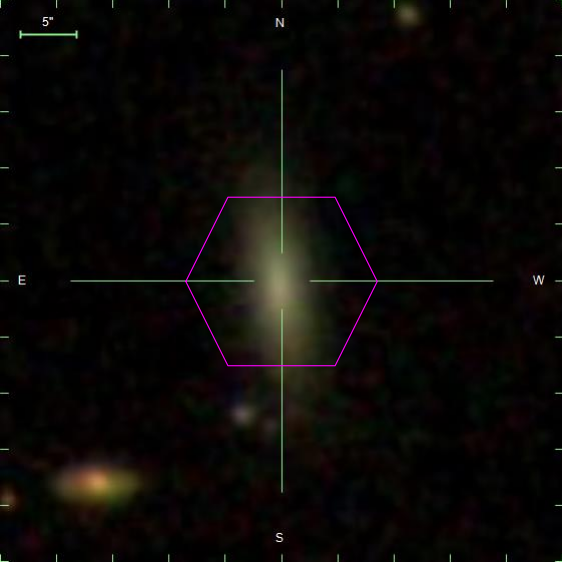}
\includegraphics[height = 0.197\textwidth]{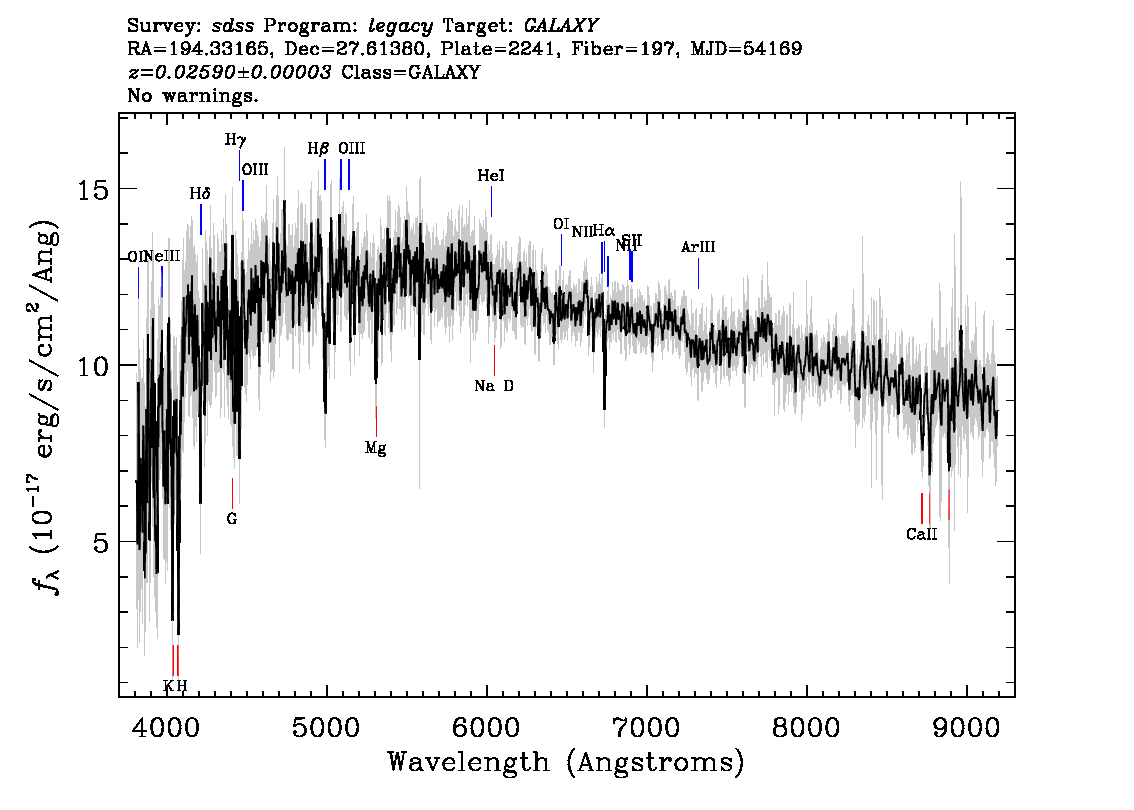}  \includegraphics[height = 0.197\textwidth]{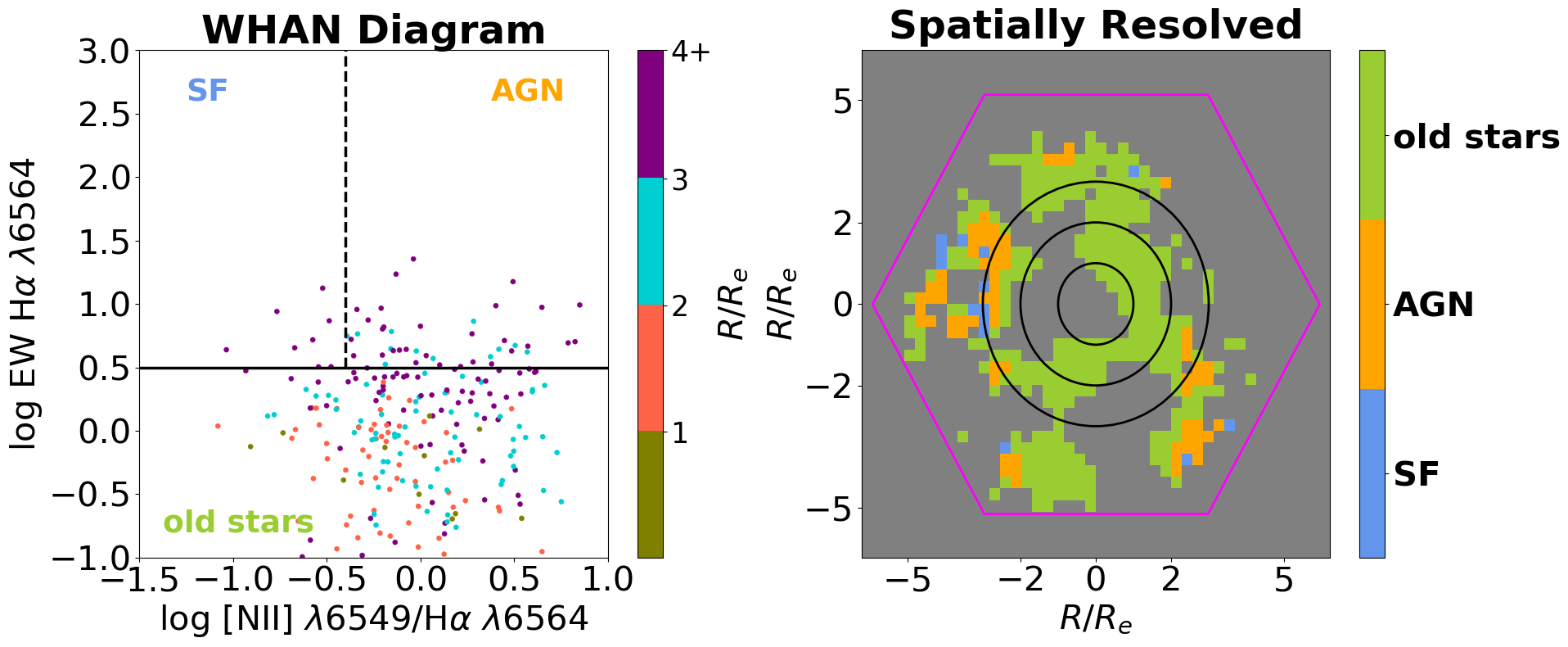} 
\bigskip
\bigskip
\ \includegraphics[height = 0.29\textwidth]{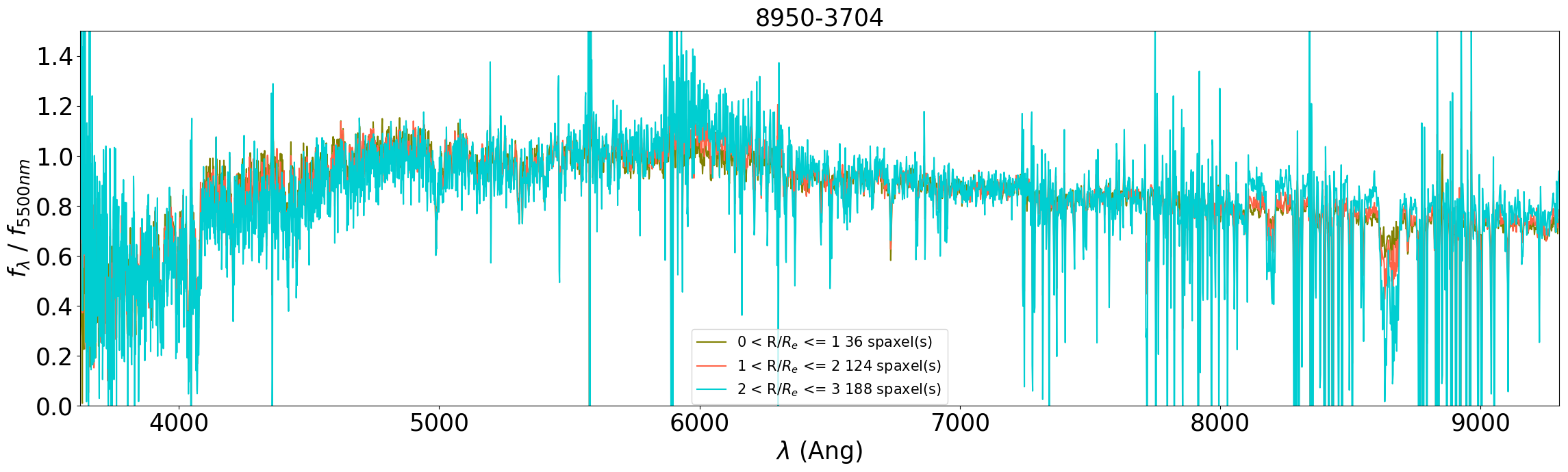}

\bigskip

\vfill\eject
\begin{center}[\textbf{MaNGA ID: | PLATE-IFU: 8950-6101}]
\end{center}
\includegraphics[height = 0.197\textwidth]{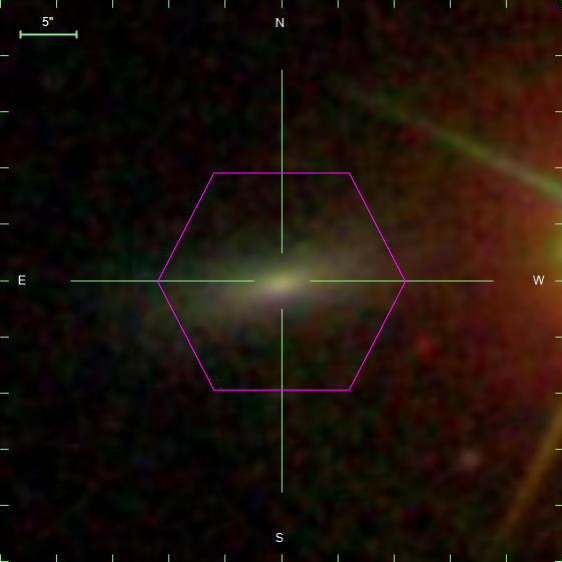}
\includegraphics[height = 0.197\textwidth]{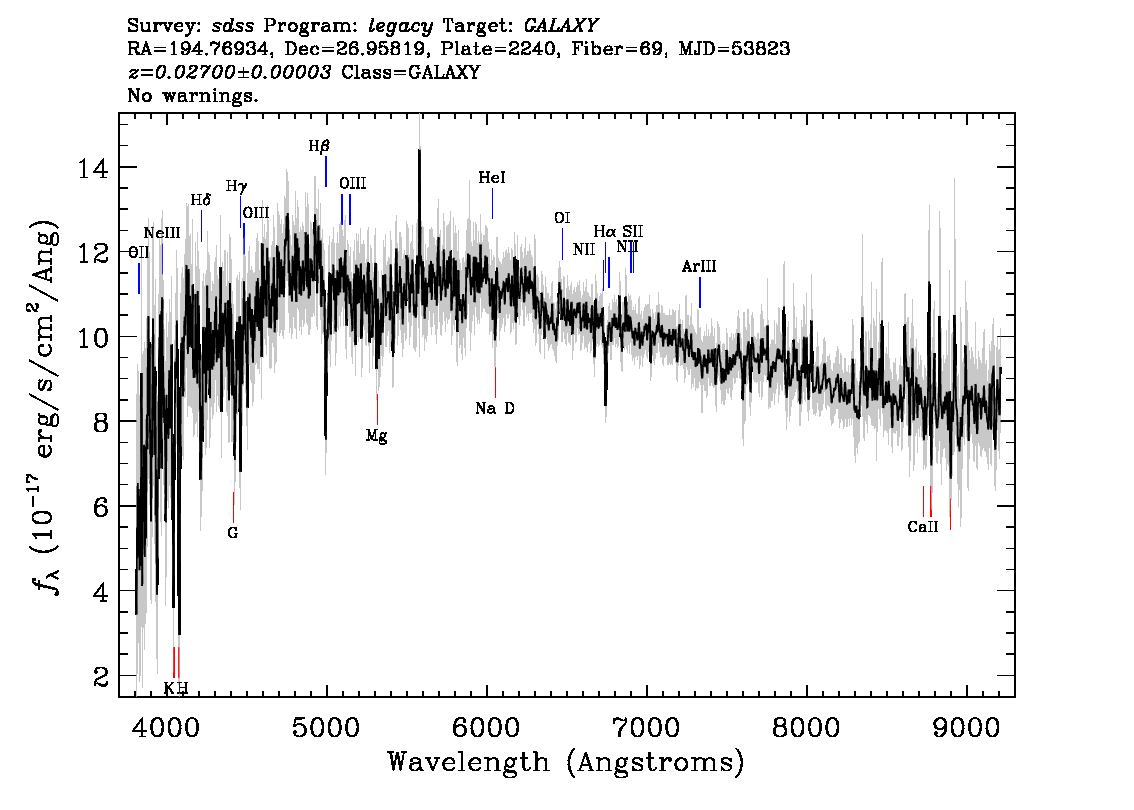}  \includegraphics[height = 0.197\textwidth]{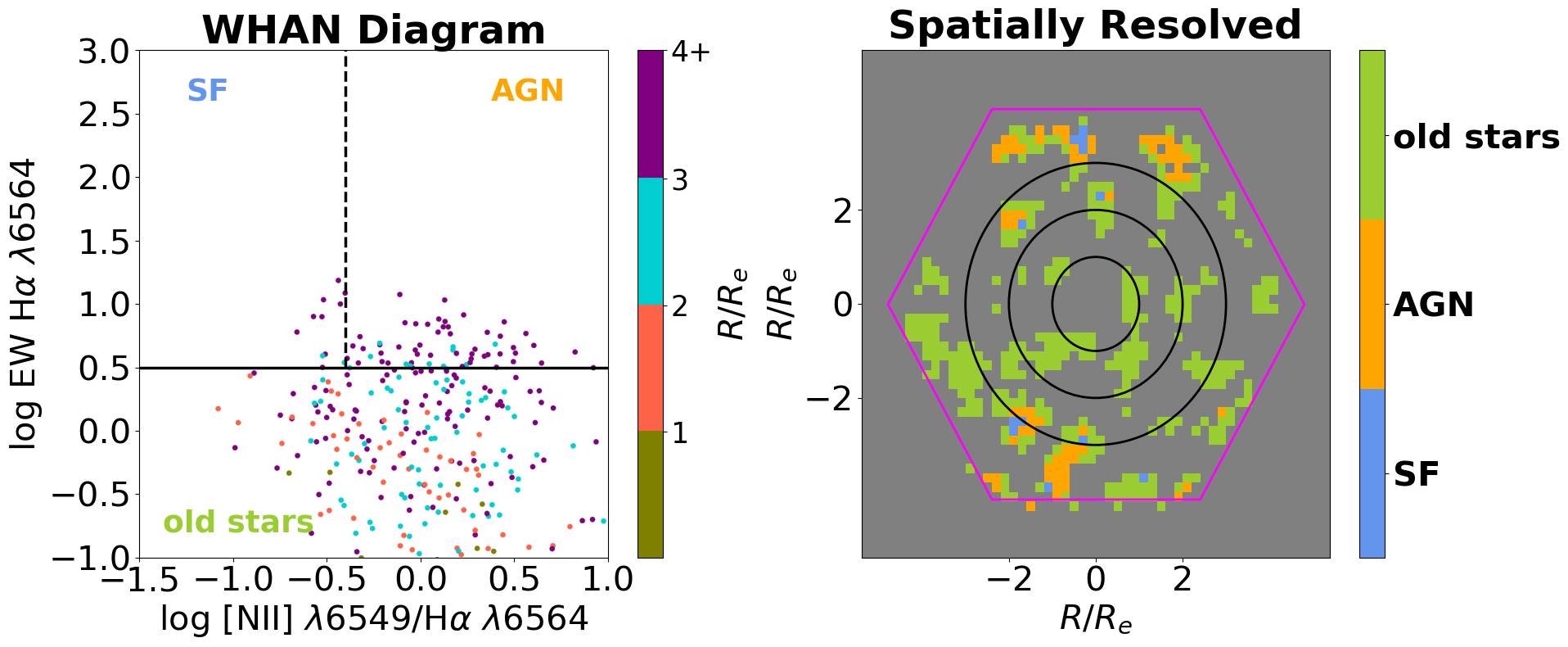} 
\bigskip
\bigskip
\ \includegraphics[height = 0.29\textwidth]{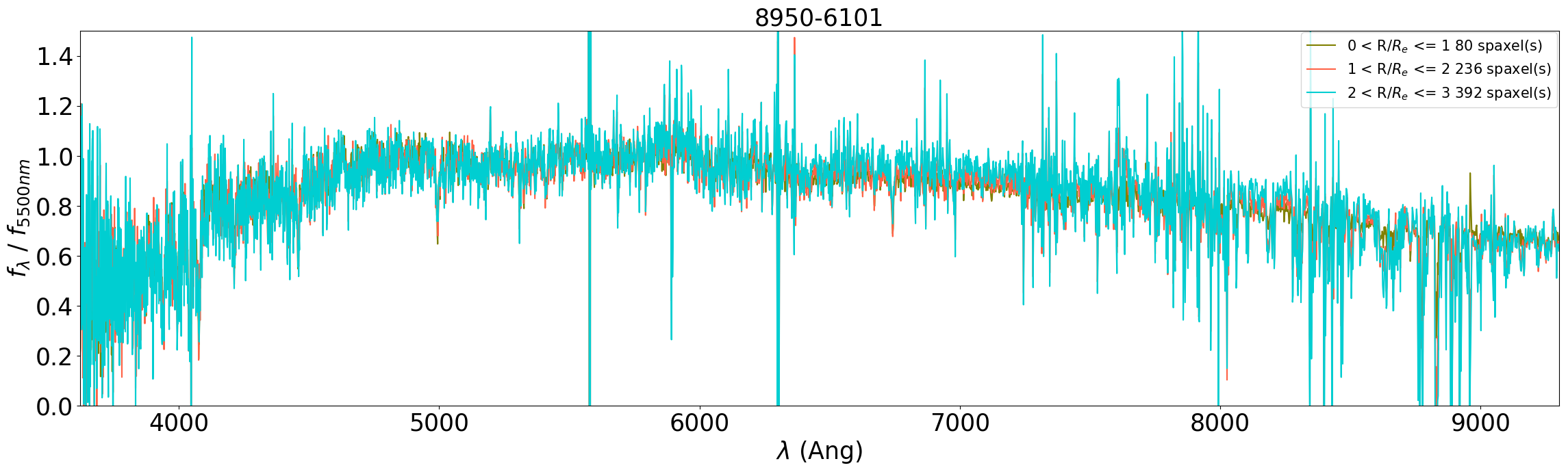}

\bigskip
\bigskip
\bigskip
\bigskip

\begin{center}[\textbf{MaNGA ID: 1-210114 | PLATE-IFU: 8979-1902}]
\end{center}
\includegraphics[height = 0.197\textwidth]{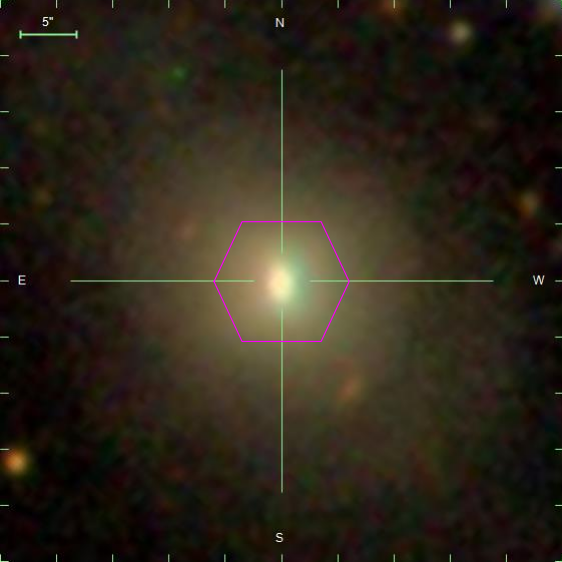}
\includegraphics[height = 0.197\textwidth]{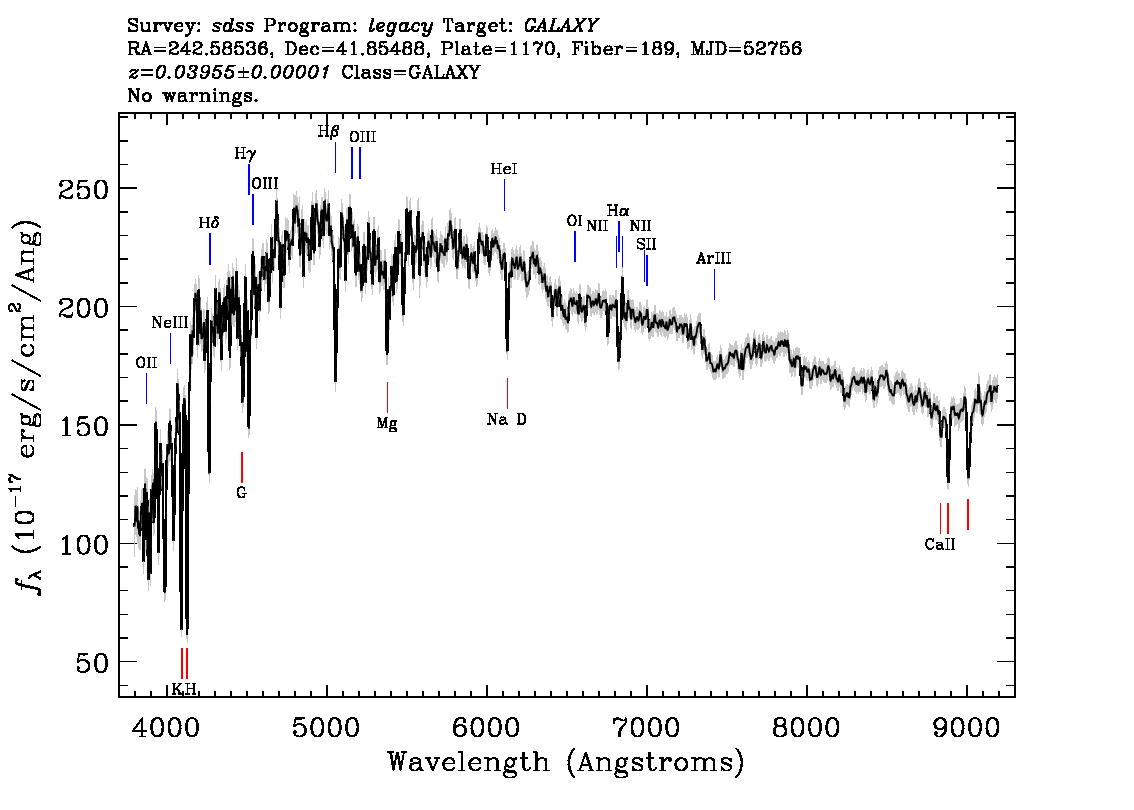}  \includegraphics[height = 0.197\textwidth]{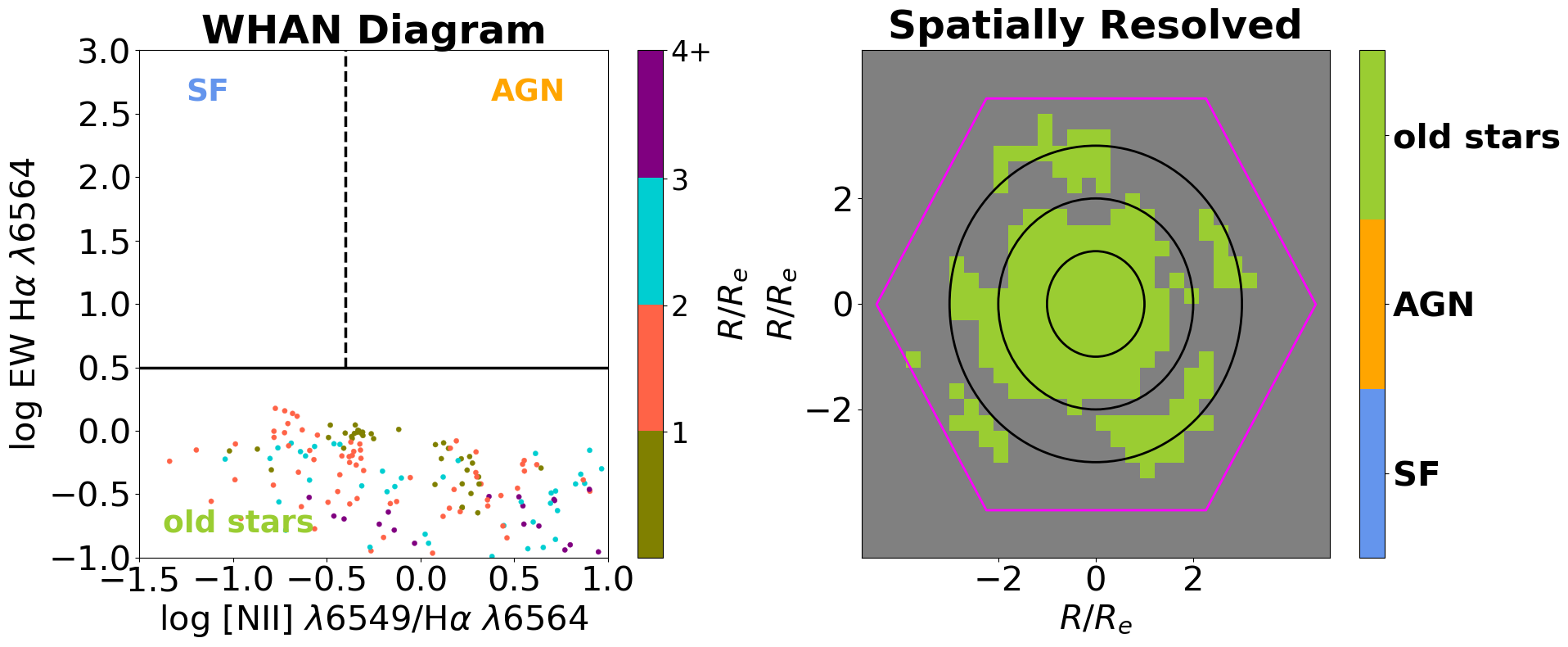}
\bigskip
\bigskip
\ \includegraphics[height = 0.29\textwidth]{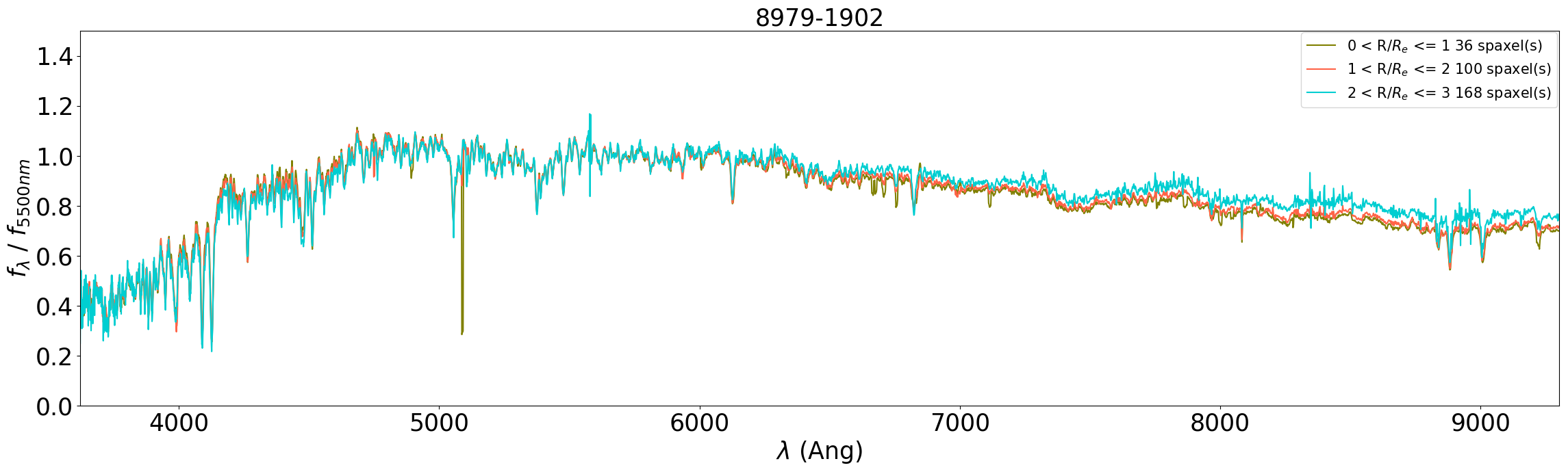}

\bigskip

\vfill\eject
\begin{center}[\textbf{MaNGA ID: 1-135235 | PLATE-IFU: 9029-1901}]
\end{center}
\includegraphics[height = 0.197\textwidth]{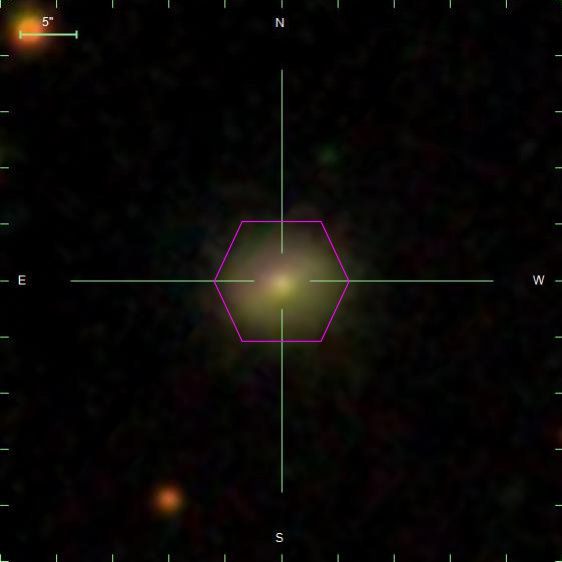}
\includegraphics[height = 0.197\textwidth]{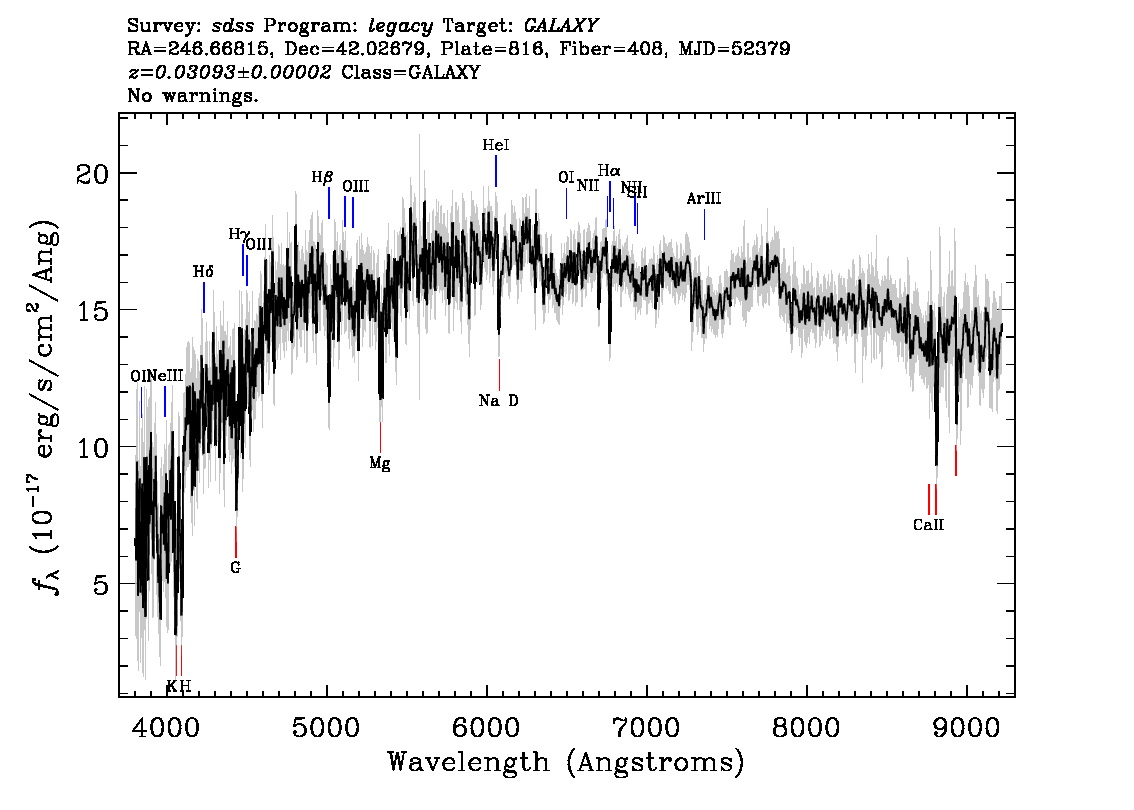}  \includegraphics[height = 0.197\textwidth]{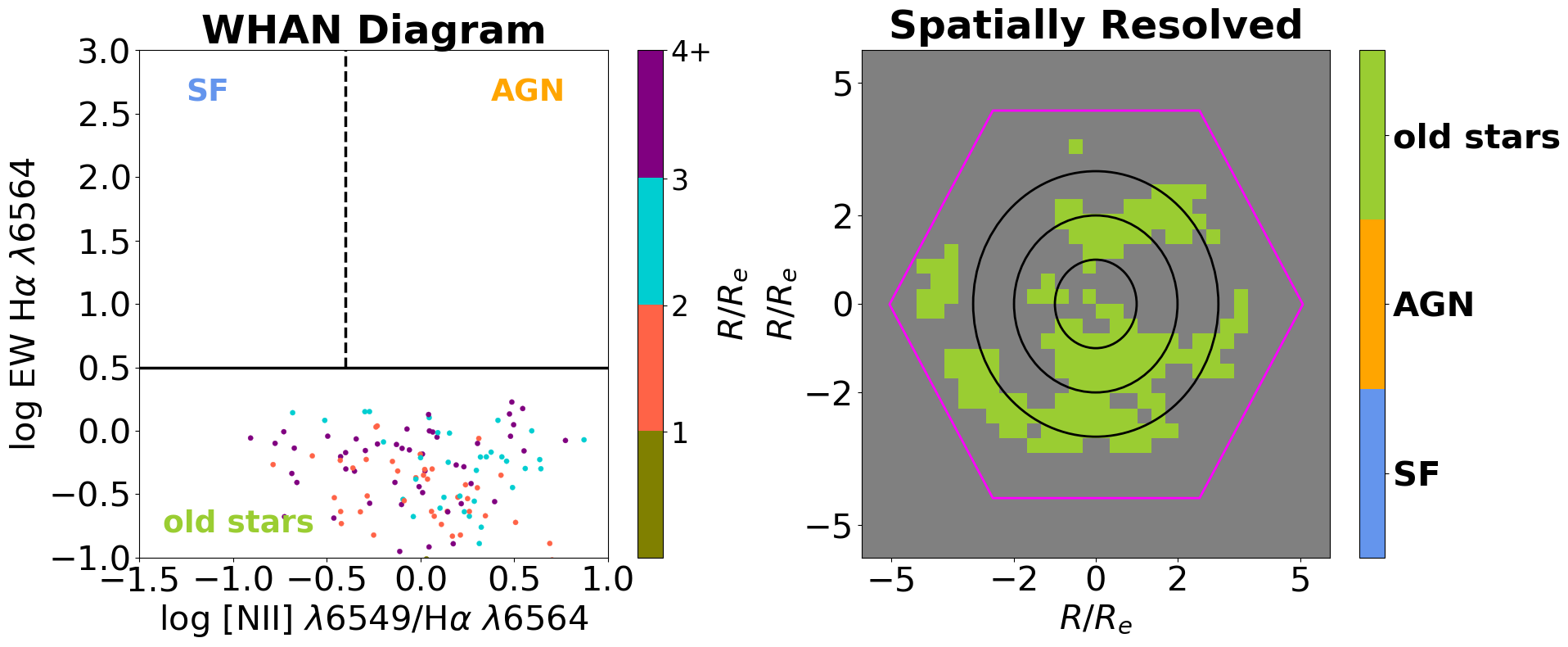} 
\bigskip
\bigskip
\ \includegraphics[height = 0.29\textwidth]{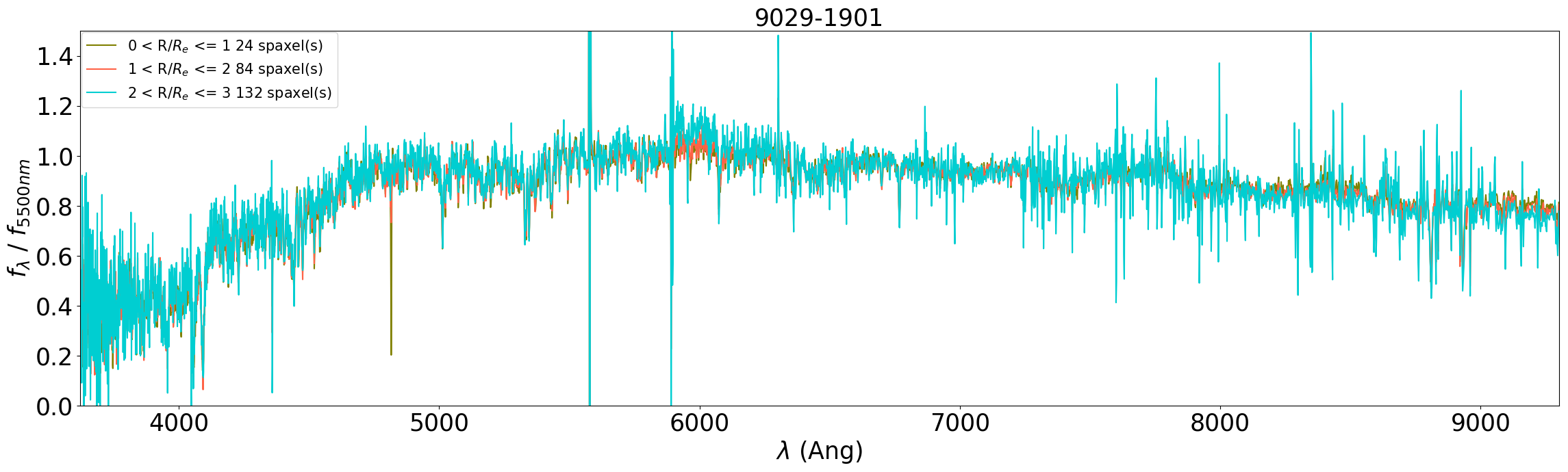}

\bigskip
\bigskip
\bigskip
\bigskip

\begin{center}[\textbf{MaNGA ID: 1-264510 | PLATE-IFU: 9041-1902}]
\end{center}
\includegraphics[height = 0.197\textwidth]{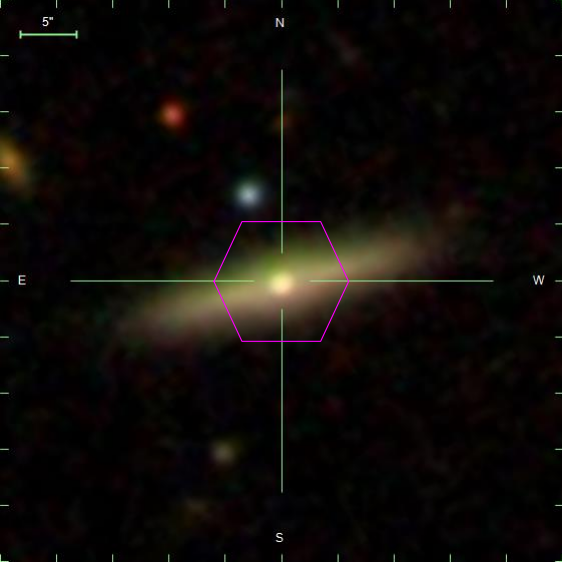}
\includegraphics[height = 0.197\textwidth]{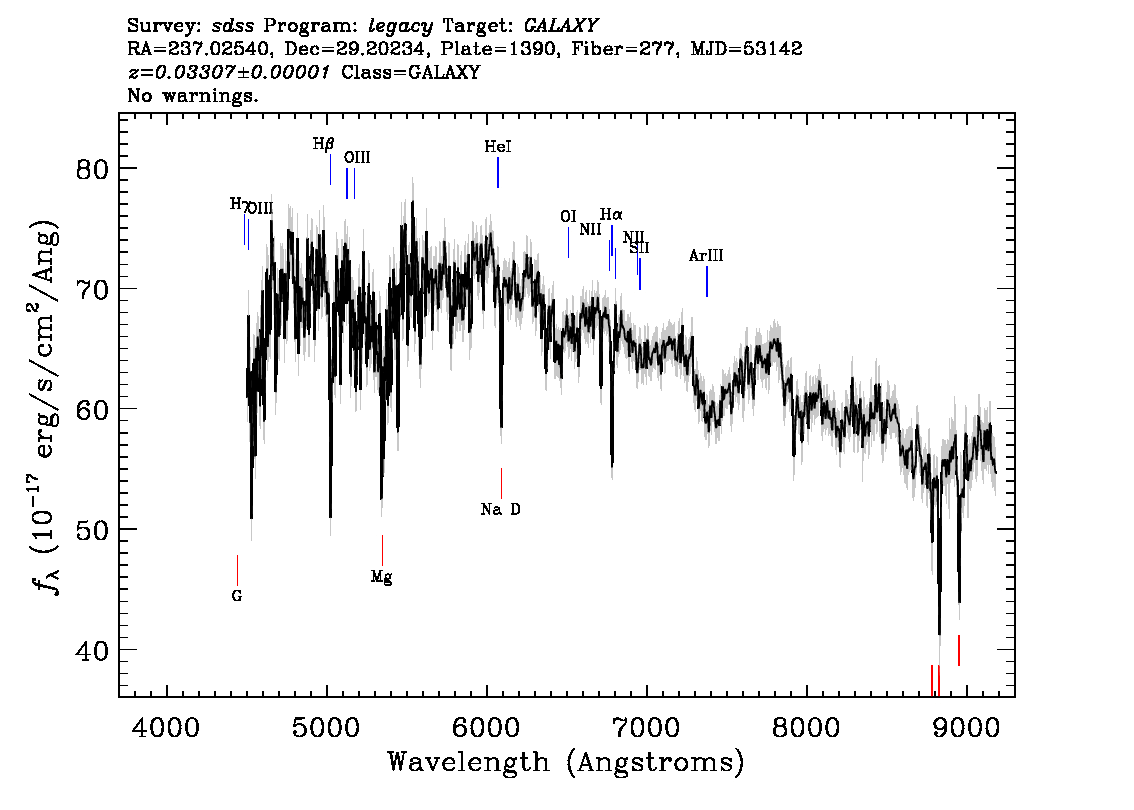}  \includegraphics[height = 0.197\textwidth]{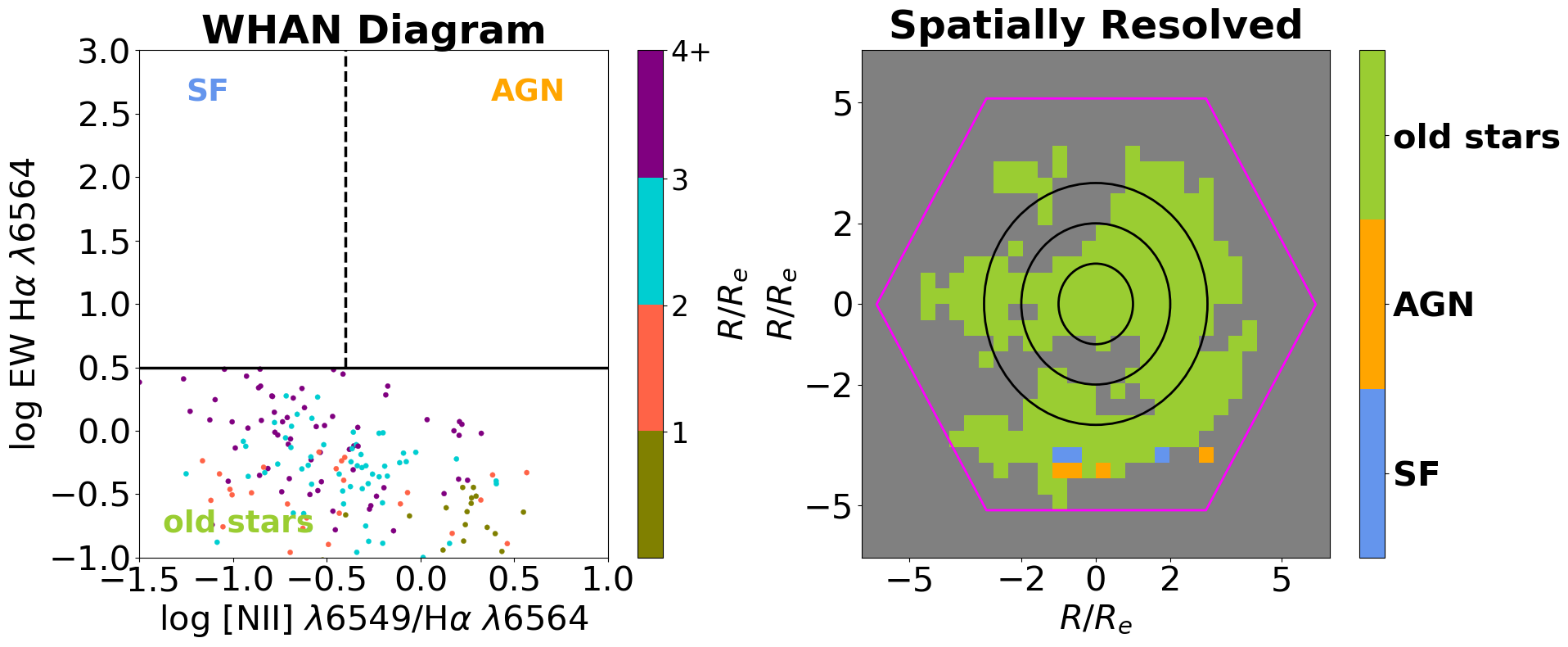} 
\bigskip
\bigskip
\ \includegraphics[height = 0.29\textwidth]{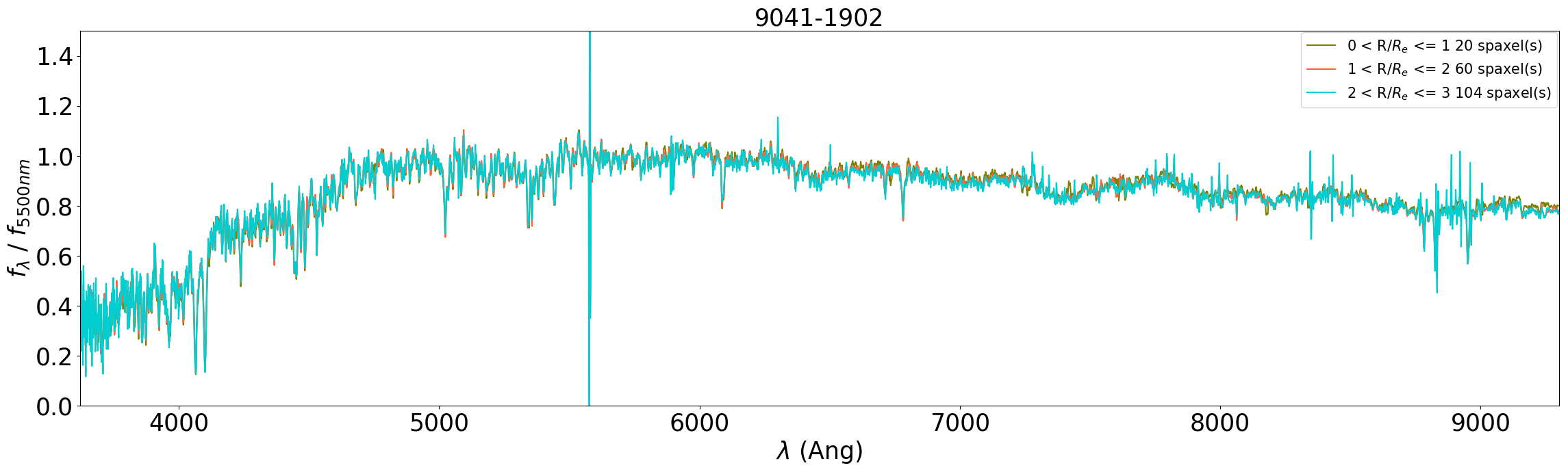}


\end{document}